\documentclass[12pt,english,a4paper]{book}
\usepackage[dvips]{graphicx}
\usepackage{fancyhdr}
\usepackage{latexsym}
\usepackage{amsfonts}
\usepackage{amssymb}
\usepackage{amsthm}
\usepackage{amsmath}
\usepackage{color}
\usepackage{colordvi}
\usepackage{axodraw}
\usepackage{array}
\usepackage{supertabular}

\fancypagestyle{plain}{ \fancyhead{}   }

\fancypagestyle{normal}{
\fancyhead[LE,RO]{\bfseries\thepage}
\fancyhead[LO]{\bfseries\rightmark}
\fancyhead[RE]{\bfseries\leftmark}

\addtolength{\headheight}{0.5pt}}

\fancypagestyle{introduccio}{
\fancyhead[LE,RO]{\bfseries\thepage}
\fancyhead[LO]{\bfseries Introducci\'o}
\fancyhead[RE]{\bfseries Introducci\'o}

\addtolength{\headheight}{0.5pt}}

\fancypagestyle{index}{
\fancyhead[LE,RO]{\bfseries\thepage}
\fancyhead[LO]{\bfseries Contents}
\fancyhead[RE]{\bfseries Contents}

\addtolength{\headheight}{0.5pt}}

\fancypagestyle{bibliografia}{
\fancyhead[LE,RO]{\bfseries\thepage}
\fancyhead[LO]{\bfseries Bibliography}
\fancyhead[RE]{\bfseries Bibliography}

\addtolength{\headheight}{0.5pt}}

\fancypagestyle{appendixa}{
\fancyhead[LE,RO]{\bfseries\thepage}
\fancyhead[LO]{\bfseries Appendix A: The Antisymmetric Tensor Formalism}
\fancyhead[RE]{\bfseries Appendix A: The Antisymmetric Tensor Formalism}

\addtolength{\headheight}{0.5pt}}

\fancypagestyle{appendixb}{
\fancyhead[LE,RO]{\bfseries\thepage}
\fancyhead[LO]{\bfseries Appendix B: Feynman Integrals}
\fancyhead[RE]{\bfseries Appendix B: Feynman Integrals}

\addtolength{\headheight}{0.5pt}}

\fancypagestyle{appendixc}{
\fancyhead[LE,RO]{\bfseries\thepage}
\fancyhead[LO]{\bfseries Appendix C: Feynman Diagrams to the Vector Form Factor}
\fancyhead[RE]{\bfseries Appendix C: Feynman Diagrams to the Vector Form Factor}

\addtolength{\headheight}{0.5pt}}

\fancypagestyle{appendixd}{
\fancyhead[LE,RO]{\bfseries\thepage}
\fancyhead[LO]{\bfseries Appendix D: Form Factors and Constraints}
\fancyhead[RE]{\bfseries Appendix D: Form Factors and Constraints}

\addtolength{\headheight}{0.5pt}}

\fancypagestyle{appendixebis}{
\fancyhead[LE,RO]{\bfseries\thepage}
\fancyhead[LO]{\bfseries Appendix~E: Dispersive Relations}
\fancyhead[RE]{\bfseries Appendix~E: Dispersive Relations}

\addtolength{\headheight}{0.5pt}}

\fancypagestyle{appendixe}{
\fancyhead[LE,RO]{\bfseries\thepage}
\fancyhead[LO]{\bfseries Appendix~F: Second-order Fluctuation of the Lagrangian}
\fancyhead[RE]{\bfseries Appendix~F: Second-order Fluctuation of the Lagrangian}

\addtolength{\headheight}{0.5pt}}

\fancypagestyle{appendixg}{
\fancyhead[LE,RO]{\bfseries\thepage}
\fancyhead[LO]{\bfseries Appendix~G: $\beta$-function Coefficients}
\fancyhead[RE]{\bfseries Appendix~G: $\beta$-function Coefficients}

\addtolength{\headheight}{0.5pt}}

\fancypagestyle{conclusions}{
\fancyhead[LE,RO]{\bfseries\thepage}
\fancyhead[LO]{\bfseries Conclusions}
\fancyhead[RE]{\bfseries Conclusions}

\addtolength{\headheight}{0.5pt}}

\usepackage[latin1]{inputenc}
\usepackage{babel}

\def\mapright#1#2{\smash{
     \mathop{-\!\!\!-\!\!\!\rightarrow}\limits^{#1}_{#2}}}

\newcommand{\no}{\nonumber}
\newcommand{\beqn}{\begin{eqnarray}}
\newcommand{\eeqn}{\end{eqnarray}}

\newcommand{\ba}{\begin{array}{c}}
\newcommand{\bat}{\begin{array}{cc}}
\newcommand{\ea}{\end{array}}

\newcommand{\nn}{\nonumber}

\newcommand{\lsim}{\stackrel{<}{_\sim}}

\newcommand{\chpt}{$\chi$PT }

\newcommand{\bea}{\begin{eqnarray}}
\newcommand{\eea}{\end{eqnarray}}
\newcommand{\beq}{\begin{equation}}
\newcommand{\eeq}{\end{equation}}

\newcommand{\cO}{{\cal O}}

\newcommand{\ket}{\,\rangle}
\newcommand{\bra}{\langle \,}

\newcommand{\mathbold}{\bf}

\newcommand{\es}{\varepsilon^{\phantom{a}}_{\mathrm{S}}}
\newcommand{\ep}{\varepsilon^{\phantom{a}}_{\mathrm{P}}}
\newcommand{\ppa}{\lambda_1^{\mathrm{PP}}}
\newcommand{\ppb}{\lambda_2^{\mathrm{PP}}}
\newcommand{\ppc}{\lambda_3^{\mathrm{PP}}}
\newcommand{\ssa}{\lambda_1^{\mathrm{SS}}}
\newcommand{\ssb}{\lambda_2^{\mathrm{SS}}}
\newcommand{\ssc}{\lambda_3^{\mathrm{SS}}}
\newcommand{\spa}{\lambda_1^{\mathrm{SP}}}
\newcommand{\spb}{\lambda_2^{\mathrm{SP}}}

\newcommand{\lam}{\lambda}
\newcommand{\bs}{b^{\mathrm{S}}}
\newcommand{\bp}{b^{\mathrm{P}}}
\newcommand{\bsp}{b^{\mathrm{SP}}}
\newcommand{\bsm}{{b^{\mathrm{S}\,\mu}}}
\newcommand{\bpm}{{b^{\mathrm{P}\,\mu}}}
\newcommand{\bspm}{{b^{\mathrm{SP}\,\mu}}}
\newcommand{\bsn}{{b^{\mathrm{S}\,\nu}}}
\newcommand{\bpn}{{b^{\mathrm{P}\,\nu}}}
\newcommand{\bspn}{{b^{\mathrm{SP}\,\nu}}}
\newcommand{\as}{{a^{\mathrm{S}}}}
\newcommand{\ap}{{a^{\mathrm{P}}}}
\newcommand{\asp}{{a^{\mathrm{SP}}}}
\newcommand{\ks}{{k^{\mathrm{S}}}}
\newcommand{\kp}{{k^{\mathrm{P}}}}

\def\mapright#1#2{\smash{
     \mathop{-\!\!\!-\!\!\!\rightarrow}\limits^{#1}_{#2}}}
\newcommand{\Int}{\displaystyle{\int}}

\newcommand{\be}{\begin{equation}}
\newcommand{\ee}{\end{equation}}
\newcommand{\bear}{\begin{eqnarray}}
\newcommand{\eear}{\end{eqnarray}}

\newcommand{\mF}{\mathcal{F}}

\begin{document}

%\hyphenation{}

% PORTADA
%
%
\thispagestyle{empty}

\begin{center}
\includegraphics[scale=0.125]{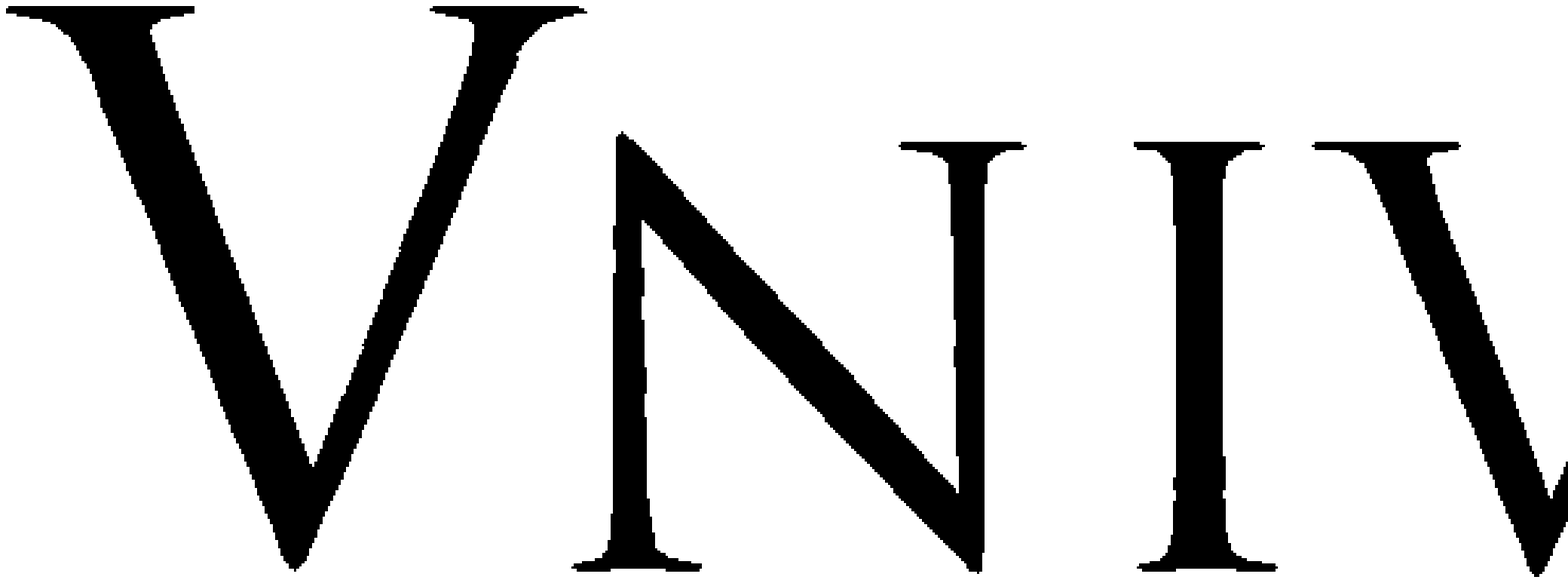}
\vspace{2.2cm}

\begin{huge}
{\mathbold
Quantum Corrections in the \\
\vspace{0.7cm}
Resonance Chiral Theory
}
\end{huge}
\vspace{7.0cm}

%\includegraphics[scale=0.4]{/afs/ific.uv.es/user/i/igroes/Trebin/logos/facultat2.ps}
%\vspace{3cm}

\begin{large}
{\mathbold
Ignasi Rosell 
}
\vspace{0.15cm}

IFIC, Departament de Física Teòrica
\vspace{0.15cm}

PhD Thesis, 
January 2007
\vspace{0.6 cm}

PhD Advisors: 
\vspace{0.15cm}

Antonio Pich and Jorge Portolés
\end{large}
\end{center}
\newpage
\thispagestyle{empty} 
\phantom{hola!}
\newpage
\thispagestyle{empty}
\phantom{hola!}
\vspace{3cm}
\begin{flushright}
To Rut. And to Pau, of course.
\end{flushright}

\newpage
\thispagestyle{empty}
\phantom{hola!}

\pagestyle{index}
\tableofcontents

\chapter{Effective Field Theories}
\section{The Magnifying Glass of the Theoretical Physicists}
\pagestyle{normal}

Who would look for a street in València with a galactic map? Or the other way around, who would want to check the position of a galaxy by using the street plan of a city? All in all, this is the key of an Effective Field Theory (EFT): the long-distance dynamics do not depend crucially on the details of the short-distance dynamics. In other words,  considering the Moon movement around the Earth in order to study our galaxy movement has no sense.

In fact, the idea of effective field theories has been always implicit when describing Nature. One takes into account the suitable degrees of freedom for the problem at hand. 

As an illustrative example the physics of the atom can be examined. An analysis considering Quantum Electrodynamics (QED) seems to be a bit useless, i.e. using quarks as degrees of freedom is not the best choice. A better approach would make use of non-relativistic electrons orbiting around the nucleus. As a first approximation, one could consider an infinite mass for the nucleus: only the electron mass and the fine structure constant would be required to describe the system. If more precision is needed, the finite mass of the proton can be taken into account, if even more precision is demanded the spin and the magnetic moment... and so on. The main idea is that the right effective theory of the system has been chosen.

From this setting one can deduce that a first question, and often not naive, is to choose the appropriate degrees of freedom at the scale under consideration. That is, effective theories are the suitable theoretical tools to describe low-energy dynamics, where the term `low' refers to a determined scale $\Lambda$. As it will be explained in the next section, only the relevant degrees of freedom, i.e. those states with $m\ll \Lambda$, are considered, while heavier excitations with $M\gg \Lambda$ have been integrated out from the action. One has to use suitable interactions among the light states, which can be organized as an expansion in powers of energy over the scale $\Lambda$.

A remarkable feature to classify effective field theories is the strength of the underlying theory in the region at hand, i.e. one can distinguish the case in which the high-energy theory is weakly or strongly coupled. In the first case the value of the effective couplings can be obtained perturbatively in terms of the underlying couplings. However, in the second case one cannot perform perturbative calculations, so this possibility is not at hand for realistic models. The effective approach we are going to use within this work is of the second kind, that is, only different restrictions coming from the underlying theory can be used, but the value of the couplings cannot be obtained directly.

It has been claimed above that the low-energy dynamics do not depend on the details of the high-energy region. This sentence should be clarified: the only effect of the high-energy theory is to fix the value of the couplings and to provide the symmetries that must be considered in order to describe the long-distance scenario.

%Being able of working at low energies without using directly the full theory is very appealing and useful. Notwithstanding, this does not mean that one cannot get information about the high-energy dynamics coming from low-energy experiments. There are always traces in the symmetries and in corrections to the results. Actually, although the best way to know the physics at short distances is usually the experiments at this regime of energies, it is not odd to learn high-energy physics from very precise low-energy physics, think for instance in the $g-2$ measurement.

To prepare this first chapter, we have made extensive use of several reviews~\cite{EFTa,EFTb}.

\section{Integration of the Heavy Modes}

We want to present from a more formal point of view what has been explained in the previous section, by following path integrals methods. Assuming that the theory at high energies is known, the effective action $\Gamma_{\mathrm{eff}}$, which encodes all the information at low energies, reads
\begin{equation}
e^{i \, \Gamma_{\mathrm{eff}}[\Phi_l]} \, = \, \int \, [\mathrm{d} \Phi_h] \,\, e^{iS[\Phi_l,\Phi_h]} \, , \label{heavymodes}
\end{equation}
where $\Phi_l$ and $\Phi_h$ refer to the light and heavy fields respectively and $S[\Phi_l,\Phi_h]$ is the action of the underlying theory. Thus the effective lagrangian is defined through the expression
\begin{equation}
\Gamma_{\mathrm{eff}}[\Phi_l] \, = \, \int \mathrm{d}^4x \,\, \mathcal{L}_{\mathrm{eff}}[\Phi_l] \, .
\end{equation}
It is possible to compute the effective action $\Gamma_{\mathrm{eff}}[\Phi_l]$, at least formally, using the saddle point technique. The heavy field $\Phi_h$ can be expanded around some field configuration $\overline{\Phi}_h$ as follows
\begin{eqnarray}\label{expansion_th}
S[\Phi_l,\Phi_h] &=& S[\Phi_l,\overline{\Phi}_h] \, + \, \int \mathrm{d}^4x \left. \frac{\delta S}{\delta \Phi_h (x)}
\right|_{\Phi_h=\overline{\Phi}_h} \Delta \Phi_h (x) \nonumber
\\&&  + \frac{1}{2} \int \mathrm{d}^4x\,\mathrm{d}^4y \left.\frac{\delta^2S}{\delta \Phi_h(x) \delta \Phi_h(y)}
\right|_{\Phi_h=\overline{\Phi}_h} \Delta \Phi_h(x) \Delta \Phi_h(y) \,+ \, \dots \, ,
\end{eqnarray}
where the definition $\Delta \Phi_h(x) \equiv \Phi_h(x)-\overline{\Phi}_h$ has been used. It can be chosen $\overline{\Phi}_h$ so that
\begin{equation}
\left.\frac{\delta S[\Phi_l,\Phi_h]}{\delta \Phi_h (x)}\right|_{\Phi_h=\overline{\Phi}_h} \, = \, 0 \, .
\end{equation}
With this choice Eq.~(\ref{heavymodes}) turns out to be
\begin{equation}
e^{i \, \Gamma_{\mathrm{eff}}[\Phi_l]} \, = \, e^{i \, S[\Phi_l,\overline{\Phi}_h]} \, \int [\mathrm{d} \Phi_h] \,
e^{i \int \mathrm{d}^4x \, \mathrm{d}^4y \left\{ \frac{1}{2} \Delta \Phi_h(x) \, A(x,y) \, \Delta \Phi_h(y)+\dots \right\}}\,,\label{tree_loop}
\end{equation}
where
\begin{equation}
A\left(x,y\right) \, \equiv \, \left.
\frac{\delta^2S}{\delta\Phi_h(x)\delta\Phi_h(y)}\right|_{\Phi_h=\overline{\Phi}_h}.
\end{equation}
By a formal Gaussian integration and assuming that the heavy field is a boson, 
\begin{equation}
\Gamma_{\mathrm{eff}}[\Phi_l] \, \equiv \, \sum_{k=0}^{\infty} \Gamma^{(k)} \, = \, S[\Phi_l,\overline{\Phi}_h[\Phi_l]]
\, + \, \frac{i}{2} \mathrm{Tr}\left(\mathrm{log}\, A[\Phi_l]\right) \,+\, \dots \, .
\end{equation} 
The expansion in Eq.~(\ref{expansion_th}) turns out to be an expansion in the number of loops, that is the first term corresponds to a tree level integration of the heavy field $\Phi_h$, as clearly seen from Eq.~(\ref{tree_loop}). 

Although the above expansion is quite general and, in principle, it can always be performed, the calculations are very often complicated or cannot be obtained perturbatively. For instance, not always the degrees of freedom of the effective theory are present in the fundamental one. However, as it has been claimed in the former section, some information for the effective low-energy action can be obtained from symmetry constraints coming from the underlying theory. 

\section{Renormalizability and Effective Theories}

Usually it is claimed that a quantum field theory should be renormalizable in order to be able to perform radiative corrections to the tree level result, i.e. that the lagrangian should contain only terms with dimension $\leq D$, with $D$ the dimension of the space-time. Otherwise one needs an infinite number of counterterms, hence an infinite number of unknown parameters, so that the theory has no predictive power.

However, an effective field theory lagrangian contains already an infinite number of terms. The lagrangian can be organized by taking into account their dimension,
\begin{equation}
\mathcal{L}_{\mathrm{eff}} \, = \, \mathcal{L}_{\le D} \,+\,\mathcal{L}_{D+1} \,+\, \mathcal{L}_{D+2} \, + \, \dots \, , \label{renorm}
\end{equation}
where $\mathcal{L}_{\le D}$ contain all terms with dimension $\leq D$, $\mathcal{L}_{D+1}$ contains terms with dimension $D+1$, and so on. The usual renormalizable lagrangian is just the first term, $\mathcal{L}_{\le D}$. Although there are an infinite number of terms in $\mathcal{L}_{\mathrm{eff}}$, the predictive power has not disappeared while one works at a given precision. As operators with higher dimensions are incorporated, a higher precision $\epsilon$ is reached, 
\begin{equation}
\epsilon \, \lesssim \, \left( {E\over \Lambda} \right)^{D_i^{\mathrm{max}}-4}\,,
\end{equation}
where $D_i^{\mathrm{max}}$ is the considered highest dimension. Accordingly, once a given precision is decided, the number of operators and thus couplings needed is finite. In other words, a non-renormalizable theory is just as good as a renormalizable theory for computations, provided one is satisfied with a finite accuracy.

With only the first term of Eq.~(\ref{renorm}) the effective lagrangian turns out to be a classical `renormalizable' theory. In fact, the Standard Model is an effective theory in which only the first piece of the expansion is considered. It is supposed to exist a more general theory where, either with the degrees of freedom already present in the usual framework or with completely new ones, there are heavier modes. In consequence it is not surprising to find corrections to the Standard Model, consequence of these new modes, i.e. New Physics or Physics Beyond the Standard Model come simply from higher scales. There are two ways to know these new scales, either experiments at very high energies or improving the precision at the present energies. 

\section{The Decoupling Theorem}

Intuitively, decoupling means that low-energy physics is ``blind'' to high-energy physics. Assuming a theory with light particles and a heavy particle of mass $M$, one can demonstrate that, under given conditions, the effects of the heavy particle in the low-energy dynamics only appears through corrections proportional to a negative power of $M$ or through renormalization. The Appelquist-Carazonne theorem is the rigorous formulation of this phenomenon~\cite{Carazzone}.

Let us consider a theory with a light field $\phi$ and a heavy field $\Phi$ with masses $m$ and $M$ respectively. $\Gamma^n (g,m,M,\mu;k_1,\dots,k_n)$ is the vertex of n light particles with momenta $k_i$, which is derived from the classical action $S[\phi,\Phi]$, where $g$ denotes the different couplings and $\mu$ is the renormalization scale. If now we consider the action $\widetilde{S}[\phi]$, which is obtained from $S[\phi,\Phi]$ by omitting the terms with heavy fields and replacing the original light particle mass and couplings by new parameters $\widetilde{m}$ and $\widetilde{g}$, the vertex of $n$ light particles can be considered again, $\widetilde{\Gamma}^n (\widetilde{g},\widetilde{m},\mu;k_1,\dots,k_n)$. Supposing some mass independent renormalization scheme, the theorem proves that
\begin{eqnarray}
\Gamma^{n}(g,m,M,\mu;k_1,\dots,k_n) & = & Z^{n/2}\, \widetilde{\Gamma}^{(n)}(\widetilde{g},\widetilde{m},\mu;k_1,\dots,k_n) \, + \, \cO({1\over M}) \, ,
\end{eqnarray}
where the new couplings, mass and scale of fields, $\widetilde{g}(g,M,\mu)$, $\widetilde{m}(g,m,M,\mu)$ and $Z(g,M,\mu)$, depend now on the heavy scale; obviously the form of these functions depend on the renormalization scheme.

As it has been indicated before, there are some conditions in order to be able to grant the validity of the theorem: the underlying theory has to be renormalizable, it should not have spontaneous symmetry breaking nor chiral fermions.

\section{Matching}

It is known that the effects of a heavy particle in the low-energy theory are present through higher-dimension operators, i.e. non-renormalizable ones which are suppressed by inverse powers of the heavy particle mass. The same physical predictions in the full and effective theories should be expected around the heavy-threshold region. Thus, both descriptions are related through a matching condition: the two theories (with and without the heavy field) should give rise to the same $S$ matrix elements for processes involving light particles. %In practice, all the one-light-particle-irreducible diagrams with light particles should be matched.

It is important to stress that while the matching conditions have not been taken into account, one is not dealing really with the effective field theory, that is, the matching procedure is a fundamental step to develop effective approaches.

Quantum Chromodynamics (QCD) is an appropriate way to understand this process. Considering the QCD lagrangian with $n_f-1$ light quark flavors plus one heavy quark of mass $M$, one assumes that at $\mu <M$ one can integrate out the heavy quark. Accepting the decoupling, the resulting effective field theory consists of the original pieces without the heavy quark plus a tower of higher-dimensional operators suppressed by powers of $1/M$. The matching conditions will relate this effective field theory to the original QCD lagrangian with $n_f$ flavors:
\begin{eqnarray}
\mathcal{L}_{\mathrm{QCD}}^{n_f}&\Longleftrightarrow & \mathcal{L}_{\mathrm{QCD}}^{n_f-1}+\sum_{i=1} \frac{c_i}{M^i} \cO_i \,.
\end{eqnarray}
At low energies these extra operators are usually neglected, being reduced the effective lagrangian to the normal QCD lagrangian with $n_f-1$ quark flavors. As it has been explained before, the two QCD theories have different renormalization properties: the running of the corresponding couplings $\alpha_s^{n_f}$ and $\alpha_s^{n_f-1}$ is different. The two effective couplings are related trough a matching condition:
\begin{eqnarray}
\alpha_s^{n_f} \left( \mu^2 \right) &=& \alpha_s^{n_f-1} \left(\mu^2 \right) \left\{ 1+ \sum_{k=} C_k \left( \mathrm{log} \frac{\mu}{M}\right) \left( \frac{\alpha_s^{n_f-1} (\mu^2)}{\pi} \right)^k \right\} \, .\label{matching}
\end{eqnarray}
Since the QCD running coupling is not a physical observable, there can be different parameters and there is no reason why they should be the same at the matching point. The physical observables are those which should be equal at the matching point: they would be the same independently from the effective field theory at hand. In fact these matching conditions require a discontinuous coupling like Eq.~(\ref{matching})

\section{Chiral Perturbation Theory}

\subsection{The QCD Lagrangian and the Running of $\alpha_s$}

With the present overwhelming experimental and theoretical evidence it is known that the $SU(3)_C$ gauge theory correctly describes the hadronic world~\cite{QCD}. Later we are going to be interested in QCD at energies between the $\rho$ mass and $2$ GeV, therefore we will start by studying the behaviour of the strong interaction at low energies. QCD describes the strong interaction between quarks and gluons through a non-Abelian $SU(N_C)$ gauge theory, with $N_C=3$. The lagrangian reads
\begin{eqnarray} \label{QCD}
\mathcal{L}_{QCD} &=& \overline{q} \left(iD\! \! \! \! / \;-\mathcal{M}\right)q\,-\,\frac{1}{4}G^a_{\mu\nu}G^{\mu\nu}_a
\,+\,\mathcal{L}_{FP}\,+\,\mathcal{L}_{GF}, \nonumber \\
D_\mu &=& \partial_\mu \, -\, ig_sG^a_\mu\frac{\lambda_a}{2} \nonumber \\
G^a_{\mu\nu} &=& \partial_\mu G^a_\nu \,-\, \partial_\nu G^a_\mu \,+\, g_sf^{abc}G^b_\mu G^c_\nu,\phantom{\frac{1}{2}}
\end{eqnarray}
where $a=1,\dots,N_C^2-1=8$, $G_\mu^a$ are the gluon fields and $g_s$ is the strong interaction coupling constant. The quark field $q$ represents a column vector in both color and flavor spaces, $\mathcal{M}$ is the quark mass matrix in flavor space. The $\lambda_a$ are the Gell-Mann matrices, so that $\lambda_a/2$ are the $SU(3)_C$ generators in the fundamental representation and $f^{abc}$ are the structure constants. The Faddeev-Popov term $\mathcal{L}_{FP}$ includes the lagrangian for the ghost fields and $\mathcal{L}_{GF}$ refers to the gauge-fixing term.

Before working perturbatively at low or high energies, one has to explore the running of the $g_s$ strong coupling, in order to confirm if it is possible to consider the coupling as a small quantity. That is, one has to renormalize the theory and, in the case of using dimensional regularization, study the dependence in $g_s$ on the scale $\mu$. Assuming that the strong coupling is small one can calculate the beta function at one-loop level,
\begin{eqnarray} \label{running}
\beta_{QCD}\,=\, \mu \frac{\partial g_s}{\partial \mu} \,=\,
-\left(11N_C-2n_f\right)\frac{g_s^3}{48\pi^2} \,,
\end{eqnarray}
so that, at least at this order, $\beta_{QCD}$ is negative for $n_f \leq 16$, with $n_f$ the number of flavors. Eq.~(\ref{running}) implies that the renormalized coupling constant varies with the scale, which is usually called the ``running'' of the coupling constant. Integrating this equation, it is obtained that
\begin{equation}
\alpha_s(q^2) \,=\, \frac{12\pi}{(11N_C-2n_f)\log(q^2/\Lambda_{QCD}^2)} \,,
\end{equation}
where $\alpha_s \,\equiv\, g_s^2/4\pi$. Written in this form, the evolution of the coupling with the scale only depends on a single parameter $\Lambda_{QCD}$, which is known as the QCD scale and is defined in terms of of $\mu$ and $\alpha_s(\mu^2)$ through
\begin{equation}
\log(\Lambda_{QCD}^2)\,=\,\log \mu^2 \,-\, \frac{12\pi}{\alpha_s(\mu^2)(33-2n_f)}\,.
\end{equation}

Eq.~(\ref{running}) allows to check the asymptotic freedom of QCD, i.e. its running coupling decreases at high energies, in contrast to the case of QED. If in QED the fact that the coupling constant decreases at long distances is interpreted as the result of the charge screening due to the presence of electron-positron virtual pairs, one thinks of an anti-screening effect in QCD, which is due to the non-Abelian nature of the gluonic interactions. 

Although Eq.~(\ref{running}) is only valid in the region where $\alpha_s$ is small, since it has been obtained by a perturbative calculation at one-loop level, one expects an increase at low energies, which leads to the confinement of QCD: the asymptotic states of QCD cannot be anymore the free quarks at this regime of energies. The phenomenology supports this idea: the confinement of quarks and gluons inside hadrons can be supposed.

As a consequence we are not able to work perturbatively at low energies by using the QCD lagrangian of Eq.~(\ref{QCD}). An effective field theory approach at long distances turns out to be the appropriate framework.

\subsection{Chiral Symmetry}

In the absence of quark masses, the QCD lagrangian of Eq.~(\ref{QCD}) turns out to be
\begin{eqnarray} \label{QCD-quiral}
\mathcal{L}_{QCD}^0 \,=\, i \overline{q}_L D\! \! \! \! / \; q_L \,+\,i \overline{q}_R D\! \! \! \! / \; q_R
\,-\,\frac{1}{4}G^a_{\mu\nu}G^{\mu\nu}_a
\,+\,\mathcal{L}_{FP}\,+\,\mathcal{L}_{GF},
\end{eqnarray}
where the `$0$' index refers to the massless case and the quark fields have been split into their chiral components. This lagrangian is invariant under independent global $G\equiv SU(n_f)_L \otimes SU(n_f)_R$ transformations of the left- and right-handed quarks in flavor space. 

Global symmetries have an influence into the spectrum, whereas local ones determine the interaction. Consistently, the global chiral symmetry, which should be approximately good in the light quark sector ($n_f=3$), should have implications in hadronic spectroscopy. Notwithstanding, it does not mean that it necessarily must be observed in the spectrum, since symmetries have always two possible realizations: either they are manifest, giving rise to a classification within the spectrum, or they are driven by a spontaneous symmetry breaking, with the resulting generation of the Goldstone bosons, according to Goldstone's theorem~\cite{Goldstone}.

Vafa and Witten~\cite{vafa-witten} proved that the lowest energy state has to be necessarily invariant under vector transformations, so that the possible spontaneous chiral symmetry breaking cannot affect the vectorial part of the chiral group.

Phenomenology is the next step. Although hadrons can be nicely classified in $SU(3)_V$ representations, degenerate multiplets with opposite parity do not exist. Moreover, the octet of pseudoscalar mesons happens to be much lighter than all the other hadronic states. This experimental evidence drives to the spontaneous $SU(3)_L \otimes SU(n_f)_R$ symmetry breaking to $SU(3)_{L+R}$. Since there are $n_f^2-1=8$ broken axial generators of the chiral group, there should be eight lightest hadronic states $J^P=0^-$ ($\pi^+,\,\pi^-,\,\pi^0,\,\eta,\,K^+,\,K^-,\,K^0$ and $\overline{K}^0$). Their small masses are generated by the quark-mass matrix, which explicitly breaks the global chiral symmetry. Taking into account this small explicit breaking, we will refer to the pion multiplet as the pseudo-Goldstone bosons.

\subsection{The Effective Chiral Lagrangian}

The general formalism to build effective lagrangians with spontaneous symmetry breaking was proposed by Callan, Coleman, Wess and Zumino~\cite{CCWZ}, who gave a suitable way to parametrize the Goldstone bosons. In the case of QCD at very low energies, it is possible to use these ideas to construct an effective lagrangian to describe the interaction among the pseudo-Goldstone bosons, the lightest pseudoscalar multiplet. Since there is a mass gap separating the pseudoscalar octet from the rest of the hadronic spectrum, one can imagine an effective field theory containing only these modes.

Thus, the basic assumption is the spontaneous chiral symmetry breaking,
\begin{equation}
G\equiv SU(3)_L \otimes SU(3)_R \longrightarrow H \equiv SU(3)_V \, .
\end{equation}
Denoting by $\phi^a$ ($a=1,\dots ,n_f^2-1=8$) the coordinates describing the pseudo-Goldstone fields in the coset space $G/H$, a coset representative $u_{R,L}(\phi)$ is chosen. The change of coordinates, carrying the pseudo-Goldstone modes, under a chiral transformation $g\equiv (g_L,g_R)\in G$ is ruled by
\begin{eqnarray}
u_L(\phi)  & \mapright{G}{} &  g_L \, u_L(\phi) \, h(g,\phi)^{\dagger} \, , \nonumber \\
u_R(\phi) & \mapright{G}{} & g_R \, u_R(\phi) \,   h(g,\phi)^{\dagger} \, ,  
\end{eqnarray}
where $h(g,\phi)\in H$. We can take the choice of a coset representative such that $u_R(\phi)=u_L^\dagger (\phi)\equiv u(\phi)$, whose explicit form in the Callan, Coleman, Wess and Zumino parameterization can be written as
\begin{eqnarray}\label{u}
u(\phi) & = & e^{\left( \frac{i}{\sqrt{2} \, F} \, \phi \right)} \,,
\end{eqnarray}
with $\phi$ defined through the following expression,
\begin{equation}\label{phi1}
\phi \, = \, \frac{1}{\sqrt{2}} \, \sum_{i=1}^{8} \, \lambda_i \, \phi_i \,=\, \left( \begin{array}{ccc}  \frac{1}{\sqrt 2}\pi^0 + \frac{1}{\sqrt 6}\eta_8  & \pi^+ & K^+ \\ \pi^- & - \frac{1}{\sqrt 2}\pi^0 + \frac{1}{\sqrt 6}\eta_8    & K^0 \\  K^- & \bar{K}^0 & - \frac{2}{\sqrt 6}\eta_8  
\end{array} \right)\,,
\end{equation}
where the normalization of the Gell-Mann matrices is given by $\langle \lambda_i \lambda_j \rangle = 2 \delta_{ij}$. 

Once the coset space is parameterized, the low-energy effective lagrangian realization of QCD for the light quark sector can be obtained, the so-called Chiral Perturbation Theory ($\chi$PT)~\cite{ChPTa,ChPTb,ChPTc}. One should write the most general lagrangian involving the matrix $u(\phi)$, which is consistent with QCD and its chiral symmetry. It is obvious that this effective approach will be useful until the resonance region, $E\ll M_\rho$, since then new degrees of freedom arise. 

$\chi$PT is worked out as a perturbative expansion in the momenta and masses of the pseudo-Goldstone bosons and it has proved to be a rigorous and fruitful scheme. Thus, the lagrangian can be organized in terms of increasing powers of momentum or, equivalently, in terms of an increasing number of derivatives,
\begin{eqnarray} \label{expansion}
\mathcal{L}_{\chi PT}&=&\sum_{n=1} \mathcal{L}^{\chi PT}_{2n} \,,
\end{eqnarray}
where the subindex, $2n$, indicates the number of derivatives. Notice that parity conservation requires an even number of these and there is no term without derivatives, since $uu^\dagger=1$.

As in any quantum field theory, quantum loops with internal lines must be explored. Taking into account the lagrangian expansion of Eq.~(\ref{expansion}) and assuming an arbitrary Feynman diagram with $N_d$ vertices of $\cO(p^d)$\footnote{The chiral order, $\cO(p^d)$, indicates the number of derivatives} and $L$ loops, it is easy to check that the chiral dimension of an amplitude is given by~\cite{ChPTa}
\begin{eqnarray} \label{contatge}
D&=&2\,+\,2L\,+\,\sum_d N_d (d-2) \,.
\end{eqnarray}
The power suppression of loop diagrams is at the basis of effective field theories. As the chiral lagrangian starts at $\cO(p^2)$, so $d\geq 2$, and all terms in Eq.~(\ref{contatge}) are positive. As a result, only a finite number of terms in the lagrangian are needed to work to a fixed chiral order, and the chiral lagrangian acts like a renormalizable field theory. For instance, the leading $D=2$ contributions are obtained with $L=0$  and $N_{d>2}=0$, i.e. tree level graphs with $\mathcal{L}_2^{\chi PT}$. Let us imagine now the calculation of amplitudes to $\cO(p^4)$, one only has two possibilities in Eq.~(\ref{contatge}), $L=0$, $N_4=1$ and $N_{d>4}=0$ or $L=1$ and $N_{d>2}=0$; that is, one only needs to consider tree level diagrams with one insertion of $\mathcal{L}^{\chi PT}_4$, or one-loop graphs with the lowest order lagrangian $\mathcal{L}_2^{\chi PT}$ to compute all scattering amplitudes to $\cO(p^4)$. 

It is clear that the chiral expansion in powers of momenta runs over some typical hadronic scale, the chiral symmetry breaking scale, $\Lambda_\chi$. In view of different arguments, as the variation of the loop contribution under a rescaling of $\mu$, one has an estimate of the scale, $\Lambda_\chi \sim 4\pi F \sim 1.2$ GeV. Furthermore, one can consider the scale related to the first heavy particles that have been integrated out, the $\rho$ multiplet, $\widetilde{\Lambda}_\chi \sim M_\rho \sim 0.77$ GeV. Notice that $\widetilde{\Lambda}_\chi < \Lambda_\chi$, so that loop contributions tend to be smaller than resonance contributions.

The effective field theory technique becomes much more powerful if couplings to external classical fields are introduced. Considering an extended QCD lagrangian, with quark couplings to external currents $v_\mu,\,a_\mu,\,s,p$:
\begin{eqnarray}\label{currents}
\mathcal{L}_{QCD}&=&\mathcal{L}_{QCD}^{0} \,+\, \overline{q}\gamma^\mu\left(v_\mu+\gamma_5a_\mu\right)q\,-\,\overline{q}\left(s-i\gamma_5p\right)q \, ,
\end{eqnarray}
the external fields will allow to compute the effective realization of general Green Functions of quark currents in a very straightforward way. Moreover, they can be used to incorporate the electromagnetic and semileptonic weak interactions, and the explicit breaking of chiral symmetry through the quark masses. Taking into account that the lagrangian of Eq.~(\ref{currents}) is to be chiral invariant, the external fields have the following chiral transformations:
\begin{equation} 
s+ip  \rightarrow  g_R (s+ip)g_L^\dagger  , \quad \ell_\mu  \rightarrow  g_L\ell_\mu
g_L^{\dagger}+ig_L\partial_\mu g_L^{\dagger} , \quad r_\mu  \rightarrow  g_Rr_\mu
g_R^{\dagger}+ig_R\partial_\mu g_R^{\dagger} ,
\end{equation}
where  $r_\mu\equiv v_\mu + a_\mu$ and $\ell_\mu\equiv v_\mu - a_\mu$ have been defined.

A very convenient way to construct the chiral invariant operators needed for the effective lagrangian is to consider tensors $X$ transforming as
\begin{eqnarray}\label{transf}
X \, \mapright{G}{} \, h(g,\phi) \, X \, h(g,\phi)^ {\dagger}  \, ,
\end{eqnarray}
since traces of products of these tensors are chiral invariant. Using the external fields and the matrix $u(\phi)$ of Eq.~(\ref{u}), the following tensors, which observe the transformations properties of Eq.~(\ref{transf}), can be constructed:
\begin{eqnarray} \label{tensors}
u_\mu &=& i\left\{ u^\dagger \left( \partial_\mu-ir_\mu \right) u - u \left( \partial_\mu-i \ell_\mu
\right)u^\dagger \right\} \, , \phantom{\frac{1}{2}}\nonumber  \\
\chi_{\pm} &=& u^\dagger \chi u^\dagger \pm u \chi^\dagger u \, , \phantom{\frac{1}{2}}\nonumber \\
f^{\mu\nu}_\pm &=& uF_L^{\mu\nu}u^\dagger \, \pm \, u^\dagger F_R^{\mu\nu}u \, , \phantom{\frac{1}{2}}
\end{eqnarray}
with $\chi=2B_0(s+ip)$ and the following tensors have been introduced,
\begin{eqnarray}
F_R^{\mu\nu} &=& \partial^\mu r^\nu \,-\, \partial^\nu r^\mu \,-\, i[r^\mu,r^\nu] \, , \nonumber \\
F_L^{\mu\nu} &=& \partial^\mu \ell^\nu \,-\, \partial^\nu \ell^\mu \,-\, i[\ell^\mu,\ell^\nu] \, .
\end{eqnarray}
$B_0$ is related to the quark condensate:
\begin{eqnarray}
\bra 0 | \overline{q}^i q^j | 0 \ket &=& - F^2 B_0 \delta^{ij}\,.
\end{eqnarray}
Besides those of Eq.~(\ref{tensors}), one can also construct tensors that follow Eq.~(\ref{transf}) by using the covariant derivative,
\begin{eqnarray} \label{covderiv}
\nabla_\mu \, X &=& \partial_\mu X \,+\, \left[ \Gamma_\mu,X \right]\,,
\end{eqnarray}
which is defined through the chiral connection,
\begin{eqnarray}
\Gamma_\mu &=&\frac{1}{2} \left\{ u^\dagger \left( \partial_\mu-ir_\mu \right) u + u \left( \partial_\mu-i \ell_\mu \right)u^\dagger \right\} \,,
\end{eqnarray}
so that if $X$ transforms as Eq.~(\ref{transf}), also does $\nabla_ \mu X$.

As it has been been indicated before, the explicit breaking of chiral symmetry through quark masses can be added by using the external currents. Taking into account that the breaking is produced in QCD due to the mass matrix,
\begin{eqnarray}
\mathcal{M}&=& \left( \begin{array}{ccc} m_u & 0 & 0 \\ 0 & m_d & 0 \\ 0 & 0 & m_s \end{array} \right) \, ,
\end{eqnarray}
the breaking is introduced in $\chi$PT with $s=\mathcal{M}$ and $p=0$ in $\chi$, see Eq.~(\ref{tensors}). Once the masses have been included, the organization of Eq.~(\ref{expansion}) turns out to be an expansion in derivatives of the pseudo-Goldstone fields and in powers of the light quark masses.

One last remark is convenient in order to understand the construction of the different pieces $\mathcal{L}_{2n}^{\chi PT}$. Taking into account that the pseudo-Goldstone masses are introduced trough $\chi$, one assumes that $\chi_\pm\sim \cO(p^2)$, and considering the definitions of $u_\mu$ and $f^{\mu\nu}_\pm$ in Eq.~(\ref{tensors}), $u_\mu \sim \cO(p), f^{\mu\nu}_\pm \sim  \cO(p^2)$.

\begin{table} 
\begin{center}
\renewcommand{\arraystretch}{1.2}
\begin{tabular}{|c|c|c|c|} 
\hline
Operator & $P$ & $C$ & h.c. \\
\hline
$u_\mu$ &  $-u^\mu$ & $u_\mu^T$ & $u_\mu$ \\
$\chi_\pm$ & $\pm \chi_\pm$ & $\chi_\pm^T$ & $\pm \chi_\pm$ \\
$f_{\mu\nu\,\pm}$ & $\pm f^{\mu\nu}_\pm$ & 
$\mp f_{\mu\nu\,\pm}^T$ & $f_{\mu\nu\, \pm}$\\
\hline
\end{tabular}
\caption{\small{Transformation properties under $C$, $P$ and hermitian conjugate of the tensors of Eq.~(\ref{tensors}).}} \label{CPT}
\end{center}
\end{table}
We only have to construct all the operators consisting of the defined tensors observing chiral and QCD symmetries. In Table \ref{CPT} the transformation properties under parity ($P$), charge conjugation ($C$) and hermitian conjugate of the tensors of Eq.~(\ref{tensors}) are shown. Employing the organization of Eq.~(\ref{expansion}), one gets that the piece of $\cO(p^2)$ reads
\begin{eqnarray} \label{p2}
\mathcal{L}_2^{\chi PT} &=& \frac{F^2}{4} \bra u_\mu u^\mu + \chi_+ \ket \, ,
\end{eqnarray}
where the brackets $\langle ... \rangle$  denote a trace of the corresponding flavour matrices. Notice that the coefficient is fixed by considering the canonical form of the kinetic piece. Taking into account the explicit chiral symmetry breaking proposed before, only two constants have been introduced in $\mathcal{L}^{\chi PT}_2$, $F$ and $B_0$, apart from masses. It is straightforward to check that $F$ is approximately the decay constant of the pion, $F\simeq 92.4$ MeV and $B_0$ can be related to the hadron masses, once the mass term of the lagrangian is obtained,
\begin{eqnarray}\label{masses}
2\,B_0\,\mathcal{M}&=&\left( \begin{array}{ccc} M_\pi^2 & 0 & 0 \\ 0 & M_\pi^2 & 0 \\ 0 & 0 & 2M_K^2-M_\pi^2 \end{array} \right) \, .
\end{eqnarray}

At $\cO(p^4)$, the most general lagrangian, invariant under parity, charge conjugation and the local chiral transformations, is given, in $SU(3)$, by~\cite{ChPTb}
\begin{eqnarray} \label{p4}
\mathcal{L}_4^{\chi PT} &=& L_1 \bra u_\mu u^\mu \ket^2 \,+\, L_2 \bra u_\mu u^\nu \ket \bra u^\mu u_\nu \ket \,+\, L_3 \bra u_\mu u^\mu u_\nu u^\nu \ket \,+\, L_4 \bra u_\mu u^\mu \ket \bra \chi_+ \ket \nonumber \\
& &  + \, L_5 \bra u_\mu u^\mu \chi_+ \ket \,+\, L_6 \bra \chi_+ \ket^2 \,+\, L_7 \bra \chi_-\ket^2
\,+\, L_8/2 \, \bra \chi_+^2 + \chi_-^2 \ket \nonumber \\ & & 
-\,i L_9 \bra f^{\mu\nu}_+ u_\mu u_\nu \ket \, +\, L_{10}/4 \, \bra f_{+ \mu\nu}f_+^{\mu\nu}-f_{-\mu\nu}f_-^{\mu\nu}\ket \nonumber \\
& & +\,iL_{11} \bra \chi_-(\nabla_\mu u^\mu + i/2\, \chi_-)\ket \,-\,L_{12}\bra (\nabla_\mu u^\mu + i/2 \, \chi_-)^2 \ket \nonumber \\ & &  
+\,H_1/2\, \bra f_{+ \mu\nu}f_+^{\mu\nu}+f_{-\mu\nu}f_-^{\mu\nu}\ket \,+\,H_2/4 \, \bra \chi_+^2-\chi_-^2 \ket \,,
\end{eqnarray}
where the terms with $L_{11}$ and $L_{12}$ vanish when the equations of motion are used and the ones with $H_1$ and $H_2$ are only needed for the renormalization. We have not included here the Wess-Zumino-Witten piece related to the chiral anomaly. In this thesis we do not deal with the odd-intrinsic parity sector of QCD.

\subsection{Renormalization}

Obviously loops are divergent and need to be renormalized. If a regularization which preserves the symmetries of the lagrangian is used, such as dimensional regularization, the needed counterterms will respect necessarily these symmetries. Since Eq.~(\ref{expansion}) contains all possible terms, the divergences can then be absorbed in a renormalization of the coupling constants of the lagrangian. At next-to-leading order, the divergences are of $\cO(p^4)$ and are thus renormalized by the low-energy couplings in Eq.~(\ref{p4}),
\begin{eqnarray}\label{gamma}
L_i&=&L_i^r(\mu)\,+\, \Gamma_i \frac{\mu^{D-4}}{32\pi^2} \left\{ \frac{2}{D-4}\,+\,C\right\}\,, \quad 
\nonumber \\
H_i&=&H_i^r(\mu)\,+\, \tilde{\Gamma}_i \frac{\mu^{D-4}}{32\pi^2} \left\{ \frac{2}{D-4}\,+\,C\right\}\,,
\end{eqnarray}
where $D$ is the space-time dimension and $C$ is the constant that fixes the renormalization scheme; notice that in $\chi$PT the modified minimal subtraction $-1$ scheme ($\overline{\mathrm{MS}}-1$) is used and one has $C=\gamma_E-\mathrm{log}4\pi-1$, with $\gamma_E\simeq 0.5772$ the Euler's constant. The explicit calculation of the one-loop generating functional gives~\cite{ChPTb}:
\begin{center}
\begin{tabular}{llllll}
$\Gamma_1=\frac{3}{32}$, & $\Gamma_2=\frac{3}{16}$, & $\Gamma_3=0$, & $\Gamma_4=\frac{1}{8}$, & 
$\Gamma_5=\frac{3}{8}$, & $\Gamma_6=\frac{11}{144}$, \\
& & & & & \\
$\Gamma_7=0$, & $\Gamma_8=\frac{5}{48}$, & $\Gamma_9=\frac{1}{4}$, & $\Gamma_{10}=-\frac{1}{4}$, &
$\tilde{\Gamma}_1=-\frac{1}{8}$, & $\tilde{\Gamma}_2=\frac{5}{24}$. \\
\end{tabular}
\end{center}
The $\mu$ dependence in the renormalized couplings $L_i^r(\mu)$ is canceled by that of the one-loop amplitude in any observable.

\chapter{Resonance Chiral Theory}

\section{Improving Phenomenological Lagrangians {\it à la} Weinberg}

Once it is accepted that the study of low-energy hadrodynamics is tampered with by our present inability to implement non-perturbative Quantum Chromodynamics fully in those processes, new ways of dealing with QCD at these regimes are required. As it has been argued in the first chapter, Effective Field Theories are one of the most appealing tool to reach this aim~\cite{EFTa,EFTb}. The success of Chiral Perturbation Theory describing the low-energy dynamics of QCD turns out to be a good proof of these ideas~\cite{review-ChPT}. There are other fruitful effective field theories of QCD that support this statement, think for instance in the Heavy Quark Effective Theory~\cite{HQET} for mesons with one heavy quark or Non-Relativistic Quantum Chromodynamics~\cite{NRQCD} in the case of mesons with both heavy quarks.

However, in the region of energies we are interested in, $M_\rho \lsim E \lsim 2$ GeV, the situation is more involved. Although chiral symmetry still provides stringent dynamical constraints, the usual $\chi$PT power counting breaks down in the presence of higher energy scales. Moreover, this regime is populated by many resonances and the absence of a mass gap in the spectrum of states makes difficult to provide a formal Effective Field Theory approach to implement QCD properly, since it is not clear which degrees of freedom are being integrated out and, anyhow, from which energy threshold it would be done the integration of heavy modes. In any case, many of the main features of Effective Field Theories will be very useful in order to carry out our procedure.

The main ingredients of our framework are the following:
\begin{enumerate}
\item We should start from the phenomenological lagrangians approach proposed by Weinberg in Ref.~\cite{ChPTa}. He suggested to construct the most general possible lagrangian, including all terms consistent with assumed symmetry principles, expecting that calculations of matrix elements would give the most general possible S-matrix consistent with analyticity, perturbative unitarity, cluster decomposition and the symmetry principles. In the case of low-energy QCD, one of the highlighted characteristics would be the introduction of the bound states, i.e. the ordinary hadrons, as the degrees of freedom. Notice that the choice of the degrees of freedom is a significant step in order to construct the effective lagrangian. 

It is important to stress that such general lagrangians, which we will call phenomenological lagrangians {\it à la} Weinberg, do not have specific dynamical content beyond the general principles of analyticity, unitarity, cluster decomposition, Lorentz invariance and assumed symmetries. This fact allows to use additional information provided by the strong interaction underlying theory to improve the description.

\item Large-$N_C$ QCD~\cite{large-NC,large-NC2} furnishes a practical scenario to work with. The limit of an infinite number of quark colors turns out to be a very useful instrument to understand many features of QCD and supplies an alternative power counting to describe the meson interaction. Assuming confinement, the $N_C \rightarrow \infty$ limit strongly constraints meson dynamics by asserting that the Green Functions of the theory are described by the tree diagrams of an effective local lagrangian with local vertices and meson fields, higher corrections in $1/N_C$ being yielded by loops described within the same lagrangian theory. The expansion in $1/N_C$ gives a good quantitative approximation scheme to the hadronic world~\cite{large-NC3}, as it will be reviewed in the next section.

\item Additional progress on our phenomenological approach is carried out by using the short-distance properties of QCD. Most of these asymptotic constraints come from matching Green Functions of QCD currents evaluated within the resonance theory with the results obtained in the leading perturbative OPE expansion. Another source of restrictions arise from form factors. %Matching hadronic effective results with the high-energy theory imposes strong constraints, as it will be explained through this chapter.
\end{enumerate}
To summarize, taking into account the difficulties of a formal EFT method in the resonance region, we are going to deal with an effective approach based on the phenomenological lagrangians' ideas of Ref.~\cite{ChPTa}. This approach can be realized by making use of the $1/N_C$ expansion and the short-distance constraints coming from QCD. All in all, we have advanced the main keys that underline the Resonance Chiral Theory (R$\chi$T)~\cite{RChTa,RChTb,RChTc}, the suggested framework in this work to handle Quantum Chromodynamics at intermediate energies, $M_\rho \lsim E \lsim 2$ GeV.

\section{The $1/N_C$ Expansion}

Dealing with QCD at intermediate energies would be handier by using an expansion parameter. The ordinary strong coupling $\alpha_s$ cannot be the solution taking into account its renormalization group equations. In the region of resonances the usual chiral counting breaks down. Accordingly, the $SU(3)$ gauge theory with very small quark bare masses has no obvious free parameter that could be used as an expansion parameter. 

't Hooft suggested that one should generalize QCD from three colours and employ an $SU(N_C)$ gauge group~\cite{large-NC}. The hope is that it may be possible to solve the theory in the large-$N_C$ limit, and that the physical $N_C=3$ case may be qualitatively and quantitatively close to the large-$N_C$ limit.

Although one might think that letting $N_C \rightarrow \infty$ would make the analysis more complicated because of the larger gauge group and consequent increase in the number of dynamical degrees of freedom, QCD simplifies as $N_C$ becomes large, and there exists a systematic expansion in powers of $1/N_C$. 
\begin{figure}
\begin{center}
\includegraphics[width=7cm]{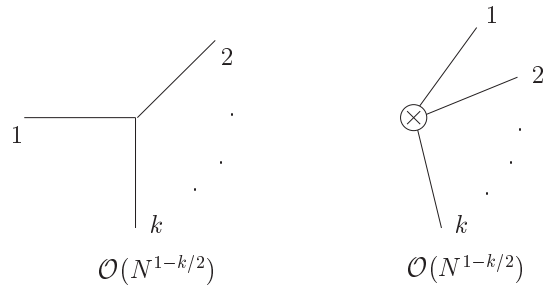}
\caption{\small{$1/N_C$ order of possible vertices.}}
\label{rule}
\end{center}
\end{figure}
%Despite the fact that this simplification has not provided the basis for a quantitative approximation scheme, the qualitative results that are available seem to be quite sufficient encouragement to justify the expansion. 

Choosing the coupling constant $g_s$ to be of $\cO(1/\sqrt{N_C})$, i.e. taking the large-$N_C$ limit with $\alpha_s N_C$ fixed, the main results are the following:
\begin{enumerate}
\item At $N_C \rightarrow \infty$ the mesons and glue states are free, stable and non-interacting. Meson masses have smooth limits and the number of meson states is infinite.
\item Meson decay amplitudes are of $\cO(1/\sqrt{N_C})$, and meson-meson elastic scattering amplitudes are of $\cO(1/N_C)$. These amplitudes follow the pattern of Figure~\ref{rule}.
\item At leading order in the $1/N_C$ expansion, meson dynamics is ruled by a sum of tree diagrams involving the exchange, not of quarks and gluons, but of infinite physical mesons. More generally, meson physics in the large-$N_C$ limit is described by the tree diagrams of an effective local lagrangian, with local vertices and local meson fields. This fact invites us quickly to think about the proper approach of the phenomenological lagrangians {\it à la} Weinberg, proposed in the last section as the suitable tool for QCD at $M_\rho \lsim E \lsim 2$ GeV.
\item Zweig's rule is exact in the large-$N_C$ limit, that is, mesons should be classified as nonets. The axial anomaly has disappeared and flavour $U(n_f)_L \otimes U(n_f)_R$ has been restored.
\item Mesons are pure $q\overline{q}$ states, that is, one finds a suppression of the $q\overline{q}$ sea at $N_C \rightarrow \infty$.
\item In the limit of large number of colours, under reasonable assumptions, $U(n_f)_R\otimes U(n_f)_L$ symmetry must spontaneously break down to $U(n_f)_V$~\cite{breakdown}.
\end{enumerate}

The preceding comments can be read in two ways. One may say that one has used the $1/N_C$ expansion to explain certain qualitative facts about the strong interactions. The other possibility is to say that one may use certain qualitative facts about the strong interactions as diagnostic tests showing that large-$N_C$ QCD is probably a good approximation to Nature~\cite{large-NC2}. Keep in mind that asserting whether the $1/N_C$ expansion is likely to be a good approximation to Nature is a very important matter from a theoretical point of view and very useful for our work, since this expansion can be used in order to justify and improve our effective approach in the resonance region. 

Notice that we have only considered the leading order terms in the $1/N_C$ expansion. It is likewise possible to show, by considering unitarity plus the diagrammatic counting rules in large-$N_C$ QCD, that the higher order corrections are sums of loop diagrams of hadrons together with subleading tree-level contributions. Just as in any theory one understands the tree approximations before trying to consider loop diagrams. In fact, the main aim of this work is to make a first step towards the knowledge of the Resonance Chiral Theory at next-to-leading order in the $1/N_C$ expansion, once the tree level contributions are under control.

On the other hand, the idea of the $1/N_C$ expansion is sometimes questioned on the grounds that $1/N_C=1/3$ is not very small. One cannot really know, theoretically, how large-$N_C$ must be for the expansion to be a good approximation except by calculating the coefficients of some of the terms that are suppressed by powers of $1/N_C$. In other words, the goodness of the expansion depend on the size of the coefficients of the expansion. The best that one can do then is to appeal to phenomenology. As it has been reviewed, there are significant phenomenological reasons to think that $1/N_C=1/3$ is small enough for the $1/N_C$ expansion to be a good approximation in QCD. In fact, it is interesting to remember why perturbation theory is successful in QED. It is not enough to say that the electric charge is small. Actually, normalized in the usual way the electric charge is approximately $e=0.302$. Perturbation theory is a good approximation in QED because when one carries out perturbative expansion, one finds that the typical expansion parameter is really $\alpha=e^2/4\pi$. If we had not yet learned how to do perturbative calculations, as in the QCD case, one would have been unable to judge, just from the value of $e$, whether this expansion would be a good approximation. If, for instance, as it is perfectly possible, the characteristic parameter in the $1/N_C$ expansion would be $1/4\pi N_C$, the next-to-leading corrections would be as tiny as electromagnetic corrections. Although this is only an extreme possibility, we want to justify that there is no reason to reject the $1/N_C$ expansion taking into account the value of $N_C$, above all considering that phenomenology seems to support the expansion~\cite{large-NC2}.

%Finally, one should consider large-$N_C$ QCD seriously because $1/N_C$ is a possible expansion parameter and is the only one the theory is known to have~\cite{large-NC2}. 

\subsection{The $1/N_C$ Expansion in Chiral Perturbation Theory}

Let us come back to the very low-energy EFT of QCD, Chiral Perturbation Theory, in order to show how this new tool we have introduced in this section, the $1/N_C$ expansion, turns out to be a useful source of dynamical information~\cite{leutwyler,polychromatic}, in the sense that it comes directly from QCD. Keep in mind that in the large-$N_C$ limit the flavour $U(n_f)_L\otimes U(n_f)_R$ has been restored.

Although formally the $\chi PT$ lagrangian of Eq.~(\ref{expansion}) could be computed from the QCD generating functional, one does not know how to calculate the values of the couplings from QCD because of its non-perturbative nature at low energies. Since it can be proved that the corresponding correlation functions of fermion bilinears are of $\cO(N_C)$, the leading-order terms in $1/N_C$ should be of $\cO(N_C)$. Moreover, they should have a single flavour trace, as terms with a single trace are of $\cO(N_C)$, while the occurrence of each additional trace reduces the order of the term by unity~\cite{kaiser}.

The leading lagrangian of Eq.~(\ref{p2}) obeys the correct $N_C$ counting rules: the different fields, the masses and momenta are all of them of $\cO(1)$, whereas $F\sim\cO(\sqrt{N_C})$. The $u(\phi)$ matrix, defined in Eq.~(\ref{u}), generates an expansion in powers of $\phi/F$, giving the required $1/\sqrt{N_C}$ suppression for each additional meson field (see Figure~\ref{rule}). Clearly, interaction vertices with $n$ mesons scale as $V_n\sim F^{2-n} \sim \cO(N_C^{1-n/2})$. Since $\mathcal{L}_2^{\chi PT}$ has an overall factor of $N_C$ and $u(\phi)$ is $N_C$-independent, the $1/N_C$ expansion is equivalent to a semiclassical expansion. Quantum corrections computed with the chiral lagrangian will have a $1/N_C$ suppression for each loop.
\begin{table}
\begin{center}
\begin{tabular}{|c| r@{$\pm$}l |c|c| }
\hline
$i$ & \multicolumn{2}{c|}{$L_i^r(M_\rho)$} & $\cO(N_C)$ & source  \\
\hline
$2L_1-L_2$  & $-0.6\,$&$\,0.6$  & $\cO(1)$ &$K_{e4}$, $\pi \pi \rightarrow \pi \pi$    \\
$L_2$       & $1.4\,$&$\,0.3$   & $\cO(N_C)$ &$K_{e4}$, $\pi \pi \rightarrow \pi \pi$  \\
$L_3$       & $-3.5\,$&$\,1.1$  & $\cO(N_C)$ &$K_{e4}$, $\pi \pi \rightarrow \pi \pi$ \\
$L_4$       & $-0.3\,$&$\,0.5$  & $\cO(1)$ & Zweig rule                            \\ 
$L_5$       & $1.4\,$&$\,0.5$   & $\cO(N_C)$ & $F_K\,:F_\pi$                           \\
$L_6$       & $-0.2\,$&$\,0.3$  & $\cO(1)$ & Zweig rule                            \\
$L_7$       & $-0.4\,$&$\,0.2$  & $\cO(1)$ & GMO, $L_5$, $L_8$                        \\
$L_8$       & $0.9\,$&$\,0.3$   & $\cO(N_C)$ &$M_\phi$, $L_5$                          \\
$L_9$       & $6.9\,$&$\,0.7$   & $\cO(N_C)$ & $\bra r^2 \ket^\pi_V$                   \\
$L_{10}$    & $-5.5\,$&$\,0.7$  & $\cO(N_C)$ &$\pi \rightarrow e \nu \gamma$           \\
\hline
\end{tabular}
\caption{\small{Phenomenological values of the couplings $L_i^r(M_\rho)$ in units of $10^{-3}$.
The fourth column shows the source used to get this information
%. Motivated by the large-$N_C$ limit we consider the $U(3)$ case
.}}
\label{Li}
\end{center}
\end{table}

More information from large-$N_C$ QCD can be obtained in the case of $\mathcal{L}_4^{\chi PT}$, shown in Eq.~(\ref{p4}). As it has been explained in Section~1.6.3, only ten additional couplings $L_i$ ($i=1,\dots,10$) are required to determine the low-energy behaviour of the Green Functions at $\cO(p^4)$. Large-$N_C$ QCD claims that terms with a single trace are of $\cO(N_C)$, while those with two traces should be of $\cO(1)$. Therefore one would say that $L_3,\,L_5,\,L_8,\,L_9$ and $L_{10}$ are of $\cO(N_C)$, while $L_4,\,L_6$, and $L_7$ are of $\cO(1)$. The case of $L_1$ and $L_2$ should be analyzed taking into account the following relation:
\begin{equation} \label{descomposicio}
\bra u_\mu u_\nu u^\mu u^\nu \ket \,=\, -2\bra u_\mu u^\mu u_\nu u^\nu \ket \,+\,\frac{1}{2}
\bra u_\mu u^\mu \ket \bra u_\nu u^\nu \ket \,+\, \bra u_\mu u_\nu \ket \bra u^\mu u^\nu \ket \, .
\end{equation}
This new operator could have been added in $\mathcal{L}_4^{\chi PT}$, but it is dependent on the terms with couplings $L_1,L_2$ and $L_3$. Therefore, the symmetries allow a new operator of $\cO(N_C)$ and once one does not include it, one could consider an additional contribution to the couplings $L_1,L_2$ and $L_3$, with the result $2\delta L_1=\delta L_2=-1/2\delta L_3 \sim \cO(N_C)$. In other words, $L_1$ and $L_2$ are really of $\cO(N_C)$, keeping $2L_1 - L_2$ of $\cO(1)$. As shown in Table~\ref{Li}, the phenomenologically determined values of those couplings~\cite{review-ChPT} follow the pattern suggested by the $1/N_C$ counting rules.

\section{The Lagrangian of Resonance Chiral Theory}

\subsection{Introduction}

We want to deal with QCD in the resonance region, $M_\rho \lsim E \lsim 2$ GeV, by following the phenomenological lagrangians {\it à la} Weinberg, which will be ruled by the $1/N_C$ expansion. One has to consider the most general lagrangian, that is, including all terms consistent with assumed symmetry principles, and considering the ordinary hadrons as degrees of freedom. %The expansion in $1/N_C$ is a powerful tool to implement non perturbative QCD.
The program to construct the lagrangian involves several tasks:

\begin{enumerate}
\item In order to be able to recover at very low energies the results of $\chi$PT, to consider chiral symmetry seems to be the best choice. On account of large-$N_C$, the mesons are put together into $U(3)$ multiplets and only operators that have one trace in the flavour space are considered~\cite{leutwyler,kaiser}.
\item It is a well known fact that, in order to make any effective description meaningful, one needs to properly match the underlying theory (QCD in this case). Notice that the QCD asymptotic behaviour sets in already at energies $E\sim 2$~GeV. Then R$\chi$T should recover the short-distance behaviour of QCD. This requirement excludes interactions with large number of derivatives, since they tend to violate the QCD ruled asymptotic behaviour of Green Functions or form factors, explaining the phenomenological success of the usual approximations, where only operators constructed with chiral tensors up to $\cO(p^2)$ are kept\footnote{The effective terms will be constructed with resonance fields and tensors which introduce the pseudo-Goldstone bosons and the sources, already introduced in Eqs.~(\ref{tensors}) and (\ref{covderiv}). We will denote as `chiral tensors' this second group. Accordingly, the operators of the lagrangian will be built by resonances and chiral tensors.}. Furthermore, this matching provides several relations between the couplings in the lagrangian, reducing the number of unknown parameters. These constraints will be analyzed in Section~2.4. 
\item Although large-$N_C$ QCD is a robust instrument to realize QCD at intermediate energies, some approximations are needed to construct the effective lagrangian. As the number of meson states is infinite at large-$N_C$, the most common one is the cut in the number of resonances, only considering the lightest states. This is known to be a good approximation since contributions from higher states are suppressed by their masses. Phenomenology supports this approximation. 
%These features are the main elements of Resonance Chiral Theory~\cite{RChTa,RChTb,RChTc}. 

In Ref.~\cite{RChTa} only contributions from the lightest resonances with non-exotic quantum numbers were taken into account, the so-called Single Resonance Approximation. Our analysis is also carried under this approximation.

Since there is an infinite number of Green Functions, it is obviously not possible to satisfy all matching conditions with a finite number of resonances and uncertainties due to truncation of the spectrum are introduced in the determination of the parameters. Eventually, one may be driven to inconsistencies in the effective parameter relations. The Minimal Hadronic Approximation (MHA) generalises the Single Resonance Approximation so the effective description includes the minimal number of resonances that allows fulfilling the QCD short-distance constraints in the considered amplitude~\cite{MHA}. Although MHA is an approximation of full large-$N_C$ QCD, it is well supported by the phenomenology of Green Functions that are order-parameter of the chiral symmetry. Deviations from the $N_C \rightarrow \infty$ limit are properly understood in some situations~\cite{MHA,juanjo}.
\item %Notice that we have a stronger reason not to include operators without resonances and of $\cO(p^4)$ or higher. 
It has been shown~\cite{RChTb} that $\mathcal{L}_4^{\chi PT}$ is largely saturated\footnote{This is much more clear in the case of vector and axial-vector resonances as their phenomenology is better known.} by the resonance exchange generated by the linear terms in the resonance field, as it will be explained in Section~2.5. Hence, the explicit introduction of the operators constructed with no resonances and chiral tensors of $\cO(p^4)$ would amount to include an overlap between both contributions. An analogous analysis at $\cO(p^6)$ has not been systematically performed but it also looks a reasonable assumption. Thus our theory stands for a complete resonance saturation of the $\chi$PT lagrangian; in other words, we are assuming that the low-energy couplings of $\mathcal{L}_n^{\chi PT}$ ($n\geq 4$) are completely determined by the resonance contributions, so one does not have to include these operators when the resonance fields are active degrees of freedom.

\item Besides the kinetic pieces, only linear couplings in the resonance fields were included in Ref.~\cite{RChTa}, since the aim of the article was to get the leading resonance contributions to the low-energy constants (LEC's) of the $\cO(p^4)$ $\chi$PT lagrangian\footnote{Solving the resonance equations of motion in an expansion in the resonance masses, the resonance fields are expressed as a series of chiral operators times inverse powers of the masses, with chiral tensors starting at $\cO(p^2)$~\cite{RChTc}. Therefore, the only possible leading resonance contributions to the LEC's of $\mathcal{L}_4^{\chi PT}$ come from operators constructed with one resonance field and one chiral tensor of $\cO(p^2)$ in the chiral counting.}, see Eq.~(\ref{p4}). In the next chapter the study of one observable to next-to-leading order in the $1/N_C$ expansion will show that in order to perform the matching with QCD operators constructed with more than one resonance will be needed~\cite{nosaltres}.

Following the path of Ref.~\cite{RChTa}, the leading resonance contributions to some $\cO(p^6)$ $\chi$PT LEC's have been studied, by considering different three-point functions~\cite{three}. A more systematic and complete approach to this issue can be found in Ref.~\cite{RChTc}.
\end{enumerate}
Notice that in contrast to many models of the resonance fields that have been widely employed in the literature, R$\chi$T only uses basic QCD symmetry features without any additional {\it ad hoc} assumptions. Its model aspect only comes from the fact that we do not include an infinite spectrum in the theory, which is one of the features of the $N_C \rightarrow \infty$ limit of QCD. 

\subsection{Constructing the Lagrangian}

As it has been pointed out above, the study is taken under the Single Resonance Approximation, where just the lightest resonances with non-exotic quantum numbers are considered. Taking into account the results at large-$N_C$, the mesons are put together into $U(3)$ multiplets. Hence, our degrees of freedom are the pseudo-Goldstone boson (the lightest pseudoscalar mesons) along with massive multiplets of the type $V(1^{--})$, $A(1^{++})$, $S(0^{++})$ and $P(0^{-+})$. With them, one constructs the most general effective action that preserves chiral symmetry invariance and QCD symmetries.  

%Since Resonance Chiral Theory must get the high-energy behaviour of QCD, only operators constructed with chiral tensors of $\cO(p^2)$ will be allowed; interactions with higher order chiral tensors tend to violate the asymptotic behaviour constrained by QCD.

Following the procedure presented in Section~1.6.3 to construct the $\chi$PT lagrangian, one considers tensors $X$ transforming as
\begin{eqnarray}\label{transf2}
X \, \mapright{G}{} \, h(g,\phi) \, X \, h(g,\phi)^ {\dagger}  \, ,
\end{eqnarray}
where now $G\equiv U(3)_L\otimes U(3)_R$. The tensors that introduce the pseudo-Goldstone bosons and the sources were already introduced in Eqs.~(\ref{tensors}) and (\ref{covderiv}), which follow the transformation properties under parity ($P$), charge conjugation ($C$) and hermitian conjugate (h.c.) of Table \ref{CPT}. Notice that the expression of $\phi$ in Eq.~(\ref{phi1}) changes in the moment one considers nonets instead of octets,
\begin{equation}\label{phi2}
\phi \, = \,  \left( \begin{array}{ccc}  \frac{1}{\sqrt 2}\pi^0 + \frac{1}{\sqrt 6}\eta_8  +\frac{1}{\sqrt{3}}\eta_0& \pi^+ & K^+ \\ \pi^- & - \frac{1}{\sqrt 2}\pi^0 + \frac{1}{\sqrt 6}\eta_8 +\frac{1}{\sqrt{3}}\eta_0   & K^0 \\  K^- & \bar{K}^0 & - \frac{2}{\sqrt 6}\eta_8  +\frac{1}{\sqrt{3}}\eta_0
\end{array} \right)\,.
\end{equation}
The resonance fields follow the same guide, so that for the vector multiplet one has,
\begin{equation} \label{vector-res}
V_{\mu\nu}=\left( \begin{array} {ccc}
\frac{1}{\sqrt{2}}\rho^{0}+\frac{1}{\sqrt{6}}\omega_8+\frac{1}{\sqrt{3}}\omega_0 & \rho^{+} & K^{*\,+} \\
\rho^{-} & -\frac{1}{\sqrt{2}}\rho^{0}+\frac{1}{\sqrt{6}}\omega_8+\frac{1}{\sqrt{3}}\omega_0 & K^{*\,0} \\
K^{*\,-} & \overline{K}^{*\,0} & -\frac{2}{\sqrt{6}}\omega_8+\frac{1}{\sqrt{3}}\omega_0 \end{array} \right)_{\mu\nu}\!\! ,
\end{equation}
where, as it is explained in Appendix~A, the antisymmetric formalism is used for spin-1 fields. The multiplets of the type $A(1^{++})$, $S(0^{++})$ and $P(0^{-+})$ are parametrized in an analogous way to Eq.~(\ref{vector-res}). The transformation properties under $P$, $C$ and hermitian conjugate of the resonance fields are shown in Table \ref{CPT2}.
\begin{table} 
\begin{center}
\renewcommand{\arraystretch}{1.2}
\begin{tabular}{|c|c|c|c|} 
\hline
Operator & $P$ & $C$ & h.c. \\
\hline
$V_{\mu\nu}$ &  $V^{\mu\nu}$ & $-V_{\mu\nu}^T$ & $V_{\mu\nu}$ \\
$A_{\mu\nu}$ & $-A^{\mu\nu}$ & $A_{\mu\nu}^T$ & $A_{\mu\nu}$ \\
$S$ & $S$ & $S^T$ & $S$\\ 
$P$ & $-P$ & $P^T$& $P$\\
\hline
\end{tabular}
\caption{\small{Transformation properties under $P$, $C$ and hermitian conjugate of the resonance fields.}} \label{CPT2}
\end{center}
\end{table}

One should now consider the most general lagrangian that preserves chiral symmetry invariance and QCD symmetries, observing the former remarks, i.e. constructed with chiral tensors up to $\cO(p^2)$ in the chiral counting and under the Single Resonance Approximation. 

In the large-$N_C$ approach, there is no limit to the number of resonances that one may include in the effective operators. One can classify the terms in the lagrangian according to the number of resonances,
\begin{eqnarray} \label{lagrangian-RChT}
\mathcal{L}_{R\chi T}&=&\mathcal{L}_{pGB}^{(2)}\,+\,\sum_{R_1}\mathcal{L}_{R_1} \,+\,\sum_{R_1,R_2}\mathcal{L}_{R_1R_2}\, +\,\sum_{R_1,R_2,R_3}\mathcal{L}_{R_1R_2R_3}\, + \, ... \,\,\,  ,
\end{eqnarray}
where the dots denote operators with four or more resonance fields, and the indexes $R_i$ run over all the different resonance fields, $V$, $A$, $S$ and $P$. However, for the purpose of this work, only operators up to three resonance fields are taken into account.

$\mathcal{L}_{pGB}^{(2)}$ keeps the $\cO(p^2)$ terms without resonances, i.e. the  lagrangian of Eq.~(\ref{p2}),
\begin{align}\label{phi}
\mathcal{L}_{pGB}^{(2)} &=\,\mathcal{L}^{\chi PT}_2\,=\, \frac{F^2}{4} \bra u_\mu u^\mu + \chi_+ \ket \, . 
\end{align}
It is important to distinguish between $\mathcal{L}_{\chi PT}$ and $\mathcal{L}_{pGB}$: although both have the same structure and operators, $\mathcal{L}_{pGB}$ differs from $\mathcal{L}_{\chi PT}$ in the value of the couplings as $\mathcal{L}_{pGB}$ belongs to the theory where the resonances are active degrees of freedom. Furthermore, notice that, as mentioned above, once a complete resonance saturation of the $\chi$PT lagrangian is supposed, no pieces of $\mathcal{L}_{pGB}^{(4)}$ or higher are added.

The second term of Eq.~(\ref{lagrangian-RChT}) corresponds to the interaction terms with one resonance field~\cite{RChTa},
\begin{align}
%\mathcal{L}_R &=\, \mathcal{L}_V + \mathcal{L}_A + \mathcal{L}_S + \mathcal{L}_P \, ,  \phantom{\frac{1}{2}} \\
%
\mathcal{L}_V &=\, \frac{F_V}{2\sqrt{2}} \bra V_{\mu\nu} f^{\mu\nu}_+ \ket \,+\, \frac{i\, G_V}{2\sqrt{2}} \bra V_{\mu\nu} [u^\mu, u^\nu] \ket \, , \label{V} \\
\mathcal{L}_A &=\, \frac{F_A}{2\sqrt{2}} \bra A_{\mu\nu} f^{\mu\nu}_- \ket\, , \label{A}\\
\mathcal{L}_S &=\, c_d \bra S u_\mu u^\mu\ket\,+\,c_m\bra S\chi_+\ket\,  \, , \phantom{\frac{1}{2}}  \label{S}%\\
\end{align}
\begin{align}
\mathcal{L}_P &=\, i\,d_m \bra P \chi_- \ket \phantom{\frac{1}{2}} \, .\phantom{holaaaaaaaaaaaaaaaaaaaaaaaaaaaa}\label{P}
\end{align}
The $\mathcal{L}_{R_1R_2}$ contain the kinetic terms and the remaining operators with two resonance fields~\cite{RChTa,RChTc}, 
\begin{align}
%\mathcal{L}_{R_1R_2}&=\mathcal{L}_{\,\mathrm{kin}\, R}\\
\mathcal{L}_{\,\mathrm{kin}\, R} &=\, \frac{1}{2} \bra \nabla^\mu R\nabla_\mu R\,-\, M_R^2 R^2 \ket  \, , \qquad  \qquad \qquad \, \, \, \, \qquad (R=S,P) \qquad \qquad \label{kinSP}\\
\mathcal{L}_{\,\mathrm{kin}\, R} &=\, -\frac{1}{2} \bra \nabla^\lambda R_{\lambda\mu} \nabla_\nu R^{\nu\mu}\,-\,
\frac{1}{2}M_R^2 R_{\mu\nu}R^{\mu\nu} \ket \, , \quad (R=V,A)    \label{kinVA} \\
\mathcal{L}_{RR}&=\, \lambda^{RR}_1 \bra RR \,u^\mu u_\mu \ket + \lambda^{RR}_2 \bra R u_\mu R u^\mu  \ket  + \lambda^{RR}_3 \bra RR\, \chi_+ \ket \,,\quad (R=S,P)\, \phantom{\frac{1}{2}}\label{SSPP}  \\
\mathcal{L}_{SP}&=\,\lambda^{SP}_1 \bra u_\alpha \{\nabla^\alpha S, P \} \ket +i \lambda^{SP}_2 \bra \{S,P\} \chi_-\ket \, , \phantom{\frac{1}{2}} \label{SP}\\
% + i \lambda^{SP}_3 \bra SP\ket\bra \chi_- \ket \, , \phantom{\frac{1}{2}} \\
%
\mathcal{L}_{SV}&=\, i \lambda^{SV}_1 \bra \{ S, V_{\mu \nu}\} u^\mu u^\nu \ket \,+\, i\lambda^{SV}_2 \bra S u_\mu V^{\mu\nu} u_\nu \ket \,+\, \lambda^{SV}_3 \bra \{ S, V_{\mu\nu}\} f^{\mu\nu}_+ \ket \, , \phantom{\frac{1}{2}} \\
\mathcal{L}_{SA}&=\, \lambda^{SA}_1 \bra \{\nabla_\mu S, A^{\mu\nu} \} u_\nu \ket \,+\,  \lambda^{SA}_2 \bra \{ S, A_{\mu\nu} \} f^{\mu\nu}_- \ket \, ,\phantom{\frac{1}{2}} \\
%
%\mathcal{L}_{PP}&=\, \lambda^{PP}_1  \bra PP \,u^\mu u_\mu \ket \,+\,\lambda^{PP}_2  \bra P u_\mu P u^\mu 
%\ket \, +\, \lambda^{PP}_3 \bra PP\, \chi_+ \ket \,,\,\phantom{\frac{1}{2}} \\
%
\mathcal{L}_{PV}&=\, i \lambda^{PV}_1\bra [\nabla^\mu P,V_{\mu\nu} ] u^\nu \ket \,+\, i  \lambda^{PV}_2 \bra [P, V_{\mu\nu} ] f^{\mu\nu}_- \ket \, , \phantom{\frac{1}{2}} \\
\mathcal{L}_{PA}&=\, i \lambda^{PA}_1  \bra [P, A_{\mu\nu} ]f^{\mu\nu}_+ \ket \,+\,\lambda^{PA}_2 \bra [ P, A_{\mu\nu} ] u^\mu u^\nu \ket \, ,\phantom{\frac{1}{2}} \\
\mathcal{L}_{VA}&=\,  \lambda^{VA}_1\bra [ V^{\mu\nu}, A_{\mu\nu} ] \chi_- \ket \,+\, i \lambda^{VA}_2 \bra  [ V^{\mu\nu}, A_{\nu\alpha} ] h^\alpha_\mu \ket \,+\, i \lambda^{VA}_3 \bra  [ \nabla^\mu V_{\mu\nu}, A^{\nu\alpha} ] u_\alpha \ket\phantom{\frac{1}{2}} \nonumber \\
&+\, i \lambda^{VA}_4 \bra  [ \nabla_\alpha V_{\mu\nu}, A^{\alpha\nu} ] u^\mu \ket \,+\, i \lambda^{VA}_5 \bra  [ \nabla_\alpha V_{\mu\nu}, A^{\mu\nu} ] u^\alpha \ket \nonumber \\& +\, i \lambda^{VA}_6 \bra  [  V_{\mu\nu}, A^{\mu}_{\,\,\alpha} ]f^{\alpha\nu}_- \ket \, ,\phantom{\frac{1}{2}} \label{VA}\!\!\!\!\!\!\!\!\\
\mathcal{L}_{RR}&=\,\lambda^{RR}_1\bra R_{\mu\nu} R^{\mu\nu} u_\alpha u^\alpha\ket \,+\,\lambda^{RR}_2 \bra R_{\mu\nu} u^\alpha R^{\mu\nu} u_\alpha \ket  \,+\,\lambda^{RR}_3 \bra R_{\mu\alpha}  R^{\nu\alpha} u^\mu u_\nu \ket  \phantom{\frac{1}{2}}   \nonumber \\
&+\,\lambda^{RR}_4 \bra R_{\mu\alpha}  R^{\nu\alpha}  u_\nu u^\mu \ket \, +\,\lambda^{RR}_5 \bra R_{\mu\alpha} \left(  u^\alpha R^{\mu\beta}  u_\beta +u_\beta R^{\mu\beta}u^\alpha\right) \ket \, \phantom{\frac{1}{2}}\nonumber \\
& +\,\lambda^{RR}_6 \bra R_{\mu\nu} R^{\mu\nu}  \chi_+ \ket\, +\,i\lambda^{RR}_7 \bra R_{\mu\alpha} R^{\alpha}_{\,\,\nu}f^{\mu\nu}_+ \ket \, .  \qquad \qquad (R=V,A)\phantom{\frac{1}{2}}
\end{align}
In the case of three resonance operators, only terms consisting of resonance fields and the covariant derivative $\nabla_\mu$ are studied, since they are the only ones that contribute to two-body form factors at tree level, see Chapter~4:
\begin{align}
\Delta \mathcal{L}_{SRR}&=\,\lambda^{SRR}_0 \bra SRR \ket \,+\, \lambda^{SRR}_1 \bra S \,\nabla_\mu R \, \nabla^\mu R \ket \, , \quad\quad (R=S,P) \phantom{\frac{1}{2}} \\
%
%\Delta \mathcal{L}_{SPP}&=\,\lambda^{SPP}_0 \bra SPP \ket \,+\, \lambda^{SPP}_1 \bra S \,\nabla_\mu P \, \nabla^\mu P \ket \, , \phantom{\frac{1}{2}} \\
%
\Delta \mathcal{L}_{SRR}&=\,\lambda^{SRR}_0 \bra S R_{\mu\nu} R^{\mu\nu} \ket \,+\, \lambda^{SRR}_1 \bra S \, \nabla_\mu R^{\mu\alpha} \, \nabla^\nu R_{\nu\alpha} \ket \,+\, \lambda^{SRR}_2 \bra S \, \nabla^\nu R^{\mu\alpha} \, \nabla_\mu R_{\nu\alpha} \ket \phantom{\frac{1}{2}}\nonumber \\
&+\, \lambda^{SRR}_3 \bra S \, \nabla_\alpha R^{\mu\nu} \, \nabla^\alpha R_{\mu\nu} \ket \,+\,\lambda^{SRR}_4 \bra S \{ R^{\mu\nu} , \nabla^2 R_{\mu\nu} \ket \phantom{\frac{1}{2}}\nonumber \\
&+\,\lambda^{SRR}_5 \bra S \{ R_{\mu\alpha} , \nabla^\mu \nabla_\nu R^{\nu\alpha}\} \ket \, ,\qquad \qquad \qquad \qquad (R=V,A)\phantom{\frac{1}{2}} \\
\Delta \mathcal{L}_{SPA}&=\,\lambda^{SPA} \bra A^{\mu\nu} \{ \nabla_\mu S, \nabla_\nu P \} \ket \, ,  \phantom{\frac{1}{2}}%\\
\end{align}
\begin{align}
\Delta \mathcal{L}_{PVA}&=\,i \lambda^{PVA}_0 \bra P [V_{\mu\nu}, A^{\mu\nu}] \ket \,+\, i\lambda^{PVA}_1 \bra P [\nabla_\mu V^{\mu\alpha}, \nabla^\nu A_{\nu\alpha}] \ket\phantom{\frac{1}{2}} \nonumber \\
&+\, i\lambda^{PVA}_2 \bra P [\nabla^\nu V^{\mu\alpha}, \nabla_\mu A_{\nu\alpha}] \ket
\,+\, i\lambda^{PVA}_3 \bra P [\nabla_\alpha V^{\mu\nu}, \nabla^\alpha A_{\mu\nu}] \ket\phantom{\frac{1}{2}} \nonumber \\
&+\, i\lambda^{PVA}_4 \bra P [ V^{\mu\nu}, \nabla^2 A_{\mu\nu}] \ket 
+\, i\lambda^{PVA}_5 \bra P [ V^{\mu\alpha}, \nabla_\mu \nabla^\nu A_{\nu\alpha}] \ket \phantom{\frac{1}{2}}\nonumber \\
&+\, i\lambda^{PVA}_6 \bra P \left[ \nabla^\nu \nabla_\mu V^{\mu\alpha},  A_{\nu\alpha}\right] \ket\, ,\phantom{\frac{1}{2}} \\
\Delta \mathcal{L}_{VRR}&=\,i\,\lambda^{VRR} \bra V^{\mu\nu} \nabla_\mu R \nabla_\nu R \ket \, ,\quad\quad (R=S,P)\phantom{\frac{1}{2}} \\
%
%\Delta\mathcal{L}_{VPP}&=\,i\,\lambda^{VPP} \bra V^{\mu\nu} \nabla_\mu P \nabla_\nu P \ket \, ,\phantom{\frac{1}{2}} \\
%
\Delta \mathcal{L}_{VVV}&=\,  i\,\lambda^{VVV}_0 \bra V^{\mu\nu} V_{\mu\alpha} V_\nu^{\,\,\alpha}\ket\,+\,
   i\,\lambda^{VVV}_1 \bra V^{\mu\nu} [\nabla_\mu V_{\alpha\beta},\nabla_\nu V^{\alpha\beta}] \ket  \phantom{\frac{1}{2}}\nonumber \\ &
 +i\,\lambda^{VVV}_2 \bra V^{\mu\nu} [\nabla^\beta V_{\mu\alpha},\nabla_\beta V_{\nu}^{\,\alpha}] \ket 
+i\,\lambda^{VVV}_3 \bra V^{\mu\nu} [\nabla_\mu V_{\beta\alpha},\nabla^\alpha V_{\nu}^{\,\beta}] \ket\phantom{\frac{1}{2}} \nonumber \\
 &
 +i\,\lambda^{VVV}_4 \bra V^{\mu\nu} [\nabla_\mu V_{\nu\alpha},\nabla_\beta V^{\alpha\beta}] \ket  +
i\,\lambda^{VVV}_5 \bra V^{\mu\nu} [\nabla^\alpha V_{\mu\nu},\nabla^\beta V_{\alpha\beta}] \ket\phantom{\frac{1}{2}} \nonumber \\ &
+i\,\lambda^{VVV}_6 \bra V^{\mu\nu} [\nabla^\alpha V_{\mu\alpha},\nabla^\beta V_{\nu\beta}] \ket  
+i\,\lambda^{VVV}_7 \bra V^{\mu\nu} [\nabla^\alpha V_{\mu\beta},\nabla^\beta V_{\nu\alpha}] \ket \phantom{\frac{1}{2}}\,, \\
%\\&+i\,\lambda^{VVV}_8 \bra V^{\mu\nu} [V_{\mu\alpha},V_\nu^{\,\alpha}]\ket \bra \chi_+ \ket +i\,\lambda^{VVV}_9 \bra \{\chi_+,V^{\mu\nu}\} [V_{\mu\alpha},V_\nu^{\,\alpha}]\ket  \, ,
%
\Delta \mathcal{L}_{VAA}&=\,i\,\lambda^{VAA}_0 \bra V^{\mu\nu} A_{\mu\alpha} A_\nu^{\,\,\alpha}\ket 
\,+i\,\lambda^{VAA}_1 \bra V^{\mu\nu} [\nabla_\mu A_{\alpha\beta},\nabla_\nu A^{\alpha\beta}] \ket\phantom{\frac{1}{2}} \nonumber \\ &
+i\,\lambda^{VAA}_2 \bra V^{\mu\nu} [\nabla^\beta A_{\mu\alpha},\nabla_\beta A_{\nu}^{\,\alpha}] \ket  
+i\,\lambda^{VAA}_3 \bra \nabla^\beta V^{\mu\nu} [ A_{\mu\alpha},\nabla_\beta A_{\nu}^{\,\alpha}] \ket \phantom{\frac{1}{2}}\nonumber \\ &
+i\,\lambda^{VAA}_4 \bra V^{\mu\nu} [\nabla_\mu A_{\beta\alpha},\nabla^\alpha A_{\nu}^{\,\beta}] \ket
+i\,\lambda^{VAA}_5 \bra \nabla_\mu V^{\mu\nu} [ A_{\beta\alpha},\nabla^\alpha A_{\nu}^{\,\beta}] \ket \phantom{\frac{1}{2}}\nonumber \\ &
 +i\,\lambda^{VAA}_6 \bra \nabla^\alpha V^{\mu\nu} [ \nabla_\mu A_{\beta\alpha}, A_{\nu}^{\,\beta}] \ket 
+i\,\lambda^{VAA}_7 \bra V^{\mu\nu} [\nabla_\mu A_{\nu\alpha},\nabla_\beta A^{\alpha\beta}] \ket\phantom{\frac{1}{2}} \nonumber \\ &
+i\,\lambda^{VAA}_8 \bra \nabla_\mu V^{\mu\nu} [ A_{\nu\alpha},\nabla_\beta A^{\alpha\beta}] \ket  
+i\,\lambda^{VAA}_9 \bra \nabla_\beta V^{\mu\nu} [\nabla_\mu A_{\nu\alpha}, A^{\alpha\beta}] \ket\phantom{\frac{1}{2}} \nonumber \\ &
+i\,\lambda^{VAA}_{10} \bra V^{\mu\nu} [\nabla^\alpha A_{\mu\nu},\nabla^\beta A_{\alpha\beta}] \ket 
+i\,\lambda^{VAA}_{11} \bra V^{\mu\nu} [\nabla^\alpha A_{\mu\alpha},\nabla^\beta A_{\nu\beta}] \ket\phantom{\frac{1}{2}} \nonumber \\ &
+i\,\lambda^{VAA}_{12} \bra \nabla^\alpha V^{\mu\nu} [ A_{\mu\alpha},\nabla^\beta A_{\nu\beta}] \ket  
+i\,\lambda^{VAA}_{13} \bra V^{\mu\nu} [\nabla^\alpha A_{\mu\beta},\nabla^\beta A_{\nu\alpha}] \ket \phantom{\frac{1}{2}}\nonumber \\ &
+i\,\lambda^{VAA}_{14} \bra \nabla^\alpha V^{\mu\nu} [ A_{\mu\beta},\nabla^\beta A_{\nu\alpha}] \ket \, .\phantom{\frac{1}{2}}\label{endlagrangian}
\end{align}
All coupling constants are real, $M_R$ are the corresponding masses of the resonances, the brackets $\langle ... \rangle$  denote a trace of the corresponding flavour matrices, and the notation defined in Ref.~\cite{RChTa,RChTc} is followed.

Keep in mind that as our lagrangian $\mathcal{L}_{R\chi T}$ satisfies the $N_C$ counting rules for an effective theory with $U(3)$ multiplets, only operators that have one trace in the flavour space are considered~\cite{leutwyler,kaiser}. The different fields, masses and momenta are of $\cO(1)$ in the $1/N_C$ expansion. Taking into account the interaction terms (see Figure~\ref{rule}), one is able to check that $F,\,F_V,\,G_V,\,F_A,\,c_d,\,c_m $ and $d_m$ are of $\cO(\sqrt{N_C})$; $\lambda_i^{R_1R_2}$ of $\cO(1)$ and $\lambda_i^{R_1R_2R_3}$ of $\cO(1/\sqrt{N_C})$. The mass dimension of these parameters is $[F]=[F_V]=[G_V]=[F_A]=[c_d]=[c_m]=[d_m]=E$, $[\lambda_i^{R_1R_2}]=E^0$ and $[\lambda_i^{R_1R_2R_3}]=E^{-1}$.

Note that the equations of motion have been used in order to reduce the number of operators. For instance, terms like $\langle P\,\nabla_\mu u^\mu \rangle$ are not present in Eq.~(\ref{P}), since using the equations of motion we would generate operators that, either have been already considered, or contain  a higher number of resonance fields.

\section{Matching with QCD}

As previously pointed out, a basic ingredient in order to take a step forward in the construction of Resonance Chiral Theory is to consider the short-distance constraints from QCD, i.e. the matching procedure between R$\chi$T and the full theory. Actually, without examining the high-energy properties of the underlying strong dynamics there are too many unknown parameters in our effective approach. Take note of the significance of the number of parameters for the predictive power of the lagrangian.

%Following the Minimal Hadronic Approximation presented in Section~2.3.1, we accept that the effective description should include the minimal number of resonances that allows matching the asymptotic constraints in the considered amplitude. 
Most of the short-distance constraints used in the literature come from considering the Green Functions of QCD currents obtained in the leading OPE expansion. The other source of information is to consider the Brodsky-Lepage behaviour of the form factors~\cite{brodsky-lepage}, that is, to demand that two-body form factors of hadronic currents vanish at high energies. This behaviour has been experimentally observed for pseudo-Goldstone bosons and photons. The doubt appears when one is considering form factors that involve resonances as asymptotic states. One of the motivations of this work is to clarify this question, relating the two-body form factors with the two-point Green Functions at next-to-leading order in the $1/N_C$ expansion~\cite{nosaltres2,nosaltres2bis}. See Chapter~4 for more information.

Another remark is needed before studying the constraints. Obviously these relations depend on the considered lagrangian. Owing to historical reasons, we start by studying the case in which only the $\mathcal{L}_{R}$ of Eq.~(\ref{lagrangian-RChT}) together with the kinetic pieces to describe the resonance interactions are included. These are the only required operators to determine the leading resonance contributions to the couplings constants of the $\cO(p^4)$ $\chi$PT lagrangian. The strong constraints are the following~\cite{polychromatic}:
\begin{enumerate}
\item Vector form factor. At leading order in the $1/N_C$ expansion, the two pseudo-Goldstone boson matrix element of the vector current reads,
\begin{eqnarray}
\mathcal{F}^v_{\pi\pi}(q^2)&=&1+\frac{F_VG_V}{F^2} \frac{q^2}{M_V^2-q^2}\,.
\end{eqnarray}
Accepting that the vector form factor should vanish at infinite momentum transfer, the resonance couplings should satisfy
\begin{eqnarray}\label{const1}
F_VG_V&=&F^2 \,.
\end{eqnarray}
\item Axial form factor. The matrix element of the axial current between one pseudo-Goldstone and one photon is parameterized by the axial form factor. From the assumed lagrangian one gets
\begin{eqnarray}
\mathcal{F}^a_{\pi \gamma}(q^2)&=&\frac{F_A^2}{M_A^2-q^2} + \frac{2F_VG_V-F_V^2}{M_V^2} \,,
\end{eqnarray}
which vanishes at $q^2\rightarrow \infty$ provided that
\begin{eqnarray}\label{const2}
2F_VG_V-F_V^2&=&0 \,.
\end{eqnarray}
\item Weinberg sum rules. The two-point function built from a left-handed and a right-handed vector quark current defines the correlator
\begin{eqnarray}
\Pi_{_{V-A}}(q^2) &=& \frac{F^2}{q^2} + \frac{F_V^2}{M_V^2-q^2} - \frac{F_A^2}{M_A^2-q^2} \,.
\end{eqnarray}
In the chiral limit it vanishes faster than $1/q^4$ at large energies~\cite{vv-aa}. This implies the conditions~\cite{vv-aa2}:
\begin{equation}\label{const3}
F_V^2-F_A^2\,=\,F^2 \,  , \qquad M_V^2F_V^2-M_A^2F_A^2 \,=\, 0 \,. 
\end{equation}
%The second relation is correct up to very small quark-mass contributions.
\item Scalar form factor. The two pseudo-Goldstone bosons matrix element of the scalar quark current contains another dynamical form factor, which for the $K\pi$ case takes the form~\cite{SFF}:
\begin{eqnarray}
\mathcal{F}^s_{K\pi}(q^2)&=&1+\frac{4c_m}{F^2} \left( c_d + (c_m-c_d) \frac{M_K^2-M_\pi^2}{M_S^2} \right) \frac{q^2}{M_S^2-q^2} \, ,
\end{eqnarray}
Requiring $\mathcal{F}^s_{K\pi}(q^2)$ to vanish at $q^2 \rightarrow \infty$, one finds that~\cite{SFF}:
\begin{equation}\label{const4}
4c_dc_m\,=\, F^2 \, , \qquad c_m-c_d\,=\,0 \, .
\end{equation}
\item $SS-PP$ sum rules. The difference of the two-point correlation functions of two scalar and two pseudoscalar currents reads
\begin{eqnarray}
\Pi_{_{S-P}}(q^2)&=&16B_0^2 \left( \frac{c_m^2}{M_S^2-q^2} - \frac{d_m^2}{M_P^2-q^2} + \frac{F^2}{8q^2}\right) \,.
\end{eqnarray}
For massless quarks, $\Pi_{_{S-P}}$ vanishes as $1/q^4$ at large energies, with a small coefficient~\cite{ss-pp}. Imposing this behaviour~\cite{ss-pp2},
\begin{equation}\label{const5}
8\left(c_m^2-d_m^2 \right) \,=\, F^2 \, , \qquad c_m^2M_S^2-d_m^2M_P^2 \simeq 0 \,. 
\end{equation}
 \end{enumerate}
Finally, assuming Eqs.~(\ref{const1}), (\ref{const2}), (\ref{const3}), (\ref{const4}) and (\ref{const5}) one has that
\begin{eqnarray} \label{matching1}
F_V\,=\,2G_V\,=\,\sqrt{2}F_A\,=\,\sqrt{2}F\,, \qquad M_A\,=\,\sqrt{2}M_V\, , \nonumber \\
c_m\,=\,c_d\,=\,\sqrt{2}d_m\,=\,\frac{F}{2} \,, \qquad M_P\,\simeq\,\sqrt{2}M_S \,,
\end{eqnarray}
that is, all the parameters of $\mathcal{L}_R$ are given in terms of the pion decay constant $F$ and the two masses of the vector and scalar multiplets, $M_V$ and $M_S$.

Considering the more general lagrangian of Eq.~(\ref{lagrangian-RChT}) all former constraints are valid except the ones coming from the axial and scalar form factor. In the case of the axial form factor, there are new contributions from Eq.~(\ref{VA}), see Eq.~(\ref{AGF}) in Appendix D. For the two pseudo-Goldstone bosons matrix element of the scalar quark current there are new contributions when one consider massive quarks. Notice that the required field redefinition of the scalar field, needed to remove the tadpole~\cite{tesis_juanjo}, would generate new contributions to the form factor coming from pieces with two resonances. So that only the first constraint of Eq.~(\ref{const4}) would be valid in the general case. The couplings of $\mathcal{L}_R$ are fixed now in terms of $F$ and the resonance masses:
\begin{align}
F_V^2&=\,F^2 \frac{M_A^2}{M_A^2-M_V^2}\, , & F_A^2&=\,F^2 \frac{M_V^2}{M_A^2-M_V^2}\, , &
G_V^2&=\,F^2  \frac{M_A^2-M_V^2}{M_A^2} \, ,& M_A^2& >\,  M_V^2\, \, \nonumber\\
c_m^2&=  \, \frac{F^2}{8}   \frac{M_P^2}{M_P^2-M_S^2} \, , &  d_m^2&=\,\frac{F^2}{8} \frac{M_S^2}{M_P^2-M_S^2}\, ,&
  c_d^2&=\,\frac{F^2}{2}\frac{M_P^2-M_S^2}{M_P^2}\, , & M_P^2& > \, M_S^2\, .
\label{opcio2}
\end{align}

\section{Leading Resonance Contributions to the $\cO(p^4)$ $\chi$PT Lagrangian}

It seems natural to expect that the lowest-mass resonances play an important role on the pseudo-Goldstone bosons dynamics, i.e. Chiral Perturbation Theory. Below the $\rho$ mass scale, the singularities associated with the pole of the resonance propagators can be replaced by the corresponding momentum expansion; the exchange of virtual resonances generates pseudo-Goldstone bosons couplings proportionals to powers of $1/M_R^2$. It can be better understood by using the EFT ideas of Chapter~1. By integrating out the lowest-mass resonances, that is, going from R$\chi$T to $\chi$PT, one would expect to obtain the largest contributions to the chiral LEC's. The so-called resonance saturation involves considering that the couplings of $\chi$PT are largely saturated by the resonance exchange. It can be justified using large-$N_C$ arguments, since tree-level resonance contributions are leading in the $1/N_C$ expansion, to be compared to other contributions related to chiral loops.

In the manner that it has been pointed out in Section~2.3.1, the only possible leading resonance contributions to the $\chi$PT LEC's of $\mathcal{L}_4^{\chi PT}$ come from operators constructed with one resonance field and one chiral tensor of $\cO(p^2)$ in the chiral counting, $\mathcal{L}_R$ of Eq.~(\ref{lagrangian-RChT}). In Ref.~\cite{RChTa} these resonance contributions were studied thoroughly. Under the Single Resonance Approximation and considering nonets for the resonance fields, as large-$N_C$ motivates, one finds the following contributions at leading order in the $1/N_C$ expansion:
\begin{align}
L_1&=\frac{G_V^2}{8M_V^2}\,,& 
L_2&=\frac{G_V^2}{4M_V^2}\,, &
L_3&=-\frac{3G_V^2}{4M_V^2}\,+\,\frac{c_d^2}{2M_S^2}\, , \nonumber \\
L_4&=0\,, &
L_5&=\frac{c_dc_m}{M_S^2}\,,&
L_6&=0\,, \nonumber \\
L_7&=0 \,, &
L_8&=\frac{c_m^2}{2M_S^2}\,-\,\frac{d_m^2}{2M_P^2} \,, &
L_9&=\frac{F_VG_V}{2M_V^2}\,, \nonumber %\\
\end{align}
\begin{align}
L_{10}&=-\frac{F_V^2}{4M_V^2}\,+\,\frac{F_A^2}{4M_A^2}\,,&
H_1&=-\frac{F_V^2}{8M_V^2}-\frac{F_A^2}{8M_A^2}\,,&
H_2&=\frac{c_m}{M_S^2}\,+\,\frac{d_m^2}{M_P^2} \,. 
\end{align}
Notice that it is not surprising to miss contributions to $L_4$, $L_6$ and $L_7$ taking into account its subleading order in the $1/N_C$ expansion, see Table \ref{Li}. 
\begin{table}
\begin{center}
\begin{tabular}{|c|r @{$\pm$} l |r@{.}l| r@{.}l | r@{.}l | r@{.}l | r@{.}l | r@{.}l|}
\hline
$i$& \multicolumn{2}{c|}{$L_i^r(M_\rho)$} & \multicolumn{2}{c|}{$V$} & \multicolumn{2}{c|}{$A$}
 & \multicolumn{2}{c|}{$S$}  & \multicolumn{2}{c|}{$\eta_1$} & 
 \multicolumn{2}{c|}{Total} & \multicolumn{2}{c|}{Total$^{\mathrm{b)}}$} \\
\hline
$1$  & $0.4\,$&$\,0.3$  & $0$&$6$   & $0$&$0$ & $0$&$0$ & $0$&$0$ & $0$&$6$   & $0$&$9$\\
$2$  & $1.4\,$&$\,0.3$  & $1$&$2$   & $0$&$0$ & $0$&$0$ & $0$&$0$ & $1$&$2$   & $1$&$8$\\
$3$  & $-3.5\,$&$\,1.1$ & $-3$&$6$  & $0$&$0$ & $0$&$6$ & $0$&$0$ & $-3$&$0$  & $-4$&$9$\\
$4$  & $-0.3\,$&$\,0.5$ & $0$&$0$   & $0$&$0$ & $0$&$0$ & $0$&$0$ & $0$&$0$   & $0$&$0$\\ 
$5$  & $1.4\,$&$\,0.5$  & $0$&$0$   & $0$&$0$ & $1$&$4^{\mathrm{a)}}$ & $0$&$0$ & $1$&$4$ & $1$&$4$\\
$6$  & $-0.2\,$&$\,0.3$ & $0$&$0$   & $0$&$0$ & $0$&$0$ & $0$&$0$ & $0$&$0$   & $0$&$0$\\
$7$  & $-0.4\,$&$\,0.2$ & $0$&$0$   & $0$&$0$ & $0$&$0$ & $-0$&$3$ & $-0$&$3$ & $-0$&$3$\\
$8$  & $0.9\,$&$\,0.3$  & $0$&$0$   & $0$&$0$ & $0$&$9^{\mathrm{a)}}$ & $0$&$0$ & $0$&$9$ & $0$&$9$\\
$9$  & $6.9\,$&$\,0.7$  & $6$&$9^{\mathrm{a)}}$& $0$&$0$ & $0$&$0$ &$0$&$0$ & $6$&$9$ & $7$&$3$\\
$10$ & $-5.5\,$&$\,0.7$ & $-10$&$0$ & $4$&$0$ & $0$&$0$ &$0$ & $0$ & $-6$&$0$ & $-5$&$5$\\
\hline
\end{tabular}
\caption{\small{Comparison between the different resonance-exchange contribution with the phenomenologically determined values of $L_i^r(M_\rho)$, in units of $10^{-3}$~\cite{EFTb}. Motivated by the large-$N_C$ limit we include $U(3)$ multiplets for the resonances. We consider only the contribution from the $\eta_1$ in the pseudoscalar channel. The superindex a) refers to an input, whereas in b) the short-distance constraints are taken into account.}}
\label{saturation}
\end{center}
\end{table}

$\eta_1$ is usually integrated out from the $\chi$PT lagrangian. Neglecting then the higher-mass $P$ resonances, the only remaining meson exchange is the one associated with this field, which generates a sizable contribution to $L_7$, 
\begin{equation}\label{eta1}
L_7=-\frac{\widetilde{d}_{\eta_1}^2}{2M_{\eta_1}^2} \, .
\end{equation}
Note that if $\eta_1$ is integrated out, $L_7$ appears naively to be of $\cO(N_C^2)$, since $M_{\eta_1}^2\sim \cO(1/N_C)$ in Eq.~(\ref{eta1}). However, the $1/N_C$ counting is not well defined in this case, since $N_C$ cannot be small ($M_{\eta_1}$ heavy) and big ($1/N_C$ expansion) at the same time.

In Table \ref{saturation} we compare the phenomenological values of these couplings together with the ones predicted by the resonance exchanges. The assumption of resonance saturation has given successful predictions for $L_i$.

A last remark is suitable. Though the scale at which the results of the integration, $\mu_0$, is known to be of the order of a typical scale of the physical system, let us say $\mu_0=M_R$, there always remains some ambiguity on the precise value of $\mu_0$ at which the resonance contributions are given. The next-to-leading order predictions would avoid this problem, as the running is under control. See Chapter~4 for more information.

\chapter{Vector Form Factor at NLO in the $1/N_C$ Expansion}

\section{Introduction}

Quantum loops including virtual resonances are a major technical challenge which still has not been properly addressed in Resonance Chiral Theory. A first step in this direction was the study of resonance loop contributions to the running of the $\chi$PT coupling $L_{10}(\mu)$, performed in Ref.~\cite{cata}, which however did not attempt an analysis of the induced ultraviolet divergences and their corresponding renormalization.

Quantum loops involving massive states have been only analysed within explicit models with additional symmetries. For instance, the gauge structure advocated in the so-called ``Hidden Local Symmetry'' description of vector resonances~\cite{HLS} implies a much simpler ultraviolet behaviour~\cite{HLSloops}. Loop corrections to some resonance parameters have also been studied~\cite{BGT:98,ChR:98} within the context of ``Heavy Vector Meson $\chi$PT''~\cite{JMW:95}, which adopts the $M_R\to\infty$ limit to guarantee a good chiral power counting; and Ref.~\cite{NJLM1} in the Nambu-Jona-Lasinio model~\cite{NJLM2}.

At the one-loop level the massive states present in R$\chi$T generate all kind of ultraviolet problems which start now to be understood. A naive chiral power counting indicates that the renormalization procedure will require higher dimensional counterterms, which presumably could generate a problematic behaviour at large momenta. Therefore, it will be necessary to perform a careful investigation of the constraints implied by the short-distance properties of QCD at the next-to-leading order in $1/N_C$.

A formal renormalization of  R$\chi$T at the one-loop level appears to be a very involved task, which requires the prior analysis of several technical ingredients, as can be seen in Chapter~5. In order to gain some understanding on the ultraviolet behaviour, it seems worth to perform first some explicit one-loop calculations of well chosen physical amplitudes. In this chapter, we present a detailed investigation of the pion vector form factor (VFF) at next-to-leading order in the $1/N_C$ expansion. This observable is defined through the two pseudo-Goldstone matrix element of the vector current:
\begin{eqnarray}\label{defchapt3}
\bra \pi^+(p_1)\, \pi^-(p_2)\,| \,\frac{1}{2}\left( \bar{u}\gamma^\mu u - \bar{d}\gamma^\mu d \right)|0 \ket & = & \mathcal{F}(q^2) \, (p_1-p_2)^\mu\, ,
\end{eqnarray}
where $q^\mu \equiv (p_1+p_2)^\mu$. At very low energies, the VFF $\mathcal{F}(q^2)$ has been studied within the $\chi$PT framework up to $\cO(p^6)$~\cite{ChPTb,VFF_ChPT}. R$\chi$T and the $1/N_C$ expansion have also been used to determine $\mathcal{F}(q^2)$ at the $\rho$ meson peak, including appropriate resummations of subleading infrared logarithms \cite{guerrero-dumm,preloop}.

We will simplify the calculation working in the two flavour theory and taking the massless quark limit. Therefore, we will assume a chiral $U(2)_L\otimes U(2)_R$ symmetry group. The small effects induced by the $U(1)_A$ anomaly will be neglected, because they are not going to be relevant in our discussion. As the isosinglet pseudoscalar can only appear within loops, and the numerical correction generated by its non-zero mass could be taken into account in a straightforward way, together with the finite quark mass effects which we are ignoring.

In the next section we will briefly resume the R$\chi$T lagrangian of interest. We will only consider the minimal set of resonance couplings (linear in the resonance fields) introduced in Ref.~\cite{RChTa}, supplemented with those counterterms required by the renormalization procedure. Notice that one of the main aims of this chapter is to justify the necessity of considering operators with more than one resonance field, in the spirit of the short-distance behaviour of our result. The renormalization of the relevant one-particle-irreducible (1PI) Feynman diagrams will be discussed in Section~3.3 and the final results of our calculation will be collected in Section~3.4. Sections~3.5 and 3.6 analyse the behaviour of the computed vector form factor at low and high energies, respectively. We will finally summarize our findings in Section~3.7. Several technical details and results have been moved to the appendices.

\section{The Lagrangian}

We are going to work within a $U(2)_L\otimes U(2)_R$ chiral theory, containing a multiplet of 4 pseudo-Goldstone bosons,
\begin{eqnarray}\label{eq:Goldstones}
\phi & =&\left(\begin{array}{cc} {1\over\sqrt 2}\pi^0 + {1\over\sqrt 2}\eta_0 & \pi^+ \\
\pi^- & - {1\over\sqrt 2}\pi^0 + {1\over\sqrt 2}\eta_0 \end{array}\right) \, ,
\end{eqnarray}
to be compared to the $U(3)_L\otimes U(3)_R$ case of Eq.~(\ref{phi2}). Under the Single Resonance Approximation, the pseudo-Goldstone bosons couple to massive $U(2)$ multiplets of the type $V(1^{--})$, $A(1^{++})$, $S(0^{++})$ and $P(0^{-+})$, with a field content analogous to the one indicated in Eq.~(\ref{eq:Goldstones}).

Our starting point is the R$\chi$T lagrangian introduced in Ref.~\cite{RChTa}, where, besides the kinetic pieces, only linear couplings in the resonance fields are included, since the intention of Ref.~\cite{RChTa} was to obtain the leading resonance contributions to the LEC's of the $\cO(p^4)$ $\chi$PT lagrangian. Therefore, $\mathcal{L}_{R\chi T}$ reads:
\begin{eqnarray}\label{lagrangian_VFF}
\mathcal{L}_{R\chi T}(\phi, \mathrm{V},\mathrm{A},\mathrm{S},\mathrm{P}) \,=\,\mathcal{L}_{pGB}^{(2)}\,+\, \sum_{R} \left( \mathcal{L}_{\mathrm{kin}\,R}  \,+\, \mathcal{L}_{R}\right)\,+\, \mathcal{L}_{R\chi T}^{NLO} \, ,
\end{eqnarray}
where $R$ runs over all the different resonance fields, $V$, $A$, $S$ and $P$. The notation of Section~2.3.2 is followed: $\mathcal{L}_{pGB}^{(2)}$ is shown in Eq.~(\ref{phi}); the different kinetic pieces are given in Eqs.~(\ref{kinSP}) and (\ref{kinVA}); and the interactive terms are defined in Eqs.~(\ref{V}), (\ref{A}), (\ref{S}) and (\ref{P}).  $\mathcal{L}_{R\chi T}^{NLO}$ refers to the subleading pieces, which will be defined below.%Notice that in the chiral limit and neglecting external scalar or pseudoscalar sources $\chi_\pm=0$.

As it has been explained in the last chapter, taking into account that only the R$\chi$T lagrangian of Eq.~(\ref{lagrangian_VFF}) is considered, one should take the usual constraints of Eq.~(\ref{matching1}):
\begin{eqnarray} \label{matching1bis}
F_V\,=\,2G_V\,=\,\sqrt{2}F_A\,=\,\sqrt{2}F\,, \qquad M_A\,=\,\sqrt{2}M_V\, , \nonumber \\
c_m\,=\,c_d\,=\,\sqrt{2}d_m\,=\,\frac{F}{2} \,, \qquad M_P\,\simeq\,\sqrt{2}M_S \,.
\end{eqnarray}

\subsection{Subleading Lagrangian}

The one loop calculation of the vector form factor with the previous lagrangian generates ultraviolet divergences which require counterterms with a higher number of derivatives. We will only include the minimal set of chiral structures needed to renormalize our calculation. We expect their corresponding couplings to be subleading in the $1/N_C$ expansion, since they are associated with quantum loop corrections.

The following $\cO(p^4)$ and $\cO(p^6)$ pseudo-Goldstone interactions will be required:
\begin{eqnarray}
\widetilde{\mathcal{L}}_{pGB}^{\,(4)} &=&\frac{i\,\widetilde{\ell}_6}{4} \bra f_+^{\mu\nu} \left[ u_\mu , u_\nu \right]  \ket  -  \widetilde{\ell}_{12} \,\bra \nabla^\mu u_\mu \nabla^\nu u_\nu\ket \,  , \label{eq.L4} \\
\widetilde{\mathcal{L}}_{pGB}^{(6)} &=& i \,\widetilde{c}_{51}\, \bra \nabla^\rho f_+^{\mu\nu} [h_{\mu\rho},u_\nu] \ket  +  i \,\widetilde{c}_{53}\, \bra  \nabla_\mu f_+^{\mu\nu} [h_{\nu\rho},u^\rho] \ket \, . \label{eq.L6}
\end{eqnarray}
%Chiral invariance forces these terms to have structures contained in the corresponding $\chi$PT lagrangian~\cite{ChPTa,ChPTb,ChPTc}.
Note that the superindex indicates the chiral order of the operator. We use a tilde to denote the R$\chi$T couplings in Eqs.~(\ref{eq.L4}) and (\ref{eq.L6}), which are different to the ones with the same names (without tilde) in $\chi$PT. For instance, the chiral coupling $\ell_6$ ($L_9$ in the three flavour case) is dominated by a contribution from vector-meson exchange and is of $\cO(N_C)$, while the corresponding resonance coupling $\widetilde{\ell}_6$ does not contain this contribution and is of $\cO(1)$.

The operator with $\widetilde{\ell}_{12}$ in Eq.~(\ref{eq.L4}) does not contribute to the tree-level calculation; nevertheless, it is needed to renormalize the pseudo-Goldstone self-energies. At $\cO(p^6)$, only the combination of couplings $\widetilde{r}^{\phantom{\, r}}_{V2}\equiv 4 F^2\,(\widetilde{c}_{53}- \widetilde{c}_{51})$ is going to be relevant for the VFF~\cite{ChPTc}. Including the lagrangians of Eqs.~(\ref{eq.L4}) and (\ref{eq.L6}), the tree-level calculation of the
vector form factor gives the result:
\begin{eqnarray}\label{eq:tree}
\mathcal{F}(q^2)& =& 1 \, +\, {F_V\, G_V\over F^2}\, {q^2\over M_V^2-q^2}\, -\, \widetilde{\ell}_6\, {q^2\over F^2}\, +\, \widetilde{r}^{\phantom{\, r}}_{V2}\, {q^4\over F^4}\, .
\end{eqnarray}

The Brodsky-Lepage requirement that the form factor should vanish at $q^2\to\infty$ implies the following conditions at leading order in $1/N_C$:
\begin{equation} \label{eq:conditions}
F_V\, G_V\, =\, F^2\qquad , \qquad
\widetilde{\ell}_6\, =\, 0\qquad , \qquad
\widetilde{r}^{\phantom{\, r}}_{V2}\,\equiv\, 4 F^2\,(\widetilde{c}_{53}- \widetilde{c}_{51})\, =\, 0\, .
\end{equation}
Therefore, the couplings $\widetilde{\ell}_6/F^2$ and $\widetilde{r}^{\phantom{\, r}}_{V2}/F^4$ are of subleading order in the $1/N_C$ expansion, i.e. $\cO(1/N_C)$, as expected on pure dimensional grounds.
\begin{figure}
\begin{center}
\includegraphics[scale=0.75]{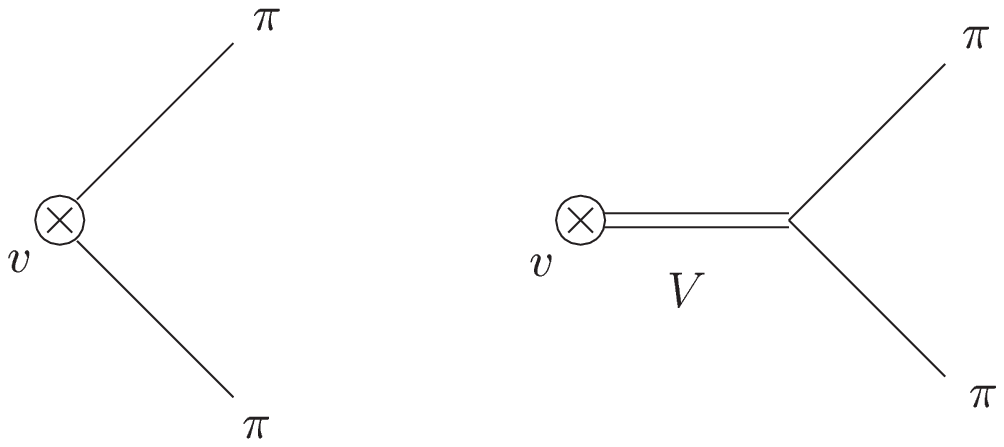}
\caption{\small{Leading-order contributions to the vector form factor of the pion. A single line stands for a pseudo-Goldstone boson while a double line indicates a vector resonance.}}
\label{fig.VFFLO}
\end{center}
\end{figure}

The renormalization of Green Functions including resonance fields forces the presence of the following additional counterterms:
\begin{eqnarray}
\mathcal{L}_{Z}^{\,(4)} &=& \frac{X_{Z_1}}{2} \bra \nabla^2 V^{\mu\nu} \left\{\nabla_\nu,\nabla^\sigma\right\} V_{\mu\sigma} \ket  + \frac{X_{Z_2}}{4} \bra \left\{\nabla_\nu,\nabla_\alpha\right\} V^{\mu\nu} \left\{\nabla^\sigma,\nabla^\alpha\right\} V_{\mu\sigma} \ket \nonumber \\
&&  +  \frac{X_{Z_3}}{4} \bra \left\{\nabla^\sigma,\nabla^\alpha\right\} V^{\mu\nu} \left\{\nabla_\nu,\nabla_\alpha\right\} V_{\mu\sigma} \ket\, , \label{eq.L4VZ} \\
\mathcal{L}_{F}^{\,(4)}&=& X_{F_1} \bra V_{\mu\nu} \nabla^2 f_+^{\mu\nu} \ket + X_{F_2} \bra V_{\mu\nu} \left\{ \nabla^\mu, \nabla_\alpha \right\} f_+^{\alpha\nu} \ket  \, , \label{eq.L4VF} \\
\mathcal{L}_{G}^{\,(4)}& =&i\, X_{G_1}\bra\left\{ \nabla^\alpha, \nabla_\mu \right\} V^{\mu\nu} \left[ u_\nu, u_\alpha\right] \ket + i\, X_{G_2} \bra V^{\mu\nu} \left[ h_{\alpha\mu} , h^\alpha_\nu \right]\ket  \, . \label{eq.L4VG}
\end{eqnarray}
The quadratic lagrangian $\mathcal{L}_{Z}^{\,(4)}$ is needed to renormalize the vector self-energy. Actually, only the sum of couplings $X_Z\equiv X_{Z_1}+X_{Z_2}+X_{Z_3}$ is relevant for this purpose. The renormalization of the vector matrix element of the vector current involves the sum of $\mathcal{L}_{F}^{\,(4)}$ couplings $X_F\equiv X_{F_1} + X_{F_2}$. Finally, the vertex with one external vector resonance and two pseudo-Goldstone legs is renormalized by $\mathcal{L}_{G}^{\,(4)}$ through the combination $X_G\equiv   X_{G_2}-  X_{G_1}/2$. The dimensions of the couplings are $[X_Z]=E^{-2}$ and $[X_{F}]=[X_{G}]= E^{-1}$.

Finally, following the notation of Eq.~(\ref{lagrangian_VFF}), one has that
\begin{eqnarray}
\mathcal{L}_{R\chi T}^{NLO} &=& \widetilde{\mathcal{L}}_{pGB}^{(4)} \,+\, \widetilde{\mathcal{L}}_{pGB}^{(6)} \,+\, \mathcal{L}_Z^{(4)}\,+\, \mathcal{L}_F^{(4)}\,+\, \mathcal{L}_G^{(4)}\,.
\end{eqnarray}
At next-to-leading order in $1/N_C$, these counterterm lagrangians only contribute through tree-level diagrams. One can then use the leading order equations of motion, 
\begin{eqnarray} \label{eq.EOM}
\nabla^\mu \nabla_\rho V^{\rho\nu} - \nabla^\nu \nabla_\rho V^{\rho\mu} &=& -  M_V^2\, V^{\mu\nu}\, -\, \frac{F_V}{\sqrt{2}}\, f_+^{\mu\nu}\,-\, \frac{i G_V}{\sqrt{2}}\,\left[u^\mu,u^\nu\right] \,,
\end{eqnarray}
to reduce the number of relevant operators. The lagrangians of Eqs.~(\ref{eq.L4VZ}), (\ref{eq.L4VF}) and (\ref{eq.L4VG}) take then the equivalent forms:
\begin{eqnarray}
\mathcal{L}_{Z}^{\,(4)}|_{\mathrm{EOM}} &=& \frac{X_Z M_V^4}{2} \bra V^{\mu\nu} V_{\mu\nu} \ket + \frac{X_Z M_V^2 F_V}{\sqrt{2}} \bra V_{\mu\nu} f_+^{\mu\nu} \ket +  \frac{i\, X_Z M_V^2 G_V}{\sqrt{2}} \bra V_{\mu\nu} \left[ u^\mu , u^\nu \right] \ket\nonumber \\
&&+ \frac{ i\, X_Z F_V G_V}{2} \bra f_+^{\mu\nu} \left[ u_\mu , u_\nu \right] \ket + \cdots\, , \label{eq.L4VZEOM} \\
\mathcal{L}_{F}^{\,(4)}|_{\mathrm{EOM}} &=& - X_{F} M_V^2  \bra V_{\mu\nu} f_+^{\mu\nu} \ket - \frac{i\, X_{F} G_V}{\sqrt{2}} \bra f^{\mu\nu}_+ \left[ u_\mu , u_\nu \right]  \ket + \cdots \, , \label{eq.L4VFEOM} \\
\mathcal{L}_{G}^{\,(4)}|_{\mathrm{EOM}} &=& - 2  i\, X_{G} M_V^2 \bra V^{\alpha\nu} \left[ u_\alpha, u_\nu \right] \ket - i\, \sqrt{2}\,  X_{G} F_V \bra f_+^{\mu\nu} \left[ u_\mu , u_\nu \right] \ket + \cdots \, , \label{eq.L4VGEOM}
\end{eqnarray}
where the dots denote other terms which are not relevant for the VFF calculation, at this order. The derivatives acting on the vector resonance fields have been traded by the heavy mass scale $M_V$ and/or derivatives acting on the pseudo-Goldstone fields, giving rise to the usual tensor structures of the $\chi$PT lagrangian. Therefore, the effect of the counterterm lagrangians $\mathcal{L}_{Z}^{\,(4)}$, $\mathcal{L}_{F}^{\,(4)}$ and $\mathcal{L}_{G}^{\,(4)}$ is just equivalent to the following shift in the couplings at next-to-leading order in $1/N_C$:
\begin{eqnarray}
\widetilde{\ell}_6^{\,\mathrm{eff}} & =& \widetilde{\ell}_6 + 2\, X_Z F_V G_V   -  2 \sqrt{2}\, X_F G_V  -  4 \sqrt{2}\, X_G F_V  \, , \nonumber\\
F_V^{\,\mathrm{eff}} &=& F_V  +  2\, X_Z M_V^2 F_V  -  2 \sqrt{2}\, X_{F} M_V^2 \, , \nonumber\\
G_V^{\,\mathrm{eff}} &=& G_V  +  2\, X_Z M_V^2 G_V  - 4 \sqrt{2}\, X_{G} M_V^2 \, , \nonumber\\
(M_V^2)^{\mathrm{eff}} &=& M_V^2  +  2\, X_Z M_V^4\, , \nonumber\\
\widetilde{r}^{\,\mathrm{eff}}_{V2}&=&  \widetilde{r}^{\phantom{\, r}}_{V2}\, . \label{eq.effcouplings}
\end{eqnarray}
Thus, since $\widetilde{\ell}_6^{\,\mathrm{eff}} \sim\widetilde{\ell}_6 \sim (M_V^2)^{\mathrm{eff}}\sim M_V^2\sim\cO(1)$ and $F_V^{\,\mathrm{eff}}\sim F_V\sim G_V^{\,\mathrm{eff}}\sim G_V\sim\cO(\sqrt{N_C})$, a consistent $1/N_C$ counting requires that $X_G$ and $X_F$ are of $\cO(1/\sqrt{N_C})$ and $X_Z$ of $\cO(1/N_C)$.

\section{Renormalization}

The renormalization procedure follows very systematic and precise steps in any well defined quantum field theory. First of all, the two-point Green Functions must be renormalized. Later the three-point Green Functions and so on. For the vector form factor up to next-to-leading order in the $1/N_C$ expansion only the two- and three-point Green Functions will contribute. The corresponding renormalizations for the one-particle-irreducible diagrams at one-loop level are given in the next subsections.

We will adopt the $\overline{MS}-1$ scheme, usually employed in $\chi$PT calculations, where one subtracts the divergent constant
\begin{eqnarray}
\lambda_\infty &=& \frac{2\, \mu^{D-4}}{D-4}+\gamma_E-\log{4\pi}-1\, ,
\end{eqnarray}
being $D$ the space-time dimension and $\gamma_E \simeq 0.5772$ the Euler's constant. However, we will impose the on-shell condition to renormalize the pion self-energy. This simplifies the calculation of physical amplitudes with external pions. Since we work in the massless quark limit, the pseudo-Goldstone tadpoles will not give any contribution. The precise definition of the relevant Feynman integrals with one, two and three propagators are relegated to Appendix~B, while the contributions from each diagram are shown in Appendix~C.

\subsection{Pion Self-energy}

\begin{figure}[t]\centering   %t!
\includegraphics[scale=0.85]{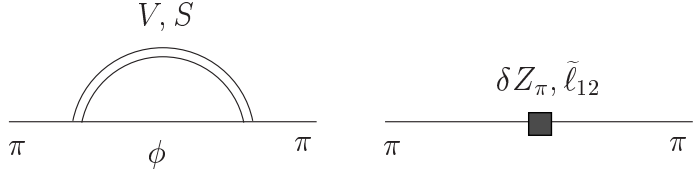}
\caption{\small One-loop diagrams and local contributions to the pion self-energy.}
\label{fig.piself}
\end{figure}

The diagrams contributing to the pion propagator are shown in Figure~\ref{fig.piself}. The kinetic lagrangians of Eqs.~(\ref{kinSP}) and (\ref{kinVA}) generate additional tadpole topologies with one resonance propagator, but they are identically zero even with massive pions. The divergences of $\cO(p^2)$ are reabsorbed through the wave-function renormalization $\pi^{b}=(1+\delta Z_\pi)^\frac{1}{2} \pi^{r}$, being $\pi^{b}$ and $\pi^{r}$ the bare and renormalized pion fields respectively. In the on-shell scheme,
\begin{equation}
\delta Z_\pi\! =\!  -\frac{2 G_V^2}{F^2} \frac{3M_V^2}{16\pi^2 F^2} \!\left\{ \lambda_\infty\! + \log{\frac{M_V^2}{\mu^2}}\! + \frac{1}{6} \right\}  +  \frac{4 c_d^2}{F^2}\frac{M_S^2}{16\pi^2 F^2} \!\left\{\lambda_\infty\! + \log{\frac{M_S^2}{\mu^2}} - \frac{1}{2}\right\}\! .\!\!\!
\end{equation}

There are also divergences of $\cO(p^4)$ which renormalize one of the couplings in $\widetilde{\mathcal{L}}_{\chi}^{\,(4)}$:
\begin{equation}
\widetilde{\ell}_{12}\,\equiv\, \widetilde{\ell}_{12}^r(\mu) + \delta \widetilde{\ell}_{12}(\mu)\, ,  \qquad\qquad \delta \widetilde{\ell}_{12}(\mu) \, = \, -\frac{G_V^2 + 2 c_d^2}{F^2} \frac{\lambda_\infty}{32\pi^2}\, .
\end{equation}
The renormalized pion self-energy takes the form
\begin{eqnarray}
-i\,\Sigma_\pi^r(p^2)& = & -i\, {p^4\over 16\pi^2 F^2}\left\{ 64\pi^2 \widetilde{\ell}_{12}^r(\mu) + \frac{2 G_V^2}{F^2}
\left[\log{\frac{M_V^2}{\mu^2}} + \phi\left(\frac{p^2}{M_V^2}\right)\right]
\right.\nonumber \\
&&\left.\hskip 3cm +  \frac{4c_d^2}{F^2}\left[\log{\frac{M_S^2}{\mu^2}} + \phi\left(\frac{p^2}{M_S^2}\right)\right]\right\} \, ,
\end{eqnarray}
where the function $\phi(p^2/M_V^2)$,
\begin{equation}
\!\!\!\phi(x) = \left(1-{1\over x}\right)^2 \! \left[ \!\left(1-{1\over x}\right)\! \log{(1-x)} -1 + {x\over 2}\right]  = - (1-x)^2 \sum_{n=0}^\infty  {x^n\over (n+2)(n+3)} ,
\end{equation}
contains finite and scale-independent contributions.

\subsection{Rho Self-energy}

The one-loop $\rho$ self-energy contains only an $\cO(p^4)$ divergence, which renormalizes the coupling $X_{Z}$ of the subleading resonance lagrangian:
\begin{equation} \label{eq:XZ_ren}
X_{Z}\,\equiv\, X_Z^r(\mu) + \delta X_{Z}(\mu) \, ,  \qquad\qquad
\delta X_{Z}(\mu)\, =\, -\frac{2 G_V^2}{F^2} \frac{\lambda_\infty}{192 \pi^2 F^2}\, .
\end{equation}
\begin{figure}  %t!
\begin{center}
\includegraphics[scale=0.80]{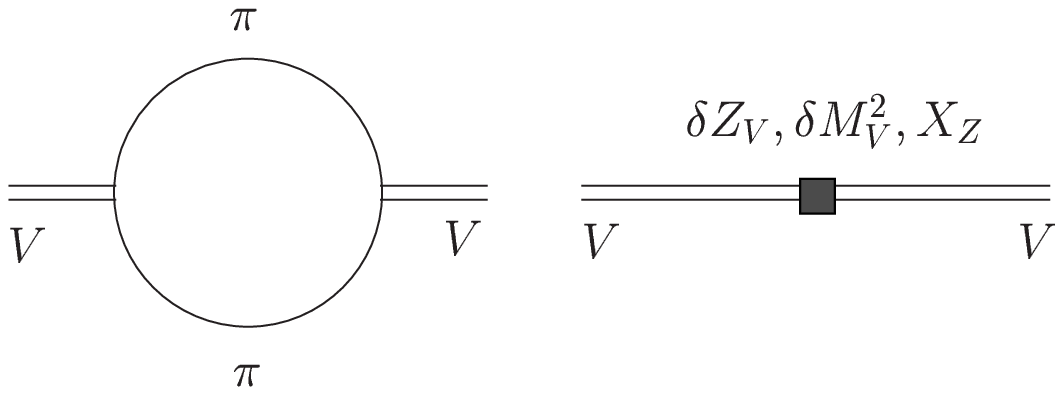}
\caption{\small{One-loop diagrams and local contributions to the $\rho$ self-energy.}}
\label{fig.rhoself}
\end{center}
\end{figure}
Thus, the vector mass and wave-function are not renormalized:
\begin{equation}
\delta M_V^2 \, =\, 0 \, ,   \qquad\qquad
\delta Z_V \, = \, 0 \, .
\end{equation}
The renormalized $\rho$ self-energy then becomes:
\begin{eqnarray}\label{eq.rhoSE}
-i\,\Sigma_V^r(q)^{\mu\nu,\rho\sigma} &=& -{i\over 2}\Omega^{L\,\mu\nu,\,\rho\sigma}(q)\,\Sigma_V^r(q^2)\, ,
\end{eqnarray}
where the antisymmetric tensor structure $\Omega^{L\,\mu\nu,\rho\sigma}(q)$ is defined in Appendix~A and 
\begin{eqnarray}\label{eq:VM_SE}
\Sigma_V^{\,r}(q^2)&=& - q^4 \left\{ 2 X^r_Z(\mu)  - \frac{2G_V^2}{F^2}\frac{1}{F^2}\left[ {1\over 6}\hat{B}_0(q^2/\mu^2) + \frac{1}{144\pi^2} \right]\right\} \, ,
\end{eqnarray}
with $\hat{B}_0(q^2/\mu^2)$ defined in Appendix~B.

\subsection{$\langle v^\mu\, V^{\rho\sigma}\rangle$ One-particle-irreducible Vertex}

\begin{figure}[t]
\begin{center}
\includegraphics[scale=0.8]{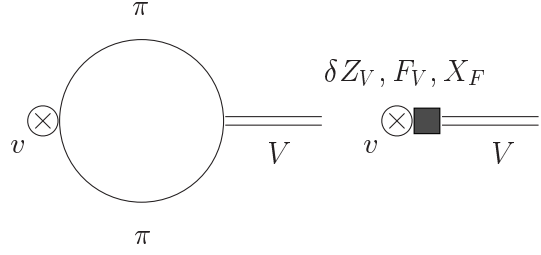}
\caption{\small{Diagrams contributing to the $\langle v^\mu\, V^{\rho\sigma}\rangle$ Green Function at NLO in $1/N_C$.}}
\label{fig.FV}
\end{center}
\end{figure}

The one-particle-irreducible amputated diagrams connecting an external vector quark current to an outgoing vector resonance are shown in Figure~\ref{fig.FV}. The one-loop contribution brings an $\cO(p^4)$ divergence which gets reabsorbed through the following renormalization of the coupling $X_F$:
\begin{equation}
X_{F}\,\equiv\, X_F^r(\mu) + \delta X_F(\mu) \, , \qquad\qquad
\delta X_F(\mu)\, =\, -\frac{\sqrt{2}G_V}{F} \frac{\lambda_\infty}{192 \pi^2 F}\, .
\end{equation}
Since there are no divergences of $\cO(p^2)$, the lowest-order coupling $F_V$ remains unchanged:
\begin{eqnarray}
\delta F_V &=& 0 \, .
\end{eqnarray}
The renormalized vertex function takes the form
\begin{equation}
i \,\Phi (q)^{\mu,\rho\sigma} \, = \, - i \, \mathcal{I}^{\rho\sigma}_{\alpha\beta}\, q^{\alpha}\, g^{\mu\beta} \left\{ F_V - 2 \sqrt{2} X_F^r(\mu)  q^2  +  \frac{2 G_V}{F^2}  q^2 \left[\frac{1}{6}\hat{B}_0(q^2/\mu^2) + \frac{1}{144\pi^2} \right]\right\} \, ,
\end{equation}
where the first term is the leading order contribution. The antisymmetric tensor structure $\mathcal{I}^{\rho\sigma}_{\alpha\beta}$ is defined in Appendix~A and the massless two-point function $\hat{B}_0(q^2/\mu^2)$ in Appendix B.

\subsection{$\langle V_{\mu\nu}\pi\pi\rangle$ One-particle-irreducible Vertex}

\begin{figure}[t]
\begin{center}
\includegraphics[scale=0.7]{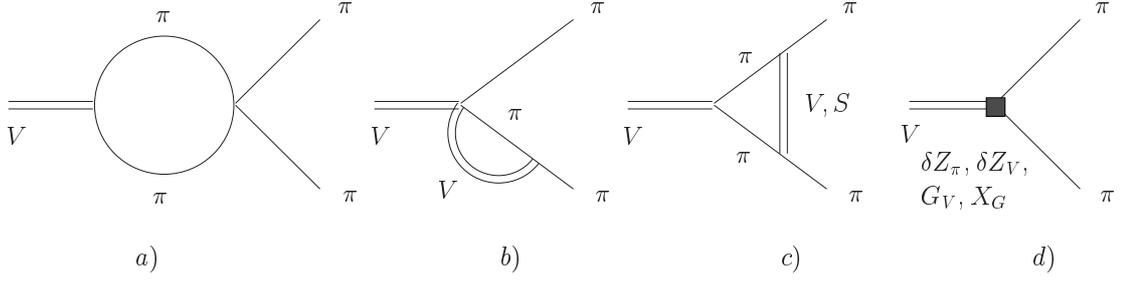}
\caption{\small{NLO diagrams contributing to the three-point Green Function $V^{\mu\nu}\to \pi\pi$.}}
\label{fig.GV}
\end{center}
\end{figure}

The one-particle-irreducible amputated diagrams connecting a vector resonance with two outgoing pseudo-Goldstone bosons at next-to-leading order in $1/N_C$ are shown in Figure~\ref{fig.GV}. The loop diagrams generate $\cO(p^2)$ and $\cO(p^4)$ divergences, which renormalize the couplings $G_V$ and $X_{G}$, respectively:
\begin{align}
G_V & \equiv  G_V^r(\mu) + \delta G_V(\mu) , \,\,\, 
\delta G_V =  G_V  \!\left[ 3 M_V^2\!\left( \frac{2 G_V^2}{F^2}-\frac{1}{2}\right) - M_S^2 \frac{4 c_d^2}{F^2} \right] \!\frac{\lambda_\infty}{16\pi^2 F^2}  ,\!\!\\
X_G&\equiv  X_G^r(\mu) + \delta X_G(\mu)  , \,\,\,
\delta X_G= \frac{\sqrt{2} G_V}{F} \! \left[ \frac{2 G_V^2}{F^2} + \frac{4 c_d^2}{F^2} - 2 \right] \frac{ \lambda_\infty}{1536\pi^2 F}  .\!\!
\end{align}
The wave-function renormalization of the external vector and pion legs amounts to a global factor $\left(\delta Z_\pi+\frac{1}{2} \delta Z_V\right)$ multiplying the lowest-order contribution (keep in mind that $\delta Z_V=0$). Taking this into account, one finally gets the finite vertex function
\begin{equation}
i \,\Gamma_{\mu\nu}^{\,r}(p_1,p_2)= \mathcal{I}^{\alpha\beta}_{\mu\nu} \,q_\alpha \,(p_1-p_2)_\beta \frac{1}{F^2}\left\{
G_V^r(\mu) \left[ 1 -\Delta\Gamma (q^2,\mu^2)\right]  - 4\sqrt{2} X_G^r(\mu) q^2\right\}\, ,
\end{equation}
where
\begin{align}
\Delta \Gamma (q^2,\mu^2) &= \, {1\over F^2}\left\{\hat{B}_0(q^2/\mu^2)\left[ \frac{2 G_V^2}{F^2} \left(\frac{M_V^4}{q^2}+2 M_V^2 + \frac{q^2}{12} \right) + \frac{4 c_d^2}{F^2}\left(\frac{M_S^4}{q^2}+\frac{q^2}{12}\right) -\frac{q^2}{6}\right] \right.\nonumber\\
&\quad+ \frac{M_V^2}{16\pi^2}\log{\frac{M_V^2}{\mu^2}} \left[ \frac{2G_V^2}{F^2}\left( \frac{M_V^2}{q^2}+5\right)
-\frac{3}{2} \right] + \frac{M_S^2}{16\pi^2}\log{\frac{M_S^2}{\mu^2}} \frac{4c_d^2}{F^2}\left(\frac{M_S^2}{q^2}-1\right) \nonumber\\
&\quad+ \frac{M_V^2}{64\pi^2} \left[ 3\frac{2G_V^2}{F^2}-1\right] +\frac{3M_S^2}{64\pi^2}\frac{4c_d^2}{F^2} +\frac{q^2}{288\pi^2}\left[\frac{2G_V^2}{F^2}+\frac{4c_d^2}{F^2}-2\right] \nonumber\\
&\quad+\frac{2 G_V^2}{F^2} C_0(q^2,0,0,M_V^2) \left[ \frac{M_V^6}{q^2}+\frac{5M_V^4}{2}+q^2 M_V^2\right] \nonumber\\
&\quad\left. + \frac{4c_d^2}{F^2} C_0(q^2,0,0,M_S^2)\left[\frac{M_S^6}{q^2}+\frac{M_S^4}{2}\right] \right\} \, .\label{eq:DGamma}
\end{align}
The three-propagator integral $C_0(q^2,M_a^2,M_b^2,M_c^2)$ is defined in Appendix~B.

\subsection{$\langle v_\mu\pi\pi\rangle$ One-particle-irreducible Vertex}

\begin{figure}
\begin{center}
\includegraphics[scale=0.7]{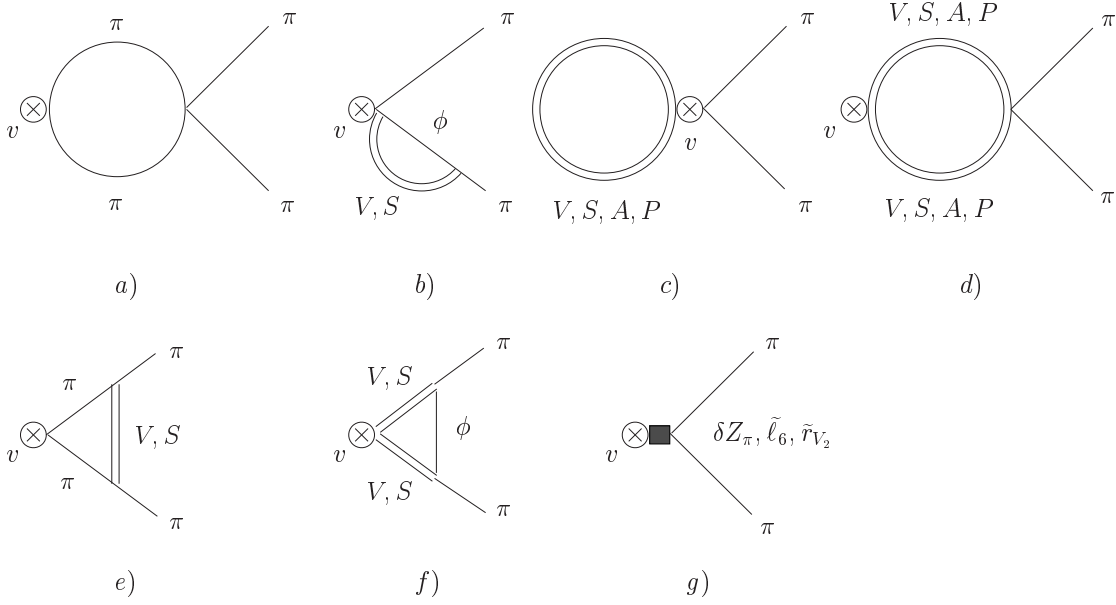}
\caption{\small{1PI diagrams
connecting an external vector current and two outgoing pions, at next-to-leading order in the $1/N_C$ expansion.}}
\label{fig.VFFl6}
\end{center}
\end{figure}

The divergences generated by the one-particle-irreducible loop diagrams shown in Figure~\ref{fig.VFFl6} get reabsorbed through the renormalization of the pion wave function $\delta Z_\pi$ and the $\cO(p^4)$ and $\cO(p^6)$ couplings $\widetilde{\ell}_6$ and $\widetilde{r}^{\phantom{\, r}}_{V2}$:
\begin{eqnarray} 
\widetilde{\ell}_6 &\equiv &\widetilde{\ell}^r_6(\mu) + \delta \widetilde{\ell}_6(\mu) \, , \quad \delta \widetilde{\ell}_6(\mu) \, =\, \left\{ 3-2\frac{2G_V^2}{F^2}+\frac{4c_d^2}{F^2} \right\} \frac{\lambda_\infty}{96\pi^2} \, ,\label{eq:l6_run} \\
\widetilde{r}^{\phantom{\, r}}_{V2}&\equiv&\widetilde{r}^{\, r}_{V2}(\mu) + \delta \widetilde{r}^{\phantom{\, r}}_{V2}(\mu) \,,\quad \delta \widetilde{r}^{\phantom{\, r}}_{V2}(\mu) \,=\, \frac{F^2\lambda_\infty}{96 \pi^2} \left\{ \frac{1}{M_V^2} + \frac{1}{M_A^2}\right\} \, . \label{eq:f1_run}
\end{eqnarray}
The resulting finite correction to the lowest-order pion form factor,
\begin{eqnarray} \label{eq:DF_1PI}
\Delta\mathcal{F}(q^2)_{\mathrm{1PI}} &=&\Delta\mathcal{F}^{\,\mathrm{ct}}  + \Delta\mathcal{F}^{\chi}+ \Delta\mathcal{F}^{V} + \Delta\mathcal{F}^A +\Delta\mathcal{F}^S +\Delta\mathcal{F}^{P}\, ,
\end{eqnarray}
contains contributions from tree-level counterterms,
\begin{eqnarray}
\Delta\mathcal{F}^{\,\mathrm{ct}} & =& -\frac{2 G_V^2}{F^2}\frac{M_V^2}{16\pi^2 F^2} \left\{ 3\log{\frac{M_V^2}{\mu^2}} + \frac{1}{2} \right\} +   \frac{4c_d^2}{F^2}\frac{M_S^2}{16\pi^2 F^2}\left\{ \log{\frac{M_S^2}{\mu^2}} - \frac{1}{2}\right\} \nonumber\\
&&- \widetilde{\ell}^r_6(\mu)\frac{q^2}{F^2}  + \widetilde{r}^{\, r}_{V2}(\mu)\frac{q^4}{F^4} \, ,\label{eq.finitol6Zpi}
\end{eqnarray}
and loop diagrams with internal pseudo-Goldstone bosons (first diagram in Figure~\ref{fig.VFFl6}),
\begin{eqnarray} 
\Delta\mF^{\chi}& =&\frac{q^2}{F^2} \left\{\frac{1}{6}\hat{B}_0(q^2/\mu^2) + \frac{1}{144\pi^2}\right\} \, ,\label{eq.finitol6L2}
\end{eqnarray}
and vector,
\begin{eqnarray}
\Delta\mathcal{F}^{V} &=& \frac{2G_V^2}{F^2} \frac{1}{F^2} \left\{ - C_0(q^2,0,0,M_V^2) \left[\frac{M_V^6}{q^2} + \frac{5 M_V^4}{2} + q^2 M_V^2\right] \right.\nonumber\\
&&\left. + C_0(q^2,M_V^2,M_V^2,0) \left[\frac{M_V^6}{q^2} + \frac{M_V^4}{2} \right] - \hat{B}_0(q^2/\mu^2) \left[\frac{M_V^4}{q^2}+2M_V^2+\frac{q^2}{12} \right]\right\}  \nonumber\\
&&-\frac{\overline{B}_0(q^2,M_V^2)}{F^2}  \left[ \left(2 M_V^2+\frac{q^2}{6}-\frac{q^4}{6 M_V^2}\right)+ \frac{2G_V^2}{F^2}\left(\frac{M_V^4}{q^2} + \frac{2 M_V^2}{3} - \frac{5 q^2}{12}\right) \right] \nonumber \\
&&+\frac{M_V^2}{16\pi^2 F^2}\log{\frac{M_V^2}{\mu^2}} \left[  \left(\frac{q^2}{2 M_V^2}-\frac{q^4}{6 M_V^4} \right)- \frac{2 G_V^2}{F^2} \left(  \frac{M_V^2}{q^2}-1 + \frac{5q^2}{12M_V^2} \right) \right] \nonumber\\
&& +\frac{M_V^2}{16\pi^2 F^2} \left[ \left(\frac{q^2}{2 M_V^2} -\frac{2 q^4}{9 M_V^4}\right)  +\frac{2G_V^2}{F^2}\left( \frac{M_V^2}{q^2} + 1 -\frac{19 q^2}{36 M_V^2}\right) \right] \, ,\label{eq.finitol6V}
\end{eqnarray}
axial-vector,
\begin{eqnarray}
\Delta\mathcal{F}^A&=&- \frac{\overline{B}_0(q^2,M_A^2)}{F^2}\left[ 2M_A^2+\frac{q^2}{6}-\frac{q^4}{6M_A^2} \right] +  \frac{M_A^2}{16\pi^2F^2} \log{\frac{M_A^2}{\mu^2}} \left[\frac{q^2}{2M_A^2}-\frac{q^4}{6M_A^4} \right] \nonumber\\ 
&& +\frac{q^2}{32\pi^2F^2} -\frac{q^4}{72\pi^2 F^2 M_A^2} \, ,\label{eq.finitol6A}
\end{eqnarray}
scalar,
\begin{eqnarray}
\Delta \mathcal{F}^S &=&\frac{4c_d^2}{F^2} \frac{1}{F^2} \left\{ - C_0(q^2,0,0,M_S^2) \left[\frac{M_S^6}{q^2}+\frac{M_S^4}{2} \right] + C_0(q^2,M_S^2,M_S^2,0) \left[\frac{M_S^6}{q^2}-\frac{M_S^4}{2}\right] \right.\nonumber\\
&& \left.  -\hat{B}_0(q^2/\mu^2) \left[\frac{M_S^4}{q^2}+\frac{q^2}{12} \right]  +\frac{M_S^4}{16\pi^2 q^2}\right\}  - \frac{q^2}{288\pi^2F^2}\left[ 1+\frac{1}{2}\frac{4c_d^2}{F^2}\right] \nonumber\\
&&- \frac{\overline{B}_0(q^2,M_S^2)}{F^2} \left[\left(\frac{2M_S^2}{3} - \frac{q^2}{6}\right) +\frac{4 c_d^2}{F^2} \left(\frac{M_S^4}{q^2}-\frac{M_S^2}{3}+\frac{q^2}{12}\right)\right] \nonumber \\
&&- \frac{M_S^2}{16\pi^2F^2}\log{\frac{M_S^2}{\mu^2}}  \left[ \frac{4c_d^2}{F^2} \left(1+\frac{M_S^2}{q^2} -\frac{q^2}{12M_S^2}\right)+ \frac{q^2}{6M_S^2} \right]\, ,\label{eq.finitol6S}
\end{eqnarray}
and pseudoscalar resonances,
\begin{eqnarray}
\Delta \mathcal{F}^{P} &=& \frac{\overline{B}_0(q^2,M_P^2)}{F^2} \left[ -\frac{2M_P^2}{3}+\frac{q^2}{6} \right] - \frac{q^2}{96\pi^2F^2} \left[ \log{\frac{M_P^2}{\mu^2}}+\frac{1}{3}\right] \, .\label{eq.finitol6P}
\end{eqnarray}
All the Feynman integrals are shown in Appendix~B.

\section{Vector Form Factor}
\begin{figure}\centering
\includegraphics[scale=0.7]{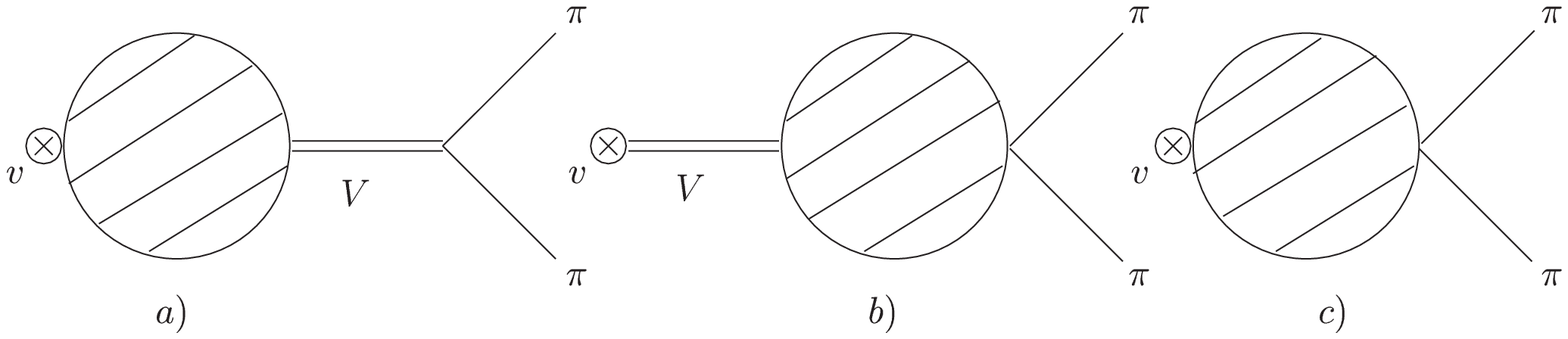}
\caption{\small{Basic topologies contributing to the Vector Form Factor at NLO.}}
\label{fig:topologies}
\end{figure}

The basic topologies contributing to the vector form factor are shown in Figure~\ref{fig:topologies}, in terms of the one-loop level 1PI diagrams computed in the previous section. The internal $\rho$ line denotes the dressed vector propagator, including the self-energy correction of Eq.~(\ref{eq:VM_SE}), which regulates the $\rho$ pole. Taking this self-energy and the subleading running of $G_V$ into account, the leading order contribution takes the form:
\begin{eqnarray}
\mathcal{F}(q^2)_{\mathrm{LO}}&=& 1 + \frac{F_VG_V^{\,r}(\mu)}{F^2} \frac{q^2}{M_V^2-q^2 - \Sigma^{\,r}_V(q^2)}\, .\label{eq:VFF_LO}
\end{eqnarray}
The topology in Figure~\ref{fig:topologies}.a generates the following subleading correction:
\begin{equation}
\Delta\mathcal{F}(q^2)_F = \frac{q^2}{M_V^2-q^2-\Sigma^{\,r}_V(q^2)} \frac{q^2}{F^2} \!\left\{ \frac{2 G_V^2}{F^2} \!\left[\frac{1}{6}\hat{B}_0(q^2/\mu^2) + \frac{1}{144\pi^2}\right]\!-2\sqrt{2} G_V X_F^r(\mu)\right\}\! .\label{eq:DFF}
\end{equation}
Figure~\ref{fig:topologies}.b brings the contribution:
\begin{equation}
\Delta \mathcal{F}(q^2)_G = - \frac{q^2}{M_V^2-q^2-\Sigma^{\,r}_V(q^2)} \frac{F_V}{\sqrt{2} F}\left\{ \frac{\sqrt{2} G_V}{F}\Delta\Gamma(q^2,\mu^2) + \frac{8 X_G^r(\mu)}{F} q^2\right\}\, ,\label{eq:DFG}
\end{equation}
where $\Delta \Gamma(q^2,\mu^2)$ is given in Eq.~(\ref{eq:DGamma}). Finally, Figure~\ref{fig:topologies}.c denotes the 1PI correction $\Delta\mathcal{F}(q^2)_{\mathrm{1PI}}$ in Eq.~(\ref{eq:DF_1PI}). Adding all contributions together, one gets the VFF at NLO:
\begin{eqnarray}
\mathcal{F}(q^2) &=&\mathcal{F}(q^2)_{\mathrm{LO}} + \Delta\mathcal{F}(q^2)_{F}  + \Delta\mathcal{F}(q^2)_G  + \Delta\mathcal{F}(q^2)_{\mathrm{1PI}}\, .
\end{eqnarray}

Using the large-$N_C$ relations of Eq.~(\ref{matching1bis}) in this result, it can be written in the form:
\begin{eqnarray} 
\mathcal{F}(q^2) &=& A(q^2){M_V^2\over M_V^2-q^2-\Sigma^{\,r}_V(q^2)}+ B(q^2)\, ,\label{eq:VFF_result}
\end{eqnarray}
where
\begin{eqnarray}
A(q^2) & = & 1 + \hat{\delta}_V + 2 M_V^2 \hat{X} - \Delta\tilde{\Gamma}(q^2)\, ,\nonumber\\
B(q^2) & = & \mathcal{G}(q^2) - \hat{\delta}_V - 2 (M_V^2+q^2) \hat{X}\, .
\end{eqnarray}
The constants
\begin{eqnarray}
\hat{\delta}_V&\equiv & {F_V G_V^r(\,\mu)\over F^2} - 1 - \Delta\Gamma (0,\mu^2)\, , \nonumber\\
\hat{X}&\equiv & X_Z^r(\mu) -{1\over F}\left[ X_F^r(\mu) + 4 X_G^r(\mu)\right]\, ,
\end{eqnarray}
and the functions $\Sigma^{\,r}_V(q^2)$,
\begin{eqnarray}
\Delta\tilde{\Gamma}(q^2)&\equiv& \Delta\Gamma (q^2,\mu^2) -\Delta\Gamma (0,\mu^2)\, ,
\end{eqnarray}
and
\begin{eqnarray}
\mathcal{G}(q^2)&\equiv& \Delta\mathcal{F}(q^2)_{\mathrm{1PI}} + \Delta\tilde{\Gamma}(q^2)\,\equiv\, G(q^2,\mu^2) -\Delta\Gamma (0,\mu^2)\,,
\end{eqnarray}
are independent of the renormalization scale $\mu$. The subleading R$\chi$T couplings $X_F^r(\mu)$ and $X_G^r(\mu)$ only appear through the constant $\hat{X}$, while $X_Z^r(\mu)$ is also present in the function $\Sigma^{\,r}_V(q^2)$. At $q^2=0$, $\Delta\tilde{\Gamma}(0) = \mathcal{G}(0) = \Sigma^{\,r}_V(0) = 0$. Therefore $\mathcal{F}(0) = 1$, as it should.

Some 1PI diagrams (Figures~\ref{fig.VFFl6}.a and \ref{fig.VFFl6}.e and the vector terms in Figures~\ref{fig.VFFl6}.b and \ref{fig.VFFl6}.c) have a corresponding reducible counterpart involving a vector propagator. The combination of both contributions can be then incorporated in $A(q^2)$. The function $G(q^2,\mu^2)$ contains the corrections generated by the other 1PI diagrams (Figures~\ref{fig.VFFl6}.d, \ref{fig.VFFl6}.f, the S term in Figure~\ref{fig.VFFl6}.b, the S, A and P terms in Figure~\ref{fig.VFFl6}.c and the $\widetilde{\ell}_6$ and $\widetilde{r}^{\phantom{\, r}}_{V2}$ pieces in Figure~\ref{fig.VFFl6}.g). Subtracting their contribution at $q^2=0$, which contains the dependence on the renormalization scale $\mu$,
\begin{equation}
G(0,\mu^2) = \Delta\Gamma (0,\mu^2)= \frac{1}{16\pi^2 F^2}\!\left\{M_V^2\!\left[\frac{3}{2}\log{\frac{M_V^2}{\mu^2}}+\frac{1}{4}\right]\!+ M_S^2\!\left[-\log{\frac{M_S^2}{\mu^2}}+\frac{1}{2}\right]\!\right\}\! ,\label{eq:G(0)}
\end{equation}
one gets:
\begin{align}
\mathcal{G}(q^2) &=\, \frac{C_0(q^2,M_V^2,M_V^2,0)}{F^2} \left[ \frac{M_V^6}{q^2}+\frac{M_V^4}{2}\right] +
\frac{C_0(q^2,M_S^2,M_S^2,0)}{F^2} \left[\frac{M_S^6}{q^2}-\frac{M_S^4}{2}\right] \nonumber \\
& +\frac{\overline{B}_0(q^2,M_V^2)}{F^2} \left[-\frac{M_V^4}{q^2}-\frac{8M_V^2}{3}+\frac{q^2}{4}+\frac{q^4}{6M_V^2}\right]
+\frac{\overline{B}_0(q^2,M_P^2)}{F^2}\left[-\frac{2M_P^2}{3}+\frac{q^2}{6}\right] \nonumber\\
&+\frac{\overline{B}_0(q^2,M_A^2)}{F^2}\left[-2M_A^2-\frac{q^2}{6}+\frac{q^4}{6M_A^2}\right]+
 \frac{\overline{B}_0(q^2,M_S^2)}{F^2} \left[-\frac{M_S^4}{q^2}-\frac{M_S^2}{3}+\frac{q^2}{12}\right]
\nonumber\\
& +\frac{1}{16\pi^2 F^2}\left\{ \frac{M_V^4 + M_S^4}{q^2} + \frac{3}{4} M_V^2 - \frac{1}{4} M_S^2 + q^2\left[\frac{1}{12}\log{\frac{M_V^2}{\mu^2}} + \frac{1}{2}\log{\frac{M_A^2}{\mu^2}} \right.\right.\nonumber \\
&\qquad \qquad \qquad \left.-\frac{1}{12}\log{\frac{M_S^2}{\mu^2}} -\frac{1}{6}\log{\frac{M_P^2}{\mu^2}} +\frac{4}{9}-16\pi^2 \widetilde{\ell}_6^{r}(\mu)\right]- \frac{q^4}{6}\left[\frac{1}{M_V^2}\log{\frac{M_V^2}{\mu^2}} \right. \nonumber\\
&\qquad \qquad \qquad \left.\left.+\frac{1}{M_A^2}\log{\frac{M_A^2}{\mu^2}} +\frac{4}{3} \left(\frac{1}{M_V^2}+\frac{1}{M_A^2}\right) - \frac{96\pi^2}{F^2}\widetilde{r}^{\, r}_{V2}(\mu)\right]\right\}\, . 
\end{align}

\section{Low-Energy Limit}

As it has been reviewed in Section~2.5, at very low energies, $q^2 \ll M_R^2$, the resonance fields can be integrated out from the effective theory. One recovers then the standard $\chi$PT lagrangian, which leads to the following result for the vector form factor of the pion~\cite{ChPTb,ChPTc}:
\begin{align}
\mathcal{F}_{\chi PT}(q^2) &=\, 1-\frac{q^2}{F^2} \left\{\ell_6^{\,r}(\mu) +\frac{1}{96\pi^2} \left[\log{\left( -\frac{q^2}{\mu^2}\right)}-\frac{5}{3}\right]\right\}+\frac{q^4}{F^4} \bigg\{ r^{\, r}_{V2}(\mu) +\frac{1}{96\pi^2}\times \nonumber \\
&  \left. \times \!\left[\log{\left( -\frac{q^2}{\mu^2}\right)}-\frac{5}{3}\right]\!\left( 2 \ell_1^{\,r}-\ell_2^{\,r} + \ell_6^{\,r}\right)(\mu) + \cO\left(N_C^0\right)\right\} +  \cO\!\left(\frac{q^6}{F^6}\right)\! .\label{eq.VFFop4CHI}
\end{align}
The Taylor expansion in powers of $q^2$ of the R$\chi$T prediction of Eq.~(\ref{eq:VFF_result}) reproduces the $\chi$PT formula, as it should. The coefficient of the $\cO\left[q^4\log{(-q^2/\mu^2)}\right]$ term satisfies the known large-$N_C$ equality~\cite{RChTa,polychromatic}
\begin{eqnarray}
2 \ell_1^r (\mu)-\ell_2^r (\mu) + \ell_6^r (\mu)  &=& F^2 \left( \frac{1}{2M_S^2} -  \frac{5}{2M_V^2}\right) \,.
\end{eqnarray}
The non-logarithmic $\cO(q^4)$ and $\cO(q^6)$ terms relate the low-energy chiral couplings $\ell_6$ and $r^{\phantom{\, r}}_{V2}$ with their R$\chi$T counterparts $\widetilde{\ell}_6$ and $\widetilde{r}^{\phantom{\, r}}_{V2}$:
\begin{align}
\ell_6^{\,r}(\mu) &=\,- \frac{F^2}{M_V^2} (1 +\hat{\delta}_V) +\widetilde{\ell}_6^{\,r}(\mu)  -\frac{1}{96\pi^2}\left[ \log{\frac{M_V^2}{\mu^2}} - \log{\frac{M_P^2}{\mu^2}} + 3 \log{\frac{M_A^2}{\mu^2}} - \frac{13}{6}\right] \nonumber\\
& = \,- \frac{F_V G_V^{\,r}(\mu)}{M_V^2} + \widetilde{\ell}_6^{\,r}(\mu)  +\frac{1}{16\pi^2}\left[ \frac{4}{3}\log{\frac{M_V^2}{\mu^2}}- \frac{1}{2}\log{\frac{M_A^2}{\mu^2}} + \frac{1}{6}\log{\frac{M_P^2}{\mu^2}}  \right.\nonumber\\
&\qquad \qquad  \left. - \frac{M_S^2}{M_V^2} \log{\frac{M_S^2}{\mu^2}}+ \frac{11}{18} + \frac{M_S^2}{2 M_V^2} \right]\, ,\label{eq:l6_rel} \\
&\nonumber \\
r^{\, r}_{V2}(\mu) & = \,\frac{F^2 F_V G_V^{\,r}(\mu)}{M_V^4} + \widetilde{r}^{\, r}_{V2}(\mu)  + \frac{2 F^4}{M_V^2} \left[\hat{X}-X_Z^r(\mu)\right] \nonumber\\
&+\frac{F^2}{96\pi^2}\left\{ \left(6\frac{M_S^2}{M_V^4}+\frac{1}{2 M_V^2} -\frac{1}{2 M_S^2}\right)\log{\frac{M_S^2}{\mu^2}} -\frac{9}{M_V^2}\log{\frac{M_V^2}{\mu^2}} - \frac{1}{M_A^2}\log{\frac{M_A^2}{\mu^2}}\right.\nonumber\\
&\qquad \qquad \left.  -\frac{167}{60 M_V^2}- \frac{17}{10 M_A^2} - \frac{3 M_S^2}{M_V^4} + \frac{17}{20 M_S^2} + \frac{1}{10 M_P^2}\right\}\, .\label{eq:f1_rel}
\end{align}
Notice that the combination of subleading R$\chi$T couplings $\hat{X}$ does not appear at $\cO(p^4)$. Therefore, the relation of Eq.~(\ref{eq:l6_rel}) adopts the same form in terms of the effective couplings defined in Eq.~(\ref{eq.effcouplings}), i.e.
\begin{eqnarray}
\widetilde{\ell}_6^{\,\mathrm{eff},r}(\mu) - \frac{F_V^{\,\mathrm{eff}} G_V^{\,\mathrm{eff},r}(\mu)}{(M_V^2)^{\mathrm{eff},r}(\mu)}& =& \widetilde{\ell}_6^{\,r}(\mu) -\frac{F_V G_V^{\,r}(\mu)}{M_V^2}\,.
\end{eqnarray}
As shown in Eq.~(\ref{eq:f1_rel}), this is no longer true at $\cO(p^6)$; nevertheless, the explicit dependence on $\hat{X}-X_Z^r(\mu)$ present in $r^{\, r}_{V2}(\mu)$ can be reabsorbed into the leading term, through the use of the effective couplings, i.e.
\begin{eqnarray}
r^{\, r}_{V2}(\mu) &=& F^2 \frac{F_V^{\,\mathrm{eff}} G_V^{\,\mathrm{eff},r}(\mu)}{(M_V^4)^{\mathrm{eff},r}(\mu)}
+\widetilde{r}^{\, \mathrm{eff},r}_{V2}+\cdots \,.
\end{eqnarray}
Eqs.~(\ref{eq:l6_rel}) and (\ref{eq:f1_rel}) contain the well known lowest-order predictions for the two $\chi$PT couplings: $\ell_6 = -M_V^2 r^{\, r}_{V2}/F^2 = - F^2/M_V^2$. Moreover, they give their dependence on the renormalization scale at the next-to-leading order. The running of the renormalized couplings $\ell_6^{\,r}(\mu),\,r^{\, r}_{V2}(\mu)$ and $\widetilde{\ell}_6^{\,r}(\mu),\,\widetilde{r}^{\, r}_{V2}(\mu)$ is different, because their corresponding effective theories have a very different particle content.

The $\mu$ dependence of a given coupling ``g'' can be characterized through the logarithmic derivative
\begin{eqnarray} \label{eq:gamma_F}
\mu\,\frac{d g}{d\mu}& =& -\frac{\gamma_g}{16\pi^2}\, .
\end{eqnarray}
From Eqs.~(\ref{eq:l6_run}) and (\ref{eq:f1_run}) one gets the running of the R$\chi$T couplings:%
\begin{equation} \label{eq:gamma_rcht}
\gamma_{\strut\,\widetilde{\ell}_6}\, =\,\frac{2}{3} \, ,\qquad\qquad \gamma_{\strut\,\widetilde{r}^{\phantom{\, r}}_{V2}}\, =\,\frac{F^2}{3} \left(\frac{1}{M_V^2}+\frac{1}{M_A^2}\right)\, =\,\frac{F^2}{2 M_V^2}\, .
\end{equation}
Eqs.~(\ref{eq:l6_rel}) and (\ref{eq:f1_rel}) give then the dependence on the renormalization scale of the corresponding $\chi$PT couplings:
\begin{equation} \label{eq:gamma_chpt}
\gamma_{\strut\ell_6}\, =\,-\frac{1}{3} \, , \qquad\qquad
\gamma_{\strut r^{\phantom{\, r}}_{V2}}\, =\,\frac{F^2}{6} \left(\frac{5}{M_V^2}-\frac{1}{M_S^2}\right)\, .
\end{equation}
These values are in perfect agreement with the low-energy results of Refs.~\cite{ChPTb,ChPTc,VFF_ChPT}. The running of the $\cO(p^6)$ coupling $r^{\phantom{\, r}}_{V2}(\mu)/F^4$ receives of course additional 2-loop contributions which are of $\cO(1/N_C^2)$.

The rigorous control of the renormalization scale dependences allows us to investigate the successful resonance saturation approximation at subleading order. The $\chi$PT couplings $\ell_6$ and $r^{\phantom{\, r}}_{V2}$ have been phenomenologically extracted from a fit to the VFF data at low momenta. This determines the scale-invariant combination~ \cite{VFF_ChPT}:
\begin{eqnarray}
\bar \ell_6&\equiv& \frac{32\pi^2}{\gamma_{\strut\ell_6}}\ell_6^{\,r}(\mu)  - \log{\frac{m_\pi^2}{\mu^2}}\, =\,
16.0\pm 0.5\pm 0.7\, ,\label{eq:l6bar}\\
r^{\, r}_{V2}(M_\rho)& =& (1.6\pm 0.5)\cdot 10^{-4}\, .\label{eq:r2Mrho}
\end{eqnarray}
Inserting these numbers in Eqs.~(\ref{eq:l6_rel}) and (\ref{eq:f1_rel}), one can estimate the corresponding scale-invariant combinations of NLO couplings in R$\chi$T:
\begin{align}
\hat{\ell}_6&\equiv\,\widetilde{\ell}_6^{\,r}(\mu) -\frac{\gamma_{\strut\,\widetilde{\ell}_6}}{32\pi^2}\log{\frac{M_V^2}{\mu^2}} -\frac{F^2}{M_V^2}\,\hat{\delta}_V\, ,\label{eq:l6hat} \\
\hat{r}^{\phantom{\, r}}_{V2}&\equiv\, \widetilde{r}^{\, r}_{V2}(\mu) + \frac{F^4}{M_V^4}\,\left(\hat{\delta}_V + 2 M_V^2 \left[\hat{X}-X_Z^r(\mu)\right]\right) - \frac{\gamma_{\strut\,\widetilde{r}^{\phantom{\, r}}_{V2}} -\frac{2 F^4}{M_V^2} \gamma_{_{X_Z}}}{32\pi^2} \log{\frac{M_V^2}{\mu^2}}\, ,\label{eq:rV2hat}
\end{align}
where $\gamma_{_{X_Z}}=-1/(6 F^2)$. Taking $F=92.4$~MeV, $M_V = 770$~MeV and $M_S = 1$~GeV, one gets $\hat{\ell}_6 = (-0.2\pm 0.9)\cdot 10^{-3}$ and $\hat{r}^{\phantom{\, r}}_{V2}= (-0.2\pm 0.5)\cdot 10^{-4}$, while a larger value of the scalar resonance mass $M_S = 1.4$~GeV shifts the $\cO(p^4)$ coupling to $\hat{l}_6 = (-0.9\pm 0.9)\cdot 10^{-3}$, without affecting $\hat{r}^{\phantom{\, r}}_{V2}$ at the quoted level of accuracy. These numbers should be compared with the large-$N_C$ predictions for the $\chi$PT couplings $\ell_6|_{N_C\to\infty} = -F^2/M_V^2 = -0.014$ and $r^{\phantom{\, r}}_{V2}|_{N_C\to\infty} =F^4/M_V^4 = 2.1\cdot 10^{-4}$. Put in a different way, the hypothesis $\hat{\ell}_6 = \hat{r}^{\phantom{\, r}}_{V2} = 0$ generates excellent predictions for $\ell_6^{\,r}(\mu)$ and $r^{\, r}_{V2}(\mu)$ at any scale $\mu$.

\section{Behaviour at Large Energies}

At large momentum transfer, the relevant renormalization scale invariant
functions take the forms:
\begin{align}
\mathcal{G}(q^2) &= \frac{1}{16\pi^2 F^2}\left\{- \,q^4\left[ \frac{1}{6}\left(\frac{1}{M_V^2}+\frac{1}{M_A^2}\right) \left(\log{\frac{-q^2}{\mu^2}}-\frac{2}{3}\right) -\frac{16\pi^2}{F^2} \widetilde{r}_{V2}^{\, r}(\mu)\right] \right.\nonumber \\
& \qquad \qquad \quad \,\,\,\left. +\, q^2\left[\frac{1}{3}\log{\frac{-q^2}{\mu^2}} +\frac{16}{9} -16\pi^2 \widetilde{\ell}_6^{\,r}(\mu)\right] + \cO\left(q^0\right)\right\} , \nonumber
\end{align}
\begin{align}
\Delta\tilde{\Gamma}(q^2) & = \frac{M_V^2}{16\pi^2 F^2}\!\left\{\log{\frac{-q^2}{M_V^2}} \!\left[\log{\frac{q^2}{M_V^2}}-2\right]\! -\!\frac{1}{2}\log^2{\frac{q^2}{M_V^2}}\! -\! \frac{\pi^2}{6}\!+\!\frac{9}{4}\!+\!\frac{M_S^2}{4 M_V^2}\right\}\! + \cO\!\left(\frac{1}{q^2}\right)\! ,\nonumber \\
\Sigma^{\,r}_V(q^2) &= \frac{-q^4}{96\pi^2 F^2}\left\{ \log{\frac{-q^2}{\mu^2}}-\frac{5}{3} + 192\pi^2 F^2 X^r_Z(\mu)\right\}\, .
\end{align}

The $\rho$ propagator makes the $A(q^2)$ piece of the VFF well behaved when $q^2\to\infty$. However, the 1PI contributions generate a wrong behaviour $\mathcal{G}(q^2)\sim q^4\log{(-q^2/\mu^2)}$ in the $B(q^2)$ term, which cannot be eliminated with a local contribution. The problem originates in the two-resonance cut which has an unphysical growing with momenta.

Although our leading R$\chi$T lagrangian of Eq.~(\ref{lagrangian_VFF}) only incorporates couplings linear in the resonance fields, the kinetic resonance lagrangian introduces some bilinear interactions through the chiral connection included in the covariant derivatives. Their couplings are fixed by chiral symmetry and give rise to the diagrams in Figures~\ref{fig.GV}.b,  \ref{fig.VFFl6}.c, \ref{fig.VFFl6}.d and \ref{fig.VFFl6}.f. Obviously, these are not the only interactions bilinear in the resonance fields even at large-$N_C$ \cite{RChTc,nosaltres2,nosaltres2bis,integral}. Therefore, it is not surprising that our calculation is unable to find the correct behaviour at large energies for those contributions with two intermediate resonances.

The contributions with an internal vector propagator in diagrams \ref{fig.VFFl6}.b and \ref{fig.VFFl6}.c give us some hint about which pieces could be missing in our calculation. These two diagrams combine with a reducible contribution of the type \ref{fig:topologies}.b: the 1PI $\langle V_{\mu\nu}\pi\pi\rangle$ vertex in Figure~\ref{fig.GV}.b. The three contributions contain identical loop functions and their sum generates a global factor $M_V^2/(M_V^2 - q^2)$, which suppresses the large-$q^2$ behaviour. Thus, these corrections have been included in the term $A(q^2)$.

It seems natural to conjecture that the remaining 1PI contributions with two-resonance cuts should combine with the corresponding reducible topologies, including $\langle V RR\rangle$ and $\langle v^\mu RR\rangle$ vertices, to generate the final propagator suppression:
\begin{eqnarray} \label{eq:conjecture}
G(q^2) &\longrightarrow & \frac{M_V^2}{M_V^2 - q^2 -\Sigma^{\,r}_V(q^2)} \,G(q^2)\, .
\end{eqnarray}
The needed lagrangian takes the form
\begin{eqnarray}
\Delta \mathcal{L}_{VRR}& =& i \, \lambda^{VSS}\,\bra V^{\mu\nu}\,\nabla_\mu S\,\nabla_\nu S \ket + i \, \lambda^{VPP}\,\bra V^{\mu\nu}\,\nabla_\mu P\, \nabla_\nu P\ket \,.
\end{eqnarray}
Our conjecture fixes the new chiral couplings in the large-$N_C$ limit. In fact, the main aim of the next chapter is to follow these ideas: once it is accepted the necessity of new terms with more than one resonance field by studying the asymptotic behaviour at large energies, we are going to analyse all the two-body form factors that can be found in the even-intrinsic-parity sector of Resonance Chiral Theory in the Single Resonance Approximation. This will be done in the spirit of correlators at next-to-leading order in the $1/N_C$ expansion. 

\section{Conclusions}

The one-loop analysis of the vector form factor of the pion has shown a series of interesting features:
\begin{enumerate}
\item As expected, loop diagrams with massive resonance states in the internal lines generate ultraviolet divergences, which require additional higher-dimensional counterterms in the R$\chi$T lagrangian. Since these counterterms give rise to tree-level contributions which grow too fast at large momenta, their corresponding couplings should be zero at leading order in the large-$N_C$ expansion. Thus, one can establish a well defined counting in powers of $1/N_C$ to organize the calculation.

The formal renormalization is completely straightforward at one loop. One can easily determine the $\mu$ dependence of all relevant renormalized couplings. Moreover, the final result is only sensitive to some combinations of the chiral couplings. In fact, using the lowest-order equations of motion, one can eliminate most of the higher-order couplings. Their effects get then reabsorbed into redefinitions of the lowest-order parameters.
\item Expanding the result in powers of $q^2/M_R^2$, one recovers the usual $\chi$PT expression at low momenta. This relates the low-energy chiral couplings $\ell_6$ and $r^{\phantom{\, r}}_{V2}$ with their corresponding R$\chi$T counterparts $\widetilde{\ell}_6$ and $\widetilde{r}^{\phantom{\, r}}_{V2}$.

The rigorous control of the renormalization scale dependences has allowed us to investigate the successful resonance saturation approximation at the next-to-leading order in $1/N_C$. The assumption $\hat{\ell}_6 = \hat{r}^{\phantom{\, r}}_{V2} = 0$ generates excellent predictions for $\ell_6^{\,r}(\mu)$ and $r^{\, r}_{V2}(\mu)$ at any scale $\mu$.

We stress again the importance of determining the resonance contributions to the chiral LEC's at next-to-leading order in $1/N_C$, since one keeps a full control of their renormalization scale dependence. Notice how the uncertainty related to the running disappears. This chapter represents a first step towards a systematic procedure to evaluate next-to-leading order contributions in the $1/N_C$ counting: in the next chapter we will present a NLO prediction of $L_8$.

\item At higher energies, we have identified an unphysical behaviour which originates in the two-resonance cuts: they generate an increase of the form factor at large values of momentum transfer. This is not surprising, since there are additional contributions generated by interaction terms with several resonances, which have not been included in the minimal R$\chi$T lagrangian. These new chiral structures should be taken into account to achieve a physical description of the VFF above the two-resonance thresholds. The short-distance QCD constraints can be used to determine their corresponding couplings.

In the next chapter we will check with several form factors the requirement of these new terms in order to fulfill a good behaviour at large energies.
\end{enumerate}

\chapter{Two-body Hadronic Form Factors From QCD}

\section{Introduction}

Once it is accepted the importance of matching the effective results evaluated within the Resonance Chiral Theory with the ones obtained with QCD, one has to study how to carry out this procedure. There are two ways of getting short-distance constraints: either to consider the Green Functions of QCD currents calculated in the leading OPE expansion or to demand that two-body form factors of hadronic currents vanish at high energies~\cite{brodsky-lepage}. Although in the first case there is no doubt about the necessity of fulfilling the asymptotic constraints in the considered amplitude, the second one is more controversial. Actually, this behaviour has only been experimentally observed for pseudo-Goldstone bosons and photons. The question appears when one is studying form factors that involve resonances as ``asymptotic states''. In this chapter we present an analysis of all two-body form factors that can be found in the even-intrinsic-parity sector of R$\chi$T in the Single Resonance Approximation~\cite{nosaltres2,nosaltres2bis}. In the spirit of correlators at next-to-leading order in the $1/N_C$ expansion, the requirement of considering the short-distance behaviour of these form factors is justified.

As a continuation of the ideas proposed in the last chapter, once these new constraints are incorporated, we expect to avoid the non-vanishing behaviour at large momentum transfer for those contributions in the vector form factor at one-loop level coming from diagrams with resonances as intermediate states. In Section~3.6 we showed the need of new operators, that is, operators with more than one resonance field, in order to generate this suppression. Notice that we propose a relation between well-behaved form factors with resonances in the final state and observables at NLO.

As soon as one is dealing with well-behaved amplitudes at large energies, a one-loop calculation provides a clear NLO prediction of the related $\chi$PT LEC's, where the scale dependence is under control. Following this path, we present a subleading prediction of $L_8$~\cite{nosaltres2}. A first step in this direction was the study of resonance loop contributions to the chiral coupling $L_{10}$~\cite{cata}. In Ref.~\cite{cata} it was suggested the importance of considering well-behaved amplitudes before studying these contributions. Sections~3.6 and 3.7~\cite{nosaltres} are a good example of these ideas in the case of $L_9$ (or $\ell_6$ in the two flavour case).

In Section~4.2, the lagrangian needed to describe all the possible two-body form factors within the Single Resonance Approximation is reviewed. Section~4.3 is devoted to clarify how to get the short-distance constraints for the form factors by relating them to  one-loop  correlators through the optical theorem; the relation between quantum loops in R$\chi$T and form factors with resonances in the final state is explained. A phenomenological example of these results is developed in Section~4.4, where a prediction of $L_8^r(\mu)$ is given, making use of dispersive relations. The study of possible inconsistencies between constraints due to the truncation of the large-$N_C$ spectrum, already suggested in former works~\cite{juanjo,ximo}, is relegated to Section~4.5. The main conclusions are summarised in Section~4.6. Some technical details and the full list of results for the form factors are collected in the Appendices~D and E.

\section{The Effective Lagrangian}

As pointed out in the introduction, the study is taken under the Single Resonance Approximation, where just the lightest resonances with non-exotic quantum numbers are considered. On account of large-$N_C$, the mesons are put together into  $U(3)$ multiplets. Since we will be interested  on the structure of the interaction at short distances, we will work under the chiral limit.

As the Resonance Chiral Theory should get the high-energy behaviour of QCD, only operators constructed with chiral tensors of $\cO(p^2)$ will be allowed; interactions with higher order chiral tensors tend to violate the asymptotic behaviour ruled by QCD.

In the large-$N_C$ approach, there is no limit to the number of resonances that one may include in the effective operators. However, as we are interested just in the two-body form factors at tree level, only operators up to three resonance fields are considered. Moreover, in the case of three resonance operators, only terms consisting of resonance fields and the covariant derivative $\nabla_\mu$ will be required.

Following these remarks the terms in the lagrangian can be classified as:
\begin{eqnarray} \label{lagrangian_formfactors}
\mathcal{L}_{R\chi T}&=&\mathcal{L}_{pGB}^{(2)} \,+\,\sum_{R_1}\mathcal{L}_{R_1}\,+\,\sum_{R_1,R_2} \mathcal{L}_{R_1R_2}\, +\,\sum_{R_1,R_2,R_3}\Delta \mathcal{L}_{R_1R_2R_3}\,   ,
\end{eqnarray}
where the indexs $R_i$ run over all the different resonance fields, $V$, $A$, $S$ and $P$. We use $\Delta$ in the last term to stress that only some terms with three resonances are added to the lagrangian. The different pieces are shown and explained in Section~2.3.2, Eqs.~(\ref{phi})~-~(\ref{endlagrangian}).

\section{Form Factors and Short-distance Constraints}

Let us consider the two-point correlation function of two QCD currents in the chiral limit:
\begin{eqnarray}
\Pi_{_{XX}}^{\mu\nu}(q)&\equiv& i\int \mathrm{d}^4x \, \mathrm{e}^{iqx}\;\langle 0|T\left(X^\mu(x)X^\nu(0)^\dagger\right)|0\rangle \, =\, \left( -g^{\mu\nu} q^2 + q^\mu q^\nu \right)\,\Pi_{_{XX}}(q^2)\, , \nonumber \\
\Pi_{_{YY}}(q)&\equiv& i\int \mathrm{d}^4x \, \mathrm{e}^{iqx}\;\langle 0|T\left(Y(x)Y(0)^\dagger\right)|0\rangle \, , \label{def1}
\end{eqnarray}
where $X^\mu(x)$ can denote the vector or axial-vector current ($X=V,A$) and $Y(x)$ the scalar or pseudo-scalar density ($Y=S,P$),
\begin{align}
V^\mu_i &=\, \bar{\psi}\,\gamma^\mu \,\frac{\lambda_i}{2}\, \psi \, ,  \qquad \qquad & S_i &=\, \bar{\psi}\, \frac{\lambda_i}{2} \, \psi \, , \nonumber \\
A^\mu_i &=\, \bar{\psi}\,\gamma^\mu \gamma_5 \,\frac{\lambda_i}{2}\, \psi \, , \qquad \qquad & P_i &=\, i\, \bar{\psi}\, \gamma_5\, \frac{\lambda_i}{2} \, \psi \, .  \label{def2}
\end{align}

The associated spectral functions are a sum of positive contributions corresponding to the different intermediate states. At large $q^2$, $\mathrm{Im}\, \Pi_{_{XX}}$  tends to a constant  whereas $\mathrm{Im}\, \Pi_{_{YY}}$ grows as $q^2$~\cite{vv-aa,ss-pp}. Therefore, since there is an infinite number of possible states, we assume a similar supression for all the absorptive contributions in the spin-1 correlators coming from each intermediate state in the $q^2 \rightarrow \infty$ limit. The high energy behaviour in the spin-0 $\mathrm{Im}\, \Pi_{_{YY}}$ is not so clear as, {\it a priori}, one could think of a constant behaviour for each intermediate cut. However, the fact  that $\Pi_{_{SS}} - \Pi_{_{PP}}$ vanishes as $1/q^4$ in the chiral limit~\cite{ss-pp}, the Brodsky-Lepage rules for the form factors~\cite{brodsky-lepage}  and the $1/q^2$ behaviour of each one-particle intermediate cut (tree-level exchanges) seems to point out  that every absorptive contribution to $\mathrm{Im}\, \Pi_{_{YY}}$ must also vanish at large momentum transfer.

The spectral functions of the correlators at next-to-leading order can be easily obtained from form factors by making use of the optical theorem. Thence, all possible two-body form factors have been calculated in order to get the imaginary part of the two-point function. In the simplest cases with just one form-factor $\mF_{m_1,m_2}(q^2)$,  one finds the relation 
\begin{eqnarray}
\left. \mathrm{Im}\, \Pi(q^2)\right|_{m_1,m_2} &= & \xi(q^2)\,\, |\mF_{m_1,m_2}(q^2)|^2\, , 
\end{eqnarray} 
with $\xi(q^2)$ a kinematic factor that depends on the considered channel. Imposing that the spectral function must vanish as $1/q^2$ at $q^2 \rightarrow \infty$ yields a specific behaviour for $\mF_{m_1,m_2}(q^2)$, depending on $\xi(q^2)$. Thus, some constraints on the effective parameters will be needed. In Appendix~D, we give the whole list of form factors in the even-intrinsic-parity sector of R$\chi$T in the Single Resonance Approximation, the exact relations between them and the spectral functions, the constraints which  are derived from the high energy analysis and the structure of the form factors after imposing the proper short-distance behaviour. Some of them can be found in former literature~\cite{polychromatic,nosaltres}.
%%: vector and scalar form factor to two
%%pseudo-Goldstones~\cite{polychromatic}, axial form factor to a photon and a
%%pseudo-Goldstone~\cite{polychromatic}, axial form factor to a pseudo-Goldstone
%%and a scalar resonance~\cite{nosaltres} and vector form factor to two scalar or
%%pseudoscalar resonances~\cite{nosaltres}.

\begin{figure}
\begin{center}
\includegraphics[scale=0.75]{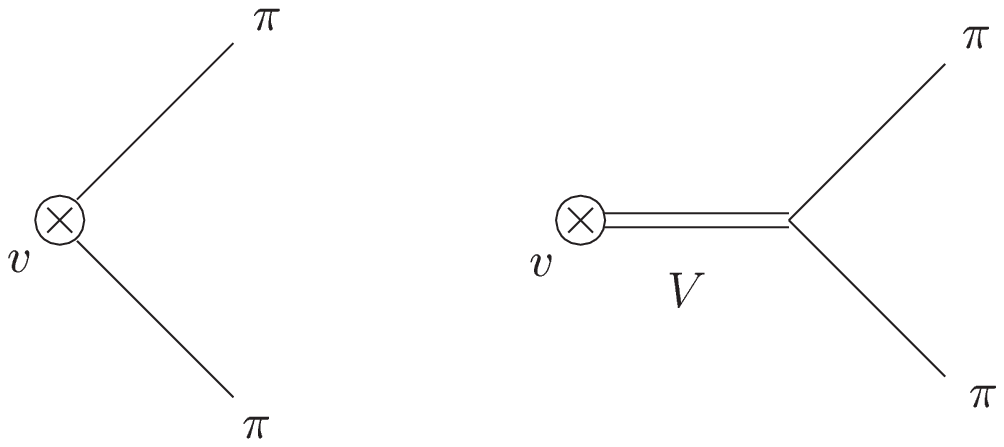}
\caption{\label{vgg}
Tree-level contributions to the vector form factor of the pion. A single line stands for a pseudo-Goldstone boson while
a double line indicates a vector resonance.}
\end{center}
\end{figure}

As an example,  we show here the case of the two pseudo-Goldstones matrix element of the vector current. The diagrams that contribute at leading order in $1/N_C$ are those depicted in Figure~\ref{vgg}. The form factor is defined through the corresponding matrix element,
\begin{eqnarray}
\bra \pi^0 (p_1) \pi^- (p_2) | \bar{d}\gamma^\mu u | 0\ket &=& \sqrt{2}\, \mathcal{F}_{\pi\pi}^{\,v} (q^2)\, (p_2-p_1)^\mu \, ,
\end{eqnarray}
where $\mathcal{F}_{\pi\pi}^{\,v}$ reads
\begin{eqnarray}
\mathcal{F}_{\pi\pi}^{\,v} (q^2)&=&1\,+\,\frac{F_VG_V}{F^2} \frac{q^2}{M_V^2-q^2} \, , \label{vffpio}
\end{eqnarray}
and it is the same form factor than the one of Eq.~(\ref{defchapt3}) in Chapter~3. Using the optical theorem, the imaginary part of the correlator is found to be
\begin{eqnarray}
\mathrm{Im}  \Pi_{_{VV}} (q^2) |_{\pi\pi}  &=& \frac{\theta(q^2)}{24\pi} |\mathcal{F}_{\pi\pi}^{\,v} (q^2)|^2 \, .
\end{eqnarray}
Imposing that $\mathrm{Im}  \Pi_{_{VV}} (q^2) |_{\pi\pi}$  vanishes in the $q^2 \rightarrow \infty$ limit leads to demanding that the form factor also does, so we find the constraint 
\begin{eqnarray}
F_V\, G_V&=&F^2 \,.
\end{eqnarray}
Taking into account this constraint, the form factor follows now the right asymptotic behaviour and reads as  
%%$\mathcal{F}_{\pi\pi}^{\,v}$ 
\begin{eqnarray}
\widetilde{\mathcal{F}}_{\pi\pi}^{\,v} (q^2)&=&\frac{M_V^2}{M_V^2-q^2} \, ,
\end{eqnarray}
as we would have obtained imposing the Brodsky-Lepage behaviour in Eq.~(\ref{vffpio}).  In this work, the tilde over a form factors denotes that the QCD short-distance constraints have already been imposed. 

\section{A next-to-leading order prediction of $L_8^r(\mu)$}

As an application of our results and because of its phenomenological importance, the observable $\Pi_{_{S-P}}(q^2)\equiv \Pi_{_{SS}}(q^2)-\Pi_{_{PP}}(q^2)$ is studied in this section in order to predict $L_8^r(\mu)$ at next-to-leading order. 

The one-loop $\chi$PT result, in the chiral limit, is
\begin{equation}
\Pi_{_{S-P}}(q^2)|_{\chi PT}=\frac{2 F^2 B_0^2}{q^2} +   32B_0^2\,L_8^r(\mu) + \frac{n_f}{2} \frac{B_0^2}{8\pi^2}  \left( 1-\log \frac{-q^2}{\mu^2} \right)  + \cO\left( q^2\right) \,,
\end{equation}
for the $U(n_f)$ case. The running in $L_8^r(\mu)$,
\begin{eqnarray}
L_8^r(\mu_2)&=&L_8^r(\mu_1)+\frac{\Gamma_8}{16\pi^2} \log \frac{\mu_1}{\mu_2},  \label{runningbis}
\end{eqnarray}
with $\Gamma_8=3/16$ for the $U(3)$ case~\cite{Herrera-Siklody:1996pm}, makes $\Pi_{_{S-P}}(q^2)$ scale independent.

A leading order prediction of the $\chi$PT coupling can be obtained easily by considering the tree-level contributions in our hadronic effective approach,
\begin{eqnarray}
\Pi_{_{S-P}}(q^2)|^{N_C\to\infty}_{R\chi T}&=&B_0^2 \left( \frac{16\, c_m^2}{M_S^2-q^2} - \frac{16\, d_m^2}{M_P^2-q^2} + \frac{2\, F^2}{q^2} \right) \,.   \label{sspplo}
\end{eqnarray}
%Imposing a good short-distance behaviour on this result~\cite{sspp,sspp2}, one %gets:
%\begin{equation}
%8\left(c_m^2-d_m^2\right) \,=\, F^2 
%%%%%%\, , \qquad c_m^2 M_S^2 -d_m^2 M_P^2 \,\simeq\, 0 
%\, .
%\end{equation}
Demanding the right high-energy behaviour ($\sim 1/q^4$) in $\Pi_{_{S-P}}(q^2)|^{N_C\to\infty}_{R\chi T}$ constraints the resonance parameters to obey the relations:
\begin{equation}\label{constraintLO}
F^2\, - \, 8\, c_m^2\, + \, 8\, d_m^2 \,\,=\,\,0\, , \quad c_m^2 \,M_S^2\, -\,d_m^2\, M_P^2\, =\, \widetilde{\delta} \, , 
\end{equation}
where $\widetilde{\delta}\equiv 3\pi \alpha_s F^4/4 \approx 0.08 \alpha_s F^2 \times (1\rm{GeV})^2$ is negligible.

In Section~2.4 it is reviewed how to fix all the low-energy couplings of $\mathcal{L}_{R\chi T}$ of Eq.~(\ref{lagrangian_formfactors}) linear in the resonance fields, by using different short-distance constraints,
\begin{align}
F_V^2&=\,F^2 \frac{M_A^2}{M_A^2-M_V^2}\, , & F_A^2&=\,F^2 \frac{M_V^2}{M_A^2-M_V^2}\, , &
G_V^2&=\,F^2  \frac{M_A^2-M_V^2}{M_A^2} \, ,& M_A^2& >\,  M_V^2\, \, \nonumber\\
c_m^2&=  \, \frac{F^2}{8}   \frac{M_P^2}{M_P^2-M_S^2} \, , &  d_m^2&=\,\frac{F^2}{8} \frac{M_S^2}{M_P^2-M_S^2}\, ,&
  c_d^2&=\,\frac{F^2}{2}\frac{M_P^2-M_S^2}{M_P^2}\, , & M_P^2& > \, M_S^2\, .
\label{opcio2bis}
\end{align}
where, at LO in $1/N_C$,  the couplings are fixed in terms of the decay constant $F$ and the resonance masses in the chiral and large-$N_C$ limit,  $M_V, \, M_A, \, M_S, \, M_P$.

The low-energy expansion of Eq.~(\ref{sspplo}) fixes the leading-order prediction of $L_8^r(\mu)$~\cite{RChTa},
\begin{eqnarray}
L_8&=&\frac{c_m^2}{2\, M_S^2} - \frac{d_m^2}{2\, M_P^2} \quad =\quad \frac{F^2}{16\, M_S^2}\, + \, \frac{F^2}{16\, M_P^2} \,,  \label{LO_pred}
\end{eqnarray}
where the constraints in Eq.~(\ref{opcio2bis}) have been considered to produce the final result. It is expected that Eq.~(\ref{LO_pred}) provides the coupling at scales of the order of the momenta involved in the processes ($\mu_0 \sim M_R$), though until now there was no information about the scale of saturation. Therefore, at LO in $1/N_C$, the uncertainty on $\mu_0$ induces an error, which for the coupling $L_8^r(\mu)$ is sizable and competes with the leading contributions.

In the large-$N_C$ limit a correlator that accepts an unsubtracted dispersive relation is determined by the position of the poles and the value of their residues. Hence, within the Single Resonance Approximation,  Eq.~(\ref{sspplo}) shows the general structure for $\Pi_{_{S-P}}$. This corresponds to the leading order saturation of the $\chi$PT $\cO(p^4)$ lagrangian by the resonance exchange.
%\footnote{$\tilde{L}_i$ are the couplings in the Resonance Chiral Theory, while $L_i$ are used for the $\chi$PT case, where the resonances have been integrated out.}. 

\subsection{Dispersive Calculation of $\Pi_{_{S-P}}$}

In this section, $\Pi_{_{S-P}}(q^2)\equiv \Pi_{_{SS}}(q^2)-\Pi_{_{PP}}(q^2)$ is computed at next-to-leading order within the Resonance Chiral Theory in the Single Resonance Approximation. By using the dispersive relations (Appendix~E), it is possible to prove that the amplitude at NLO in $1/N_C$ shows the structure
\begin{equation}\label{dispa}
\Pi_{_{S-P}}(q^2) = \frac{2 F^2B_0^2}{q^2}  +  \frac{16 c_{m}^{r\,\, 2}B_0^2 }{M_{S}^{r\,\,2}  - q^2}  -  \frac{16 d_{m}^{r\,\,2}B_0^2 }{M_{P}^{r\,\, 2}  - q^2}   +  \sum_{m_1,m_2} \Delta \Pi_{_{S-P}}(q^2)|_{m_1,m_2}\, ,
\end{equation}
where the contributions $\Delta \Pi_{_{S-P}}(q^2)|_{m_1,m_2}$ are given by the two meson absorptive cut $m_1,m_2$. Their imaginary part is related to the corresponding two-meson form factors through the optical theorem (the precise relations are given in Appendix~E), so the functions are given by the dispersive integral
\begin{align}
\Delta \Pi_{_{S-P}}(q^2)|_{m_1,m_2} &=\, \lim_{\epsilon\to 0} \left[ \int_0^{M_R^2-\epsilon} \!\!\! \mathrm{d}t\, \frac{1}{\pi}\frac{\mbox{Im}\Pi_{_{S-P}}(t)|_{m_1,m_2}}{t\, -\, q^2} \, + \,  \int_{M_R^2+\epsilon}^\infty \!\!\! \mathrm{d}t\, \frac{1}{\pi}\frac{\mbox{Im}\Pi_{_{S-P}}(t)|_{m_1,m_2}}{t\, -\, q^2}\right. \nonumber \\
&\qquad \left. \,-\, \frac{2}{\pi\epsilon} \, \lim_{t\to M_R^2} \left\{(M_R^2-t)^2\, \frac{\mbox{Im}\Pi_{_{S-P}}(t)|_{m_1,m_2}}{t\,-\,q^2}\right\}\,\, \right]\, , \label{correNLO}
\end{align} 
where $M_R$ is the mass of the intermediate resonance produced in the $m_1,m_2$ form-factor. It obeys  the properties
\begin{equation} 
\lim_{t\to M_R^2} \mathrm{Re} \overline{D}(t)|_{m_1,m_2}=0\, , \qquad 
\lim_{t\to M_R^2} \frac{\mathrm{d}}{\mathrm{d}t} \mathrm{Re} \overline{D}(t)|_{m_1,m_2}=0\,,
\end{equation} 
with $\overline{D}(t)|_{m_1,m_2}\equiv (M_R^2-t)^2 \Delta\Pi_{_{S-P}}(t)|_{m_1,m_2}$. 

Notice that the dispersive integrals are convergent because the form-factors are well behaved at infinite momentum. This ensures the absence of non-vanishing contributions in the part of the amplitude that comes from unitarity. The remaining terms in the correlator do not contain cuts and are analytical. These polynomial terms must vanish, remaining only the pole+unitarity structure in Eq.~(\ref{dispa}). This fixes any possible $\widetilde{L}_8$ arising at NLO, since the full polynomial must be zero. Furthermore, we will impose the $1/q^4$ behaviour prescribed by the OPE for  $\Pi_{_{S-P}}(q^2)$ up to NLO in $1/N_C$. 

For the first absorptive cut one gets the contributions 
\begin{align}
\Pi_{_{S-P}} (q^2)|_{tree}& =\, B_0^2  \left\{ \frac{2 F^2}{q^2} +  \frac{16  c_m^{r\,\,2} }{M_S^{r\,\, 2} - q^2}   -  \frac{16 d_m^{r\,\, 2}}{M_P^{r\,\, 2} - q^2}  \right\} \, , \label{totc0} \\
\Delta \Pi_{_{S-P}} (q^2)|_{\eta\pi} & =\, \frac{n_f}{2} \frac{B_0^2}{8 \pi^2}  \left(\frac{M_S^2}{M_S^2 - q^2}\right)^2  \left[ -1 + \frac{q^2}{M_S^2} - \log{\left(\frac{-q^2}{M_S^2}\right)} \right] \,, \label{totc1} \\
\Delta \Pi_{_{S-P}} (q^2)|_{V\pi}& =\,  \frac{n_f}{2}  \frac{B_0^2}{8 \pi^2}\frac{2G_V^2}{F^2}  \left(\frac{M_P^2}{M_P^2 - q^2}\right)^2 \left[ \left(1-\frac{q^2}{M_P^2}\right) \left(-\frac{ M_V^4}{q^4} - \frac{  M_V^4}{q^2 M_P^2} \right.  \right. \nn \\
 & \left. + \frac{5 M_V^2}{2q^2} + 1 - \frac{ 9 M_V^2}{2M_P^2} + \frac{ 3 M_V^4}{M_P^4} \right)- \left( 1-\frac{4M_V^2}{M_P^2}+\frac{3M_V^2q^2}{M_P^4} \right) \times \nn \\
 & \times  \!\left(1-\frac{M_V^2}{M_P^2}\right)^2 \!  \! \log \frac{M_P^2-M_V^2}{M_V^2} \! +\! \! \left( 1 -\frac{M_V^2}{q^2}\right)^3\! %\times   \\ &\times
\!\left.\log \!{\left(1-\frac{q^2}{M_V^2}\right)\!} \right]   ,     \label{totc4}\\
\Delta \Pi_{_{S-P}} (q^2)|_{A\pi}& =\,0\,, \phantom{\frac{1}{2}}\label{totc2}\\
\Delta \Pi_{_{S-P}} (q^2)|_{S\pi}& =\,\frac{n_f}{2}\frac{B_0^2}{8 \pi^2}   \frac{4 c_d^2}{F^2} \left(\frac{M_P^2}{M_P^2 - q^2}\right)^2 \left\{ \left(\frac{F^2}{2 c_d^2}-1\right)^2 \left(1-\frac{M_S^2}{M_P^2}\right)^2 \times \right. \nn \\
 &   \times \left[ 1-\frac{q^2}{M_P^2}  + \left(1-\frac{2 M_S^2}{M_P^2} +\frac{M_S^2 q^2}{M_P^4}\right) \log\frac{M_S^2}{M_P^2-M_S^2}+ \left(1-\frac{M_S^2}{q^2}\right)\times \right. \nn\\
 & \times \left. \log{\!\left(1-\frac{q^2}{M_S^2}\right)\!} \right]\! + \!\left(\frac{F^2}{2 c_d^2}-1\right)\! \left(1-\frac{M_S^2}{M_P^2}\right) \!  \left[  \frac{4 M_S^2}{M_P^2} -\frac{2 M_S^2}{q^2} -\frac{ 2 M_S^2 q^2}{M_P^4} \right. \nn \\
 &  +\left( \frac{2 M_S^2}{M_P^2}  - \frac{ 2 M_S^4}{M_P^4} -\frac{ 2 M_S^2 q^2}{M_P^4} + \frac{ 2 M_S^4 q^2}{M_P^6}   \right)  \log \frac{M_S^2}{M_P^2-M_S^2}-  2   \left(\frac{M_S^2}{M_P^2} \right. \nn \\
 &\left.\left.- \frac{M_S^2}{q^2}\right)   \left(1-\frac{M_S^2}{q^2}\right)  \log{\left(1-\frac{q^2}{M_S^2}\right)}  \right]- \frac{M_S^2}{M_P^2} -\frac{M_S^4}{M_P^4} -\frac{M_S^4}{q^4} +\frac{M_S^2}{2 q^2}  \nn \\
 & + \!\frac{2 M_S^4}{M_P^2 q^2}\! + \!\frac{M_S^2 q^2}{2M_P^4} \! \left.+\! \left(\frac{M_S^2}{M_P^2}\!-\!\frac{M_S^2}{q^2}\right)^2\! \left(1-\!\frac{M_S^2}{q^2}\right) \! \log{\!\left(1-\!\frac{q^2}{M_S^2}\right)\!} \right\} \! ,\label{totc5}\\
\Delta \Pi_{_{S-P}} (q^2)|_{P\pi}& =\, \frac{n_f}{2} \frac{B_0^2}{8 \pi^2} \!\frac{16d_m^2}{F^2} \!\left(\frac{M_P^2-M_S^2}{M_S^2 - q^2}\right)^2\! \left[ -1 +\frac{q^2}{M_S^2} + \!\left( 1-\!\frac{2 M_P^2}{M_S^2} + \!\frac{M_P^2 q^2}{M_S^4} \!\right)\! \times \right. \nn \\ 
 & \left. \times  \log\frac{M_P^2-M_S^2}{M_P^2} -  \left( 1 - \frac{M_P^2}{q^2}\right)  \log{\left(1-\frac{q^2}{M_P^2}\right)} \right] , \label{totc3}
\end{align}

It is possible to show that states with higher thresholds turn out to be more an more suppressed (Appendix E.2). Only contributions from cuts that contain up to one resonance field are taking into account: the $\pi\eta$, the $A\pi$ and the $P\pi$ cut of the scalar correlator (Sections~D.3.1, D.3.2 and D.3.3 respectively) and the $V\pi$ and the $S\pi$ cut of the pseudoscalar correlator (Sections~D.4.1 and D.4.2). All the results from Appendix~D have been multiplied by a factor $n_f/2$ in order to go from $2$ to $n_f$ light flavours. The results in Eqs.~(\ref{totc4})-(\ref{totc3}) include also a factor 2 that accounts the two possible absorptive structure, e.g., in the case of Eq.~(\ref{totc4}) it is possible $\rho^0 \pi^-$ and $\rho^- \pi^0$. The pion scalar form factor constraint from Eq.~(\ref{SGG}) has been used in Eq.~(\ref{totc5}).

\subsection{Short-distance Constraints at One-loop}

At high $q^2$ the first absorptive contribution vanishes as 
\begin{align}
  \Pi_{_{S-P}} (q^2)|_{tree}& = \frac{B_0^2}{q^2} \! \Bigg\{ 2 F^2 \! -\!  16  c_m^{r\,\,2}\! + \! 16 d_m^{r\,\, 2}\!+\!\frac{16}{q^2}\! \bigg[ d_m^{r\,\,2}M_P^2\!-\!c_m^{r\,\,2}M_S^2 \bigg]\! \Bigg\} \!  +\! \cO\!\left(\!\frac{1}{q^6}\!\right)\! , \!\!\label{sdc0}\\
\Delta \Pi_{_{S-P}} (q^2)|_{\eta\pi} &= \frac{n_f}{2}  \frac{B_0^2}{8\pi^2 q^2}\,M_S^2\,\Bigg\{ 1 +\frac{M_S^2}{q^2}  \bigg[ 1 - \log \frac{-q^2}{M_S^2} \bigg] \Bigg\} 
 + \cO\!\left(\! \frac{1}{q^6}\!\right)  , \label{sdc1}\\
\Delta \Pi_{_{S-P}} (q^2)|_{V\pi}&=  \frac{n_f}{2}  \frac{B_0^2}{8\pi^2 q^2}\frac{2G_V^2}{F^2} \, M_P^2 \,
 \Bigg\{ -1 +\frac{9M_V^2}{2M_P^2} - \frac{3 M_V^4}{M_P^4}  -  \frac{3M_V^2}{M_P^2} \left(1-\frac{M_V^2}{M_P^2}\right)^2\times  \nn \\
&  \left.\times \log \frac{M_P^2-M_V^2}{M_V^2}+\frac{M_P^2}{q^2} \bigg[ -1 -\frac{2M_V^4}{M_P^4} +\frac{2M_V^2}{M_P^2}+ \log \frac{-q^2}{M_V^2} \right. \nn \\
& - \left(1+\frac{2M_V^2}{M_P^2} \right) \left(1-\frac{M_V^2}{M_P^2} \right)^2 \log \left( \frac{M_V^2}{M_V^2} -1 \right) \bigg] \Bigg\} + \cO\!\left( \!\frac{1}{q^6}\!\right)  , \label{sdc4}\\
\Delta \Pi_{_{S-P}} (q^2)|_{A\pi}&=0\,, \phantom{\frac{1}{2}}\label{sdc2}\\
\Delta \Pi_{_{S-P}} (q^2)|_{S\pi}&=\frac{n_f}{2} \frac{B_0^2}{8\pi^2 q^2}   \frac{4 c_d^2}{F^2} \, M_P^2 \,\Bigg\{ \left(\frac{F^2}{2 c_d^2}-1\right)^2 \left(1-\frac{M_S^2}{M_P^2}\right)^2 \left(-1+\frac{M_S^2}{M_P^2}\times \right.\nn \\
&\left.\times \log\frac{M_S^2}{M_P^2-M_S^2}\right)  + \frac{M_S^2}{2M_P^2} \, + \frac{2M_S^2}{M_P^2} \, \left(\frac{F^2}{2\, c_d^2}-1\right) \left(1-\frac{M_S^2}{M_P^2}\right) \bigg( -1  \nn \\
& \left. +\left(-1 +\frac{M_S^2}{M_P^2}\right)  \log \frac{M_S^2}{M_P^2-M_S^2}\right)+\frac{M_P^2}{q^2} \bigg[ \left( \frac{F^2}{2c_d^2}-1\right)^2 \left(1-\frac{M_S^2}{M_P^2}\right)^2 \times \nn\\
&  \times \! \left(-1+\log \frac{-q^2}{M_P^2-M_S^2} \right)\! +\!\frac{M_S^4}{M_P^4}\! \left(-1 + \log \frac{-q^2}{M_S^2} \right)\!+\!\frac{2M_S^2}{M_P^2}\! \left( \frac{F^2}{2c_d^2}-1 \right)\!\times \nn \\
& \times \!\left(1-\frac{M_S^2}{M_P^2} \right) \!\left( \frac{M_S^2}{M_P^2} \log \frac{M_S^2}{M_P^2- M_S^2} - \log \frac{-q^2}{M_P^2-M_S^2 } \right)\! \bigg]  \Bigg\}+ \cO\!\left(\! \frac{1}{q^6}\!\right)  ,\label{sdc5}\\
\Delta \Pi_{_{S-P}} (q^2)|_{P\pi}&= \frac{n_f}{2}  \frac{B_0^2}{8\pi^2 q^2 } \frac{16d_m^2}{F^2} \left(1-\frac{M_S^2}{M_P^2}\right)^2 \,M_P^2 \, \Bigg\{   \frac{M_P^2}{M_S^2} + \frac{M_P^4}{M_S^4}\log\!\frac{M_P^2-M_S^2}{M_P^2} + \nn \\
& +\frac{M_P^2}{q^2} \bigg[1-\log \frac{-q^2}{M_P^2-M_S^2 } \bigg]\Bigg\} + \cO\!\left( \!\frac{1}{q^6}\!\right)  , \label{sdc3}
\end{align}

Once the leading-order relations in Eq.~(\ref{opcio2bis}) have been used, imposing the vanishing of the logarithm $\ln(-q^2)/q^4$ gives the constraint 
\begin{equation}\label{log}
\left(1-\frac{M_V^2}{M_A^2}\right) \, =\,
\frac{M_S^2}{M_P^2}\left(1\, - \, \frac{M_S^2}{2\, M_P^2}\right)\, ,
\end{equation}
which requires $M_A \leq \sqrt{2} M_V$. Imposing the right short-distance behaviour ($\sim 1/q^4$) in $\Pi(t)$, one gets
\begin{eqnarray}
F^2\, (1+\delta_{_{\rm NLO}}^{(2)})\, - \, 8 c_m^{r\, 2} \, +\, 8 \,
d_m^{r\, 2}\, = \, 0 \, , \label{eq:NLO_rela}
\\
F^2 \,M_S^2 \,\delta_{_{\rm NLO}}^{(4)}\, - \, 8 c_m^{r\, 2}\, M_S^{r\, 2} \,
+\, 8 \, d_m^{r\, 2}\, M_P^{r\, 2} \, = \, -8\,\widetilde{\delta} \, , \label{eq:NLO_relb}
\end{eqnarray}
where the corrections
\begin{equation} 
\delta_{_{\mathrm{NLO}}}^{(m)} = \frac{3 M_S^2}{32\pi^2F^2}
\!\left\{ 1 + \!\left(1-\frac{M_S^2}{M_P^2}\right)\!
\xi_{S\pi}^{(m)} \! + 2\!\left(\frac{M_P^2}{M_S^2}-1\right)\! \xi_{P\pi}^{(m)} 
%\right.\no\\&& \hskip -.3cm\left. 
- \frac{2
M_P^2}{M_S^2}\!\left(1-\frac{M_V^2}{M_A^2}\right)\! \xi_{V\pi}^{(m)}
\right\} \end{equation}
are known functions of the resonance masses:
\beqn 
\xi_{S\pi}^{(2)} &\!\!\! = &\!\!\! 1 - \frac{6 M_S^2}{M_P^2} +
\left(\frac{4 M_S^2}{M_P^2}-\frac{6 M_S^4}{M_P^4}\right)
\ln{\left(\frac{M_P^2}{M_S^2}-1\right)} ,
\no\\
\xi_{P\pi}^{(2)} &\!\!\! = &\!\!\!  1 +\frac{M_P^2}{M_S^2}
\ln{\left(1-\frac{M_S^2}{M_P^2}\right)} ,\no
\\
\xi_{V\pi}^{(2)} &\!\!\! = &\!\!\! 1 + \frac{3 M_V^2}{M_P^2}\!\left[
\frac{M_V^2}{M_P^2}-\frac{3}{2}+
\!\left(1-\frac{M_V^2}{M_P^2}\right)^2\!
\ln{\!\left(\frac{M_P^2}{M_V^2}-1\right)}\!\right]\!  ,
\no \\
\xi_{S\pi}^{(4)} &\!\!\! = &\!\!\! -4  + \left(2-\frac{4
M_S^2}{M_P^2}\right) \ln{\left(\frac{M_P^2}{M_S^2}-1\right)} ,
\\
\xi_{P\pi}^{(4)} &\!\!\! = &\!\!\!  1 +
\ln{\left(\frac{M_P^2}{M_S^2}-1\right)} ,
\no\\
\xi_{V\pi}^{(4)} &\!\!\! = &\frac{M_P^2}{M_S^2}\left(1- \ln \frac{M_S^2}{M_V^2}\right) - \frac{2M_V^2}{M_S^2} \left(1-\frac{M_V^2}{M_P^2} \right) 
\no \\&\!\!\!&\!\!\!\!\!\
   + \left(\frac{M_P^2}{M_S^2}+\frac{2M_V^2}{M_S^2} \right) \left(1-\frac{M_V^2}{M_P^2}\right)^2 \ln \left(\frac{M_P^2}{M_V^2}-1 \right)  .\no
\eeqn

Note that from Eqs.~(\ref{eq:NLO_rela}) and (\ref{eq:NLO_relb}) one determines the effective couplings $c_m^r$ and $d_m^r$:
\begin{eqnarray}
c_m^{r\,\,2}&=& \frac{F^2}{8} \frac{M_P^{r\,\,2}}{M_P^{r\,\,2}-M_S^{r\,\,2}} \left(1+\delta_{_{\rm NLO}}^{(2)}-\frac{M_S^2}{M_P^2}\delta_{_{\rm NLO}}^{(4)}-\frac{8}{M_P^2 F^2} \widetilde{\delta} \right) \, ,\label{cmr} \\
d_m^{r\,\,2}&=& \frac{F^2}{8} \frac{M_S^{r\,\,2}}{M_P^{r\,\,2}-M_S^{r\,\,2}} \left(1+\delta_{_{\rm NLO}}^{(2)}-\delta_{_{\rm NLO}}^{(4)} -\frac{8}{M_S^2 F^2} \widetilde{\delta}\right) \, .\label{dmr}
\end{eqnarray}

\subsection{Saturation of $L_8^r(\mu)$ at Next-to-leading Order in $1/N_C$}

Once we have extracted information from short distance QCD, we are ready to study the low energy limit of the theory. One finds the contributions:
\begin{align}
 \Pi_{_{S-P}} (q^2)|_{tree} \,=&\, B_0^2 \, \left( \frac{2\, F^2}{q^2} \, \,\, + \,\, \, \frac{16\, c_m^{r\,\,2}}{M_S^{r\,\,2}} \,\,\, - \,\,\, \frac{16\, d_m^{r\,\,2}}{M_P^{r\,\,2}}  \right)  \, + \cO\left( q^2\right)  \,, \label{c0}\\
\Delta \Pi_{_{S-P}} (q^2)|_{\eta\pi} \,=&\,\frac{n_f}{2}  \frac{B_0^2}{8\pi^2}  \left[ -1-\log \frac{-q^2}{M_S^2} \right] + \cO\left( q^2\right)  \,, \label{c1}\end{align}\begin{align}%\\
\Delta \Pi_{_{S-P}} (q^2)|_{V\pi}\,=&\, -  \frac{n_f}{2}\,  \frac{B_0^2}{8\pi^2}\frac{2G_V^2}{F^2} \bigg\{ -\frac{17}{6} +7\frac{M_V^2}{M_P^2} -4\frac{M_V^4}{M_P^4} + \left( 1-\frac{4M_V^2}{M_P^2} \right) \times \nonumber \\ & 
\times \left( 1-\frac{M_V^2}{M_P^2} \right)^2 \log \frac{M_P^2-M_V^2}{M_V^2} \bigg\} + \cO\left( q^2\right)  , \label{c4}\\
\Delta \Pi_{_{S-P}} (q^2)|_{A\pi}\,=&\,0\,, \phantom{\frac{1}{2}}\label{c2}\\
\Delta \Pi_{_{S-P}} (q^2)|_{S\pi}\,=&\,- \frac{n_f}{2}\frac{B_0^2}{8\pi^2}   \frac{4 c_d^2}{F^2} 
\left\{\left( \frac{F^2}{2 c_d^2}-1\right)^2  \left(1-\frac{M_S^2}{M_P^2}\right)^2  \left[ -2 + \left(1-\frac{2 M_S^2}{M_P^2}\right)\times   \right. \right. \nn \\
& \left.\times \log \frac{M_P^2-M_S^2}{M_S^2} \right]+ \, \frac{1}{6}\, + \left( \frac{F^2}{2\, c_d^2}-1\right) \left(1-\frac{M_S^2}{M_P^2}\right) \,  \times \, \nn \\
&  \left. \times \! \left[ 1-\frac{2 M_S^2}{M_P^2} + \!\left(\frac{2 M_S^2}{M_P^2}-\frac{2 M_S^4}{M_P^4}\right)\!   \log \frac{M_P^2-M_S^2}{M_S^2} \right]\!  \right\}\! + \!\cO\!\left( q^2\right)  ,\label{c5}\\
\Delta \Pi_{_{S-P}} (q^2)|_{P\pi}\,=&\, \frac{n_f}{2} \frac{B_0^2}{8\pi^2} \frac{16d_m^2}{F^2} \left( \frac{M_P^2-M_S^2}{M_S^2} \right)^2 \left[ -2 + \frac{2M_P^2-M_S^2}{M_S^2}\times \nn\right. \\
&\times \left. \log \frac{M_P^2}{M_P^2-M_S^2} \right] + \cO\left( q^2\right) \, . \label{c3}
\end{align}
It is interesting to remark that the non-analytic $\log{(-q^2)}$ structure that arises in $\chi$PT from the $\pi\eta$ loop is exactly reproduced at low energies by the $\pi\eta$ cut within the resonance theory; working within a chiral invariant framework ensures the proper low energy behaviour. The remaining cuts with resonances are absent in $\chi$PT and they only produce analytical contributions that go to the low-energy constants. 

This produces for $L_8^r(\mu)$ within $U(n_f)$ at any renormalization scale $\mu$, 
\begin{align}
L_8^r(\mu)& =\,  \frac{F^2}{16} \left(\frac{1}{M_S^{r\, 2}}+\frac{1}{M_P^{r\,2}}\right) \left\{ 1+  \delta_{_{\rm NLO}}^{(2)} - \frac{M_S^{r\, 2}\delta_{_{\rm NLO}}^{(4)} + 8\widetilde{\delta}/F^2 }{M_S^{r\, 2}+M_P^{r\, 2}}\right\}\nn \\
% \frac{F^2\,(1\, +\, \delta_{NLO})}{16\, M_S^{r\,\,2} }  \,\, + \,\,  \frac{d_m^{r\,\,2}}{2\,M_P^{r\,\,2}} \,\,\left(\frac{ \, M_P^{r\,\,2}-M_S^{r\,\,2}}{M_{S}^{r\,\, 2}}\right) \, +\phantom{aaaaaaaaaaaaaaa}\nn \\
&\, +\, \frac{n_f}{2}  \frac{1}{256\, \pi^2} \left[ -2-\log \frac{\mu^2}{M_S^2} \right]  \, +  \nn \\
&\, + \,  \frac{n_f}{2}\,  \frac{1}{128\, \pi^2} \left(\frac{M_A^2-M_V^2}{M_A^2}\right)  \bigg[ \frac{17}{6} -7\frac{M_V^2}{M_P^2} +4\frac{M_V^4}{M_P^4}\nonumber \\
 & \qquad \qquad \qquad \, -\,  \left( 1-\frac{4M_V^2}{M_P^2} \right) \left( 1-\frac{M_V^2}{M_P^2} \right)^2\,  \log \frac{M_P^2-M_V^2}{M_V^2} \bigg] \, +\nn \\
&\, + \, \frac{n_f}{2}\,  \frac{1}{128\, \pi^2} \left(\frac{M_P^2-M_S^2}{M_P^2}\right)  \bigg[ -\frac{1}{6} - \frac{M_S^2}{M_P^2} +4\frac{M_S^4}{M_P^4}\nonumber \\
 & \qquad \qquad \qquad \, +\, \frac{M_S^4}{M_P^4}\,  \left( -3+\frac{4M_S^2}{M_P^2} \right)   \log \frac{M_P^2-M_S^2}{M_S^2} \bigg]\nn \,+\, \\
&\, + \, \frac{n_f}{2} \, \frac{1}{128\, \pi^2} \left(\frac{M_P^2-M_S^2}{M_S^2}\right)  \left[ -2 + \frac{2M_P^2-M_S^2}{M_S^2} \log \frac{M_P^2}{M_P^2-M_S^2} \right]\, .\label{predictionL8}
\end{align}
In the first line we have the tree-level contribution, where the NLO relation from Eqs.~(\ref{cmr}) and (\ref{dmr}) have been used. The next lines contain the one-loop contributions, respectively from $\pi\eta$, $V\pi$, $S\pi$ and $P\pi$, and where the LO constraints from Eq.~(\ref{opcio2bis}) have been employed. 

A last remark is required: the calculation has been done within the $U(3)$ case, whereas the usual $\chi$PT results are obtained in the $SU(3)$ framework. Therefore,  we have to take into account the matching  between the $U(3)$ and $SU(3)$ Chiral Perturbation Theories~\cite{leutwyler}. The difference between the value of $L_8$ in the two versions of the effective theory is related to the difference between the corresponding coefficients $\Gamma_8$, that is, the different running. Accordingly, the leading order prediction of $L_8$ is the same in both cases~\cite{kaiser}, since the running is a next-to-leading order effect. One gets~\cite{leutwyler}
\begin{eqnarray}
L_8^{SU(3)} (\mu) &=& L_8^{U(3)} (\mu) + \frac{ \Gamma_8^{SU(3)}-\Gamma_8^{U(3)}}{16\pi^2} \log \frac{M_0}{\mu} \,,
\end{eqnarray}
where $\Gamma_8^{U(3)}=3/16$~\cite{Herrera-Siklody:1996pm}, $\Gamma_8^{SU(3)}=5/48$~\cite{ChPTb},  and $M_0= 850\pm 50$~MeV~\cite{kaiser2} is the mass of the $\eta'$ in the chiral limit. 

\subsection{Phenomenology}

At this point we have the chiral coupling $L_8^r(\mu)$ expressed in terms of the resonance masses $M_V$, $M_A$, $M_S\simeq M_S^r$, $M_P\simeq M_P^r$, the decay constant $F$ and the $U(3)-SU(3)$ matching contribution, given by $M_0$, the mass of the $\eta'$ in the chiral limit. 
%In addition, the low energy coupling depends on the renormalized resonance coupling $d_m^r$.  Although our knowledge of this parameter is rather poor, we expect deviation at the order of the 30\% level with respect to the leading order, so we take  
%\begin{eqnarray}
%d_m^{r\,\, 2}& = & d_m^2\, \,  \left\{1\, \pm \, \frac{1}{N_C}\right\}\, , 
%\end{eqnarray}
%with $d_m$ provided by Eq.~(\ref{opcio2bis}) and a relative error of $1/3$~\footnote{Since the $1/N_C$ expansion actually refers to the amplitudes (where one finds $d_m^{r\,\, 2}$) than directly to the couplings, we consider that this is the proper size for the NLO corrections rather than $d_m^{r}= d_m\,\, \left\{1\pm\frac{1}{N_C}\right\}$.}. Thus, the second term in the right-hand-side of Eq.~(\ref{predictionL8}) turns out to be
%\begin{eqnarray}
%\frac{d_m^{r\,\,2}}{2\, M_P^{r\,\,2}}\, \left(\frac{M_P^{r\,\,2}-M_S^{r\,\,2}}{M_S^{r\,\,2}}\right)&=& \frac{F^2}{16\, M_P^{r\,\,2}} \,\,\, \left\{1\, \pm \,\frac{1}{N_C}\right\}\, , 
%\end{eqnarray}
%generating for $L_8^r(\mu)$ a similar structure to that found in the leading order saturation in Eq.~(\ref{LO_pred}), together with new NLO corrections.

The different input parameters are defined in the chiral limit. We take the ranges \cite{ChPTb,polychromatic,kaiser2,PDG,MVsplit} $M_V=(770\pm
5)$~MeV, %$M_A=(1.23\pm 0.04)$ GeV,
$M_S^r=(1.14\pm 0.16)$~GeV,
%%% (varied in the range
%%% $980$~MeV$\leq M_S^r\leq M_P^r\simeq 1300$~MeV~\cite{,PDG}),
$M_P^r=(1.3\pm 0.1)$~GeV, $M_{0}=(0.85 \pm 0.05)$~GeV
and $F=(89 \pm 2)$~MeV, and use the relation of Eq.~(\ref{log}) to fix $M_A$, keeping the constraint $M_P\geq M_S$ from Eq.~(\ref{opcio2bis}) and imposing $M_A \geq 1$ GeV. The correction $\widetilde{\delta}$ turns out to be negligible. For the renormalization scale $\mu_0=770$~MeV, one obtains the following contributions
%For the masses in the chiral limit we consider  $M_V=776\pm 16$~MeV~\cite{PDG,MVsplit}, $M_A=1.23\pm 0.04$~GeV~\cite{PDG}, $M_S^r=1.14\pm 0.16$~GeV (varied in the range $980$~MeV$\leq M_S^r\leq M_P^r\simeq 1300$~MeV~\cite{polychromatic,PDG}), $M_P^r=1.3\pm 0.1$~GeV~\cite{PDG}. We estimate $M_0=900\pm 50$~MeV~\cite{kaiser} and the pion decay constant in the chiral limit  $F=89 \pm 2$~MeV is considered~\cite{ChPTb,PDG}. 
\begin{equation}
10^3\, \cdot \, L_8^r(\mu_0)\,=\, \underbrace{\,0.33\,}_{tree} - \underbrace{\,0.05\,}_{U(3)\to SU(3)}  - \underbrace{\,0.72\,}_{\pi\pi} + \underbrace{\,0.55\,}_{V\pi}+ \underbrace{\,0.38\,}_{S\pi}+ \underbrace{0.00}_{A\pi} + \underbrace{\,0.12\,}_{P\pi} \pm\, 0.4 \, , 
\end{equation}
where one finds the expected suppression of heavier thresholds.% This measurement carries the uncertainties 
%\begin{equation}
%10^3 \cdot  \Delta L_8^r(\mu_0)|_{\mathrm{R \chi T}}\,=\, \underbrace{\,^{+0.3}_{-0.5}\,}_{M_S^r}   \,\underbrace{\,^{+0.25}_{-0.40}\,}_{M_P^r} \, \pm  \underbrace{\,0.08\,}_{F} \pm  \underbrace{\,0.03\,}_{M_0}  \,\underbrace{\,^{+0.0010}_{-0.0040}\,}_{M_V}\, \pm   \underbrace{\,0.12\,}_{\mathrm{truncation}}\, .
%\end{equation}

The largest uncertainties originate in the badly known values of $M_S^r$ and $M_P^r$, which already appear in the leading order prediction. The keypoint is the fact that the rest are purely  NLO errors in $1/N_C$ and they remain small, validating the perturbative expansion in $1/N_C$.
%The uncertainty associated with the $d_m^{r\, 2}$ value is already a factor of three smaller.
To account for the higher-mass intermediate states which have been neglected, we have added an additional truncation error equal to $0.12\cdot 10^{-3}$, the size of the heaviest included channel ($P\pi$). Note that the smallness of the truncation error ensures that the Single Resonance Approximation is fair within this framework. All errors have been added in quadrature. Therefore we arribe to
\be L_8^r(\mu_0) = (0.6\pm 0.4)\cdot 10^{-3}\, , \ee
to be compared with the value $L_8^r(\mu_0)=(0.9\pm 0.3)\cdot10^{-3}$, usually adopted in phenomenological analyses.
%We set the truncation error as the contribution from the last channel, $P\pi$, and it accounts for the cuts with higher thresholds, neglected in our approach. The main uncertainties come from our lack of knowledge on $M_S^r$ and $M_P^r$, which already enter at leading order. The keypoint is the fact that the rest are purely  NLO errors in $1/N_C$ and they remain small, validating the perturbative expansion in $1/N_C$. To end with, the smallness of the truncation error ensures that the Single Resonance Approximation is fair within this framework. Summing up  the errors quadratically,  one  finally gets 
%\begin{eqnarray}
%L_8^r(\mu_0)|_{\mathrm{R \chi T}}&=& \left( 0.6\,\pm 0,4  \right)\cdot 10^{-3}  \,,
%\end{eqnarray}
%to be compared to the value coming from $\chi PT$ analyses $L_8^r(\mu_0)=\left(0.9\pm 0.3 \right) \cdot 10^{-3}$~\cite{RChTa}.

It is interesting to recall that for the considered scale $\mu_0=770$~MeV, our NLO prediction for $L_8^{r}(\mu_0)$ suffers small deviations with respect to its value at leading order, $L_8^{N_C\to\infty}= 0.8 \cdot 10^{-3}$, given by Eq.~(\ref{LO_pred}). Through a simple $\chi$PT analysis one finds that varying the renormalization scale between $\mu_1=0.5$~GeV and $\mu_2=1$~GeV  produces a variation on the renormalized coupling of the order of $|L_8^r(\mu_2)-L_8^r(\mu_1)|\sim 0.5\cdot 10^{-3}$. The outcome of our $1/N_C$ calculation  shows a perfect agreement with these considerations, being the possible deviations between LO and NLO in $1/N_C$ of the order of the expected renormalization scale uncertainties in $L_8$. 

\section{Conflict between High-energy Constraints}

The Resonance Chiral Theory is an effective approach of QCD that models large-$N_C$ by cutting the tower of resonances, that is, an infinite number of meson fields is not considered. However, it is known that really an infinite tower of resonances is needed to recover the large-$N_C$ behaviour within QCD. Therefore, it is not surprising to find some conflicts between the constraints of Appendix~D. In fact, it should have been expected, since our approach does not fully recover QCD and, eventually, it may lead to inconsistencies: not all behaviours of QCD can be satisfied at the same time within the MHA.

In Ref.~\cite{ximo} it was claimed that there exists in general a problem between QCD short-distance constraints for Green Functions and those coming from form factors and cross-sections following from the quark counting rule~\cite{brodsky-lepage}. However, from the general analysis of two-body form factors, developed in detail in Appendix~D, we find that the spin-0 sector does not lead to contradictions. On the other hand, the form factors related to spin-1 mesons drive us in some cases to constraints which do not agree with those coming from other form factors.   

This incompatibility can be solved by including a second multiplet, being this idea supported by large-$N_C$. Note that we follow the Minimal Hadronic Approximation~\cite{MHA}, therefore a second multiplet should be incorporated if there exists a conflict between the short-distance constraints of the problem at hand.
  
For instance,  in the left-right correlator, if the analysis is taken up to next-to-leading order in the $1/N_C$ expansion, all the constraints related to the vector and axial form factors (Appendix~D) should be considered. However, the restrictions for $\lambda_i^{VA}$ in Eq.~(\ref{VAG}) from the vector form factor to an axial resonance field and a pion, and those in Eq.~(\ref{AVG}) from the axial form factor to a vector resonance field and a pion are incompatible. The proposed solution is the inclusion of a second multiplet for the  $V(1^{--})$ and $A(1^{++})$ resonances, but only for internal lines. Then one should add new pieces to the lagrangian of Eq.~(\ref{lagrangian_formfactors}):
\begin{align}
\mathcal{L}_{V'}&=\,\frac{F_V'}{2\sqrt{2}}\bra V'_{\mu\nu} f^{\mu\nu}_+ \ket + \frac{iG_V'}{2\sqrt{2}} \bra V'_{\mu\nu}[ u^\mu, u^\nu] \ket   \, , \\ 
\mathcal{L}_{A'}&=\,\frac{F_A'}{2\sqrt{2}}\bra A'_{\mu\nu} f^{\mu\nu}_- \ket \, , \end{align}\begin{align}%\\ 
\mathcal{L}_{V'A}&=\,\lambda^{V'A}_1 \bra [ V'_{\mu\nu},A^{\mu\nu}] \chi_-\ket + i\,\lambda^{V'A}_2 \bra [V'^{\mu\nu},A_{\nu\alpha}]h^\alpha_\mu \ket  \phantom{\frac{1}{2}} \nonumber \\
& +i\,\lambda^{V'A}_3 \bra [\nabla^\mu V'_{\mu\nu}, A^{\nu\alpha}]u_\alpha \ket 
 +i\,\lambda^{V'A}_4 \bra [\nabla_\alpha V'_{\mu\nu},A^{\alpha\nu}]u^\mu \ket\phantom{\frac{1}{2}}\nonumber \\
&
+ i\,\lambda^{V'A}_5 \bra [\nabla_\alpha V'_{\mu\nu}, A^{\mu\nu}] u^\alpha \ket
+i\,\lambda^{V'A}_6 \bra [V'_{\mu\nu},A^{\mu}_{\,\alpha}]  f^{\alpha\nu}_- \ket 
 \,,\phantom{\frac{1}{2}} \\ 
\mathcal{L}_{VA'}&=\,\lambda^{VA'}_1 \bra [ V_{\mu\nu},A'^{\mu\nu}] \chi_-\ket + i\,\lambda^{VA'}_2 \bra [V^{\mu\nu},A'_{\nu\alpha}]h^\alpha_\mu \ket \phantom{\frac{1}{2}}\nonumber \\
&+  i\,\lambda^{VA'}_3 \bra [\nabla^\mu V_{\mu\nu}, A'^{\nu\alpha}]u_\alpha \ket 
 +i\,\lambda^{VA'}_4 \bra [\nabla_\alpha V_{\mu\nu},A'^{\alpha\nu}]u^\mu \ket\phantom{\frac{1}{2}}\nonumber  \\
&+ i\,\lambda^{V'A}_5 \bra [\nabla_\alpha V_{\mu\nu}, A'^{\mu\nu}] u^\alpha \ket
+i\,\lambda^{VA'}_6 \bra [V_{\mu\nu},A'^{\mu}_{\,\alpha}]  f^{\alpha\nu}_- \ket  \,.\phantom{\frac{1}{2}}
\end{align}

From the results from Eq.~(\ref{VAG}) and Eq.~(\ref{AVG}) is now obvious that the new constraints are, respectively:
\begin{align}
F_V(2\lambda^{VA}_2\!-2\lambda^{VA}_3\!+\lambda^{VA}_4\!+2\lambda^{VA}_5)+
F_V'(2\lambda^{V'A}_2\!-2\lambda^{V'A}_3\!+\lambda^{V'A}_4\!+2\lambda^{V'A}_5)&=F_A\,,\phantom{\frac{1}{2}} \nonumber \\ 
F_V(-2\lambda^{VA}_2\!+\lambda^{VA}_3)+
F_V'(-2\lambda^{V'A}_2\!+\lambda^{V'A}_3) &= 0 \, , \phantom{\frac{1}{2}}\nonumber \\
F_A(2\lambda^{VA}_2\!-\lambda^{VA}_4\!-2\lambda^{VA}_5)+
F_A'(2\lambda^{VA'}_2\!-\lambda^{VA'}_4\!-2\lambda^{VA'}_5)&=-F_V+2G_V   ,\phantom{\frac{1}{2}} \nonumber \\
F_A(-2\lambda^{VA}_2\!+\lambda^{VA}_3)+
F_A'(-2\lambda^{VA'}_2\!+\lambda^{VA'}_3)&=-G_V \, ,\phantom{\frac{1}{2}}
\end{align}
so the incompatibility is not present any longer. In this way, the incompatibilities in the lightest resonances couplings can be carried to the couplings of higher states that produce mild effects on the region of validity of our effective description. 

\section{Conclusions}

Resonance Chiral Theory is an effective framework to handle QCD at energies where one has hadronic resonances and pseudo-Goldstones from the chiral symmetry breaking. The expansion in powers of $1/N_C$ provides a key in order to construct the effective action. In addition to embed $\chi$PT at low energies, this theory must recover perturbative QCD and the OPE at short distances. 

Several constraints on the R$\chi$T couplings are derived from the study of  Green Functions of QCD currents at large-$N_C$. The other source of information is the consideration of the Brodsky-Lepage behaviour of the form factors, e.g. the pion vector form factor. This work  shows  the necessity of also taking into account the form factors with resonances in the final state.  They are related to two-point Green Functions at next-to-leading order in the $1/N_C$ expansion and rule their asymptotic behaviour at one-loop. All two-body form factors that can be found in the even-intrinsic-parity sector of Resonance Chiral Theory (Single Resonance Approximation) have been analysed, producing  the constraints and form factor structures shown in Appendix~D.

It is important to remark that there are no new constraints coming from the short-distance analysis of form factors with  one photon and one meson in the final state, that is, two-meson form factor analysis provides the most stringent set of constraints. This is not a surprise taking into account the relation between the vector resonances and the photon because of their quantum numbers.

The need of taming the resonance form factors at high energies was hinted in the last chapter, in order to improve the short distance behaviour of  the pion vector form factor at the  one-loop level~\cite{nosaltres}. This immediately leads to demand operators with more than one resonance field. Thus,  one must study amplitudes with resonances as external states at LO in $1/N_C$ whenever a calculation is carried at the loop level. In our case, the optical theorem tells us that the relevant amplitudes are just the two-body form factors. 

We have illustrated the analysis showing the case of the $\Pi_{_{S-P}}(q^2)$ correlator and the R$\chi$T prediction of the corresponding low energy coupling $L_8^r(\mu)$ at NLO in $1/N_C$.  From the Weinberg sum rules for $\Pi_{_{S-P}}(q^2)$ and  the pion scalar form factor one gets its expression at leading order.  Dispersive integrals show that the correlator up to NLO in $1/N_C$ is just given   by terms proportional to the squared modulus of form factors and renormalized resonance parameters.  Furthermore, the local $\chi$PT operators are shown to be absent within our present realization of the resonance lagrangian. The modified parameters can be partially fixed by taking the Weinberg sum rule analysis of  $\Pi_{_{S-P}}(q^2)$ up to NLO. This produces a slight modification to the leading relation, which can be read as:
\begin{eqnarray}
F^2\, (1+\delta_{_{\rm NLO}}^{(2)})\, - \, 8 c_m^{r\, 2} \, +\, 8 \,
d_m^{r\, 2}\, = \, 0 \, , \nn
\\
F^2 \,M_S^2 \,\delta_{_{\rm NLO}}^{(4)}\, - \, 8 c_m^{r\, 2}\, M_S^{r\, 2} \,
+\, 8 \, d_m^{r\, 2}\, M_P^{r\, 2} \, \simeq \, 0 \, . \nn
\end{eqnarray}

The chiral invariance in R$\chi$T leads to the recovering of the $\chi$PT structure at low energies. The $\pi\eta$ cut in R$\chi$T reproduces the long distance non-analytic term $\log(-q^2)$ from  one-loop $\chi$PT. This keeps the control on the renormalization scale $\mu$  appearing in $\chi$PT within $\log (-q^2/ \mu^2)$ and $L_8^r(\mu)$.   The remaining absorptive cuts generate analytic terms in $q^2$ and they only contribute to the 
LEC's. All this provides the determination for $\mu_0=770$~MeV,  
$$
L_8^r(\mu_0)\quad =\quad (\, 0.6\,\pm 0.4\, ) \, \cdot \, 10^{-3} \, .
$$
which can be compared to $\chi$PT value $L_8(\mu_0)|_{\mathrm{exp}} =\left(0.9 \pm 0.3\right) \cdot 10^{-3}$~\cite{RChTa}. Since the dependence on the scale is always exactly controlled, this problematic uncertainty disappears in our picture. On the other hand, the bulk of the error is due to the current ignorance on the values of the masses of the scalar and pseudoscalar multiplets in the chiral limit. The reduction of their relative uncertainties below the 5\% level would drastically improve the result. Until then, purely NLO errors in $1/N_C$ and the Single Resonance Approximation produce just a subdominant  contribution to the global error. This validates the perturbative expansion in $1/N_C$ and points out the way to proceed in order in increase the accuracy of the determination. 

To end with, we have commented some problems that appear in the spin-1 sector due to the truncation of the large--$N_C$ spectrum of infinite resonances. Because of this cut in the tower of resonances, it is clear that QCD cannot be exactly  recovered through our effective approach and that some conflicts between constraints may eventually arise. In our case, it is shown that this incompatibility can be solved by including a second multiplet.

\chapter{One-loop Renormalization: Scalar and Pseudoscalar Resonances}

\section{Introduction}

Since its inception Resonance Chiral Theory has been applied both to the study of resonance contributions in weak interaction processes (radiative and non--leptonic kaon decays) \cite{weak} and to the study of form factors of mesons \cite{guerrero-dumm}, where only the R$\chi$T lagrangian at tree level has been used and, accordingly, the leading contribution in the large-$N_C$ approach we are describing has been obtained. 

The next-to-leading order in the $1/N_C$ expansion arises from one loop calculations within the theory and its control starts to be necessary both on grounds of the convergence of the predictions and to straighten our knowledge of non-perturbative QCD. A Dyson-Schwinger resummation of subleading orders is required to describe the amplitudes near the resonance peak~\cite{preloop}, leading, eventually,  to systematic one-loop calculations~\cite{nosaltres,nosaltres2,nosaltres2bis,cata,integral}. Improving the phenomenological determinations  of non-perturbative QCD quantities is needed in order to distinguish new physics effects. As it has been pointed out in the previous chapter, it also allows getting the resonance contributions to the $\chi$PT LEC's at next-to-leading order, keeping the dependence on the renormalization scale under control. Furthermore, quantum loops are essential to find the quantum field theory description and to properly understand the hadronic interactions  beyond {\it ad hoc} modelings.

R$\chi$T is non-renormalizable. Moreover the lack of an expansion parameter in the lagrangian does not make feasible the application of a perturbative renormalization program based on a well defined power-counting scheme analogous to the one in $\chi$PT. Nevertheless from a practical point of view the situation is similar to the $\chi$PT case \cite{buchler}. As shown in Chapter~3~\cite{nosaltres}, where the vector form factor of the pion was calculated at one-loop level in R$\chi$T, it is possible to construct a finite number of operators, within the theory, whose couplings can absorb the divergences coming from one loop diagrams. The only requirement is, of course, that the regularization procedure of the loop divergences respects the symmetries of the lagrangian.

In the present chapter we have studied the full one-loop generating functional that arises from R$\chi$T when one multiplet of scalar and pseudoscalar resonances are considered and only up to bilinear couplings in the resonances are included. The divergent contributions have been evaluated and, consequently, the full set of operators needed to renormalize the theory properly has been obtained. The conceptual differences with the $\chi$PT renormalization program will also be stressed.

In Section~5.2 we describe shortly the content of R$\chi$T that is of interest in our case and its main features. Section~5.3 is devoted to explain the procedure and hints which are followed to perform the evaluation of the generating functional, whose results are given in Section~5.4 and commented in Section~5.5. In Section~5.6 we point out the conclusions and summarize. Some technical details are relegated to Appendix~F and most of the results to Appendix~G.

\section{R$\chi$T with Scalars and Pseudoscalars Resonances}

We consider the R$\chi$T lagrangian constituted by pseudo-Goldstone bosons and one multiplet of both scalar and pseudoscalar resonances. Motivated by the large-$N_C$ limit we include $U(3)$ multiplets for the spectrum though we limit ourselves to $SU(3)$ external currents as we are not interested in anomaly related issues. Our lagrangian reads:
\begin{equation}\label{eq:lagr1}
\mathcal{L}_{R\chi T}(\phi, \mathrm{S},\mathrm{P}) \,=\,\mathcal{L}_{pGB}^{(2)}\,+\, \mathcal{L}_{\mathrm{kin}\,S} \,+\,\mathcal{L}_{\mathrm{kin}\,P}  \,+\, \mathcal{L}_{S}\,+\,  \mathcal{L}_{P}\,+\, \mathcal{L}_{SS}\,+\,\mathcal{L}_{PP}\,+\,\mathcal{L}_{SP} \, ,
\end{equation}
where the notation of Section~2.3.2 is followed. The different pieces of Eq.~(\ref{eq:lagr1}) are given in Eqs.~(\ref{phi}), (\ref{kinSP}), (\ref{kinSP}), (\ref{S}), (\ref{P}), (\ref{SSPP}), (\ref{SSPP}) and (\ref{SP}) respectively. In other words, we have considered all terms observing chiral and QCD symmetries which are constructed with scalar and pseudoscalar resonances together with chiral tensors of $\cO(p^2)$, up to bilinear couplings in the resonance fields and under the Single Resonance Approximation.

Several comments on our lagrangian theory are suitable here:
\begin{enumerate}
\item[-] The R$\chi$T lagrangian satisfies, by construction, the structures of chiral dynamics at very low-energies ($E \ll M_R$). Notwithstanding, it is clear that there is no small coupling or kinematical parameter that could allow us to perform a perturbative expansion in order to solve the effective action of the theory, as it happens in $\chi$PT. We stress again that  the large-$N_C$ limit guides a loop perturbative expansion, not in the lagrangian, but in the observables evaluated with it. 

It has also been proposed \cite{Bruns:2004tj} that, due to the fact that the chiral counting is spoiled when resonances are included in loops, it could be possible to keep the chiral counting by disentangling the \lq \lq {\em hard}" modes that could be absorbed in the renormalization program. In this way one gets a chiral expansion even if resonance contributions in the loop are considered. This procedure can be useful but only if one is interested in the application at very low energies out of the resonance region. 

\item[-] Short-distance constraints on the asymptotic behaviour of form factors and Green Functions provide, in the $1/N_C$ expansion, different relations between the couplings. Assuming the usual constraints of Eq.~(\ref{matching1})~\cite{polychromatic}, one has for the $\mathcal{L}_S$, $\mathcal{L}_P$, $\mathcal{L}_{\mathrm{kin}\,S}$ and $\mathcal{L}_{\mathrm{kin}\,P}$ couplings:
\begin{equation} \label{eq:largecd}
c_m\,=\,c_d\,=\,\sqrt{2}d_m\,=\,\frac{F}{2} \, , \qquad M_P\, \simeq \, \sqrt{2} M_S\, ,
\end{equation}
as it has been explained in Section~2.4. High-energy constraints on the $\lambda_i^{RR}$ couplings in the $N_C \rightarrow \infty$ are shown in Appendix~D, see Chapter~4 for more information. Taking into account that no terms with three resonance fields are considered, the following relations are found~\cite{integral}:
\begin{eqnarray} \label{eq:largecd2}
\lambda_3^{SS} \, = \, \lambda_3^{PP} & = & 0 \, , \nonumber \\
\lambda_1^{SP} \, = \, 4 \, \lambda_2^{SP} \, = \, - \, \frac{d_m}{c_m} \,=\, \frac{-2c_m+c_d}{2d_m}
& = & - \, \frac{1}{\sqrt{2}} \; , 
\end{eqnarray}
where we have used Eq.~(\ref{eq:largecd}). From Appendix~D these results can be obtained easily, by neglecting the couplings with three resonances:
\begin{enumerate}
\item From the scalar form factor $\bra P^i | s^j |\pi^k \ket$, see Eq.~(\ref{SGP}), it is obtained that
\begin{eqnarray}
\lambda_1^{SP}&=&-\frac{d_m}{c_m}\,.
\end{eqnarray}
\item The asymptotic behaviour of the scalar form factor $\bra S^i | s^j |S^k \ket$ gives, see Eq.~(\ref{SSS}),
\begin{eqnarray}
\lambda_3^{SS}&=&0\,.
\end{eqnarray} 
\item Studying the high-energy behaviour of the scalar form factor $\bra P^i | s^j |P^k \ket$, see Eq.~(\ref{SPP}), one gets
\begin{eqnarray}
\lambda_3^{PP}&=&0\,.
\end{eqnarray}
\item From the ultraviolet limit of the pseudoscalar form factor $\bra S^i | p^j |\pi^k \ket$, see Eq.~(\ref{PGS}), it is found that
\begin{eqnarray}
\lambda_1^{SP}&=&\frac{-2c_m+c_d}{2d_m} \,.
\end{eqnarray}
\item The pseudoscalar form factor $\bra S^i | p^j |P^k \ket$, see Eq.~(\ref{PSP}), relates $\lambda_1^{SP}$ and $\lambda_2^{SP}$:
\begin{eqnarray}
\lambda_1^{SP} & = & 4 \, \lambda_2^{SP}\,.
\end{eqnarray}
\end{enumerate}
Though the relations shown in Eqs.~(\ref{eq:largecd}) and (\ref{eq:largecd2}) could be used to simplify the outcome of the calculations presented in this chapter, we will give the full results without short-distance constraints built-in so as not to lose generality.
\item[-] From the R$\chi$T lagrangian in Eq.~(\ref{eq:lagr1}), the equations of motion for the pseudo-Goldstone and resonance fields are obtained as the system of coupled equations:
\begin{align}
\nabla^\mu u_\mu \,=&\, \frac{i}{2}  \chi_- 
-\frac{2c_d}{F^2} \nabla^\mu \left\{u_\mu ,S \right\} 
+\frac{i\,c_m}{F^2}  \left\{\chi_- ,S\right\} 
-\frac{1}{2F^2} \big[u_\mu,[\nabla^\mu S,S] \big]\nonumber \\ &
-\frac{2 \ssa}{F^2} \nabla_\mu \left\{ u^\mu, SS\right\} 
 -\frac{4\ssb}{F^2} \nabla_\mu \left( S\, u^\mu S\right)\,+\,
\frac{i\,\ssc}{F^2} \left\{ \chi_- ,SS\right\} \nonumber \\ &
-\frac{d_m}{F^2}  \left\{\chi_+ , P\right\} \,-\,\frac{1}{2F^2} 
\big[u_\mu,[\nabla^\mu P,P] \big] 
-\,\frac{2 \ppa}{F^2} \nabla_\mu \left\{ u^\mu, PP\right\} \nonumber \\&
-\,\frac{4\ppb}{F^2} \nabla_\mu \left( P\, u^\mu P\right)\,
+\,\frac{i\,\ppc}{F^2} \left\{ \chi_- ,PP\right\}
-\frac{2\spa}{F^2} \nabla^\mu \{ \nabla_\mu S, P \} \nonumber \\
& +\frac{\spa}{2F^2} \Big[ u_\mu , \big[ S , \{ P, u^\mu \} \big] \Big]  
-\frac{\spb}{F^2} \big\{ \chi_+ , \{ S, P \} \big\} \, ,\label{eom1} \\ \nonumber \\
\nabla^\mu \nabla_\mu S \,=&\, -M_S^2 \,S \,+\, c_m \, \chi_+ 
 \,+\,c_d \, u_\mu u^\mu  \, 
+\, \ssa \left\{ S, u_\mu u^\mu \right\}\,+\, 2 \ssb u_\mu S u^\mu 
\phantom{\frac{1}{2}}\nonumber \\
&+\, \ssc \left\{ S, \chi_+ \right\}\,  -\, \lam_1^{\mathrm{SP}} 
\nabla_\mu \{ P , u^\mu \}
\,+\, i \lam_2^{\mathrm{SP}} \{ P , \chi_- \}  \, , \phantom{\frac{1}{2}}\label{eom2}\\
& \nonumber \\
\nabla^\mu \nabla_\mu P \,=&\, -M_P^2 \,P \,+\, i \,d_m \, \chi_- \,+\,
 \ppa \left\{ P, u_\mu u^\mu \right\}\,+\, 2 \ppb u_\mu P u^\mu 
 \phantom{\frac{1}{2}} \nonumber \\
& +\, \ppc \left\{ P, \chi_+ \right\} \, +\, \lam_1^{\mathrm{SP}}  
\{ \nabla_\mu S , u^\mu \}
\,+\, i \lam_2^{\mathrm{SP}} \{ S , \chi_- \}\, \,.\phantom{\frac{1}{2}}\label{eom3}
\end{align}
\end{enumerate}
Like it has been stressed previously, the lack of an expansion coupling or parameter in R$\chi$T hinders a perturbative renormalization like the one applied in $\chi$PT. By studying the vector form factor of the pion at next-to-leading order, in Chapter~3 it was shown that, using dimensional regularization, all the divergences could be absorbed by the introduction of local operators fulfilling the symmetry requirements. This is a particular case of the well known fact that all divergences are local in a quantum field theory \cite{Collins}, and are given by a polynomial in the external  momenta or masses. Hence it is reasonable to consider the construction of the full set of operators that renders our ${\cal L}_{R \chi T}(\phi,S,P)$ theory finite up to one-loop. Accordingly we perform the one-loop generating functional of our lagrangian theory to evaluate the full set of divergences that arise. This we pursue in the rest of the chapter.

\section{Generating Functional at One Loop}

The generating functional of the connected Green Functions, $W[J]$, is the logarithm of the vacuum-to-vacuum transition amplitude in the presence of external sources $J(x)$ coupled to bilinear quark currents:
\begin{eqnarray}
e^{\,i \, W[J]} & = & \frac{1}{{\cal N}} \, \int  \, [\, \mathrm{d} \psi \, ] \, \, \, e^{\, i \, S_0[\psi,J]} \; ,
\end{eqnarray}
where the normalization is such that $W[0]=0$ and the field $\psi$ is, in our case, short for the pseudo-Goldstone and resonance mesons. The evaluation of the generating functional of our lagrangian theory ${\cal L}_{R\chi T}(\phi,S,P)$, is readily done with the background field method \cite{bfm,bfm2}, where the action is expanded around the classical fields $\psi_{cl}$. By defining the quantum field as $\Delta \psi = \psi - \psi_{cl}$, the expansion up to one loop ($L=1$) is given by:
\begin{align}
W[J]_{L=1} \, = \, S_0[\psi_{cl},J] \, - \, i \, \log \left[ \,  \int \, [\, \mathrm{d} \Delta \psi \, ] \, \exp \left( \phantom{\frac{1}{2}} \right. \right. & 
\! \! \! \!   i
\int \, \mathrm{d}^4x_1 \, \frac{\delta \, S_0[\psi,J]}{\delta \psi_i(x_1)} \, 
\Big|_{\psi_{cl}}
\, \Delta \psi_i(x_1) \, \nonumber \\ 
& 
\! \! \! \!\! \! \! \! \! \! \!\! \! \! \!\! \! \! \!\! \! \! \!\! \! \! \! \! \! \!\! \! \! \! \! \! \!\! \! \! \!\! \! \! \!\! \! \! \!\! \! \!  \! \! \! \!\! \! \! \! \! \! \!\! \! \! \!\! \! \! \!\! \! \! \!\! \! \!  \!\! \! \! \!\! \! \! \!\! \! \! \!\! \! \!
\left. \left.+ \, \frac{i}{2!} \, \int \,
 \mathrm{d}^4x_1 \, \mathrm{d}^4x_2 \, \, \Delta \psi_i(x_1) \, 
\frac{\delta^2 \, S_0[\psi,J]}{\delta \psi_i(x_1) \, \delta \psi_j(x_2)} \,
 \Big|_{\psi_{cl}} \, \Delta \psi_j(x_2) \, \right) \, 
 \right]   ,
\end{align}
but for an irrelevant constant. The $i,j$ indices run over all the different fields and are summed over. The classical field $\psi_{cl}$ is, by definition, the solution of:
\begin{eqnarray} \label{eq:eom}
\frac{\delta S_0[\psi,J]}{\delta \psi_i(x)} \, \Bigg|_{\psi_{cl}} \, = \, 0 \, ,
\end{eqnarray}
that provides the implicit relation $\psi_{cl} = \psi_{cl}[J]$ and the equations of motion for the classical fields.  Solving the remaining gaussian integral in the Euclidean spacetime and coming back to Minkowsky we have finally:
\begin{eqnarray} \label{eq:sl1}
W[J]_{L=1}  &=& S_0[\psi_{cl},J] \, + \, S_1[\psi_{cl},J]\, ,  \\
S_1[\psi_{cl},J] &=& \frac{i}{2} \,\log  \,\mbox{det} \, {\cal D}(\psi_{cl},J) 
 \, \;, 
\end{eqnarray}
where $ {\cal D}(\psi_{cl},J)$ is the quadratic differential operator specified by:
\begin{equation}
\langle \, x \,  | \, {\cal D}(\psi_{cl},J) \, | \,  y \, \rangle_{ij} \, = \,
 \frac{\delta^2 \, S_0[\psi,J]}{\delta \psi_i(x) \, \delta \psi_j(y)}  \Bigg|_{\psi_{cl}} \; .
\end{equation}
The action at one loop needs regularization and, following the use within $\chi$PT, we will proceed by working in $D$ spacetime dimensions, a procedure that preserves the relevant symmetries of our theory. Divergences in the functional integration are local and, within dimensional regularization, can be absorbed through local operators that satisfy the same symmetries than the original theory \cite{Collins}. The one-loop renormalized lagrangian is thus defined by:
\begin{eqnarray} \label{eq:renoo}
 {\cal L}_{1}[\psi,J] \, = \,  
\mu^{D-4} \, \left( \, {\cal L}_{1}^{\mathrm{ren}}[\psi,J; \mu] \, + \, 
 \frac{1}{(4\pi)^2} \, \frac{1}{D-4} \, 
 {\cal L}_{1}^{\mathrm{div}}[\psi,J] \, \right) \, .
\end{eqnarray}
In Eq.~(\ref{eq:renoo}) we have split the one-loop bare lagrangian into a renormalized and  a divergent part, and the scale $\mu$ is introduced in order to restore the correct dimensions in the renormalized lagrangian for $D\ne 4$. The divergent part ${\cal L}_{1}^{\mathrm{div}}$ contains the counterterms which exactly cancel the divergences found in the result for the one-loop generating functional of Eq.~(\ref{eq:sl1}).

Up to one loop ${\cal L}_{1}[\psi,J]$ can be written in terms of a minimal basis of $N$ operators $ {\cal O}_i[\psi,J]$. For a non-renormalizable theory, such as R$\chi$T, $N$ grows with the number of loops. Accordingly we expect to find in our evaluation of $ S_1[\psi,J]$ many more operators that those in the original tree level theory $S_0[\psi,J]$. The structure of these obeys the same construction principles (symmetries) that gave ${\cal L}_{R\chi T}(\phi,S,P)$ in Eq.~(\ref{eq:lagr1}), though we foresee  that higher-order chiral tensors may be involved. A detailed study of the functional integration shows that the new terms have the structure $\chi^{(4)}$, $ R \, \chi^{(4)} $ or $R \, R \, \chi^{(4)}$ (with a single or multiple traces) and $\chi^{(2)}$, $R\, \chi^{(2)}$ and $ R \, R \, \chi^{(2)}$ (with multiple traces)\footnote{As it will be emphasized later, in the procedure and due to a necessary field redefinition, terms with more than two resonances will be generated. We attach to our initial scheme and only will keep terms with up to two resonances.}.

\subsection{Expansion Around the Classical Solutions}

Following the aforementioned procedure we expand the action associated to our lagrangian ${\cal L}_{R \chi T}(\phi,S,P)$ in Eq.~(\ref{eq:lagr1}) around the solutions of the classical equations of motion: $u_{cl}(\phi)$, $S_{cl}$ and $P_{cl}$. The fluctuations of the pseudoscalar Goldstone fields $\Delta_i$ ($i=0,...,8$), and of the scalar and pseudoscalar resonances $\varepsilon_{S_i}$ and $\varepsilon_{P_i}$, are parameterized as\footnote{This is a convenient choice for the pseudoscalar fluctuation variables in order to simplify several cumbersome expressions. Notice that, once the  ``gauge'' $u_R = u_L^{\dagger} \equiv u$ is enforced, it implies that the classical and the quantum pseudo-Goldstone fields commute: $u_{cl} \, \exp(i \Delta /2) = \exp(i \Delta / 2) \, u_{cl}$.}:
\begin{align} \label{eq:fluc1}
u_R&=\,u_{cl}\,e^{i \Delta / 2}\, ,  &u_L&=\,u_{cl}^\dagger  \, 
e^{-i\Delta / 2}\,, \nonumber \\
S&=\,S_{cl}\,+\,\frac{1}{\sqrt{2}}{\es} \,, 
&P&=\,P_{cl}\,+\,\frac{1}{\sqrt{2}}{\ep}\,,
\end{align}
with 
\begin{align} \label{eq:fluc2}
\Delta &=\,\Delta_i \lam_i / F \,, 
&\es&=\,{\es}_{i}\,\lam_i\,, 
& \ep&=\,{\ep}_{i}\,\lam_i\,.
\end{align}
In the following we will drop the subindex ``$cl$'' for simplicity.
\par
Expanding the lagrangian using Eqs.~(\ref{eq:fluc1}) and (\ref{eq:fluc2}) up to terms quadratic in the fields ($\Delta_i ,\,\varepsilon_{S_i} ,\,\varepsilon_{P_i}$) and using the EOM of Eqs.~(\ref{eom1}), (\ref{eom2}) and (\ref{eom3}), we obtain the second-order fluctuation lagrangian, that takes the form\footnote{The intricacies of this evaluation are explained in detail in Appendix~F}:
\begin{align} \label{eq:expansion}
\Delta \mathcal{L}_{\mathrm{R}\chi\mathrm{T}}&=\, -\frac{1}{2}\,\Delta_i 
\left( d'_\mu d'^\mu + \sigma \right)_{ij} \Delta_j 
- \frac{1}{2}\,{\es}_{i} \left( d^\mu d_\mu + \ks \right)_{ij} {\es}_{j} 
- \frac{1}{2}\,{\ep}_{i} \left( d^\mu d_\mu + \kp \right)_{ij} {\ep}_{j} 
\nonumber \\
&+ \,{\es}_{i}\, \as_{ij}\, \Delta_j 
+ \,{\ep}_{i}\, \ap_{ij}\, \Delta_j 
+\,{\ep}_i\, \asp_{ij}\, {\es}_j  \phantom{\frac{1}{2}} \nonumber \\ 
&
+ \,{\es}_{k}\, \bs_{\mu \, ki}\, d^\mu_{ij} \Delta_j 
+ \,{\ep}_{k}\, \bp_{\mu \, ki}\, d^\mu_{ij} \Delta_j  
+\, {\ep}_k \, \bsp_{\mu \, ki}\, d^\mu_{ij} {\es}_j   \,\, . 
\phantom{\frac{1}{2}} 
\end{align}
Derivatives and matrices are defined in Appendix~F where it is also shown that in order to write $\Delta \mathcal{L}_{R \chi T}$ in the form displayed above we need to perform two field redefinitions. This procedure generates operators with multiple resonance fields. However our theory, as specified in Section~5.2, does not include operators with more than two resonances and, for consistency, we shall keep this structure in the fluctuation lagrangian, thus disregarding operators with three or more resonance fields in the following. We will comment later on the consequences of this feature. It is customary to write the second-order fluctuation lagrangian as:
\begin{equation} \label{eq:gausso}
\Delta {\cal L}_{\mathrm{R} \chi \mathrm{T}} \, = \, 
- \, \frac{1}{2} \, \eta \, \left( \, \Sigma_{\mu} \, \Sigma^{\mu} \, + \, 
\Lambda \, \right) \, \eta^{\top} \; , 
\end{equation}
where $\eta$ collects the fluctuation fields, $\eta=\left(\Delta_i,{\es}_j,{\ep}_k\right)$, $i,j,k = 0,...,8$, $\eta^{\top}$ is its transposed and the rest of definitions are given in Appendix~F.

\subsection{Divergent Part of the Generating Functional at One Loop}

After we have performed the second-order fluctuation on our lagrangian theory we come back to our discussion at the beginning of this section in order to identify the one-loop generating functional, specified now by the action:
\begin{equation}
S_{1} \, = \, \frac{i}{2} \, \log \, \mbox{det} \, 
\left( \, \Sigma_{\mu} \, \Sigma^{\mu} \, + \, \Lambda \, \right) \; .
\end{equation}
We use dimensional regularization to extract the divergence of this expression. As emphasized in the literature \cite{Barvinsky:1985an} it is convenient to employ the Schwinger-DeWitt proper-time representation, embedded in the heat-kernel formalism, in order to extract the residue at the $D-4$ pole. Ref.~\cite{bfm} shows that, in fact, symmetry considerations can also provide this information (at least up to one loop).

Hence we get:
\begin{equation}\label{eq:oneloop}
S_{1}=-\frac{1}{(4\pi)^2} \, \frac{1}{D-4} 
\int \mathrm{d}^4x \,\, \mbox{Tr}\,  \left( \, \frac{1}{12} \, Y_{\mu\nu}\, 
Y^{\mu\nu} \, + \,  \frac{1}{2} \Lambda^2 \, \right) \, + \, 
S_1^{\mathrm{finite}}\,  , 
\end{equation}
where $\mbox{Tr}$ is short for the trace in the flavour space, $Y_{\mu \nu}$ denotes the field strength tensor of $Y_{\mu}$ in Eq.~(\ref{eq:y}):
\begin{eqnarray}
Y_{\mu \nu} = \partial_{ \mu} Y_{\nu} - \partial_{\nu} Y_{\mu} + [Y_{\mu}, Y_{\nu}]\,.
\end{eqnarray}
The finite remainder $S_1^{\mathrm{finite}}$ cannot be simply expressed as a local lagrangian, but can be worked out for a given transition~\cite{ChPTb,Unterdorfer:2002zg}.
\par
Finally we get the one-loop divergence as:
\begin{eqnarray} \label{eq:s1div}
{S}_{1}^{\mathrm{div}}
&= & - \, \frac{1}{(4 \, \pi)^2} \, \frac{1}{D-4} \, \int \mathrm{d}^4 x \, {\cal L}_{1}^{\mathrm{div}} \; , 
\end{eqnarray}
where
\begin{eqnarray}
{\cal L}_{1}^{\mathrm{div}} & = &   
\,\frac{1}{12} \langle \gamma_{\mu\nu}^{\prime} \gamma^{\prime\,\mu\nu} 
+ 2 \gamma_{\mu\nu}\gamma^{\mu\nu}\rangle \,
+\frac{1}{2} \langle \sigma^2 + \kp^2 + \ks^2 \rangle 
+\langle \as {\as}^\top + \ap {\ap}^\top + \asp {\asp}^\top \rangle
\nonumber \\ 
&&
- \frac{1}{12} \langle  \gamma^{\prime\,\mu\nu} \left( {\bs_{\mu}}^\top 
\bs_{\nu} +{\bp_{\mu}}^\top \bp_{\nu} \right) \rangle \,
 - \, \frac{1}{12} \langle \gamma^{\mu\nu}
\left( \bs_{\mu}{\bs_\nu}^{\top}+\bp_{\mu}{\bp_\nu}^\top+ 
\bsp_{\mu}{\bsp_\nu}^{\top}+{\bsp_{\mu}}^\top \bsp_{\nu} \right)
\rangle  \nonumber \\ 
&&
-\langle {\as}^\top \big( \bar{d}^{\mu}_+ \bs_\mu + \frac{1}{2}
 {\bsp_\mu}^\top \bpm \big) 
 +{\ap}^\top \big( \bar{d}^{\mu}_+ \bp_\mu - \frac{1}{2} {\bsp_\mu} 
 \bsm \big)
% \rangle - \langle 
+{\asp}^\top \big( \hat{d^\mu} \bsp_\mu + \frac{1}{2} {\bpm} 
{\bs_\mu}^\top \big) \rangle \nonumber \\
&&
+ \frac{1}{4} \langle \sigma \big( {\bs_\mu}^\top \bsm +{\bp_\mu}^\top
 \bpm \big)
+\ks \big(  \bsm {\bs_\mu}^\top+{\bsp_\mu}^\top \bspm \big) 
+ \kp \big(  \bpm {\bp_\mu}^\top+  \bspm{\bsp_\mu}^\top \big) 
%\rangle \, +\frac{1}{4} \langle 
\rangle 
\nonumber \\
&&
 +\frac{1}{4} \langle \tilde{d}^{\mu}_- {\bs_\mu}^\top \bar{d}^{\nu}_+ 
\bs_\nu + 
\tilde{d}^{\mu}_- {\bp_\mu}^\top \bar{d}^{\nu}_+ \bp_\nu +
\hat{d^\mu} {\bsp_\mu}^\top \hat{d^\nu} \bsp_\nu \rangle \nonumber \\
&&
 -\frac{1}{12} \langle \tilde{d}_{+\mu} {\bs_\nu}^\top \bar{d}^{[\mu}_-
 \bsn^{]} + 
\tilde{d}_{+\mu} {\bp_\nu}^\top \bar{d}^{[\mu}_- \bpn^{]} +
\hat{d_\mu} {\bsp_\nu}^\top \hat{d}^{[\mu} \bspn^{]} \rangle \nonumber \\
&&
+\frac{1}{4} \langle \tilde{d}^{\mu}_- {\bs_\mu}^\top {\bsp_\nu}^\top 
\bpn
-\tilde{d}^{\mu}_- {\bp_\mu}^\top \bsp_\nu \bsn
+\hat{d^\mu} {\bsp_\mu}^\top \bpn {\bs_\nu}^\top \rangle \nonumber \\
&&
-\frac{1}{12} \langle \tilde{d}^{\mu}_+ {\bsn}^\top {\bsp_{[\mu}}^\top 
\bp_{{\nu]}}
-\tilde{d}^{\mu}_+ {\bpn}^\top {\bsp_{[\mu}} \bs_{{\nu]}}
+\hat{d^\mu} {\bspn}^\top {\bp_{[\mu}} {\bs_{{\nu]}}}^\top
 \rangle \nonumber \\
&&
+\frac{1}{48} \langle \big( {\bs_\mu}^\top \bsm{\bs_\nu}^\top\bsn+ 
{\bs_\mu}^\top \bsn{\bs_\nu}^\top\bsm+{\bs_\mu}^\top \bs_\nu {\bsm}^\top 
\bsn\big)\nonumber \\
&& \qquad
 +\big( {\bp_\mu}^\top \bpm{\bp_\nu}^\top\bpn+ {\bp_\mu}^\top 
\bpn{\bp_\nu}^\top\bpm+{\bp_\mu}^\top \bp_\nu {\bpm}^\top \bpn\big)\nonumber \\
&& \qquad
+\big( {\bsp_\mu}^\top \bspm{\bsp_\nu}^\top\bspn+ {\bsp_\mu}^\top 
\bspn{\bsp_\nu}^\top\bspm 
\, \, +\,  {\bsp_\mu}^\top \bsp_\nu {\bspm}^\top \bspn\big)
\rangle\nonumber \\
&&
+\frac{1}{24} \langle \big( {\bs_\mu}^\top \bsm{\bp_\nu}^\top\bpn+ 
{\bs_\mu}^\top \bsn{\bp_\nu}^\top\bpm+{\bs_\mu}^\top \bs_\nu {\bpm}^\top 
\bpn\big)\nonumber \\
&& \qquad
+\big( {\bsp_\mu}^\top \bspm{\bsn}{\bs_\nu}^\top+ {\bsp_\mu}^\top
 \bspn{\bs_\nu}{\bsm}^\top+{\bsp_\mu}^\top \bspn {\bsm} {\bs_\nu}^\top 
 \big)\nonumber \\
&& \qquad
 +\big(  \bpm {\bp_\mu}^\top{\bspn}{\bsp_\nu}^\top+ \bpm 
{\bp_\nu}^\top {\bspn}{\bsp_\mu}^\top  +\bp_\mu{\bp_\nu}^\top 
 {\bspm} 
{\bspn}^\top \big)
\rangle \, ,
\end{eqnarray}
where derivatives and matrices are defined in Appendix~F and $\gamma_{\mu \nu} = \partial_{\mu} \gamma_{\nu} - 
\partial_{\nu} \gamma_{\mu} + [ \gamma_{\mu}, \gamma_{\nu} ]$ (correspondingly for $\gamma_{\mu \nu}'$). Moreover for two vectors $A_{\mu}$, $ B_{\mu}$ we write $A_{[\mu}B_{\nu]}= A_\mu B_\nu - A_\nu B_\mu$. This result is completely general for the second-order fluctuation lagrangian in Eq.~(\ref{eq:expansion}). However, and as explained in Appendix~F, the expressions given there are valid only for operators with up to two resonances as we limit ourselves in this article.

\subsection{Result}

When worked out, $S_1^{\mathrm{div}}$ in Eq.~(\ref{eq:s1div}) can be expressed in a basis of operators that satisfy the same symmetry requirements than our starting lagrangian ${\cal L}_{\mathrm{R} \chi \mathrm{T}}(\phi, S,P)$. A minimal basis of $R\chi T$ operators that, upon integration of the resonances, contributes to the ${\cal O}(p^6)$ $\chi PT$ lagrangian, in $SU(3)$, can be found in Ref.~\cite{RChTc}. However, up to now, a basis for the one-loop $R \chi T$ has still not been worked out. This is precisely our result generated by $S_1^{\mathrm{div}}$. Hence, at one loop, the $R\chi T$ lagrangian needed to renormalize our theory reads:
\begin{equation} \label{eq:main}
{\cal L}_{1} \, = \,   \sum_ {i=1}^{18} \, \alpha_i \, 
{\cal O}_i \, + \, \sum_{i=1}^{66} \, \beta_i^R \, {\cal O}_i^R \, + \, 
\sum_{i=1}^{379} \, \beta_i^{RR} \, {\cal O}_i^{RR} \; .
\end{equation}
The ${\cal O}_i$ operators correspond to those up to ${\cal O}(p^4)$ in $U(3)_L \otimes U(3)_R$ $\chi PT$ \cite{Herrera-Siklody:1996pm}. ${\cal O}_i^{R}$ and ${\cal O}_i^{RR}$ involve one and two resonance fields, respectively, together with $\chi^{(2)}$ and $\chi^{(4)}$ chiral tensors. The couplings in the bare lagrangian $ {\cal L}_{1}$ read, in accordance with Eq.~(\ref{eq:renoo}):
\begin{eqnarray} \label{eq:rge}
\alpha_i & = & \mu^{D-4} \, \left( \, \alpha_i^r(\mu) \, + \, 
\frac{1}{(4\pi)^2} \, \frac{1}{D-4} \, \gamma_i  \right) \; , \nonumber \\
\beta_i^R & = & \mu^{D-4} \, \left( \, \beta_i^{R,r}(\mu) \, + \, 
\frac{1}{(4\pi)^2} \, \frac{1}{D-4} \, \gamma_i^R  \right) \; , \nonumber \\
\beta_i^{RR} & = & \mu^{D-4} \, \left( \, \beta_i^{RR,r}(\mu) \, + \, 
\frac{1}{(4\pi)^2} \, \frac{1}{D-4} \, \gamma_i^{RR}  \right) \; ,
\end{eqnarray}
where  $\gamma_i$, $\gamma_i^R$ and $\gamma_i^{RR}$ are the divergent coefficients given by $S_1^{\mathrm{div}}$ that constitute the $\beta$-function of our lagrangian (we use the terminology of Ref.~\cite{buchler}). The determination of the latter though straightforward involves a long calculation. In order to diminish the possibility of errors we have performed two independent evaluations. One of them has been carried out with the help of the FORM~3 program \cite{Vermaseren:2000nd} and the other with Mathematica \cite{Wolfram}. In Table~\ref{table_result} we show the $\cO_i$ operators, together with their $\beta$-function. The operators in this table constitute a minimal basis. The rest of the result is rather lengthy and is relegated to Appendix~G.  

\begin{table} 
\begin{center}
\begin{tabular}{|>{$}r<{$} | >{$}c<{$} | >{$}p{12cm}<{$} |}
\hline & & \\
i & \cO_i &\qquad \qquad \qquad \qquad \qquad \qquad \quad \gamma_i  \\ & & \\ \hline \hline 

1&       \bra  u \cdot u \ket &   - 2    N        {\ppa}    {M_P^2} + 1/2    N        {M_P^2}    (\spa)^2 - 2    N        {\ssa}    {M_S^2}
          + 1/2    N        {M_S^2}    (\spa)^2 + N  F^{-2}  c_d^2    {M_S^2} \\ \hline

2&       \bra  {\chi_+} \ket &   - 2    N        {\ppc}    {M_P^2} - 2    N        {\ssc}    {M_S^2} \\ \hline

3&      - \bra  u_\mu \ket^2 &    2        {\ppb}    {M_P^2} - 1/2        {M_P^2}    (\spa)^2 + 2        {\ssb}    {M_S^2}
          - 1/2        {M_S^2}    (\spa)^2 - c_d^2  F^{-2}  {M_S^2} \\ \hline

4&       \bra  u_\mu   u_\nu u^\mu u^\nu   \ket &  1/6    N    F^{-4}    c_d^4 - 1/12    N        (\spa)^2 + 1/24    N
                 (\spa)^4 + 1/16    N     + 1/6    N  F^{-2}  c_d^2    (\spa)^2 - 1/6    N  F^{-2}  c_d^2 \\ \hline

5&       \bra  u \cdot u\ket^2 &  1/2    F^{-4}    c_d^4 - 1/2        {\ppa}    (\spa)^2 +     (\ppa)^2 - 1/2        {\ssa}
             (\spa)^2 +     (\ssa)^2 + 1/8        (\spa)^4 + 1/16     - F^{-2}c_d^2    {\ssa} + 1/2  F^{-2}  c_d^2
         (\spa)^2 - 1/4  F^{-2}  c_d^2 \\ \hline

6&       \bra  u_\mu   u_\nu \ket^2 &  F^{-4}    c_d^4 -     {\ppb}    (\spa)^2 + 2        (\ppb)^2 - 
             {\ssb}    (\spa)^2 + 2        (\ssb)^2 + 1/4        (\spa)^4 + 1/8     - 2  F^{-2}  c_d^2    {\ssb} +F^{-2} c_d^2
         (\spa)^2 - 1/2  F^{-2}  c_d^2 \\ \hline

7&       \bra  u \cdot u  u \cdot u \ket &  1/3    N    F^{-4}    c_d^4 - 1/2    N        {\ppa}    (\spa)^2 + N        (\ppa)^2 - 1/2
             N        {\ssa}    (\spa)^2 + N        (\ssa)^2 + 1/12    N        (\spa)^2 + 1/12    N        (\spa)^4 -
         N  F^{-2}  c_d^2    {\ssa} + 1/3    N  F^{-2}  c_d^2    (\spa)^2 - 1/12    N  F^{-2}  c_d^2 \\ \hline

8&     \bra  {\chi_+} \ket \bra u \cdot u \ket &  F^{-4}    c_d^3    c_m + 2        {\ppa}    {\ppc} - 1/2        {\ppc}    (\spa)^2 + 2
                 {\ssa}    {\ssc} - 1/2        {\ssc}    (\spa)^2 + 1/8     +F^{-2} d_m    c_d    {\spa} +F^{-2} c_d    c_m    (\spa)^2 - 1/
         2  F^{-2}  c_d    c_m -F^{-2} c_d^2    {\ssc} - 1/4  F^{-2}  c_d^2 \\ \hline

9&       \bra  {\chi_+}  u \cdot u \ket &  N    F^{-4}    c_d^3    c_m + 2    N        {\ppa}    {\ppc} - 1/2    N        {\ppc}    (\spa)^2
          + 2    N        {\ssa}    {\ssc} - 1/2    N        {\ssc}    (\spa)^2 + 1/8    N     + N F^{-2}   d_m    c_d    {\spa} + NF^{-2}
         c_d    c_m    (\spa)^2 - 1/2    N  F^{-2}  c_d    c_m - N  F^{-2}  c_d^2    {\ssc} - 1/4    N F^{-2}   c_d^2 \\ \hline

10&       \bra  {\chi_+}\ket^2 &  F^{-4}    c_d^2    c_m^2 +     (\ppc)^2 +     (\ssc)^2 + 1/16     + 2  F^{-2}  d_m
             c_m    {\spa} + F^{-2}d_m^2 - 1/2   F^{-2} c_d    c_m + F^{-2}c_m^2    (\spa)^2 \\ \hline

11&      \bra  {\chi_-}\ket^2 &      {\spa}    {\spb} - 1/8        (\spa)^2 - 2        (\spb)^2 + F^{-2}d_m    c_d    {\spa}
          - 2  F^{-2}  d_m    c_m    {\spa} - F^{-2}d_m^2    (\spa)^2 +F^{-2} c_d    c_m - 1/4  F^{-2}  c_d^2 - F^{-2}c_m^2 \\ \hline

12&    1/2  \bra  \chi_+^2 +\chi_-^2 \ket &N    F^{-4}    c_d^2    c_m^2 + N        (\ppc)^2 + N        (\ssc)^2 + 1/16  N               + N  F^{-2}  d_m^2 + 1/2    N   F^{-2} c_d    c_m + N  F^{-2}    c_m^2    (\spa)^2 
+N        {\spa}    {\spb} - 1/8    N        (\spa)^2 - 2    N        (\spb)^2 + NF^{-2}
             d_m    c_d    {\spa} - N  F^{-2}  d_m^2    (\spa)^2  - 1/4    N  F^{-2}  c_d^2 - N
         F^{-2} c_m^2 \\ \hline

13&     - i \,\bra  f_+^{\mu\nu}  u_\mu  u_\nu \ket & - 1/6    N          (\spa)^2 + 1/4    N
                - 1/3    N   F^{-2}   c_d^2 \\ \hline

14&    1/4    \bra  f^{+\,2}_{\mu\nu}  -f^{-\,2}_{\mu\nu}  \ket &   - 1/4    N   + 1/6    N        (\spa)^2 + 1/3    N   F^{-2} c_d^2   \\ \hline

15&     1/2    \bra  f^{+\,2}_{\mu\nu}  +f^{-\,2}_{\mu\nu}  \ket  &  - 1/8    N  - 1/12    N        (\spa)^2 - 1/6    N   F^{-2} c_d^2
          \\ \hline

16&  1/4  \bra  \chi_+^2   -\chi_-^2  \ket & 2 N    F^{-4}    c_d^2    c_m^2 + 2 N        (\ppc)^2 +2 N        (\ssc)^2 + 1/8    N
              + 8  F^{-2}  N    d_m    c_m    {\spa} +2 N  F^{-2}  d_m^2 - 3    N   F^{-2} c_d    c_m + 2N F^{-2}   c_m^2    (\spa)^2
-2N        {\spa}    {\spb} + 1/4    N        (\spa)^2 + 4    N        (\spb)^2 -2 NF^{-2}
             d_m    c_d    {\spa}  +2 N  F^{-2}  d_m^2    (\spa)^2  + 1/2    N  F^{-2}  c_d^2 + 2N     F^{-2} c_m^2 \\ \hline

17&  -    \bra  u_\mu  \ket \bra u^\mu   u \cdot u \ket & - 4        {\ppa}    {\ppb} +     {\ppa}    (\spa)^2 +     {\ppb}   
         (\spa)^2 - 4        {\ssa}    {\ssb} +     {\ssa}    (\spa)^2 +     {\ssb}    (\spa)^2 - 1/2        (\spa)^4 + 1/
         4     + 2 F^{-2}   c_d^2    {\ssa} + 2  F^{-2}  c_d^2    {\ssb} - F^{-2} c_d^2 \\ \hline

18&       \bra  u_\mu  \ket \bra u^\mu   {\chi_+} \ket &   - 2    F^{-4}    c_d^3    c_m + 4        {\ppb}    {\ppc} - 
             {\ppc}    (\spa)^2 + 4        {\ssb}    {\ssc} -     {\ssc}    (\spa)^2 - 1/4     - 2  F^{-2}  d_m    c_d    {\spa} - 2
          F^{-2}   c_d    c_m    (\spa)^2 + F^{-2}c_d    c_m - 2  F^{-2}  c_d^2    {\ssc} + 1/2  F^{-2}  c_d^2  \\  \hline
\end{tabular}
\caption{\small{Operators involving only pseudo-Goldstone bosons and external currents and their $\beta$-function coefficients at one loop, when both scalar and pseudoscalar resonances are included. }} \label{table_result}
\end{center}
\end{table}

\section{Features and Use of the Renormalized R$\chi$T Lagrangian}

In order to understand the aspects and use of the renormalized $R \chi T$ lagrangian that we have obtained above, we would like to emphasize here several of its features:
\begin{enumerate}
\item In Table~\ref{table_result} we have collected the full basis of ${\cal O}(p^2)$ and ${\cal O}(p^4)$ $U(3)_L \otimes U(3)_R$ $\chi$PT operators generated in the functional integration at one loop. We should recover the result first obtained in Ref.~\cite{Herrera-Siklody:1996pm}. After the comparison is made\footnote{Notice that the notation of Ref.~\cite{Herrera-Siklody:1996pm} is different to ours though, to ease the comparison, the order chosen is the same. We always quote our notation for the operators.} we agree indeed with their results. Notice though that in order to disentangle the resonances, it is not enough to withdraw all the resonance couplings. This is because the derivative terms in ${\cal L}_{\mathrm{kin}}(S,P)$, which do not carry any resonance coupling, also contribute through the functional integration to several of the operators, namely ${\cal O}_4$, ${\cal O}_7$, ${\cal O}_{13}$, ${\cal O}_{14}$ and ${\cal O}_{15}$ in Table~\ref{table_result}. We have confirmed that ${\cal L}_{\mathrm{kin}}(S,P)$ gives precisely the difference between our coefficients $\gamma_4$, $\gamma_7$, $\gamma_{13}$, $\gamma_{14}$, $\gamma_{15}$ and those of Ref.~\cite{Herrera-Siklody:1996pm} once the resonance couplings have been switched off.
\begin{figure}
\begin{center}
\includegraphics[scale=0.63]{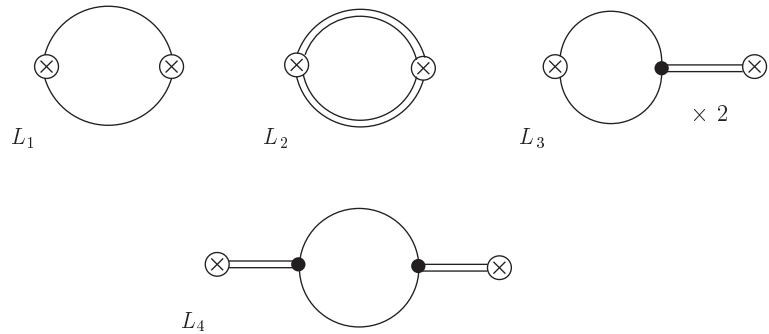}
\caption{\label{fig:loop}
One-loop contributions to the $\Pi_{_{SS}}^{ij}(q^2)$ correlator in the chiral limit when only scalar resonances are included. A single line stands for a pseudo-Goldstone boson while a double line indicates a scalar resonance. Their result is divergent.}
\end{center}
\end{figure}
\item In the procedure we have employed to evaluate the functional integration of ${\cal L}_{\mathrm{R} \chi \mathrm{T}}$ up to one loop we have withdrawn those operators with three or more resonance fields and kept up to two resonances. A cut in the number of resonances is necessary because to reach the Gaussian expression in Eq.~(\ref{eq:gausso}) we need to perform several field transformations (see Appendix~F) that generate operators with more resonance fields which in turn require additional field transformations and so on. One of the differences of R$\chi$T with respect to \chpt (in the strong \cite{ChPTa,ChPTb,ChPTc} or electroweak interaction \cite{Kambor:1989tz} form of the latter) is that we do not have an expansion parameter into the lagrangian that can provide a natural cut for higher order terms in these field transformations. Notice that the cut in the number of resonances seems to hinder our result, as it does not allow us to renormalize divergent one loop diagrams with three or more resonance fields as external legs. However we would not expect to treat these loops as we are not including, in our leading order lagrangian, interacting terms with three or more resonance fields.
\end{enumerate}
To end this section we would like to show a simple example of the application of our result. We consider the one-loop renormalization of the two-point function of scalar currents:
\begin{equation}
\Pi_{_{SS}}^{ij} (q^2) \, = \, i \, \int  \mathrm{d}^4 x \, e^{i q \cdot x} \, \langle 0 |T\{ S^i(x) S^j(0)\} | 0 \rangle \, , \qquad \, \;  \; 
S^i(x) = \overline{q}(x) \lambda^i q(x) \; , \label{SS}
\end{equation}
in the chiral limit and when only scalar resonances are considered. The divergent loop diagrams contributing are those depicted in Fig.~\ref{fig:loop}. In order to cancel the divergences one needs to add the counterterm contributions in Figure~\ref{fig:counter}, where diagram $C_1$ is given by ${\cal O}_{12} + 2 {\cal O}_{16} = \langle \chi_+^2 \rangle$ in Table~\ref{table_result}, $C_2$ by ${\cal O}_{4}^{R}=\langle S \chi_{+} \rangle $ in Table~G.1 of Appendix~G and $C_3$ by ${\cal O}_{1}^{RR}= \langle S S \rangle$ in Table~G.2 of Appendix~G, once the pseudoscalar resonance couplings are disconnected. The cancellation works as follows: one part of the contribution of $C_1$ cancels completely the divergence in the loops $L_1+L_2$. Another piece of $C_1$ together with $C_2$ eliminates the divergence coming from $L_3$ and, finally, all remaining contributions of $C_1$ and $C_2$ add to $C_3$ in order to render $L_4$ finite. Notice that, as there are no nonlocal divergences, the contributions of one-particle-reducible diagrams are brought finite once one-particle-irreducible diagrams have been properly renormalized.
\begin{figure}
\begin{center}
\vspace*{1.4cm}
\includegraphics[scale=0.68]{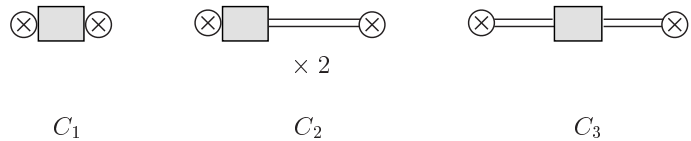}
\caption{\label{fig:counter}
Counterterm contributions that renormalize the one-loop result of 
~\ref{fig:loop}. A double line stands for a scalar resonance.}
\end{center}
\vspace*{-0.4cm}
\end{figure}
\section{Running of the couplings and short-distance behaviour}

\subsection{Running of the couplings}

Our result provides the running of the $\alpha_i$, $\beta_i^R$ and $\beta_i^{RR}$ couplings through the renormalization group equations. From Eq.~(\ref{eq:rge}) we get:
\begin{equation} \label{eq:rgeq}
\mu \, \frac{d}{d \mu} \, \alpha_i^r(\mu) \, = \, - \, \frac{\gamma_i}{16\,\pi^2}  \, , 
\end{equation}
and, analogously, for $\beta_i^R$ and $\beta_i^{RR}$. This result can be potentially useful if we are interested in the evaluation of the resonance couplings at this order. Though $\mu$ is known to be of the order of a typical scale of the physical system, let us say $\mu_0 = M_S$ or $\mu_0 = M_P$, there always remains some ambiguity on the precise value of $\mu_0$ at which the low-energy couplings are determined at leading-order in the $1/N_C$ expansion. The running provides an estimate of the reliance of such determinations. If the coupling under request varies drastically with the scale it is clear that the value obtained has a large uncertainty, while if it has a smooth dependence on the scale the determination is more reliable. Note that the running is a next-to-leading order effect.

In the case of R$\chi$T with only scalar and pseudoscalar resonance fields the Weinberg's dimensional analysis of Eq.~(\ref{contatge}) holds\footnote{This is not longer true in the case of vector and axial-vector resonances due to the propagator structure. Keep in mind the necessity of $\widetilde{\mathcal{L}}_{pGB}^{(6)}$ in order to renormalize the vector form factor at next-to-leading order in Chapter~3. In Eq.~(\ref{eq:f1_run}) one can see that only contributions from spin-1 resonances are responsible for these kind of divergences.}, once one assumes that the masses of the resonances are of $\cO(p)$ in the chiral counting and they appears explicitly in the lagrangian, unlike the pseudo-Goldstone bosons, whose masses are taking into account trough the chiral tensors $\chi_\pm$. Therefore, as it has been pointed out in Section~5.3, the terms needed to renormalize the theory at subleading order in the $1/N_C$ expansion are constructed with up to two resonances and chiral tensors up to $\cO(p^4)$. Notice that in the case of chiral tensors of $\cO(p^2)$ and $\cO(p^0)$ a $M_R^2$ and $M_{R_1}^2 M^2_{R_2}$ factor respectively are required in the divergent pieces in order to fulfill the ``generalized'' chiral counting. 

This has an {\it a priori} surprising consequence: there are counterterms associated with the operators of $\mathcal{L}_{pGB}^{(2)}$. In other words, the structure of $\mathcal{L}_{pGB}^{(2)}$ changes when one goes beyond the leading order:
\begin{eqnarray}\label{phibis}
\mathcal{L}_{pGB}^{(2)} &=&  \alpha_1 \bra u_\mu u^\mu \ket  + \alpha_2 \bra \chi_+ \ket \, , 
\end{eqnarray}
where we have followed the notation of Eq.~(\ref{eq:main}). Therefore, 
\begin{eqnarray} 
\alpha_1&=& \frac{F^2}{4} \sum_{n=0} \alpha_1^{(n)} \left(\frac{M_R}{F}\right)^{2n} \, , \nonumber\\
\alpha_2&=&\frac{F^2}{4} \sum_{n=0} \alpha_2^{(n)} \left(\frac{M_R}{F}\right)^{2n} \, , \label{notation}
\end{eqnarray}
where the coefficients have been defined in such a way that $\alpha_1^{(0)}=\alpha_2^{(0)}=1$. Notice that the suppression of higher terms in the $1/N_C$ expansion is explicitly shown, since $F\sim \cO(\sqrt{N_C})$ and $M_R \sim \cO(1)$. At next-to-leading order only the coefficients until $n=1$ must be considered.

\subsection{Vanishing $\beta$-functions and short-distance behaviour}

Within this issue it is interesting to take a closer look to the running of the couplings in $\mathcal{L}^{(2)}_{pGB}$ and $\mathcal{L}^{(4)}_{pGB}$, corresponding to the $\cO_i$ operators involving only pseudo-Goldstones. The corresponding $\beta$-function coefficients are $\gamma_i$, see Table 5.1.

%First of all note that the running, as expected, is a next-to-leading order effect in the $1/N_C$ expansion. This fact is guided by the $M_R^2/F^2$ factors taking into account that $M_R \sim {\cal O}(1)$ and $F \sim {\cal O}(\sqrt{N_C})$.
An interesting aspect is the interval over which $\mu$ runs. It is well known \cite{Collins} that the couplings are only relevant at the scale of the momenta involved in the processes (in order to diminish the role of the logarithms). In our case $\mu \sim M_S, \, M_P$. Thus we do not expect a large running for the scale, namely a few hundreds of MeV. This last conclusion brings us to the next point. At next to leading order in $1/N_C$ we can ignore the running on the couplings appearing in $\gamma_i$, since the running, as expected, is a next-to-leading order effect in the $1/N_C$ expansion\footnote{This fact is clear taking into account that $F$, $F_V$, $G_V$, $F_A$, $c_d$, $c_m$ and $d_m$ are of $\cO(\sqrt{N_C})$ and $\lambda_i^{R_1R_2}$ and $M_R$ of $\cO(1)$.}. Hence we can input the leading order values for the couplings, given by Eqs.~(\ref{eq:largecd}) and (\ref{eq:largecd2}), to obtain the leading logarithm in the evolution of these couplings. It is remarkable that, at this order, Eq.~(\ref{eq:rgeq}) predict a vanishing $\beta$-function for ${\cal O}_2$, ${\cal O}_8$, ${\cal O}_9$, ${\cal O}_{10}$, ${\cal O}_{11}$, ${\cal O}_{12}$, ${\cal O}_{16}$ and ${\cal O}_{18}$, i.e. all those operators involving $\chi_+$ and/or $\chi_{-}$. If the Large-$N_C$ estimates for the couplings are to be reliable we come to the conclusion that the predictions for these couplings are rather robust.

This feature must be explained. Notice that this behaviour was predicted for $\widetilde{L}_8$ in Chapter~4, relating it to short-distance constraints.

This result can be understood by following the optical theorem and taking into account the short-distance constraints that have been used. The imaginary part of any Feynman diagram can be obtained by cutting through the diagrams in all possible ways such that the cut propagators can simultaneously be put on shell, that is, by replacing $1/(p^2-m^2+i\varepsilon)\rightarrow -2\pi i \delta (p^2-m^2)$ in each cut propagator. Then one should perform the loop integrals and finally sum the contributions of all possible cuts.

To understand it we can start by considering again the one-loop renormalization of the two-point function of scalar currents, defined in Eq.~(\ref{SS}). We can use the cutting rules to calculate the spectral function of the correlator, as it was explained in Chapter~4. There are only four possible cuts: two pseudo-Goldstone, one pseudo-Goldstone and one pseudoscalar resonance, and two scalar or pseudoscalar resonance fields. The optical theorem allows to use the constrained form factors reviewed in Section~5.2 and analyzed in great detail in Section~D.3 of Appendix~D. Taking into account that for these contributions the highest behaviour at large energies could be $\cO(q^0)$, the suppression ruled by the constrained form factors leads to an $\cO(q^{-4})$ behaviour at large energies.

The following step consists of relating the spectral function to the divergent part of the contributions, which is responsible of the $\mu$ dependence on our couplings. We see that the relevant discontinuities can come only from two-point Feynman integrals. In Appendix~B it can be seen that the divergent piece and the imaginary part have always the same asymptotic behaviour. In other words, the same suppression must happen for the divergent piece. The one-point Feynman integral is not important for this purpuse, because although it has  not discontinuities, its behaviour at large energies is always lower than the two-point functions, as it does not depend on $q^2$. 

The needed counterterms contributions that renormalize the one-loop result are depicted in Figure~\ref{fig:counter}. As it was explained at the end of Section~5.4, diagram $C_1$ is given by ${\cal O}_{12} + 2 {\cal O}_{16} = \langle \chi_+^2 \rangle$, $C_2$ by ${\cal O}_{4}^{R}=\langle S \chi_{+} \rangle $ and $C_3$ by ${\cal O}_{1}^{RR}= \langle S S \rangle$. In Table~\ref{table_result} and in Appendix~G the $\beta$-function coefficients of the corresponding vertices are available. $C_1$, $C_2$ and $C_3$ give a behaviour at large energies of $\cO(q^0)$, $\cO(q^{-2})$ and $\cO(q^{-4})$ respectively. Considering the suppression explained before, $C_1$ is not needed to renormalize the process. So, as we have obtained and following our notation, $\gamma_{12}+2\gamma_{16}$ vanish once the short-distance contrainsts of Eqs.(\ref{eq:largecd}) and (\ref{eq:largecd2}) are implemented. 

Therefore, the process is very easy. The commented suppression will be observed in those processes where the scalar and/or pseudoscalar form factor play a role, understanding now why the operators involving $\chi_+$ and/or $\chi_-$ of Table~\ref{table_result} do not run at one loop. 
\begin{figure}
\begin{center}
%\vspace{+1.65cm}
\includegraphics[scale=0.6]{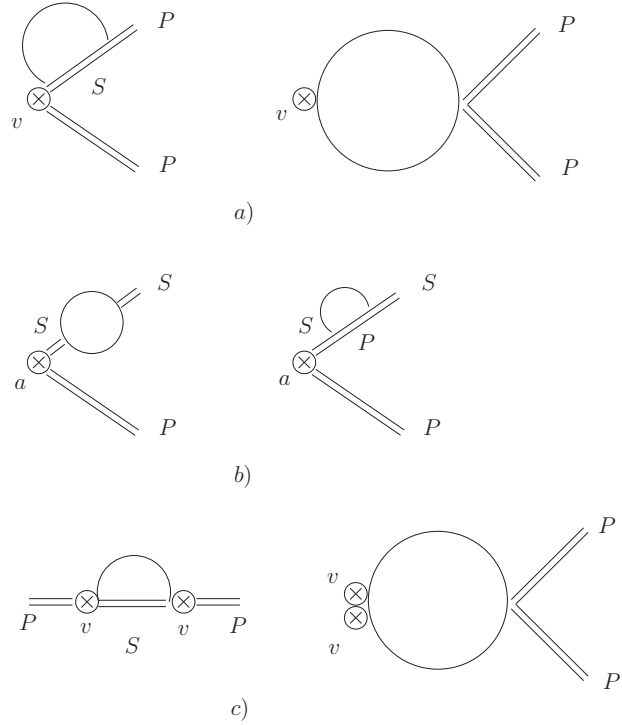}
%\vspace{+1cm}
\caption{\label{vanishing}
One-loop diagrams which are renormalized with $\cO^{RR}_{180}$ (first line), $\cO^{RR}_{316-318}$ (second line) and $\cO^{RR}_{190-191}$ (third line).}
\end{center}
\end{figure}

Following these ideas one can understand most of the found suppressions:
\begin{enumerate}
\item Two-point function of scalar current. The affected counterterms by the suppression are the following: $\cO_{10}=\bra \chi_+ \ket^2$ and ${\cal O}_{12} + 2 {\cal O}_{16} = \langle \chi_+^2 \rangle$. Then, one gets $\gamma_{10}=\gamma_{12}+2\gamma_{16}=0$.
% with an asymptotic $\cO(q^0)$ behaviour; ${\cal O}_{4}^{R}=\langle S \chi_{+} \rangle $ and $\cO_5^R= \bra S \ket \bra \chi_+ \ket$ with an asymptotic $\cO(q^{-2})$ behaviour. Then, $\gamma_{10}=\gamma_{12}+2\gamma_{16}=\gamma_4^R=\gamma_5^R=0$.
\item Two-point function of pseudoscalar current. Affected counterterms by the suppression: $\cO_{11}=\bra \chi_- \ket^2$ and $\cO_{12}-2\cO_{16}=\bra\chi_-^2\ket$. Then, $\gamma_{11}=\gamma_{12}-2\gamma_{16}=0$.
% with an asymptotic $\cO(q^0)$ behaviour; $\cO_2=\bra \chi_+\ket,\,\cO^R_{44}=i\bra P \chi_-\ket$ and $\cO^R_{45}=i\bra P\ket\bra \chi_- \ket$ with an asymptotic $\cO(q^{-2})$ behaviour. Then, $\gamma_{11}=\gamma_{12}-2\gamma_{16}=\gamma_2=\gamma_{44}^R=\gamma_{45}^R=0$.
\item Scalar Form Factor $\bra \pi^i | s^j | \pi^k\ket$. Affected counterterms by the suppression: $\cO_8=\bra \chi_+ \ket \bra u_\mu u^\mu \ket,\, \cO_9 = \bra \chi_+ u_\mu u^\mu \ket$ and $\cO_{18}=\bra u_\mu \ket \bra u^\mu \chi_+ \ket$. Then, $\gamma_8=\gamma_9=\gamma_{18}=0$.
\item Scalar Form Factor $\bra P^i | s^j | \pi^k \ket$. Affected counterterms by the suppression: $\cO^R_{52}=\bra \chi_+ \{ u_\mu, \nabla^\mu P \} \ket,\,\cO^R_{53}\bra u^\mu \chi_+ \ket \bra \nabla_\mu P \ket, \cO^R_{54}=\bra \chi_+ \ket \bra u_\mu \nabla^\mu P\ket$ and $\cO^R_{55}=\bra \chi_+ \nabla^\mu P \ket \bra u_\mu \ket$. Then, $\gamma^R_{52}=\gamma^R_{53}=\gamma^R_{54}=\gamma^R_{55}=0$.
\item Pseudoscalar Form Factor $\bra S^i | p^j | \pi^k \ket$. Affected counterterms by the suppression: $\cO^R_{32}=i\bra \chi_- \{ u_\mu, \nabla^\mu S \} \ket,\, \cO^R_{33}= i\bra u_\mu \chi_-\ket \bra \nabla^\mu S \ket, \, \cO^R_{34}=i\bra \chi_- \ket \bra u_\mu \nabla^\mu S\ket$ and $\cO^R_{35}=i\bra \chi_- \nabla^\mu S \ket \bra u_\mu \ket$. Then, $\gamma^R_{32}=\gamma^R_{33}=\gamma^R_{34}=\gamma^R_{35}=0$.
\end{enumerate}
We cannot consider the form factors to two resonance fields because they involve other cuts that have been not analyzed asymptotically.

Following these ideas we have been able to explain $15$ vanishing $\beta$-functions. In total $28$ vanishing ones have been found, so $13$ are not so clear. In any case, in $6$ of these $13$ we can conjecture that an accidental suppression happens because the structure of the loops is very similar to the loops that appeared in the constrained form factors, so that the high-energy constraints can lead to the found suppression: $\cO^{RR}_{180}=i\bra f^{\mu\nu}_+ \nabla_\mu P \nabla_\nu P \ket$ is a counterterm for $\bra P^i | v^{\mu\,j} | P^k \ket$; $\cO^{RR}_{316} = \bra f^{\mu\nu}_- \{\nabla_\mu P, \nabla_\nu S \} \ket, \, \cO^{RR}_{317}=\bra \nabla_\mu P \ket \bra f^{\mu\nu}_- \nabla_\nu S \ket$ and $\cO^{RR}_{318}=\bra \nabla_\mu S \ket \bra f^{\mu\nu}_- \nabla_\nu P \ket$ are counterterm for $\bra S^i | a^{\mu\,j} | P^k \ket$; $\cO^{RR}_{190}= \bra P f^{\mu\nu}_+ P f_{+\, \mu\nu} \ket$ and $\cO^{RR}_{191}= \bra PP f^{\mu\nu}_+ f_{\mu\nu\, +} \ket$ are counterterms of $\bra P^i | v^{\mu\,j} v^{\nu\,k} | P^l \ket$. The corresponding loop contibutions are shown in Figure~\ref{vanishing} and the cancellation between these pairs of loops can be checked in Table~G.2 of Appendix~G, taking into account the relevant couplings for each diagram.

In the case of $6$ of the other $7$ vanishing $\beta$-function we can conclude nothing following the same procedure, since not all the operators with the same structure have a vanishing $\beta$-function. For instance, the coupling related to the operator $\cO^{R}_{26}=\bra u_\nu S u^\nu \chi_+\ket$ has not running, while it does not happen the same with $\cO^{R}_{25}$ and $\cO^{R}_{27-31}$, all of them constructed with the same operators. 

The case of $\cO_2=\bra \chi_+ \ket$ is a different question. As a consequence of imposing the correct short-distance behaviour of the scalar form factors $\bra S^i | s^j | S^k \ket $ and $\bra P^i | s^j | P^k \ket $, one has $\lambda_3^{SS}=\lambda_3^{PP}=0$, so that, for instance, there are not divergences of $\bra 0|s^i | 0\ket $ that can be renormalized by the counterm $\cO_2$. Therefore, following the notation of Eq.(\ref{notation}), $\alpha_2$ does not run at one-loop level.

As we have pointed out above, the case of vector and axial-vector resonances is different because the structure of the propagators breaks down the Weinberg chiral counting. In any case, if the procedure of Chapter~4 is followed, there is no problem because the interest is directly in the imaginary part and not in the form factors, so the needed suppression is obtained. Therefore, extrapolating this behaviour to all the resonance fields and studying all form factors, these vanishing $\beta$-functions will be obtained in many more cases. Furthermore, if the behaviour of different scatterings were studied at large energies, the same could happen with operators that are not related to external currents. Eventually one could expect to obtain $\gamma_i=0$ for all $i$, that is, all operators involving only pseudo-Goldstone bosons and external currents would have vanishing $\beta$-function, what would be very interesting in order to understand the saturation, since $\widetilde{L}_i$ could vanish without problems, allowing an easy resonance saturation of the couplings of $\mathcal{L}_4^{\chi PT}$ at one loop, as it was suggested in Ref.~\cite{cata}.

\appendix
\chapter*{Appendix A \newline \newline The Antisymmetric Tensor Formalism}
\addcontentsline{toc}{chapter}{Appendix A: The Antisymmetric Tensor Formalism}
\newcounter{alpha}
\renewcommand{\thesection}{\Alph{alpha}}
\renewcommand{\theequation}{\Alph{alpha}.\arabic{equation}}
\renewcommand{\thetable}{\Alph{alpha}}
\setcounter{alpha}{1}
\setcounter{equation}{0}
\setcounter{table}{0}

Although the antisymmetric tensor formalism for spin $1$ massive fields was already proposed at the end of 60's~\cite{anti1}, its use was not regular until it was rediscovered in Ref.~\cite{ChPTb} in order to introduce the $\rho$ resonance field in the chiral lagrangian, Ecker {\it et al.} turned it into the usual way to work with spin-$1$ resonances in R$\chi$T~\cite{RChTa}.

In Ref.~\cite{anti2} it was proved that for antisymmetric tensor fields with mass there are (up to multiplicative factors and a total four divergence) only two possible lagrangians of second order in derivatives, if one assumes the existence of a Klein-Gordon divisor. They correspond to having either the Lorentz condition or else a Bianchi identity satisfied by the fields. In the case of describing spin $1$ particles, one has these two possibilities, where $\phi_{\mu\nu}=-\phi_{\nu\mu}$,
\begin{enumerate}
\item The subsidiary condition is the Bianchi identity, i.e.  $\epsilon^{\mu\lambda\rho\sigma}\partial_\lambda\phi_{\rho\sigma}=0$, and $\phi_{ik}$ are frozen, so the dynamical degrees of freedom are $\phi_{i0}$, where $i$ runs over $i=1,2,3$. Notice that there are $3$ degrees of freedom, as it should be.
\item The subsidiary condition is now the Lorentz condition, that is, $\partial^\rho \phi_{\rho\nu}=0$, and $\phi_{i0}$ are frozen, so the three degrees of freedom are $\phi_{ij}$.
\end{enumerate}
Because of historical reasons, the first option has been chosen. In this case the free lagrangian is proved to be
\begin{eqnarray}\label{free}
\mathcal{L}&=&-\frac{1}{2}\partial^\mu \phi_{\mu\nu} \, \partial_\rho \phi^{\rho\nu} +\frac{1}{4} M^2 \phi_{\mu\nu} \phi^{\mu\nu}\,,
\end{eqnarray}
from where the equations of motion are
\begin{eqnarray}\label{eomfree}
\partial^\mu \partial_\sigma \phi^{\sigma \nu} -\partial^{\nu}\partial_\sigma \phi^{\sigma\mu} + M^2 \phi^{\mu\nu} &=&0\,.
\end{eqnarray}
With the definition $\phi_\mu=\partial^\nu \phi_{\nu\mu}/M$  one obtains from Eq.~(\ref{eomfree}) the familiar Proca equation
\begin{eqnarray}
\partial_\rho \left( \partial^\rho \phi^\mu -\partial^\mu \phi^\rho \right) + M^2 \phi^\mu &=& 0 \,.
\end{eqnarray}
From the lagrangian of Eq.~(\ref{free}) one derives the free propagator
\begin{equation} \label{propagador}
\bra 0|T\left\{\phi_{\mu\nu}(x),\phi_{\rho\sigma}(y)\right\}|0\ket \,=\,
\int
\frac{\mathrm{d}^4k}{(2\pi)^4} e^{-ik(x-y)} 
\left\{\frac{2i}{M^2-q^2}\Omega^L_{\mu\nu,\rho\sigma} \,+\, \frac{2i}{M^2}
\Omega_{\mu\nu,\rho\sigma}^T\right\}\,,
\end{equation} 
where the following antisymmetric tensors have been defined
\begin{eqnarray}\label{omega1}
\Omega^L_{\mu\nu,\rho\sigma}(q)&=&\frac{1}{2q^2}\left( g_{\mu\rho}q_\nu q_\sigma\,-\,g_{\rho\nu}q_\mu
q_\sigma\,-\,(\rho\leftrightarrow\sigma)\right)\,,\nonumber\\
\Omega_{\mu\nu,\rho\sigma}^T(q)&=&-\frac{1}{2q^2}\left( g_{\mu\rho}q_\nu q_\sigma\,-\,g_{\rho\nu}q_\mu
q_\sigma\,-\,q^2g_{\mu\rho}g_{\nu\sigma}\,-\,(\rho\leftrightarrow\sigma)\right)\,,
\end{eqnarray}
and superindexs $L$ and $T$ refer to longitudinal and transversal respectively. Let us consider as ``generalized identity'' the  $\mathcal{I}_{\mu\nu,\rho\sigma}$ tensor,
\thispagestyle{appendixa}
\begin{eqnarray}\label{omega2}
\mathcal{I}_{\mu\nu,\rho\sigma}&=&\frac{1}{2}\left(g_{\mu\rho}g_{\nu\sigma}\,
-\,g_{\mu\sigma}g_{\nu\rho}\right)\,,
\end{eqnarray}
since any antisymmetric tensor $\mathcal{A}_{\mu\nu}=-\mathcal{A}_{\nu\mu}$ satisfies that
\begin{equation}
\mathcal{A}\cdot\mathcal{I}\,=\,\mathcal{I}\cdot\mathcal{A}\,=\,\mathcal{A} \,.
\end{equation}
$\Omega^T_{\mu\nu}(q)$ and $\Omega^L_{\mu\nu}(q)$ satisfy the properties of the projectors:
\begin{align}
\Omega^T\,+\,\Omega^L\,&=\,\mathcal{I}&
\Omega^T\cdot\Omega^L\,=\,\Omega^L\cdot\Omega^T&=\,0\, , \nonumber \\
\Omega^T\cdot\Omega^T\,&=\,\Omega^T\, , &
\Omega^L\cdot\Omega^L&=\,\Omega^L \, .
\end{align}
The propagator of Eq.~(\ref{propagador}) has the normalization
\begin{eqnarray}
\bra 0 | \, \phi_{\mu\nu} \,|\phi,p\ket &=&  
\frac{i}{M}[p_\mu\epsilon_\nu(p)\,-\,p_\nu\epsilon_\mu(p)] \, ,
\end{eqnarray}
where $\epsilon_\mu(p)$ is the polarization vector.

\subsubsection{Advantages Using the Antisymmetric Formalism}

There are different ways to include massive spin-$1$ fields in effective lagrangians, mainly the Proca and the antisymmetric tensor formalisms. In Ref.~\cite{RChTb} it was analysed this ambiguity in the context of $\chi$PT to $\cO(p^4)$. It was shown that, provided the consistency with QCD asymptotic behaviour is incorporated, the structure of the effective couplings induced by vector and axial-vector exchange is model independent. 

In the case of the antisymmetric tensor formalism no local terms, of the $\mathcal{L}_{pGB}$, constructed with chiral tensors of $\cO(p^4)$ or higher are required to fulfill the short-distance behaviour of QCD, that is, $\widetilde{L}_i = 0$, where $\widetilde{L}_i$ are the couplings in Resonance Chiral Theory, while $L_i$ are used for the $\chi$PT case, when resonances have been integrated out. 

Notwithstanding, for the Proca formalism business is not so easy. Considering the Proca lagrangian,
\begin{eqnarray}
\mathcal{L}^{\mathrm{Proca}}&=&\sum_{R=V,A}\left( \mathcal{L}^{\mathrm{Proca}}_{\mathrm{kin}}(R) + \mathcal{L}^{\mathrm{Proca}}_{\mathrm{int}}(R) \right)\, ,
\end{eqnarray}
where the lagrangian has been split in a kinetic and an interaction piece,
\begin{eqnarray}
\mathcal{L}^{\mathrm{Proca}}_{\mathrm{kin}}(R)&=&-\frac{1}{4}\bra  \hat{R}_{\mu\nu}\hat{R}^{\mu\nu}\,-\,2M_R^2\hat{R}_\mu\hat{R}^\mu \ket  \qquad (R=V,A) \, , \nonumber \\
\mathcal{L}^{\mathrm{Proca}}_{\mathrm{int}}(V)&=&-\,\frac{1}{2\sqrt{2}}\left(
f_V\bra\hat{V}^{\mu\nu}f^+_{\mu\nu}\ket\,+\,ig_V\bra\hat{V}_{\mu\nu}[u^\mu,u^\nu]\ket \right) \,+\, \dots \, , \phantom{\frac{1}{2}}\nonumber\\
\mathcal{L}^{\mathrm{Proca}}_{\mathrm{int}}(A)&=&f_A\bra\hat{A}^{\mu\nu}f^-_{\mu\nu}\ket \, +\, \dots \,, 
\end{eqnarray}
and the dots refer to terms not relevant for the Green Functions that are analysed at large energies. Finally the following definition has been used,
\begin{eqnarray}
\hat{R}_{\mu\nu}&=&\nabla_\mu \hat{R}_\nu \,-\, \nabla_\nu \hat{R}_\mu \, .
\end{eqnarray}
Imposing a reasonable short-distance behaviour for the two-point function built from a left- and a right-handed vector quark current, the pion form factor and the elastic meson-meson scattering, one finds the following constraints:
\begin{eqnarray}
\widetilde{L}_1^{\mathrm{Proca}}\,=\,\frac{1}{8}g_V^2 \,,\qquad
\widetilde{L}_2^{\mathrm{Proca}}\,=\,\frac{1}{4}g_V^2 \,,\qquad
\widetilde{L}_3^{\mathrm{Proca}}\,=\,-\frac{3}{4}g_V^2 \,,\nonumber\\
\widetilde{L}_9^{\mathrm{Proca}}\,=\,\frac{1}{2}f_Vg_V \,,\qquad
\widetilde{L}_{10}^{\mathrm{Proca}}\,=\,-\frac{1}{4}f_V^2 +\frac{1}{4}f_A^2\,,
\end{eqnarray}
while the rest vanish. Therefore the convenience of the antisymmetric formalism is manifest.
\thispagestyle{appendixa}
\newpage
\thispagestyle{appendixa}

\appendix
\chapter*{Appendix B \newline \newline Feynman Integrals}
\addcontentsline{toc}{chapter}{Appendix B: Feynman Integrals}
\newcounter{feyn}
\renewcommand{\thesection}{\Alph{feyn}}
\renewcommand{\theequation}{\Alph{feyn}.\arabic{equation}}
\renewcommand{\thetable}{\Alph{feyn}}
\setcounter{feyn}{2}
\setcounter{equation}{0}
\setcounter{table}{0}

The calculation of Chapter~3 involves the following Feynman Integrals:
\begin{align}
A_0(M^2)& \equiv \int \frac{\mathrm{d}k^D}{i(2\pi)^D} \frac{1}{k^2+i\epsilon - M^2}\,=\, - \frac{M^2}{16\pi^2} \left[ \lambda_\infty +  \log{\frac{M^2}{\mu^2}}\right] \, , \\
B_0(q^2,M_a^2,M_b^2) &\equiv  \int \frac{\mathrm{d}k^D}{i(2\pi)^D} \frac{1}{(k^2+i\epsilon  -  M_a^2) [(q-k)^2+i\epsilon  - M_b^2]} \nonumber \\
\quad =- \frac{1}{16\pi^2}\!\bigg[ \lambda_\infty &\left. + \frac{M_a^2}{M_a^2 - M_b^2}\log{\frac{M_a^2}{\mu^2}} - \frac{M_b^2}{M_a^2 - M_b^2}\log{\frac{M_b^2}{\mu^2}}\right]\! +  \bar{J}(q^2,M_a^2,M_b^2) ,
\end{align}
and the finite function
\begin{align}
&C_0(q^2,M_a^2,M_b^2,M_c^2) \,\equiv\, \nonumber \\ &\qquad\quad \int \frac{\mathrm{d}k^D}{i(2\pi)^D} \frac{1}{[(p_1-k)^2+i\epsilon - M_a^2] [(p_2+k)^2+i\epsilon - M_b^2] (k^2+i\epsilon -M_c^2)} ,
\end{align}
with $D$ the space-time dimension, $q\equiv p_1+p_2$ and, with massless outgoing pions, $p_1^2=p_2^2=0$. The divergences are collected in the factor
\begin{eqnarray}
\lambda_\infty &\equiv&   \frac{2 \mu^{D-4}}{D-4}+\gamma_E - \log 4\pi - 1\, ,
\end{eqnarray}
being $\gamma_E\simeq 0.5772$ the Euler's constant and $\mu$ the renormalization scale.

The two-propagator integral contains the finite function
\begin{align}
&\bar{J} (q^2,M_a^2,M_b^2) \, = \, \frac{1}{32 \pi^2} \left\{ 2 + \left[ \frac{M_a^2-M_b^2}{q^2} - \frac{M_a^2+M_b^2}{M_a^2-M_b^2}\right] \log \frac{M_b^2}{M_a^2} \right.\nonumber\\
&\qquad \left.-\frac{\lambda^{1/2} (q^2,M_a^2,M_b^2)}{q^2} \log{\left( \frac{[q^2 + \lambda^{1/2} (q^2,M_a^2,M_b^2)]^2 - (M_a^2-M_b^2)^2}{[q^2 -\lambda^{1/2} (q^2,M_a^2,M_b^2)]^2 - (M_a^2-M_b^2)^2} \right) } \right\}\, ,
\end{align}
with $\lambda (x,y,z)\equiv x^2 + y^2 + z^2 - 2 x y - 2 x z - 2 y z$. 

Some useful particular cases are:
\begin{eqnarray}
B_0(q^2,0,0)&=& - \frac{\lambda_\infty}{16\pi^2} +  \hat{B}_0(q^2/\mu^2)\, ,\\
B_0(q^2,M^2,M^2)& = & - \frac{1}{16\pi^2}\left\{\lambda_\infty +\log{\frac{M^2}{\mu^2}} + 1 \right\} + \overline{B}_0(q^2,M^2) \, , \\
B_0(q^2,0,M^2)& = & - \frac{1}{16\pi^2}\left\{\lambda_\infty +\log{\frac{M^2}{\mu^2}} \right\}  +  \bar{J}(q^2,0,M^2) \, , 
\end{eqnarray}
with the finite parts
\begin{eqnarray}
\hat{B}_0(q^2/\mu^2)& = & \frac{1}{16\pi^2}\left\{1-\log{\left(-\frac{q^2}{\mu^2}\right)}  \right\} \, ,\\
\overline{B}_0(q^2,M^2)&=& \bar{J}(q^2,M^2,M^2) \, =\, \frac{1}{16\pi^2}
\left\{2 - \sigma_M \log{  \left(\frac{\sigma_M+1}{\sigma_M -1}\right) } \right\}  \, ,\\
\bar{J}(q^2,0,M^2)&=& \frac{1}{16\pi^2}\left\{ 1 -\left( 1 -{M^2\over q^2}\right) \log{ \left(1 -{q^2\over M^2}\right)}\right\}\, ,
\end{eqnarray}
where $\sigma_M=\sqrt{1-4 M^2/q^2}$.

The relevant three-propagator integrals are:
\begin{eqnarray}
C_0(q^2,0,0,M^2) &=& -\frac{1}{16 \pi^2 q^2} \left\{ \mbox{Li}_2\left(1+\frac{q^2}{M^2}\right) - \mbox{Li}_2(1)\right\} \, ,\\
C_0(q^2,M^2,M^2,0) &=& \frac{1}{16 \pi^2 q^2}\log^2{\left({\sigma_M-1\over\sigma_M+1}\right)}\, ,
\end{eqnarray}
where
\begin{eqnarray}
\mbox{Li}_2(y)&\equiv&-   \int_0^1 \frac{\mathrm{d}x}{x} \log{ (1-xy) } =  -  \Int_0^y\, \frac{\mathrm{d}x}{x} \log{ (1-x) }
\end{eqnarray}
is the usual dilogarithmic function.
\thispagestyle{appendixb}

\appendix
\chapter*{Appendix C \newline \newline Feynman Diagrams for the Vector Form Factor}
\addcontentsline{toc}{chapter}{Appendix C: Feynman Diagrams for the Vector Form Factor}
\newcounter{appendixc}
\renewcommand{\thesection}{\Alph{appendixc}}
\renewcommand{\theequation}{\Alph{appendixc}.\arabic{equation}}
\renewcommand{\thetable}{\Alph{appendixc}}
\setcounter{appendixc}{5}
\setcounter{equation}{0}
\setcounter{table}{0}

We show the contributions from the different Feynman diagrams to the vector form factor of the pion at next-to-leading order in the $1/N_C$ expansion. As it has been pointed out in Chapter~3, a $U(2)_L \otimes U(2)_R$ chiral theory is used and we work in the massless limit. Keep in mind that only the lagrangian of Section~3.2 is employed.

In the following figures a single line stands for a pseudo-Goldstone boson while a double line indicates a resonance field; notice that the resonance in the s-channel is always a $\rho^0$.

\subsubsection{Wave-function Renormalization}

\begin{table}[h!,20pt]
\[
\begin{array}{ll}
\lower15pt
\hbox{\epsfbox{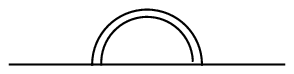}} & 
{\displaystyle 
=i\frac{2G_V^2}{F^4}\left\{-(p^2+M_V^2)A_0(M_V^2)+(p^2-M_V^2)^2B_0(p^2,0,M_V^2)\right\} }\\
&{\displaystyle\,\,\,\,
+i\frac{4c_d^2}{F^4}
\left\{(3p^2-M_S^2)A_0(M_S^2)+(p^2-M_S^2)^2B_0(p^2,0,M_S^2)\right\}\,,
}\\ \\

\lower15pt
\hbox{\epsfbox{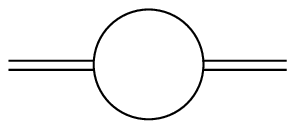}} &
{\displaystyle =
-i\frac{2G_V^2}{F^4}
\frac{(q^2)^2}{2}\Omega^L_{\mu\nu,\rho\sigma}(q)
\left\{\frac{1}{6}B_0(q^2,0,0) + \frac{1}{144\pi^2}
\right\}. }
\phantom{hoooooooooollllllllllllla}
\end{array}
\]
\end{table}
\vspace{-0.5cm}

\subsubsection{Contributions without Resonance Fields}

\begin{table}[h!,20pt]
\[
\begin{array}{lcl}
\lower15pt
\hbox{\epsfbox{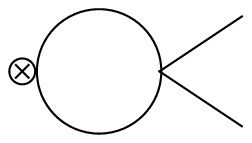}} & 
%\phantom{xxx} &
\rightarrow \quad &
{\displaystyle \frac{q^2}{F^2}\left\{\frac{1}{6}B_0(q^2,0,0)+\frac{1}{144\pi^2} \right\} \,.}
\phantom{holaaaaaaaaaaaaaaaaaaaaaaaa} 
\end{array}
\]
\end{table}

\newpage
\thispagestyle{appendixc}
\subsubsection{Contributions with Vector Resonance Fields}

\begin{table}[h!,20pt]
\[
\begin{array}{lcl}
\lower15pt
\hbox{\epsfbox{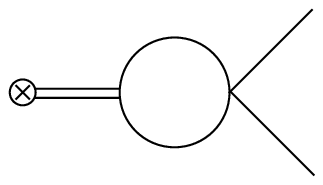}} & 
%\phantom{xxx} &
\rightarrow  &
{\displaystyle \frac{F_VG_V}{F^2}\frac{q^2}{M_V^2-q^2}
\frac{q^2}{F^2}\left\{\frac{1}{6}B_0(q^2,0,0)+\frac{1}{144\pi^2} \right\}\,, } \\ \vspace{-0.015cm} \\
%\phantom{holaaaaaaaaaaaaaaaaaaaaaaaa} 

\lower15pt
\hbox{\epsfbox{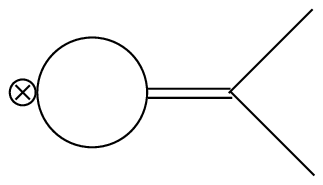}} & 
%\phantom{xxx} &
\rightarrow  &
{\displaystyle \frac{2G_V^2}{F^2}\frac{q^2}{M_V^2-q^2}
\frac{q^2}{F^2}\left\{\frac{1}{6}B_0(q^2,0,0)+\frac{1}{144\pi^2} \right\} \,,} \\ \vspace{-0.015cm} \\

\lower15pt
\hbox{\epsfbox{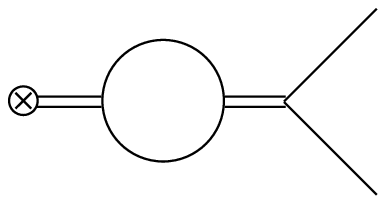}} & 
%\phantom{xxx} &
\rightarrow  &
{\displaystyle \frac{F_VG_V}{F^2}\frac{2G_V^2}{F^2}\left(\frac{q^2}{M_V^2-q^2}\right)^2
\frac{q^2}{F^2}\left\{\frac{1}{6}B_0(q^2,0,0)+\frac{1}{144\pi^2} \right\} \,,} \\ \vspace{-0.015cm} \\

\lower15pt
\hbox{\epsfbox{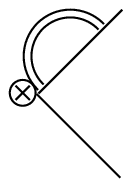}} & 
%\phantom{xxx} &
\rightarrow  &
{\displaystyle \frac{2G_V^2}{F^4}
\left\{-3A_0(M_V^2)+\frac{M_V^2}{32\pi^2} \right\} \,,} \\ \vspace{-0.015cm}\\

\lower15pt
\hbox{\epsfbox{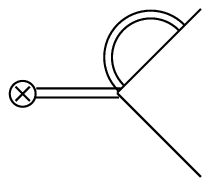}} & 
%\phantom{xxx} &
\rightarrow  &
{\displaystyle \frac{F_VG_V}{F^4}\frac{q^2}{M_V^2-q^2}
\left\{-\frac{3}{2}A_0(M_V^2)+\frac{M_V^2}{64\pi^2} \right\} \,,} \\ \vspace{-0.015cm}\\

\lower15pt
\hbox{\epsfbox{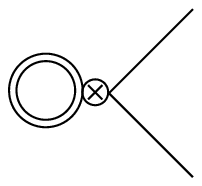}} & 
%\phantom{xxx} &
\rightarrow  &
{\displaystyle \frac{1}{F^2}
\left\{\frac{3}{2}A_0(M_V^2)-\frac{M_V^2}{64\pi^2} \right\}\,, } \\ \vspace{-0.015cm}\\

\lower15pt
\hbox{\epsfbox{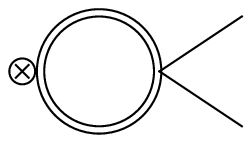}} & 
%\phantom{xxx} &
\rightarrow  &
{\displaystyle \frac{1}{F^2} \left\{B_0(q^2,M_V^2,M_V^2,0)\left[ -2M_V^2-\frac{q^2}{6}+
\frac{q^4}{6M_V^2}\right] \right. }\\
& & {\displaystyle\left.+A_0(M_V^2)\left[
\frac{1}{2}-\frac{q^2}{3M_V^2}\right]-\frac{7M_V^2}{64\pi^2}+\frac{q^2}{48\pi^2}-
\frac{q^4}{288\pi^2M_V^2}\right\} \,,} \\ \vspace{-0.015cm}\\

\lower15pt
\hbox{\epsfbox{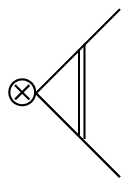}} & 
%\phantom{xxx} &
\rightarrow  &
{\displaystyle \frac{2G_V^2}{F^4}\left\{ C_0(q^2,0,0,M_V^2)\left[
-\frac{M_V^6}{q^2}-\frac{5M_V^4}{2}-q^2M_V^2\right]\right. } \\
& & {\displaystyle \left.+B_0(q^2,0,0)\left[-\frac{M_V^4}{q^2}
-2M_V^2-\frac{q^2}{12} \right] \right. } \\
& & {\displaystyle \left. +A_0(M_V^2)\left[
\frac{M_V^2}{q^2}+2\right]-\frac{M_V^2}{64\pi^2}-\frac{q^2}{288\pi^2} \right\} \,,} 

\end{array}
\]
\end{table}

\newpage

\begin{table}[h!,20pt]
\[
\begin{array}{lcl}

\lower15pt
\hbox{\epsfbox{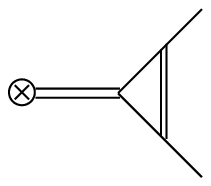}} & 
%\phantom{xxx} &
\rightarrow  &
{\displaystyle \frac{2G_V^2}{F^4}\frac{F_VG_V}{F^2}\frac{q^2}{M_V^2-q^2}\left\{ C_0(q^2,0,0,M_V^2)\left[
-\frac{M_V^6}{q^2}-q^2M_V^2\right.\right. }\\
& & {\displaystyle \left.\left.-\frac{5M_V^4}{2}\right]
+B_0(q^2,0,0)\left[-\frac{M_V^4}{q^2}-
2M_V^2-\frac{q^2}{12} \right]\right. } \\
& & {\displaystyle \left. +A_0(M_V^2)\left[
\frac{M_V^2}{q^2}+2\right]
-\frac{M_V^2}{64\pi^2}-\frac{q^2}{288\pi^2} \right\}\,, } \\ \\

\lower15pt
\hbox{\epsfbox{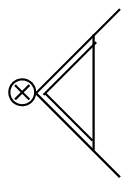}} & 
%\phantom{xxx} &
\rightarrow  &
{\displaystyle \frac{2G_V^2}{F^4}\left\{ C_0(q^2,M_V^2,M_V^2,0)\left[
\frac{M_V^6}{q^2}+\frac{M_V^4}{2}\right] \right. } \\
& & {\displaystyle \left.+B_0(q^2,M_V^2,M_V^2)\left[-\frac{M_V^4}{q^2}
-\frac{2M_V^2}{3}+\frac{5q^2}{12} \right] \right. } \\ 
& & {\displaystyle \left. +A_0(M_V^2)\left[
\frac{M_V^2}{q^2}+\frac{2}{3}\right]+\frac{M_V^2}{192\pi^2}-\frac{q^2}{288\pi^2} \right\} \,.} \\

\end{array}
\]
\end{table}
\vspace{-0.5cm}

\subsubsection{Contributions with Scalar Resonance Fields}
\thispagestyle{appendixc}
\begin{table}[h!,20pt]
\[
\begin{array}{lcl}

\lower15pt
\hbox{\epsfbox{as.eps}} & 
%\phantom{xxx} &
\rightarrow  &
{\displaystyle \frac{4c_d^2}{F^4}\left\{A_0(M_S^2)+\frac{M_S^2}{32\pi^2}\right\} \,, } \\ \\

\lower15pt
\hbox{\epsfbox{tadpole.eps}} & 
%\phantom{xxx} &
\rightarrow  &
{\displaystyle \frac{1}{F^2}A_0(M_S^2) \,, } \\ \\

\lower15pt
\hbox{\epsfbox{dobl.eps}} & 
%\phantom{xxx} &
\rightarrow  &
{\displaystyle \frac{1}{F^2}\left\{ B_0(q^2,M_S^2,M_S^2)\left[-\frac{2M_S^2}{3}+\frac{q^2}{6}\right]
\right.}\\
& & {\displaystyle \left.
-\frac{1}{3}A_0(M_S^2)-\frac{M_S^2}{24\pi^2}+\frac{q^2}{144\pi^2} \right\}\,, }\\ \\

\lower15pt
\hbox{\epsfbox{triang.eps}} & 
%\phantom{xxx} &
\rightarrow  &
{\displaystyle 
\frac{4c_d^2}{F^4}\left\{C_0(q^2,0,0,M_S^2)\left[-\frac{M_S^6}{q^2}-\frac{M_S^4}{2}\right]
+B_0(q^2,0,0)\left[-\frac{M_S^4}{q^2} \right.\right.} \\
& & {\displaystyle
\left.\left.-\frac{q^2}{12}\right]+\frac{M_S^2}{q^2}A_0(M_S^2)-\frac{M_S^2}{64\pi^2}-\frac{q^2}{288\pi^2} \right\}
\,,} 

\end{array}
\]
\end{table}

\newpage

\begin{table}[h!,20pt]
\[
\begin{array}{lcl}

\lower15pt
\hbox{\epsfbox{triangb.eps}} & 
%\phantom{xxx} &
\rightarrow  &
{\displaystyle 
\frac{4c_d^2}{F^4}\frac{F_VG_V}{F^2}\frac{q^2}{M_V^2-q^2}\left\{C_0(q^2,0,0,M_S^2)\left[-\frac{M_S^6}{q^2}
-\frac{M_S^4}{2}\right] -\frac{M_S^2}{64\pi^2}\right. } \\
& & {\displaystyle \left. +B_0(q^2,0,0)\left[-\frac{M_S^4}{q^2}-\frac{q^2}{12}\right] 
+\frac{M_S^2}{q^2}A_0(M_V^2)-\frac{q^2}{288\pi^2} \right\}
\,,} \\ \vspace{-0.1cm}\\

\lower15pt
\hbox{\epsfbox{btriang.eps}} & 
%\phantom{xxx} &
\rightarrow  &
{\displaystyle  \frac{4c_d^2}{F^4}\left\{C_0(q^2,M_S^2,M_S^2,0)\left[	\frac{M_S^6}{q^2}- 
\frac{M_S^4}{2}\right]\right. } \\
& & {\displaystyle +B_0(q^2,M_S^2,M_S^2)\left[-\frac{M_S^4}{q^2}+\frac{M_S^2}{3}-\frac{q^2}{12}\right] }\\
& & {\displaystyle \left.+A_0(M_S^2)
\left[\frac{M_S^2}{q^2}-\frac{1}{3}\right]+\frac{M_S^2}{192\pi^2}-\frac{q^2}{288\pi^2}\right\} \,.}
\phantom{holaaaaaa}
%\lower15pt
%\hbox{\epsfbox{/afs/ific.uv.es/user/i/igroes/Trebin/Feynman/btriang.eps}} & 
%\phantom{xxx} &
%\rightarrow  &
%{\displaystyle 0 \,.} 

\end{array}
\]
\end{table}
\vspace{-0.5cm}
\thispagestyle{appendixc}

\subsubsection{Contributions with Axial Resonance Fields}

\begin{table}[h!,20pt]
\[
\begin{array}{lcl}

\lower15pt
\hbox{\epsfbox{tadpole.eps}} & 
%\phantom{xxx} &
\rightarrow  &
{\displaystyle \frac{1}{F^2}
\left\{\frac{3}{2}A_0(M_A^2)-\frac{M_A^2}{64\pi^2} \right\}\,,\phantom{holaaaaaaaaaaaaaaaaaa} } \\ \vspace{-0.1cm}\\

\lower15pt
\hbox{\epsfbox{dobl.eps}} & 
%\phantom{xxx} &
\rightarrow  &
{\displaystyle \frac{1}{F^2} \left\{B_0(q^2,M_A^2,M_A^2,0)\left[ -2M_A^2-\frac{q^2}{6}+
\frac{q^4}{6M_A^2}\right] \right. }\\
& & {\displaystyle\left.+A_0(M_A^2)\left[
\frac{1}{2}-\frac{q^2}{3M_A^2}\right]-\frac{7M_A^2}{64\pi^2}+\frac{q^2}{48\pi^2}-
\frac{q^4}{288\pi^2M_A^2}\right\} \,.} \phantom{holaaaaaaaaaaaaaaaaaaaaaaaaaaaaaaaaaaa} 

\end{array}
\]
\end{table}
\vspace{-0.5cm}

\subsubsection{Contributions with Pseudoscalar Resonance Fields}

\begin{table}[h!,20pt]
\[
\begin{array}{lcl}

\lower15pt
\hbox{\epsfbox{tadpole.eps}} & 
%\phantom{xxx} &
\rightarrow  &
{\displaystyle \frac{1}{F^2}A_0(M_P^2) \,, } \\ \vspace{-0.1cm}\\

\lower15pt
\hbox{\epsfbox{dobl.eps}} & 
%\phantom{xxx} &
\rightarrow  &
{\displaystyle \frac{1}{F^2}\left\{ B_0(q^2,M_P^2,M_P^2)\left[-\frac{2M_P^2}{3}+\frac{q^2}{6}\right]
\right.}\\
& & {\displaystyle \left.
-\frac{1}{3}A_0(M_P^2)-\frac{M_P^2}{24\pi^2}+\frac{q^2}{144\pi^2} \right\}\,. }
\phantom{holaaaaaaaaaaaaaaaaaaaaaaaa}

\end{array}
\]
\end{table}

\appendix
\chapter*{Appendix D \newline \newline Form Factors and Constraints}
\addcontentsline{toc}{chapter}{Appendix D: Form Factors and Constraints}
\newcounter{apd}
\renewcommand{\thesection}{\Alph{apd}}
\renewcommand{\theequation}{\Alph{apd}.\arabic{equation}}
\renewcommand{\thetable}{\Alph{apd}}
\renewcommand{\thefigure}{\Alph{apd}.\arabic{figure}}
\setcounter{apd}{4}
\setcounter{equation}{0}
\setcounter{table}{0}
\setcounter{figure}{0}

In this appendix all two-body form factors that can be found in the even-intrinsic-parity sector of the Resonance Chiral Theory in the Single Resonance Approximation are analysed, following the ideas of Section~4.3. 

The following items are presented for each form factor:
\begin{enumerate}
\item The form factor(s) is(are) defined through the corresponding matrix element.
\item The expression of the form factor(s) is(are) shown.
\item Using the optical theorem, the spectral function is given 
 in terms of the form factors.
\item The constraints found by imposing a good high-energy behaviour of the spectral function.
\item  Once the constraints are imposed, the 
well behaved form factor(s) is(are) presented again and quoted with a tilde. 
\end{enumerate}

Notice that when $R^0_{I=0}$  or $\eta$  
is written, we refer to the singlet in the $U(2)$ case. 
The following usual notation is employed throughout the section :
\begin{equation}
\lambda\left(a,b,c\right) = a^2+b^2+c^2-2ab-2ac-2bc  \, , \ \sigma_M = \lambda^{1/2}(q^2,M^2,M^2)/q^2 =  
\sqrt{1-4M^2/q^2}  
\, .\nonumber
\end{equation}
\subsection{Vector Form Factors}

\begin{figure}
\begin{center}
\includegraphics[scale=0.8]{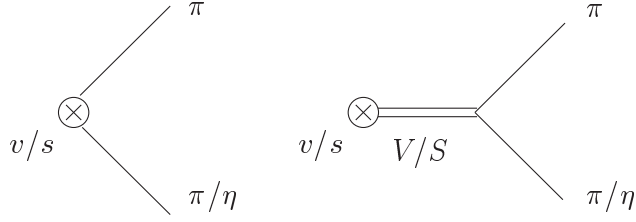}
\caption{\label{AR0F0}
Tree-level contributions to the vector/scalar form factor to two pseudo-Goldstones.}
\end{center}
\end{figure}

\subsubsection{Vector Form Factor to $\pi\pi$ (Figure \ref{AR0F0})}

\begin{eqnarray}
\bra \pi^0 (p_1) \pi^- (p_2) | \bar{d}\gamma^\mu u | 0\ket &=& \sqrt{2}\, \mathcal{F}_{\pi\pi}^{\,v} (q^2)\, (p_2-p_1)^\mu \, ,\phantom{\frac{1}{2}} \\
\mathcal{F}_{\pi\pi}^{\,v} (q^2)&=&1\,+\,\frac{F_VG_V}{F^2} \frac{q^2}{M_V^2-q^2} \, , \\
\mathrm{Im} \Pi_{_{VV}} (q^2) |_{\pi\pi}  &=& \frac{\theta(q^2)}{24\pi} |\mathcal{F}_{\pi\pi}^{\,v} (q^2)|^2 \, , \\
F_V\, G_V&=&F^2 \,,\phantom{\frac{1}{2}}
\end{eqnarray}
\begin{eqnarray}
\mathcal{\tilde{F}}_{\pi\pi}^{\,v} (q^2)&=& \frac{M_V^2}{M_V^2-q^2} \, .
\end{eqnarray}

\begin{figure}
\begin{center}
\includegraphics[scale=0.8]{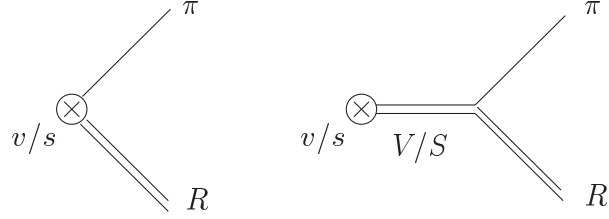}
\caption{\label{AR1F0}
Tree-level contributions to the vector/scalar form factor to one resonance field and one pseudo-Goldstone.}
\end{center}
\end{figure}

%\vspace{0.15cm}
%
%\newpage
\subsubsection{Vector Form Factor to A$\pi$ (Figure \ref{AR1F0})}
\thispagestyle{appendixd}
\begin{eqnarray}
\bra A^0_{I=1} (p_A,\varepsilon) \pi^- (p_\pi) | \bar{d}\gamma^\mu u | 0\ket \,&=& \frac{i\, \sqrt{2}}{M_A} \Big\{
( q\varepsilon^* \, p_A^\mu \, -\, qp_A\, {\varepsilon^*}^\mu )\, \mathcal{F}^{\,v}_{A\pi} (q^2) \nonumber \\
&&+( q\varepsilon^* \, p_\pi^\mu \, -\, qp_\pi {\varepsilon^*}^\mu )\, \mathcal{G}^{\,v}_{A\pi} (q^2) \Big\} \, ,
\end{eqnarray}
\begin{eqnarray}
\mathcal{F}^{\,v}_{A\pi} (q^2) &=& \frac{F_A}{F} \,+ \,\frac{F_V}{F}  \frac{M_A^2-q^2}{M_V^2-q^2}\Big[
-2 \lambda^{VA}_2+2 \lambda^{VA}_3- \lambda^{VA}_4-2\lambda^{VA}_5 \Big] \, ,\nonumber \\ \nonumber \\
\mathcal{G}^{\,v}_{A\pi} (q^2) &=& \frac{2F_V}{F}  \frac{M_A^2}{M_V^2-q^2}\Big[-2\lambda^{VA}_2 + \lambda^{VA}_3 \Big] \, ,
\end{eqnarray} 
\begin{align}
&\mathrm{Im}  \Pi_{_{VV}} (q^2) |_{A\pi} = \theta(q^2-M_A^2) \frac{1-M_A^2/q^2}{48\pi} \!\Bigg\{ \!\!\left( \frac{M_A^2}{q^2}\!+4+\!\frac{q^2}{M_A^2} \right)\! |\mathcal{F}^{\,v}_{A\pi}|^2\!\! +\! (1-M_A^2/q^2)^2\times\nonumber \\
& \qquad \times \left( \frac{q^2}{M_A^2} + \frac{q^4}{2M_A^4}\right) |\mathcal{G}^{\,v}_{A\pi}|^2 
%\nonumber \\&&
+2 (1-M_A^2/q^2) \left( 1+\frac{2q^2}{M_A^2} \right)\mathrm{Re} \{ \mathcal{F}^{\,v}_{A\pi} {\mathcal{G}^{\,v}_{A\pi}}^* \}  \Bigg\} \, ,
\end{align} 
\begin{equation}
2\lambda^{VA}_2-2\lambda^{VA}_3+\lambda^{VA}_4+2\lambda^{VA}_5\,=\,F_A/F_V \, , \qquad \qquad
-2\lambda^{VA}_2+\lambda^{VA}_3 \,=\, 0 \, ,\label{VAG}
\end{equation}
\begin{equation}
\mathcal{\tilde{F}}^{\,v}_{A\pi} (q^2) \,=\,\frac{F_A}{F}\frac{M_V^2-M_A^2}{M_V^2-q^2} \, , \qquad \qquad \mathcal{\tilde{G}}^{\,v}_{A\pi} (q^2)\,=\, 0 \, .
\end{equation}

\subsubsection{Vector Form Factor to P$\pi$ (Figure \ref{AR1F0})}
\begin{figure}
\begin{center}
\includegraphics[scale=0.8]{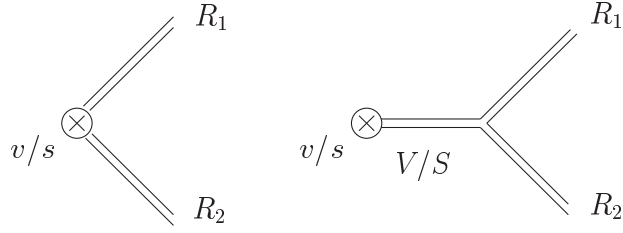}
\caption{\label{AR2F0}
Tree-level contributions to the vector/scalar form factor 
to two resonance fields.}
\end{center}
\end{figure}

\begin{eqnarray}
\bra P^- (p_P) \pi^0 (p_\pi) | \bar{d}\gamma^\mu  u | 0\ket \,&=& \sqrt{2} \left( q p_\pi \, p_P^\mu - q p_P\,  p_\pi^\mu \right)\, \mathcal{F}^{\,v}_{P\pi} (q^2)  \, ,\phantom{\frac{1}{2}} \\
\mathcal{F}^{\,v}_{P\pi} (q^2) &=& \frac{2 \lambda^{PV}_1 F_V}{F} \frac{1}{M_V^2-q^2} \,,\phantom{\frac{1}{2}}  
\end{eqnarray}
\begin{eqnarray}
 \mathrm{Im}  \Pi_{_{VV}} (q^2) |_{P\pi}  &=& \theta(q^2-M_P^2) \, \frac{\left(1-M_P^2/q^2\right)^3}{96\pi} q^4   |\mathcal{F}^{\,v}_{P\pi}|^2 \, ,\phantom{\frac{1}{2}} \\
\lambda_1^{PV}&=&0\, ,\phantom{\frac{1}{2}}\\
\mathcal{\tilde{F}}^{\,v}_{P\pi} (q^2) &=& 0 \,.\phantom{\frac{1}{2}}
\end{eqnarray}
\subsubsection{Vector Form Factor to VV (Figure \ref{AR2F0})}
\thispagestyle{appendixd}
\begin{align}
&\bra \! V^0_{I=1} (p_1,\varepsilon_1) V^- (p_2,\varepsilon_2) | \bar{d}\gamma^\mu u | 0\!\ket =\sqrt{2} \Big( \varepsilon_1^* \varepsilon_2^* \,(p_2-p_1)^\mu - ( q\varepsilon_1^*\, {\varepsilon_2^*}^\mu - q\varepsilon_2^* \,{\varepsilon_1^*}^\mu ) \Big) \! \mathcal{F}^{\,v}_{VV} (q^2) \nonumber \\
&+\sqrt{2} ( q\varepsilon_1^*\, {\varepsilon_2^*}^\mu - q\varepsilon_2^* \,{\varepsilon_1^*}^\mu )\mathcal{G}^{\,v}_{VV} (q^2)  
%\nonumber \\& 
+\sqrt{2} \frac{(p_2-p_1)^\mu}{M_V^2} ( q\varepsilon_1^*\,q\varepsilon_2^*  \,-\, p_1 p_2\, \varepsilon_1^* \varepsilon_2^* )  \mathcal{H}^{\,v}_{VV} (q^2)   ,
\end{align}
\vspace{-0.2cm}
\begin{eqnarray}
\mathcal{F}^{\,v}_{VV} (q^2) &\!\!\!=&\!\!\! -1\,+\,2\lambda^{VV}_7 \,+\, \frac{F_V}{\sqrt{2}(M_V^2-q^2)}\Big[6 \lambda^{VVV}_0 + (4M_V^2+2q^2) \lambda^{VVV}_2    \Big. \nonumber \\
&\!\!\!&\!\!\! \Big. +(4M_V^2-2q^2) \left( -2\lambda^{VVV}_1+ \lambda^{VVV}_3 + \lambda^{VVV}_4-2 \lambda^{VVV}_5 \right)+ 4 q^2 \lambda^{VVV}_6  \Big. \nonumber \\
&\!\!\!&\!\!\! \Big. + 8M_V^2 \lambda^{VVV}_7    \Big] \, ,\nonumber \\ \nonumber \\
\mathcal{G}^{\,v}_{VV} (q^2) &\!\!\!=&\!\!\!  \frac{4 \, F_V\, M_V^2}{\sqrt{2}(M_V^2-q^2)}\Big[  -2\lambda^{VVV}_1  +\lambda^{VVV}_3 + \lambda^{VVV}_4-2\lambda^{VVV}_5 -\lambda^{VVV}_6 + \lambda^{VVV}_7    \Big] \, , \phantom{\frac{1}{2}}\nonumber 
\\ \nonumber \\
\mathcal{H}^{\,v}_{VV} (q^2) &\!\!\!=&\!\!\! -2\lambda^{VV}_7\,+\, \frac{F_V}{\sqrt{2}(M_V^2-q^2)}\Big[-6 \lambda^{VVV}_0+(4M_V^2+2q^2) \big( 2\lambda^{VVV}_1 - \lambda^{VVV}_2  \big. \Big. \nonumber \\
&& \Big.\big. -\lambda^{VVV}_3  - \lambda^{VVV}_4+2 \lambda^{VVV}_5 -2\lambda^{VVV}_7   \big) \Big] \, , 
\end{eqnarray} 
\begin{align}
\mathrm{Im}  \Pi_{_{VV}} (q^2) |_{VV}  &=\, \theta(q^2-4M_V^2) \frac{\sigma^3_{M_V}}{24\pi} \Bigg\{ \left( 3+\frac{q^2}{M_V^2}\right) |\mathcal{F}^{\,v}_{VV}|^2
+ \left( \frac{q^2}{M_V^2}+\frac{q^4}{4M_V^4}\right) |\mathcal{G}^{\,v}_{VV}|^2  \nonumber \\
& +\left( 3-\frac{2q^2}{M_V^2}+\frac{q^4}{2M_V^4} \right) |\mathcal{H}^{\,v}_{VV}|^2 
 -\frac{3q^2}{M_V^2}\mathrm{Re} \{ \mathcal{F}^{\,v}_{VV} {\mathcal{G}^{\,v}_{VV}}^* \}
\nonumber 
\end{align}
\begin{equation} 
 + \left( 6-\frac{2q^2}{M_V^2} \right)\mathrm{Re} \{ \mathcal{F}^{\,v}_{VV} {\mathcal{H}^{\,v}_{VV}}^* \} - \frac{q^2}{M_V^2}  \mathrm{Re} \{ \mathcal{G}^{\,v}_{VV} {\mathcal{H}^{\,v}_{VV}}^* \} \Bigg\} \,,
\end{equation} 
\begin{align}
2\lambda^{VVV}_1+\lambda^{VVV}_2-\lambda^{VVV}_3-\lambda^{VVV}_4+2\lambda^{VVV}_5+2\lambda^{VVV}_6
&=-\frac{1}{\sqrt{2}F_V}+\frac{\sqrt{2}}{F_V}\lambda^{VV}_7  \, , \nonumber \\
-2\lambda^{VVV}_1+\lambda^{VVV}_3+\lambda^{VVV}_4-2\lambda^{VVV}_5-\lambda^{VVV}_6+\lambda^{VVV}_7&= 0\,, \nonumber \\
2\lambda^{VVV}_1-\lambda^{VVV}_2-\lambda^{VVV}_3-\lambda^{VVV}_4+2\lambda^{VVV}_5 -2\lambda^{VVV}_7&= -\frac{\sqrt{2}}{F_V}\lambda^{VV}_7  \, , \nonumber \\
-\frac{3}{2M_V^2} \lambda^{VVV}_0\!+ 2\lambda^{VVV}_1\! -\lambda^{VVV}_2\! -\lambda^{VVV}_3\! -\lambda^{VVV}_4\! + 2\lambda^{VVV}_5\! &-2\lambda^{VVV}_7\!=\frac{\lambda^{VV}_7}{\sqrt{2}F_V}
 \, , \label{VVV}
\end{align}
\begin{equation}
\mathcal{\tilde{F}}^{\,v}_{VV} (q^2) \,=\,-\frac{M_V^2}{M_V^2-q^2} \, , \qquad \qquad \mathcal{\tilde{G}}^{\,v}_{VV} (q^2)\,=\,\mathcal{\tilde{H}}^{\,v}_{VV} (q^2)\,=\, 0 \, .
\end{equation}

\subsubsection{Vector Form Factor to AA (Figure \ref{AR2F0})}
\thispagestyle{appendixd}
\begin{align}
&\bra \!A^0_{I=1} (p_1,\varepsilon_1) A^- (p_2,\varepsilon_2) | \bar{d}\gamma^\mu u | 0\!\ket =\sqrt{2}\! \Big( \varepsilon_1^* \varepsilon_2^* \,(p_2-p_1)^\mu - ( q\varepsilon_1^*\, {\varepsilon_2^*}^\mu - q\varepsilon_2^* \,{\varepsilon_1^*}^\mu ) \Big) \! \mathcal{F}^{\,v}_{AA} (q^2) \nonumber \\
&+\sqrt{2}\, ( q\varepsilon_1^*\, {\varepsilon_2^*}^\mu - q\varepsilon_2^* \,{\varepsilon_1^*}^\mu )\mathcal{G}^{\,v}_{AA} (q^2) 
% \nonumber \\&
 +\,\sqrt{2} \frac{(p_2-p_1)^\mu}{M_V^2} ( q\varepsilon_1^*\,q\varepsilon_2^*  \,-\, p_1 p_2\, \varepsilon_1^* \varepsilon_2^* )  \mathcal{H}^{\,v}_{AA} (q^2)  \, ,
\end{align}
\begin{align}
\mathcal{F}^{\,v}_{AA} (q^2) &=\, -1 \,+\,2\lambda^{AA}_7\,+\, \frac{F_V}{\sqrt{2}(M_V^2-q^2)}\Big[2 \lambda^{VAA}_0+ 2q^2( \lambda^{VAA}_3+\lambda^{VAA}_8) \Big. \nonumber \\
&\Big. +(2M_A^2-q^2)\left( 2\lambda^{VAA}_2  +\lambda^{VAA}_7-\lambda^{VAA}_9-2\lambda^{VAA}_{10}+\lambda^{VAA}_{12}+2 \lambda^{VAA}_{13}-\lambda^{VAA}_{14}  \right) \Big. \nonumber \\
&\Big. +(-q^2-2M_A^2)\lambda^{VAA}_6 \Big] \, , \nonumber 
\\ \nonumber \\
\mathcal{G}^{\,v}_{AA} (q^2) &=  \frac{\sqrt{2}F_V M_A^2}{M_V^2-q^2}\Big[
-  \lambda^{VAA}_6\!+  \lambda^{VAA}_7\! 
- \lambda^{VAA}_9\! -2 (\lambda^{VAA}_{10} \!+ \lambda^{VAA}_{11}\!) + \lambda^{VAA}_{12}\! 
 - \lambda^{VAA}_{14}\! \Big]  ,\! \nonumber 
%\\ \nonumber \\
\end{align}
\begin{align}
\mathcal{H}^{\,v}_{AA} (q^2) &=\,-2\lambda^{AA}_7\,+\, \frac{F_V}{\sqrt{2}(M_V^2-q^2)}\Big[
-2 \lambda^{VAA}_0 +4q^2\lambda^{VAA}_1
+(-4M_A^2+2q^2) \lambda^{VAA}_2 \nonumber \\
&-2q^2 (\lambda^{VAA}_3+\lambda^{VAA}_4-  \lambda^{VAA}_5) +(2M_A^2+2q^2) \lambda^{VAA}_6
  \nonumber \\
&+2M_A^2( -\lambda^{VAA}_7+\lambda^{VAA}_9+2\lambda^{VAA}_{10}-\lambda^{VAA}_{12}-2\lambda^{VAA}_{13}+\lambda^{VAA}_{14}) \Big] \, ,
\end{align} 
\begin{align}
\mathrm{Im}  \Pi_{_{VV}} (q^2) |_{AA}  \,&=\, \theta(q^2-4M_A^2) \frac{\sigma^3_{M_A}}{24\pi} \Bigg\{ \left( 3+\frac{q^2}{M_A^2}\right) |\mathcal{F}^{\,v}_{AA}|^2
+ \left( \frac{q^2}{M_A^2}+\frac{q^4}{4M_A^4}\right) |\mathcal{G}^{\,v}_{AA}|^2  \nonumber \\
& +\left( 3-\frac{2q^2}{M_A^2}+\frac{q^4}{2M_A^4} \right) |\mathcal{H}^{\,v}_{AA}|^2 
 -\frac{3q^2}{M_A^2}\mathrm{Re} \{ \mathcal{F}^{\,v}_{AA} {\mathcal{G}^{\,v}_{AA}}^* \}\nonumber \\
&  + \left( 6-\frac{2q^2}{M_A^2} \right)\mathrm{Re} \{ \mathcal{F}^{\,v}_{AA} {\mathcal{H}^{\,v}_{AA}}^* \} - \frac{q^2}{M_A^2}  \mathrm{Re} \{ \mathcal{G}^{\,v}_{AA} {\mathcal{H}^{\,v}_{AA}}^* \} \Bigg\} \,,
\end{align} 
\begin{eqnarray}
-2\lambda^{VAA}_2\! +2\lambda^{VAA}_3 \!  -\lambda^{VAA}_6\!-\lambda^{VAA}_7\!+ 2\lambda^{VAA}_8\! +&& \nonumber \\
+\lambda^{VAA}_9\!+2\lambda^{VAA}_{10}\!- \lambda^{VAA}_{12} \!-2\lambda^{VAA}_{13}+\lambda^{VAA}_{14}\!&=&-\frac{\sqrt{2}}{F_V}+\frac{2\sqrt{2}}{F_V}\lambda^{AA}_{7}  \, , \nonumber \\
-\lambda^{VAA}_6\!+\lambda^{VAA}_7\!-\lambda^{VAA}_9\!-2\lambda^{VAA}_{10}\!-2\lambda^{VAA}_{11} \!+\lambda^{VAA}_{12}\! -\lambda^{VAA}_{14} \!&=& 0 \, , \nonumber \\
2\lambda^{VAA}_1\!+\lambda^{VAA}_2\!-\lambda^{VAA}_3\!-\lambda^{VAA}_4\!+\lambda^{VAA}_5\!+\lambda^{VAA}_6\!&=&-\frac{\sqrt{2}}{F_V}\lambda^{AA}_{7} \, , \nonumber \\
-\frac{1}{M_A^2}\lambda^{VAA}_{0}\ -2\lambda^{VAA}_2\!+\lambda^{VAA}_6\!-\lambda^{VAA}_7\!+\lambda^{VAA}_9\! + && \nonumber \\
+2\lambda^{VAA}_{10}\!-\lambda^{VAA}_{12}\!-2\lambda^{VAA}_{13}\!+\lambda^{VAA}_{14}\!&=&\frac{\sqrt{2}M_V^2 \lambda^{AA}_7}{F_V M_A^2}  \,,\label{VAA}
\end{eqnarray}
\begin{equation}
\mathcal{\tilde{F}}^{\,v}_{AA} (q^2) \,=\,-\frac{M_V^2}{M_V^2-q^2} \, , \qquad \qquad \mathcal{\tilde{G}}^{\,v}_{AA} (q^2)\,=\,\mathcal{\tilde{H}}^{\,v}_{AA} (q^2)\,=\, 0 \, .
\end{equation}

\subsubsection{Vector Form Factor to RR (R=S,P) (Figure \ref{AR2F0})}
\thispagestyle{appendixd}
\begin{eqnarray}
\bra R^0_{I=1} (p_1) R^- (p_2) | \bar{d}\gamma^\mu u | 0\ket &=& \sqrt{2}\, \mathcal{F}_{RR}^{\,v} (q^2)\, (p_2-p_1)^\mu \, , \phantom{\frac{1}{2}}\\
\mathcal{F}_{RR}^{\,v} (q^2)&=&1\,+\,\frac{F_V}{\sqrt{2}} \, \lambda^{VRR} \frac{q^2}{M_V^2-q^2} \, , \\
\mathrm{Im}  \Pi_{_{VV}} (q^2) |_{RR}  &=& \theta(q^2-4M_R^2)\,\frac{\sigma^3_{M_R}}{24\pi} \,|\mathcal{F}_{RR}^{\,v} (q^2)|^2 \, , \\
\lambda^{VRR}&=&\frac{\sqrt{2}}{F_V}  \, , \\
\mathcal{\tilde{F}}_{RR}^{\,v} (q^2)&=& \frac{M_V^2}{M_V^2-q^2} \, .
\end{eqnarray}
\vspace{0.15cm}
\subsubsection{Vector Form Factor to SV (Figure \ref{AR2F0})}
\thispagestyle{appendixd}
\begin{eqnarray}
\bra S^0_{I=0} (p_S) V^- (p_V,\varepsilon) | \bar{d}\gamma^\mu  u | 0\ket \,&=& \frac{\sqrt{2}}{M_V} \Big\{
( q\varepsilon^* \, p_V^\mu \, -\, qp_V\, {\varepsilon^*}^\mu )\, \mathcal{F}^{\,v}_{SV} (q^2) \nonumber \\
&&+( q\varepsilon^* \, p_S^\mu \, -\, qp_S {\varepsilon^*}^\mu )\, \mathcal{G}^{\,v}_{SV} (q^2) \Big\} \, ,
\end{eqnarray}
\begin{eqnarray}
\mathcal{F}^{\,v}_{SV} (q^2) &=& 4\lambda^{SV}_3  +\frac{\sqrt{2}F_V}{M_V^2-q^2} \bigg[- 2\lambda^{SVV}_0-M_V^2\lambda^{SVV}_1-\frac{q^2+M_V^2-M_S^2}{2}\times  \bigg. \nonumber \\
&& \bigg. \times (\lambda^{SVV}_2+2\lambda^{SVV}_3) +(M_V^2+q^2)(2\lambda^{SVV}_4+\lambda^{SVV}_5)\bigg] \,,  \nonumber \\
\mathcal{G}^{\,v}_{SV} (q^2) &=& -\frac{\sqrt{2}F_VM_V^2}{M_V^2-q^2} \lambda^{SVV}_1    \, , 
\end{eqnarray} 
\begin{align}
 \mathrm{Im}  \Pi_{_{VV}} &(q^2) |_{SV}  \,=\, \theta(q^2-(M_S+M_V)^2) \, \frac{\lambda^{1/2}(q^2,M_S^2,M_V^2)}{48\pi q^2}\, \, \Bigg\{\frac{1}{M_V^2 q^2} \Big[ (M_V^2-M_S^2)^2    \nonumber \\
&  -2q^2(M_S^2-2M_V^2)+q^4\Big]  |\mathcal{F}^{\,v}_{SV}|^2 + \frac{1}{ 2M_V^4q^2} \Big[2M_V^2(M_V^2-M_S^2)^2 \nonumber \\
& \Big. + q^2(-3M_V^4+6M_S^2M_V^2+ M_S^4)-2M_S^2q^4 +q^6\Big]  |\mathcal{G}^{\,v}_{SV}|^2 \phantom{\frac{1}{2}} \nonumber \\
& +\frac{1}{M_V^2 q^2} \!\Big[\! -2(M_V^2-M_S^2)^2-2q^2(M_V^2+M_S^2)+4q^4\! \Big]\! \mathrm{Re} \!\{ \mathcal{F}^{\,v}_{SV} {\mathcal{G}^{\,v}_{SV}}^* \} \! \Bigg\}  ,
\end{align} 
\begin{equation}
\lambda_2^{SVV}+2\lambda_3^{SVV}-4\lambda_4^{SVV}-2\lambda_5^{SVV}\,=\,-\frac{4\sqrt{2}}{F_V} \lambda_3^{SV}   \, , \qquad \qquad
\lambda_1^{SVV}\,=\,0\, , \label{VSV}
\end{equation}
\begin{equation}
\mathcal{\tilde{F}}^{\,v}_{SV} (q^2) = \frac{\sqrt{2} F_V}{M_V^2-q^2} \left[  \frac{8M_V^2}{\sqrt{2}F_V}\lambda_3^{SV} -2\lambda_0^{SVV}+\frac{M_S^2}{2} \left( \lambda_2^{SVV}+2\lambda_3^{SVV}\right) \right] \,,
\ \mathcal{\tilde{G}}^{\,v}_{SV} (q^2) =0 \, .
\end{equation}
%\newpage
\subsubsection{Vector Form Factor to PA (Figure \ref{AR2F0})}
\begin{figure}
\begin{center}
\includegraphics[scale=0.70]{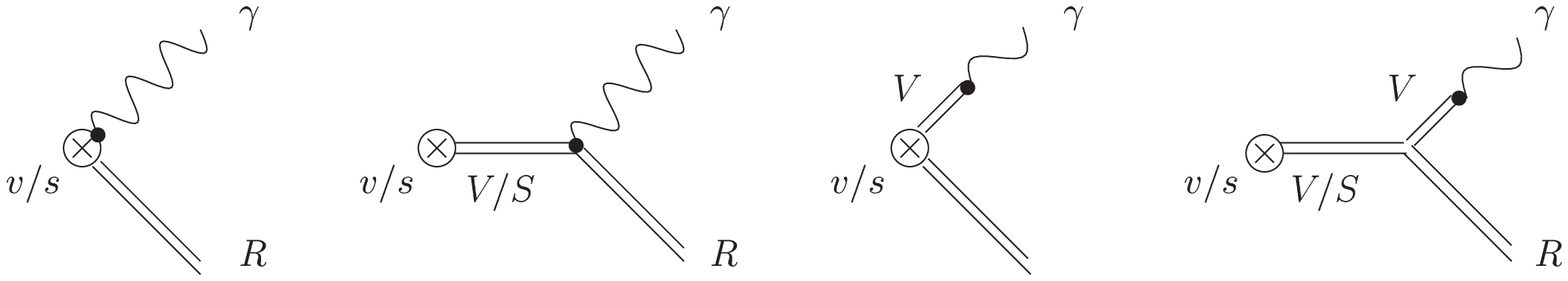}
\caption{\label{AR1F1}
Tree-level contributions to the vector/scalar form factor 
to one resonance field and one photon.}
\end{center}
\end{figure}

\thispagestyle{appendixd}
\begin{eqnarray}
\bra P^0_{I=1} (p_P) A^- (p_A,\varepsilon) | \bar{d}\gamma^\mu  u | 0\ket \,&=& \frac{i\,\sqrt{2}}{M_A} \Big\{
( q\varepsilon^* \, p_A^\mu \, -\, qp_A\, {\varepsilon^*}^\mu )\, \mathcal{F}^{\,v}_{PA} (q^2) \nonumber \\
&&+( q\varepsilon^* \, p_P^\mu \, -\, qp_P {\varepsilon^*}^\mu )\, \mathcal{G}^{\,v}_{PA} (q^2) \Big\} \, ,
\end{eqnarray}
\begin{eqnarray}
\mathcal{F}^{\,v}_{PA} (q^2) &=& 4\lambda^{PA}_1  +\frac{\sqrt{2}F_V}{M_V^2-q^2} \bigg[ 2\lambda^{PVA}_0+M_A^2\lambda^{PVA}_1+\frac{q^2+M_A^2-M_P^2}{2} (\lambda^{PVA}_2+2\lambda^{PVA}_3)\bigg. \nonumber \\
&& \bigg.  -M_A^2(2\lambda^{PVA}_4+\lambda^{PVA}_5)-q^2\lambda^{PVA}_6\bigg] \,,  \nonumber \\
\mathcal{G}^{\,v}_{PA} (q^2) &=& \frac{\sqrt{2}F_VM_A^2}{M_V^2-q^2} \lambda^{PVA}_1    \, , 
\end{eqnarray} 
\begin{align}
 \mathrm{Im}  \Pi_{_{VV}} &(q^2) |_{PA}  \,=\, \theta(q^2-(M_P+M_A)^2) \, \frac{\lambda^{1/2}(q^2,M_P^2,M_A^2)}{48\pi q^2}\, \, \Bigg\{\frac{1}{M_A^2 q^2} \Big[ (M_A^2-M_P^2)^2    \nonumber \\
&  -2q^2(M_P^2-2M_A^2)+q^4\Big]  |\mathcal{F}^{\,v}_{PA}|^2 + \frac{1}{ 2M_A^4q^2} \Big[2M_A^2(M_A^2-M_P^2)^2 \nonumber \\
& \Big. + q^2(-3M_A^4+6M_P^2M_A^2+ M_P^4)-2M_P^2q^4 +q^6\Big]  |\mathcal{G}^{\,v}_{PA}|^2 \nonumber \\
&+ 
\frac{1}{M_A^2 q^2} \!\Big[ -2(M_A^2-M_P^2)^2 -2q^2(M_A^2+M_P^2)+4q^4 \Big]\! \mathrm{Re} \!\{ \mathcal{F}^{\,v}_{PA} {\mathcal{G}^{\,v}_{PA}}^* \} \! \Bigg\}  ,
\end{align} 
\begin{equation}
\lambda_2^{PVA}+2\lambda_3^{PVA}-2\lambda_6^{PVA}\,=\,\frac{4\sqrt{2}}{F_V} \lambda_1^{PA}   \, ,
\qquad \qquad \lambda_1^{PVA}\,=\,0\, , \label{VPA}
\end{equation}
\begin{eqnarray}
\mathcal{\tilde{F}}^{\,v}_{PA} (q^2) &=& \frac{\sqrt{2}F_V}{M_V^2-q^2} \bigg[\frac{4M_V^2}{\sqrt{2}F_V} \lambda_1^{PA} +2\lambda_0^{PVA} +\frac{M_A^2-M_P^2}{2}\left(\lambda_2^{PVA}+2\lambda_3^{PVA}\right)\nonumber \\ && -M_A^2 \left( 2\lambda_4^{PVA}+\lambda_5^{PVA}\right)\bigg] \, , \qquad \qquad 
\mathcal{\tilde{G}}^{\,v}_{PA} (q^2) \,=\, 0 \,.
\end{eqnarray}
%\newpage
\subsubsection{Vector Form Factor to V$\gamma$ (Figure \ref{AR1F1})}
\thispagestyle{appendixd}
\begin{align}
&\bra \gamma (p_\gamma, \varepsilon_\gamma) V^- (p_V,\varepsilon_V) | \bar{d}\gamma^\mu  u | 0\ket \,=\, \frac{\sqrt{2} \, e \,F_V}{M_V} \times \nonumber \\
&\times\!\! \bigg[\frac{1}{q^2}\Big\{ M_V^2 \, q\varepsilon_V^* (q p_\gamma \, {\varepsilon^*_\gamma}^\mu - q \varepsilon_\gamma^*  \, p_\gamma^\mu ) 
+ (qp_V \, p_\gamma^\mu - qp_\gamma \, p_V^\mu ) (qp_\gamma \, \varepsilon_V^* \varepsilon_\gamma^* - q\varepsilon_\gamma^* \, q \varepsilon_V^* ) \Big\} \mathcal{F}^{\,v}_{V\gamma} (q^2)
 \nonumber \\
&+\Big\{ M_V^2 \, q\varepsilon_V^* \, {\varepsilon_\gamma^*}^\mu -\varepsilon_\gamma^* \varepsilon_V^* (qp_V \, p_\gamma^\mu - qp_\gamma \, p_V^\mu )  + q\varepsilon_V^* \, q\varepsilon_\gamma^* (p_\gamma-p_V)^\mu \Big\} \mathcal{G}^{\,v}_{V\gamma} (q^2)     +\Big\{ q\varepsilon_\gamma^* \, {\varepsilon_V^*}^\mu \!           \nonumber \\
&- q\varepsilon_V^* \, {\varepsilon_\gamma^*}^\mu\!   + \varepsilon_V^* \varepsilon_\gamma^*\, (p_\gamma^\mu -p_V^\mu) +\frac{2}{M_V^2-q^2}\! \left(q\varepsilon_\gamma^* \,qp_V \, {\varepsilon_V^*}^\mu \!- q\varepsilon_V^* \, q\varepsilon_\gamma^* \, p_V^\mu   \right)\! \Big\}
   \bigg] \, ,
\end{align}
\begin{align}
\mathcal{F}^{\,v}_{V\gamma} (q^2) \,=\,& \frac{2\sqrt{2} \, F_V\, q^2}{(M_V^2-q^2)M_V^2} \big[ 2 \lambda^{VVV}_1\!\!- \lambda^{VVV}_3\!\! - \lambda^{VVV}_4\!\! +2 \lambda^{VVV}_5\!\! + \lambda^{VVV}_6\!\! - \lambda^{VVV}_7\! \big] \, ,\nonumber 
\end{align}
\begin{align}
\mathcal{G}^{\,v}_{V\gamma} (q^2) \,=\,& \frac{\sqrt{2} \, F_V}{(M_V^2-q^2)M_V^2} \big[ 3\lambda^{VVV}_0\!\!+2qp_V\left(\lambda^{VVV}_2\!\! +\lambda^{VVV}_6\!\!   + \lambda^{VVV}_7\!\right)\big] \nonumber \\
&\, + \frac{2\lambda^{VV}_7}{M_V^2} + \frac{1}{M_V^2-q^2} \big[  2\lambda^{VV}_7-1 \big] \, , 
\end{align} 
\begin{align}
& \mathrm{Im}  \Pi_{_{VV}} (q^2) |_{V\gamma}  \,\propto\,  \Big[ 
\left(\frac{q^4}{8}-\frac{3M_V^2q^2}{8}+\frac{M_V^4}{4} \right)  |\mathcal{F}^{\,v}_{V\gamma}|^2 +
\left( \frac{q^4}{2}-\frac{M_V^2q^2}{2}-\frac{M_V^4}{2} \right)  |\mathcal{G}^{\,v}_{V\gamma}|^2 \nonumber \\
&+\left( -\frac{q^4}{2} +2M_V^2q^2 -3M_V^4 \right) \mathrm{Re} \{ \mathcal{F}^{\,v}_{V\gamma} {\mathcal{G}^{\,v}_{V\gamma}}^* \} + 
\left( \frac{q^2}{2}-\frac{3M_V^2}{2} \right) \mathrm{Re} \{ \mathcal{F}^{\,v}_{V\gamma} \}\nonumber \\
& +
\left(- 3 q^2 +6M_V^2 \right)  \mathrm{Re} \{ \mathcal{G}^{\,v}_{V\gamma} \} +\left(\frac{q^2}{2M_V^2} +1\right) +\cO\left(\frac{1}{q^2}\right)  \Big] \, ,
\end{align} 
%Following these results we have to constraint that
%\begin{eqnarray}
%\mathcal{F}^{\,v}_{V\gamma}-\frac{1}{M_V^2} \propto \cO\left(\frac{1}{q^2} \right) \, , \nonumber \\
%\mathcal{G}^{\,v}_{V\gamma}-\frac{1}{2M_V^2} \propto \cO\left(\frac{1}{q^2} \right) \, , \nonumber \\
%q^2 \left( \mathcal{F}^{\,v}_{V\gamma} -2\mathcal{G}^{\,v}_{V\gamma}\right) +3 \propto \cO\left(\frac{1}{q^2} \right) \,  \nonumber 
%\end{eqnarray}
%
%\vspace{0.2cm}
\begin{align}
-2\lambda^{VVV}_1+\lambda^{VVV}_3+\lambda^{VVV}_4-2\lambda^{VVV}_5-\lambda^{VVV}_6+\lambda^{VVV}_7&= \frac{1}{2\sqrt{2}F_V}\,,\, \mathrm{[cf\,\ref{VVV}]}  \nonumber \\
\lambda^{VVV}_2+\lambda^{VVV}_6+\lambda^{VVV}_7&=\frac{\sqrt{2}}{F_V}\lambda^{VV}_7 -\frac{1}{2\sqrt{2}F_V}\, , \, \mathrm{[cf\,\ref{VVV}]}   \nonumber \\
\lambda^{VV}_7 &=-\frac{F_V\, \lambda^{VVV}_0}{\sqrt{2} \,M_V^2} \, ,\, \mathrm{[cf\,\ref{VVV}]} \label{VVF}
\end{align}
\begin{equation}
\mathcal{\tilde{F}}^{\,v}_{V\gamma}\,=\,-\frac{q^2}{(M_V^2-q^2)M_V^2}\,, \qquad \mathcal{\tilde{G}}^{\,v}_{V\gamma}\,=\,-\frac{3M_V^2+q^2}{2(M_V^2-q^2)M_V^2} \, .
\end{equation}
\subsubsection{Vector Form Factor to S$\gamma$ (Figure \ref{AR1F1})}
\thispagestyle{appendixd}
\begin{eqnarray}
\bra \gamma (p_\gamma, \varepsilon) S^- (p_S) | \bar{d}\gamma^\mu  u | 0\ket \,&=& \frac{\sqrt{2}\, e \,F_V }{3} 
\left( q\varepsilon^* \, p_\gamma^\mu \, -\, qp_\gamma\, {\varepsilon^*}^\mu \right)\, \mathcal{F}^{\,v}_{S\gamma} (q^2)  \, ,
\end{eqnarray}
\vspace{-0.4cm}
\begin{eqnarray}
\mathcal{F}^{\,v}_{S\gamma} (q^2) &=&4 \lambda^{SV}_3 \left( \frac{1}{M_V^2-q^2}+\frac{1}{M_V^2}\right)  +\frac{\sqrt{2}F_V}{M_V^2(M_V^2-q^2)} \bigg[ -2\lambda^{SVV}_0\bigg. \nonumber \\
&& \bigg. -\frac{q^2-M_S^2}{2} (\lambda^{SVV}_2+2\lambda^{SVV}_3) +q^2(2\lambda^{SVV}_4+\lambda^{SVV}_5)\bigg] \,,  
\end{eqnarray} 
\begin{align}
 \mathrm{Im}  \Pi_{_{VV}} (q^2) |_{S\gamma}  &=\, \theta(q^2-M_S^2) \, F_V^2\, e^2 \,\frac{(1-M_S^2/q^2)^3}{432\pi} \,q^2\, |\mathcal{F}^{\,v}_{S\gamma}|^2 ,
\end{align} 
\begin{eqnarray}
\lambda_2^{SVV}+2\lambda_3^{SVV}-4\lambda_4^{SVV}-2\lambda_5^{SVV}&=&-\frac{4\sqrt{2}}{F_V} \lambda_3^{SV}   \, , \, \qquad \, \mathrm{[cf\,\ref{VSV}]} \label{VSF}
\end{eqnarray}
\begin{align}
\mathcal{\tilde{F}}^{\,v}_{S\gamma} (q^2) &=\,
\frac{\sqrt{2}F_V}{(M_V^2-q^2)M_V^2} 
\left[
\frac{8\, M_V^2}{\sqrt{2}\, F_V}\,  \lambda_3^{SV}  
- 2\lambda_0^{SVV}+\frac{M_S^2}{2}\left(\lambda_2^{SVV}+2\lambda_3^{SVV}\right) \right] \, . 
\end{align}

\newpage
\subsection{Axial Form Factors}
\begin{figure}
\begin{center}
\includegraphics[scale=0.8]{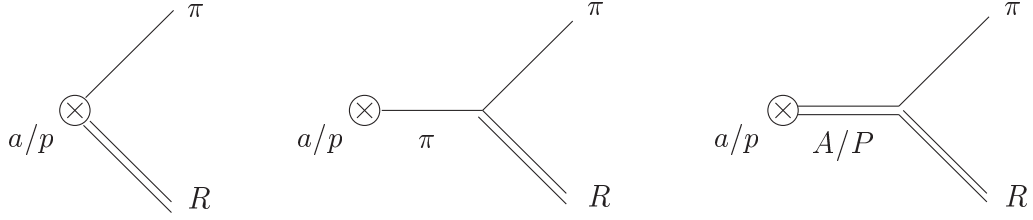}
\caption{\label{BR1F0}
Tree-level contributions to the axial/pseudoscalar form factor to one resonance field and one pseudo-Goldstone.}
\end{center}
\end{figure}

\thispagestyle{appendixd}
\subsubsection{Axial Form Factor to V$\pi$ (Figure \ref{BR1F0})}
\begin{eqnarray}
\bra V^0_{I=1} (p_V,\varepsilon) \pi^- (p_\pi) | \bar{d}\gamma^\mu \gamma_5 u | 0\ket \,&=& \frac{i\, \sqrt{2}}{M_V} \Big\{
( q\varepsilon^* \, p_V^\mu \, -\, qp_V\, {\varepsilon^*}^\mu )\, \mathcal{F}^{\,a}_{V\pi} (q^2) \nonumber \\
&&+( q\varepsilon^* \, p_\pi^\mu \, -\, qp_\pi {\varepsilon^*}^\mu )\, \mathcal{G}^{\,a}_{V\pi} (q^2) \Big\} \, ,
\end{eqnarray}
\begin{eqnarray}
\mathcal{F}^{\,a}_{V\pi} (q^2) &=& -\frac{F_V}{F}+\frac{2G_V}{F}-\frac{2G_V}{F}\frac{M_V^2}{q^2} + \frac{F_A}{F}\frac{q^2}{M_A^2-q^2}\Big[(-\frac{2M_V^2}{q^2}+2) \lambda^{VA}_2\nonumber \\
&&+(\frac{M_V^2}{q^2}-1) \lambda^{VA}_4+ (\frac{2M_V^2}{q^2}-2) \lambda^{VA}_5 \Big] \, , \nonumber \\ \nonumber \\
\mathcal{G}^{\,a}_{V\pi} (q^2) &=& -\frac{2G_V}{F}\frac{M_V^2}{q^2}+ \frac{2F_A}{F}\frac{M_V^2}{M_A^2-q^2} \Big[ -2\lambda^{VA}_2+\lambda^{VA}_3 \Big] \, , 
\end{eqnarray} 
\begin{eqnarray}
\mathrm{Im}  \Pi_{_{AA}} (q^2) |_{V\pi}  &=& \theta(q^2-M_V^2) \frac{1-M_V^2/q^2}{48\pi} \Bigg\{ \left( \frac{M_V^2}{q^2}+4+\frac{q^2}{M_V^2} \right) |\mathcal{F}^{\,a}_{V\pi}|^2 \nonumber \\
&&+ (1-M_V^2/q^2)^2 \left( \frac{q^2}{M_V^2} + \frac{q^4}{2M_V^4}\right) |\mathcal{G}^{\,a}_{V\pi}|^2 \nonumber \\
&&+2 (1-M_V^2/q^2) \left( 1+\frac{2q^2}{M_V^2} \right)\mathrm{Re} \{ \mathcal{F}^{\,a}_{V\pi} {\mathcal{G}^{\,a}_{V\pi}}^* \}  \Bigg\} \, ,
\end{eqnarray} 
\begin{equation}
2\lambda^{VA}_2-\lambda^{VA}_4-2\lambda^{VA}_5\,=\,-\frac{F_V}{F_A}+\frac{2G_V}{F_A} \,  , \qquad
-2\lambda^{VA}_2+\lambda^{VA}_3\,=\,-\frac{G_V}{F_A} \, ,\label{AVG}
\end{equation}
\begin{eqnarray}
\mathcal{\tilde{F}}^{\,a}_{V\pi} (q^2) &=&\left(\frac{F_V}{F}-\frac{2G_V}{F}\right) \frac{M_V^2-M_A^2}{M_A^2-q^2} -\frac{2G_V}{F}\frac{M_V^2}{q^2} \, , \nonumber \\
\mathcal{\tilde{G}}^{\,a}_{V\pi} (q^2) &=&-\frac{2G_V}{F}\frac{M_V^2 M_A^2}{(M_A^2-q^2)q^2} \, .
\end{eqnarray}

\newpage
\subsubsection{Axial Form Factor to S$\pi$ (Figure \ref{BR1F0})}
\begin{figure}
\begin{center}
\includegraphics[scale=0.8]{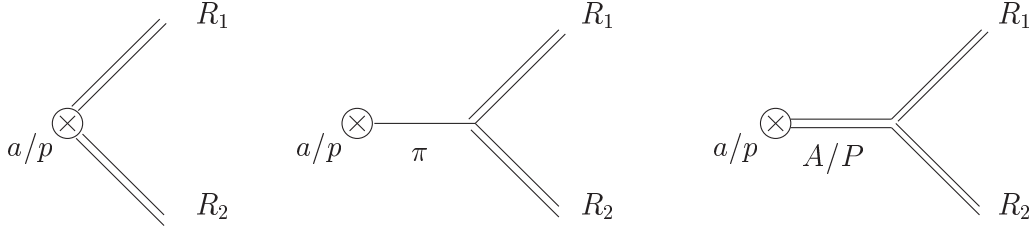}
\caption{\label{BR2F0}
Tree-level contributions to the axial/pseudoscalar form factor to two resonances.}
\end{center}
\end{figure}

\thispagestyle{appendixd}
\begin{eqnarray}
\bra S^0_{I=0} (p_S) \pi^- (p_\pi) | \bar{d}\gamma^\mu \gamma_5 u | 0\ket \,&=&-2\, i \,\mathcal{F}^{\,a}_{S\pi} (q^2) \left( g^{\mu\nu}-\frac{q^\mu q^\nu}{q^2} \right) {p_\pi}_\nu \, , \\
\mathcal{F}^{\,a}_{S\pi} (q^2) &=& \frac{2c_d}{F}-\frac{\sqrt{2}\,F_A}{F} \frac{q^2}{M_A^2-q^2}\lambda^{SA}_1 \, , \\
\mathrm{Im}  \Pi_{_{AA}} (q^2) |_{S\pi}  &=& \theta(q^2-M_S^2) \frac{(1-M_S^2/q^2)^3}{48\pi} |\mathcal{F}^{\,a}_{S\pi} (q^2)|^2 \, , \\
\lambda^{SA}_1&=&-\frac{\sqrt{2}c_d}{F_A} \, , \\
\mathcal{\tilde{F}}^{\,a}_{S\pi} (q^2) &=&\frac{2c_d}{F}\frac{M_A^2}{M_A^2-q^2} \, .
\end{eqnarray}
\subsubsection{Axial Form Factor to VA (Figure \ref{BR2F0})}
\begin{align}
&\bra V^0_{I=1} (p_V,\varepsilon_V) A^- (p_A,\varepsilon_A) | \bar{d}\gamma^\mu \gamma_5 u | 0\ket \,=\, \frac{\sqrt{2}}{M_VM_A}\frac{1}{2q^2} \times  \nonumber \\
& \qquad \times \Big\{2\left( qp_A p_V^\mu - qp_V p_A^\mu \right) \Big[ p_A p_V \varepsilon_A^* \varepsilon_V^*  -q\varepsilon_A^* q\varepsilon_V^*\, \Big] \mathcal{F}^{\,a}_{VA} (q^2) \nonumber \\
&\qquad \qquad+ 2M_V^2 \Big[ \left( qp_A p_V^\mu - qp_V p_A^\mu \right)\varepsilon_A^* \varepsilon_V^*-\left( p_V^\mu+p_A^\mu  \right)q\varepsilon_A^* q\varepsilon_V^* +q^2 q\varepsilon_V^* {\varepsilon_A^*}^\mu\,\Big]  \mathcal{G}^{\,a}_{VA} (q^2) \nonumber \\
&\qquad \qquad +2M_A^2 \Big[ \left( qp_A p_V^\mu - qp_V p_A^\mu \right)\varepsilon_A^* \varepsilon_V^*+\left( p_V^\mu+p_A^\mu  \right)q\varepsilon_A^* q\varepsilon_V^* -q^2 q\varepsilon_A^* {\varepsilon_V^*}^\mu\,\Big]  \mathcal{H}^{\,a}_{VA} (q^2) \nonumber \\
&\qquad \qquad +\Big[\left(M_V^2+M_A^2 \right) \left( qp_A p_V^\mu - qp_V p_A^\mu \right)\varepsilon_A^* \varepsilon_V^*+ \left( M_V^2+M_A^2 \right) \left(p_A^\mu
-p_V^\mu \right)  q\varepsilon_A^* q\varepsilon_V^* \Big. \nonumber \\
&\qquad \qquad \qquad \quad \Big.\Big. + \left( M_V^2-M_A^2 \right)  \left( M_A^2 \, q \varepsilon_A^* {\varepsilon_V^*}^\mu +M_V^2 \, q \varepsilon_V^* {\varepsilon_A^*}^\mu\right)\Big]  \mathcal{I}^{\,a}_{VA} (q^2) \Big\} \, ,
\end{align}
\vspace{-0.25cm}
\begin{align}
&\mathcal{F}^{\,a}_{VA} (q^2) = 2 \lambda^{VA}_4\!+4\lambda^{VA}_5\!+4\lambda^{VA}_6\! -\frac{\sqrt{2}F_A}{M_A^2-q^2} \Big[
(M_A^2-M_V^2)(-2\lambda^{VAA}_1\!+\lambda^{VAA}_4\!- \lambda^{VAA}_5\!)\nonumber \\
&-2 \lambda^{VAA}_0\!\! -4qp_A \lambda^{VAA}_2\!\! -2M_V^2\lambda^{VAA}_3\!\!+(q^2+M_A^2/2+3M_V^2/2)\lambda^{VAA}_6\!\! -(M_A^2+M_V^2) \lambda^{VAA}_8\!\!\nonumber \\
& +(q^2+M_A^2/2-M_V^2/2)\left( - \lambda^{VAA}_7\!+ \lambda^{VAA}_9\! + 2 \lambda^{VAA}_{10}\!-  \lambda^{VAA}_{12}\! -2 \lambda^{VAA}_{13}\! +\lambda^{VAA}_{14} \right)  \Big]\, , \nonumber
\end{align}
\begin{align}
&\mathcal{G}^{\,a}_{VA} (q^2) =-2 \lambda^{VA}_2\!\!+2\lambda^{VA}_3\!\!+2\lambda^{VA}_6\!\! -\frac{\sqrt{2}F_A}{M_A^2-q^2}\! \Big[
-\lambda^{VAA}_0\!\! -2qp_A \lambda^{VAA}_2\!\! +(M_A^2-M_V^2)/4 \times \nonumber \\
& \!\left( -\lambda^{VAA}_7\!\! +\! \lambda^{VAA}_9\!\!  +\! 2\lambda^{VAA}_{10}\!\!  -\! \lambda^{VAA}_{12}\!\!  -\!
2\lambda^{VAA}_{13}\!\! + \!\lambda^{VAA}_{14}\!\!  \right)\! -M_V^2\lambda^{VAA}_3\!\! + (M_A^2/4 +3M_V^2/4)\lambda^{VAA}_6\! \!\nonumber \\
&+(-q^2-M_A^2/2-M_V^2/2) \lambda^{VAA}_{8}\!\! 
+(q^2\!+M_A^2/2\!-M_V^2/2) \left( -2\lambda^{VAA}_1\!\!+ \lambda^{VAA}_4\!\!- \lambda^{VAA}_5\!\right) \! \Big]  ,\!\! \!\! \!\! \!\! \nonumber\\
&\mathcal{H}^{\,a}_{VA} (q^2) \,=\,2 \lambda^{VA}_2+2\lambda^{VA}_6 -\frac{\sqrt{2}F_A}{M_A^2-q^2} \Big[
- \lambda^{VAA}_0\! -2qp_A  \lambda^{VAA}_2\!   -M_V^2\lambda^{VAA}_3\! \nonumber \\
&+(M_A^2-M_V^2)/2 \left( -2\lambda^{VAA}_1\! + \lambda^{VAA}_4\! - \lambda^{VAA}_5\! \right)  
+(-2q^2+3M_V^2+M_A^2) \lambda^{VAA}_6\! /4 \nonumber \\
&-2q^2\lambda^{VAA}_{11}\!+(2q^2\!\!+M_V^2\!\!-M_A^2)/4\! \left( \lambda^{VAA}_7\!\! - \lambda^{VAA}_9\!\! -2 \lambda^{VAA}_{10}\!\! +\lambda^{VAA}_{12}\!\! - \lambda^{VAA}_{14}\! \right)\! \nonumber \\
&- (M_V^2\!+M_A^2) \lambda^{VAA}_8\! /2-(q^2+M_A^2/2-M_V^2/2)\lambda^{VAA}_{13}\!\Big]  ,\!\! \!\! \!\! \nonumber \\  
&\mathcal{I}^{\,a}_{VA} (q^2) \,=\,-\frac{F_A\,q^2}{\sqrt{2}(M_A^2-q^2)} \Big[ +4 \lambda^{VAA}_{1}\! -2\lambda^{VAA}_{4}\! +2\lambda^{VAA}_{5}\! +\lambda^{VAA}_{6}\! - \lambda^{VAA}_{7}\! +2\lambda^{VAA}_{8}\!  \nonumber \\
& + \lambda^{VAA}_{9}\!+2 \lambda^{VAA}_{10}\! - \lambda^{VAA}_{12}\! -2\lambda^{VAA}_{13}\! + \lambda^{VAA}_{14}\! \Big] \, ,
\end{align} 
%\vspace{-0.5cm}
\thispagestyle{appendixd}
\begin{align}
& \mathrm{Im}  \Pi_{_{AA}} (q^2) |_{VA}  \,\propto\,  \Big[ 
\left( q^4/8 + \cO (q^2)\right)  |\mathcal{F}^{\,a}_{VA}|^2 +\cO (q^2)\, |\mathcal{G}^{\,a}_{VA}|^2 +\cO (q^2)\, |\mathcal{H}^{\,a}_{VA}|^2 \nonumber \\
&+\left( (M_A^4+4M_V^2M_A^2+M_V^4)/8+\cO (q^{-2})\right)  |\mathcal{I}^{\,a}_{VA}|^2 
+\cO (q^2)\,\mathrm{Re} \{ \mathcal{F}^{\,a}_{VA} {\mathcal{G}^{\,a}_{VA}}^* \} \nonumber \\ &
+\cO (q^2)\,\mathrm{Re} \{ \mathcal{F}^{\,a}_{VA} {\mathcal{H}^{\,a}_{VA}}^* \} 
+\left( q^2(M_A^2+M_V^2)/4  + \cO (q^0)\right) \mathrm{Re} \{ \mathcal{F}^{\,a}_{VA} {\mathcal{I}^{\,a}_{VA}}^* \} \nonumber \\ &
+\cO (q^0)\,\mathrm{Re} \{ \mathcal{G}^{\,a}_{VA} {\mathcal{H}^{\,a}_{VA}}^* \} 
+\cO (q^0)\,\mathrm{Re} \{ \mathcal{G}^{\,a}_{VA} {\mathcal{I}^{\,a}_{VA}}^* \} 
+\cO (q^0)\,\mathrm{Re} \{ \mathcal{H}^{\,a}_{VA} {\mathcal{I}^{\,a}_{VA}}^* \}  \Big] \, ,
\end{align} 
\begin{align}
& -2\lambda^{VAA}_{2}\!   + \lambda^{VAA}_{6}\! -\lambda^{VAA}_{7}\! + \lambda^{VAA}_{9}\! +2\lambda^{VAA}_{10}\! 
- \lambda^{VAA}_{12}\! -2\lambda^{VAA}_{13}\! +\lambda^{VAA}_{14}\! \,
\nonumber \\ &  \qquad 
=\,\frac{1}{\sqrt{2}F_A} \Big\{ -2\lambda^{VA}_{4}\! -4\lambda^{VA}_{5 }\! -4\lambda^{VA}_{6 }\! \Big\} \, ,  \nonumber \\
\nonumber \\
&M_V^2 \Big\{ 4\lambda^{VAA}_{1}\! +4\lambda^{VAA}_{2}\! -4\lambda^{VAA}_{3}\! -2\lambda^{VAA}_{4}\! +2 \lambda^{VAA}_{5}\! +3 \lambda^{VAA}_{6}\! +\lambda^{VAA}_{7}\! -2\lambda^{VAA}_{8}\! - \lambda^{VAA}_{9}\! 
\nonumber \\ &  \qquad -2\lambda^{VAA}_{10}\! 
+ \lambda^{VAA}_{12}\! +2 \lambda^{VAA}_{13}\! -\lambda^{VAA}_{14}\! \Big\} +
M_A^2 \Big\{ -4\lambda^{VAA}_{1}\! -4\lambda^{VAA}_{2}\!  +2 \lambda^{VAA}_{4}\!
\nonumber \\ &  \qquad  -2 \lambda^{VAA}_{5}\!+\lambda^{VAA}_{6}\! -\lambda^{VAA}_{7}\! -2 \lambda^{VAA}_{8}\! +\lambda^{VAA}_{9}\! 
+ 2\lambda^{VAA}_{10}\! - \lambda^{VAA}_{12}\! -2\lambda^{VAA}_{13}\! + \lambda^{VAA}_{14}\! \Big\} 
\nonumber \\ & \qquad -4\lambda^{VAA}_{0}\!\,=\,
\frac{2\sqrt{2}M_A^2}{F_A} \Big\{ \lambda^{VA}_{4}\! +2\lambda^{VA}_{5 }\! +2\lambda^{VA}_{6 }\! \Big\} \, ,  \nonumber \\
\nonumber \\
&-2\lambda^{VAA}_{1}\! -\lambda^{VAA}_{2}\!  +\lambda^{VAA}_{4}\! - \lambda^{VAA}_{5}\!  -\lambda^{VAA}_{8}\!  \,=\,\frac{\sqrt{2}}{F_A} \Big\{ \lambda^{VA}_{2}\! -\lambda^{VA}_{3 }\! -\lambda^{VA}_{6}\! \Big\} \, ,  \nonumber \\
\nonumber \\
& -2\lambda^{VAA}_{2}\! - \lambda^{VAA}_{6}\! +\lambda^{VAA}_{7}\!  - \lambda^{VAA}_{9}\! -2\lambda^{VAA}_{10}\! 
\nonumber \\ & \qquad 
-4\lambda^{VAA}_{11}\! + \lambda^{VAA}_{12}\! -2\lambda^{VAA}_{13}\! -\lambda^{VAA}_{14}\! \,=\,-2\sqrt{2}/F_A \left\{ \lambda^{VA}_{2}\! +\lambda^{VA}_{6 }\! \right\} \, ,  \nonumber 
\end{align}
\begin{align}
&4\lambda^{VAA}_{1}\! -2\lambda^{VAA}_{4}\! +2 \lambda^{VAA}_{5}\! + \lambda^{VAA}_{6}\! -\lambda^{VAA}_{7}\! +2\lambda^{VAA}_{8}\! + \lambda^{VAA}_{9}\! +2\lambda^{VAA}_{10}\!  
\nonumber \\ &\qquad  - \lambda^{VAA}_{12}\! -2\lambda^{VAA}_{13}\!+\lambda^{VAA}_{14}\! \,=\, 0 \, ,\label{AVA}
\end{align}
%\vspace{0.5cm}
\begin{align}
&\mathcal{\tilde{F}}^{\,a}_{VA} (q^2) =\mathcal{\tilde{I}}^{\,a}_{VA} (q^2) \,=\,0\,,\nonumber\phantom{\frac{1}{2}}\\
&\mathcal{\tilde{G}}^{\,a}_{VA} (q^2) =-\frac{\sqrt{2}F_A}{M_A^2-q^2}\! \Bigg\{
\frac{\sqrt{2}M_A^2}{F_A} \!\left(  \lambda^{VA}_2\!-\lambda^{VA}_3\!-\lambda^{VA}_6 \!\right)\!
-\lambda^{VAA}_0\!\!  +(M_A^2-M_V^2)/4 \times \nonumber \\
&\qquad  \times \left(-4\lambda^{VAA}_1\!\!+ 2\lambda^{VAA}_4\!\!-2 \lambda^{VAA}_5\! -\lambda^{VAA}_7\!\! + \lambda^{VAA}_9\!\!  + 2\lambda^{VAA}_{10}\!\!  - \lambda^{VAA}_{12}\!\!  -2\lambda^{VAA}_{13}\!\! + \lambda^{VAA}_{14}\!  \right) \nonumber\phantom{\frac{1}{2}} \\&\qquad-M_A^2 \lambda^{VAA}_2\!\!
-M_V^2\lambda^{VAA}_3\! + (M_A^2/4 +3M_V^2/4)\lambda^{VAA}_6\!
-(M_A^2/2+M_V^2/2) \lambda^{VAA}_{8}\! \Bigg\} \, ,\phantom{\frac{1}{2}} \nonumber\\
&\mathcal{\tilde{H}}^{\,a}_{VA} (q^2) =\mathcal{\tilde{G}}^{\,a}_{VA} (q^2)
 \quad + \quad 
\frac{2M_A^2}{M_A^2-q^2} \left[ 2\lambda^{VA}_2-\lambda^{VA}_3\right] . 
\end{align}
\subsubsection{Axial Form Factor to PV (Figure \ref{BR2F0})}
\thispagestyle{appendixd}
\begin{eqnarray}
\bra P^0_{I=1} (p_P) V^- (p_V,\varepsilon) | \bar{d}\gamma^\mu \gamma_5 u | 0\ket \,&=& \frac{\sqrt{2}\,i}{M_V} \Big\{
( q\varepsilon^* \, p_V^\mu \, -\, qp_V\, {\varepsilon^*}^\mu )\, \mathcal{F}^{\,a}_{PV} (q^2) \nonumber \\
&&+( q\varepsilon^* \, p_P^\mu \, -\, qp_P {\varepsilon^*}^\mu )\, \mathcal{G}^{\,a}_{PV} (q^2) \Big\} \, ,
\end{eqnarray}
\vspace{-0.5cm}
\begin{align}
\mathcal{F}^{\,a}_{PV} (q^2) &=\, 2\lambda^{PV}_1 \left( \frac{M_V^2}{q^2} -1 \right) -4 \lambda^{PV}_2 +\frac{\sqrt{2}F_A}{M_A^2-q^2} \bigg[ 2\lambda^{PVA}_0+M_V^2\lambda^{PVA}_1\bigg. \nonumber \\
& \bigg. +\frac{q^2+M_V^2-M_P^2}{2} (\lambda^{PVA}_2+2\lambda^{PVA}_3) -q^2 (2\lambda^{PVA}_4+\lambda^{PVA}_5)-M_V^2  \lambda^{PVA}_6 \bigg] \,,  \nonumber \\
\mathcal{G}^{\,a}_{PV} (q^2) &= \frac{2M_V^2}{q^2}\lambda^{PV}_1+\frac{\sqrt{2}F_A}{M_A^2-q^2}\left(M_V^2\lambda^{PVA}_1\right)    \, , 
\end{align} 
%
%\vspace{-0.3cm}
\begin{align}
 \mathrm{Im}  \Pi_{_{AA}} &(q^2) |_{PV}  = \theta(q^2-(M_P+M_V)^2)  \frac{\lambda^{1/2}(q^2,M_P^2,M_V^2)}{48\pi q^2} \Bigg\{\frac{1}{M_V^2 q^2} \Big[ (M_V^2-M_P^2)^2    \nonumber \\
&  -2q^2(M_P^2-2M_V^2)+q^4\Big]  |\mathcal{F}^{\,a}_{PV}|^2 + \frac{1}{ 2M_V^4q^2} \Big[2M_V^2(M_V^2-M_P^2)^2 \nonumber \\
& \Big. + q^2(-3M_V^4+6M_P^2M_V^2+ M_P^4)-2M_P^2q^4 +q^6\Big]  |\mathcal{G}^{\,a}_{PV}|^2  \phantom{\frac{1}{2}}\nonumber \\
&+ \!\frac{1}{M_V^2 q^2} \!\Big[ \! -2(M_V^2-M_P^2)^2 \!-\! 2q^2(M_V^2+M_P^2)\!+\! 4\,q^4  \Big]  \mathrm{Re} \{ \mathcal{F}^{\,a}_{PV} {\mathcal{G}^{\,a}_{PV}}^* \} \! \Bigg\}  ,
\end{align} 
\vspace{-0.3cm}
\begin{equation}
\lambda_2^{PVA}\!\!+2\lambda_3^{PVA}\!\!-4 \lambda_4^{PVA}\!\!-2\lambda_5^{PVA}=-\frac{2\sqrt{2}}{F_A}\left(\lambda_1^{PV}\!+2\lambda_2^{PV}\! \right)  \, , \
\lambda_1^{PVA}=\frac{\sqrt{2} \lambda_1^{PV}}{F_A}\, , \label{APV}
\end{equation}
\vspace{-0.3cm}
\begin{eqnarray}
\mathcal{\tilde{F}}^{\,a}_{PV} (q^2) &=&\frac{\sqrt{2}F_A}{M_A^2-q^2} \left[ \frac{\sqrt{2}}{F_A} \left( \frac{M_A^2M_V^2}{q^2}-M_A^2 \right)\lambda_1^{PV}-\frac{2\sqrt{2}M_A^2}{F_A} \lambda_2^{PV}+2 \lambda_0^{PVA}+\right. \nonumber \\
&&\left.\frac{M_V^2-M_P^2}{2} \left(\lambda_2^{PVA}+2\lambda_3^{PVA}\right) -M_V^2 \lambda_6^{PVA}\right] \, , \nonumber \\
\mathcal{\tilde{G}}^{\,a}_{PV} (q^2) &=&\frac{2M_V^2M_A^2}{(M_A^2-q^2)q^2} \lambda_1^{PV} \,.
\end{eqnarray}

\subsubsection{Axial Form Factor to SA (Figure \ref{BR2F0})}
\thispagestyle{appendixd}
\begin{eqnarray}
\bra S^0_{I=0} (p_S) A^- (p_A,\varepsilon) | \bar{d}\gamma^\mu \gamma_5 u | 0\ket \,&=& \frac{\sqrt{2}}{M_A} \Big\{
( q\varepsilon^* \, p_A^\mu \, -\, qp_A\, {\varepsilon^*}^\mu )\, \mathcal{F}^{\,a}_{SA} (q^2) \nonumber \\
&&+( q\varepsilon^* \, p_S^\mu \, -\, qp_S {\varepsilon^*}^\mu )\, \mathcal{G}^{\,a}_{SA} (q^2) \Big\} \, ,
\end{eqnarray}
\begin{eqnarray}
\mathcal{F}^{\,a}_{SA} (q^2) &=& 2\lambda^{SA}_1 \left( \frac{M_A^2}{q^2} -1 \right) -4 \lambda^{SA}_2 +\frac{\sqrt{2}F_A}{M_A^2-q^2} \bigg[ 2\lambda^{SAA}_0+M_A^2\lambda^{SAA}_1\bigg. \nonumber \\
&& \bigg. +\frac{q^2+M_A^2-M_S^2}{2} (\lambda^{SAA}_2+2\lambda^{SAA}_3) -(M_A^2+q^2)(2\lambda^{SAA}_4+\lambda^{SAA}_5)\bigg] \,,  \nonumber \\
\mathcal{G}^{\,a}_{SA} (q^2) &=& \frac{2M_A^2}{q^2}\lambda^{SA}_1+\frac{\sqrt{2}F_A}{M_A^2-q^2}\left(M_A^2\lambda^{SAA}_1\right)    \, , 
\end{eqnarray} 
\begin{align}
 \mathrm{Im}  \Pi_{_{AA}}& (q^2) |_{SA}  = \theta(q^2-(M_S+M_A)^2) \, \frac{\lambda^{1/2}(q^2,M_S^2,M_A^2)}{48\pi q^2}\, \, \Bigg\{\frac{1}{M_A^2 q^2} \Big[ (M_A^2-M_S^2)^2    \nonumber \\
&  -2q^2(M_S^2-2M_A^2)+q^4\Big]  |\mathcal{F}^{\,a}_{SA}|^2 + \frac{1}{ 2M_A^4q^2} \Big[2M_A^2(M_A^2-M_S^2)^2 \nonumber \\
& \Big. + q^2(-3M_A^4+6M_S^2M_A^2+ M_S^4)-2M_S^2q^4 +q^6\Big]  |\mathcal{G}^{\,a}_{SA}|^2\nonumber \\
&  + \frac{1}{M_A^2 q^2} \! \Big[\! -\!2(M_A^2-M_S^2)^2 \!-\!2q^2(M_A^2+M_S^2)\!+\!4\,q^4 \Big] \mathrm{Re} \{ \mathcal{F}^{\,a}_{SA} {\mathcal{G}^{\,a}_{SA}}^* \} \! \Bigg\}  ,
\end{align} 
\begin{equation}
\lambda_2^{SAA}\!\!+2\lambda_3^{SAA}\!\!-4\lambda_4^{SAA}\!\!-2\lambda_5^{SAA}=-\frac{2\sqrt{2}}{F_A}\left(\lambda_1^{SA}\!+2\lambda_2^{SA} \!\right)  \, , \ 
\lambda_1^{SAA}=\frac{\sqrt{2} \lambda_1^{SA}}{F_A}\, , 
\end{equation}
\begin{eqnarray}
\mathcal{\tilde{F}}^{\,a}_{SA} (q^2) &=& \frac{\sqrt{2}F_A}{M_A^2-q^2} \left[ -\frac{\sqrt{2}}{F_A} \left(-\frac{M_A^4}{q^2} +3M_A^2 \right) \lambda_1^{SA} -\frac{4\sqrt{2} M_A^2}{F_A} \lambda_2^{SA} \right. \nonumber \\
&& +2\lambda_0^{SAA}+M_A^2\lambda_1^{SAA}-\frac{M_S^2}{2} \left( \lambda_2^{SAA}+2\lambda_3^{SAA}\right) \left. \right] \, , \nonumber \\
\mathcal{\tilde{G}}^{\,a}_{SA} (q^2) &=&\frac{2M_A^4}{(M_A^2-q^2)q^2} \lambda_1^{SA}\, .
\end{eqnarray}
\subsubsection{Axial Form Factor to SP (Figure \ref{BR2F0})}
\thispagestyle{appendixd}
\begin{eqnarray}
\bra S^0_{I=0} (p_S) P^- (p_P) | \bar{d}\gamma^\mu \gamma_5 u | 0\ket \,&=&-2\, i \,\mathcal{F}^{\,a}_{SP} (q^2) \left( g^{\mu\nu}-\frac{q^\mu q^\nu}{q^2} \right) {p_P}_\nu \, , \\
\mathcal{F}^{\,a}_{SP} (q^2) &=& \sqrt{2} \lambda_1^{SP}-  \frac{q^2}{M_A^2-q^2}F_A \lambda^{SPA}   \, , 
\end{eqnarray}
\begin{eqnarray}
\mathrm{Im}  \Pi_{_{AA}} (q^2) |_{SP}  &=& \theta(q^2-(M_S+M_P)^2) \frac{\lambda^{3/2}(q^2,M_S^2,M_P^2)}{48\pi q^6} |\mathcal{F}^{\,a}_{SP} (q^2)|^2 \, , 
\end{eqnarray}
\vspace{-0.4cm}
\begin{eqnarray}
\lambda^{SPA}&=&-\frac{\sqrt{2}\lambda_1^{SP}}{F_A} \,  , \\
 \mathcal{\tilde{F}}^{\,a}_{SP} (q^2) &=&\frac{\sqrt{2}M_A^2}{M_A^2-q^2} \lambda_1^{SP}\,.
\end{eqnarray}

\begin{figure}
\begin{center}
\includegraphics[scale=0.8]{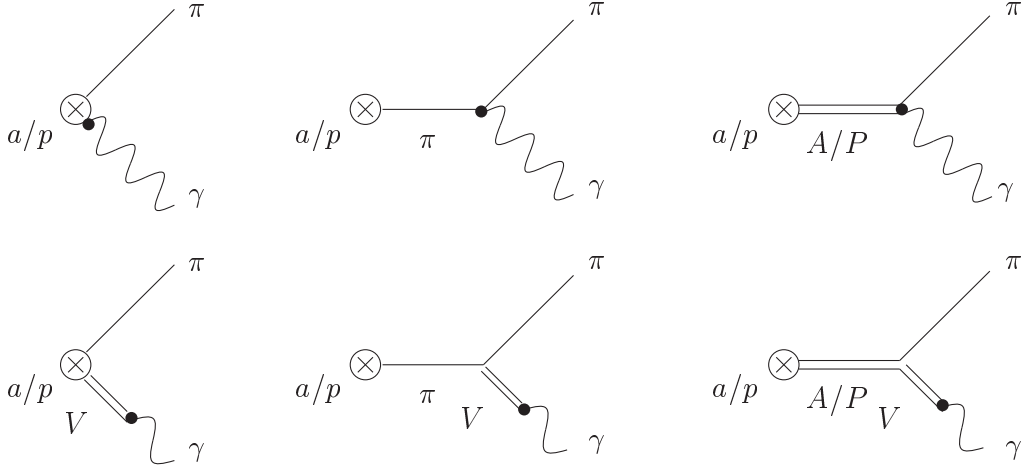}
\caption{\label{BR0F1}
Tree-level contributions to the axial/pseudoscalar form factor to one pseudo-Goldstone and one photon.}
\end{center}
\end{figure}

\subsubsection{Axial Form Factor to $\pi\gamma$ (Figure \ref{BR0F1})}
\thispagestyle{appendixd}
\begin{equation}
\bra\! \gamma (p_\gamma, \varepsilon ) \pi^- (p_\pi) | \bar{d}\gamma^\mu \gamma_5 u | 0\!\ket = 
i\,\sqrt{2}e F \!\left(\! {\varepsilon^*}^\mu \!-\! 2 q\varepsilon^* \, \frac{q^\mu}{q^2} \right)\!+\!
\frac{i\,\sqrt{2}  e}{F}( q\varepsilon^* \, p_\gamma^\mu  - qp_\gamma\, {\varepsilon^*}^\mu ) \mathcal{F}^{\,a}_{\pi\gamma} (q^2)   ,
\end{equation}
\begin{equation}
\mathcal{F}^{\,a}_{\pi\gamma} (q^2) =\frac{F_A^2}{M_A^2-q^2}\! + \!\frac{2F_VG_V-F_V^2}{M_V^2}\!+\! \frac{F_AF_V}{M_V^2} \frac{q^2}{M_A^2-q^2} \!\left( 2 \lambda_2^{VA}\!-\lambda_4^{VA}\!-2\lambda_5^{VA}\right)  ,%
\end{equation} 
\begin{align}
 \mathrm{Im}  \Pi_{_{AA}} (q^2) |_{\pi\gamma}  &=\,\frac{e^2}{F^2} \, \frac{q^2}{48\pi}\, |\mathcal{F}^{\,a}_{\pi\gamma}|^2\, - \frac{e^2}{12\pi} \mathrm{Re} \{ \mathcal{F}^{\,a}_{\pi\gamma}\} \,+\, \frac{e^2\,F^2}{12\pi q^2} \, ,
\end{align} 
\begin{eqnarray}
2\lambda^{VA}_2-\lambda^{VA}_4-2\lambda^{VA}_5&=&-\frac{F_V}{F_A}+\frac{2G_V}{F_A} \, , \qquad  \mathrm{[cf\,\ref{AVG}]} \label{AGF}
\end{eqnarray}
\begin{eqnarray}
\mathcal{\tilde{F}}^{\,a}_{\pi\gamma} (q^2) &=&\frac{1}{M_A^2-q^2} \left[ F_A^2+\frac{M_A^2}{M_V^2}\left(2F_VG_V-F_V^2\right) \right] \, .
\end{eqnarray}

\begin{figure}
\begin{center}
\includegraphics[scale=0.8]{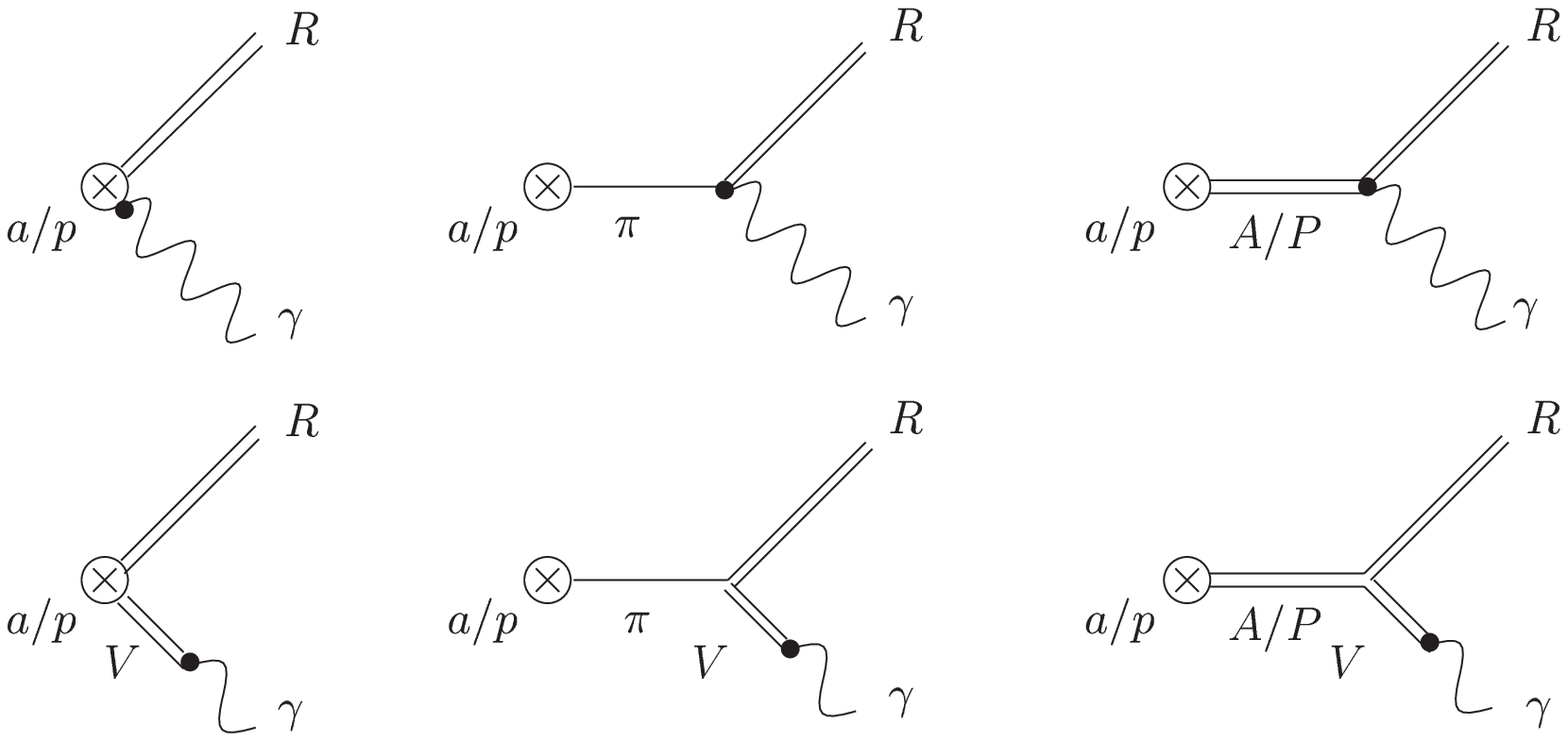}
\caption{\label{BR1F1}
Tree-level contributions to the axial/pseudoscalar form factor 
to one resonance field and one photon.}
\end{center}
\end{figure}

\subsubsection{Axial Form Factor to A$\gamma$ (Figure \ref{BR1F1})}
\thispagestyle{appendixd}
\begin{align}
&\bra \gamma (p_\gamma,\varepsilon_\gamma) A^- (p_A,\varepsilon_A) | \bar{d}\gamma^\mu \gamma_5 u | 0\ket \,=\, \frac{e}{\sqrt{2} M_A}\frac{1}{q^2} \times  \nonumber \\
& \times \bigg\{2/M_V^2 \left( qp_A p_\gamma^\mu - qp_\gamma p_A^\mu \right) \Big[ p_A p_\gamma \varepsilon_A^* \varepsilon_\gamma^*  -q\varepsilon_A^* q\varepsilon_\gamma^*\, \Big] \mathcal{F}^{\,a}_{A\gamma} (q^2) \nonumber \\
& \ +2 M_A^2/M_V^2\Big[ \left( qp_A p_\gamma^\mu - qp_\gamma p_A^\mu \right)\varepsilon_A^* \varepsilon_\gamma^*+\left( p_\gamma^\mu+p_A^\mu  \right)q\varepsilon_A^* q\varepsilon_\gamma^* -q^2 q\varepsilon_A^* {\varepsilon_\gamma^*}^\mu\,\Big]  \mathcal{G}^{\,a}_{A\gamma} (q^2) \nonumber \\
& \ +2 M_A^2 \, F_A \Big[\varepsilon_A^* \varepsilon_\gamma^* \left(  p_\gamma^\mu+p_A^\mu  \right)+ \frac{2}{M_A^2-q^2}\left(\left( p_\gamma^\mu+p_A^\mu  \right)q\varepsilon_A^* q\varepsilon_\gamma^*-q^2 q\varepsilon_\gamma^* {\varepsilon_A^*}^\mu \right) \Big] \bigg\} \,  ,
\end{align}
%\vspace{-0.4cm}
\begin{align}
&\mathcal{F}^{\,a}_{A\gamma} (q^2) = 2F_V \left(  \lambda^{VA}_4\!\!\!+2\lambda^{VA}_5\!\!\!+2\lambda^{VA}_6\!\right) 
+\frac{4F_A}{M_A^2-q^2} \bigg\{ M_V^2\lambda^{AA}_7\!\!\!-\frac{F_V}{\sqrt{2}} \Big[- \lambda^{VAA}_0\!\!\!\nonumber \\
&+qp_A \left(-2\lambda^{VAA}_2\!\!\!+\lambda^{VAA}_6\!\!\! - \lambda^{VAA}_7\!\!\!+ \lambda^{VAA}_9\! \!\!+ 2 \lambda^{VAA}_{10}\!\!\!-  \lambda^{VAA}_{12}\!\!\! -2 \lambda^{VAA}_{13}\!\!\! +\lambda^{VAA}_{14}\!\right) \Big] \bigg\} \, , \nonumber \\
&\mathcal{G}^{\,a}_{A\gamma} (q^2) =2F_V \!\left(  \lambda^{VA}_2\!+\lambda^{VA}_6 \right)\!
+\!\frac{F_A}{M_A^2-q^2} \bigg\{\!\!-M_V^2+2M_V^2\lambda^{AA}_7\!\!\!+\sqrt{2}F_V\!\Big[\lambda^{VAA}_0\!\!\!
+\!2q^2\lambda^{VAA}_{11}\! \nonumber \\
&+\!qp_A \!\left(2 \lambda^{VAA}_2\!\!+2\lambda^{VAA}_{13}\! \right)\!\!
+\!qp_\gamma \left(\lambda^{VAA}_6\!\!-\!\lambda^{VAA}_7\!\! + \!\lambda^{VAA}_9\!\! +\!2 \lambda^{VAA}_{10}\!\! -\!\lambda^{VAA}_{12}\!\!+\! \lambda^{VAA}_{14}\! \right)\! \Big]\! \bigg\}  ,
\end{align} 
%\vspace{-0.2cm}
\begin{align}
& \mathrm{Im}  \Pi_{_{AA}} (q^2) |_{VA}  \propto  \Big[ 
 \cO (q^4)  \, |\mathcal{F}^{\,a}_{A\gamma}|^2  +\cO (q^2)\, |\mathcal{G}^{\,a}_{A\gamma}|^2+\cO (q^2)\,\mathrm{Re} \{ \mathcal{F}^{\,a}_{A\gamma} {\mathcal{G}^{\,a}_{A\gamma}}^* \} 
 +\cO (q^{-4}) \Big]  ,
\end{align} 
%
%Following these results we have to constraint that
%\begin{eqnarray}
%\mathcal{F}^{\,a}_{A\gamma} \propto \cO (\frac{1}{q^4}) \, , \nonumber \\
%\mathcal{G}^{\,a}_{A\gamma} \propto \cO (\frac{1}{q^2}) \, .\nonumber 
%\end{eqnarray}
\begin{align}
& -2\lambda^{VAA}_{2}\!   + \lambda^{VAA}_{6}\! -\lambda^{VAA}_{7}\! + \lambda^{VAA}_{9}\! +2\lambda^{VAA}_{10}\! 
- \lambda^{VAA}_{12}\! -2\lambda^{VAA}_{13}\! +\lambda^{VAA}_{14}\! \,
\nonumber \\ & \qquad 
=\,\frac{1}{\sqrt{2}F_A} \Big\{ -2\lambda^{VA}_{4}\! -4\lambda^{VA}_{5 }\! -4\lambda^{VA}_{6 }\! \Big\} \, , 
\, \mathrm{[cf\,\ref{AVA}]}  \nonumber \\
\nonumber \\
&
 -2\lambda^{VAA}_{2}\!\!  +\lambda^{VAA}_{6}\!\! -\lambda^{VAA}_{7}\!\!  +\lambda^{VAA}_{9}\!\! + 2\lambda^{VAA}_{10}\!\! - \lambda^{VAA}_{12}\!\! -2\lambda^{VAA}_{13}\! + \lambda^{VAA}_{14}\!  -\frac{2}{M_A^2}\lambda^{VAA}_{0}\!\,
\nonumber \\
& \qquad =\,
\frac{\sqrt{2}}{F_A} \Big\{ \lambda^{VA}_{4}\! +2\lambda^{VA}_{5 }\! +2\lambda^{VA}_{6 }\! \Big\}
+\frac{2\sqrt{2}M_V^2}{F_VM_A^2}  \lambda^{AA}_7  \,, \, \mathrm{[cf\,\ref{VAA},\ref{AVA}]}\nonumber \\
\nonumber \\
& 2\lambda^{VAA}_{2}\! + \lambda^{VAA}_{6}\! -\lambda^{VAA}_{7}\!  + \lambda^{VAA}_{9}\! +2\lambda^{VAA}_{10}\! 
\nonumber \\ & \qquad 
+4\lambda^{VAA}_{11}\! - \lambda^{VAA}_{12}\! +2\lambda^{VAA}_{13}\! +\lambda^{VAA}_{14}\! \,=\,\frac{2\sqrt{2}}{F_A} \Big\{ \lambda^{VA}_{2}\! +\lambda^{VA}_{6 }\! \Big\} \, ,\, \mathrm{[cf\,\ref{AVA}]} \label{AAF}\\&\nonumber
\end{align}
\begin{equation}
\mathcal{\tilde{F}}^{\,a}_{A\gamma} (q^2) = 0  , \quad
\mathcal{\tilde{G}}^{\,a}_{A\gamma} (q^2) = \frac{F_A}{M_A^2-q^2} \bigg\{-M_V^2 
+\frac{M_A^2F_V}{F_A} \left( 2\lambda^{VA}_2- \lambda^{VA}_4-2 \lambda^{VA}_5\right)
 \bigg\}  ,
\end{equation}

\subsubsection{Axial Form Factor to P$\gamma$  (Figure \ref{BR1F1})}
\thispagestyle{appendixd}
\begin{eqnarray}
\bra \gamma (p_\gamma, \varepsilon ) P^- (p_P) | \bar{d}\gamma^\mu \gamma_5 u | 0\ket \,&=&i\, \sqrt{2} \, e
( q\varepsilon^* \, p_\gamma^\mu \, -\, qp_\gamma\, {\varepsilon^*}^\mu )\, \mathcal{F}^{\,a}_{P\gamma} (q^2)  \, ,
\end{eqnarray}
\begin{align}
\mathcal{F}^{\,a}_{P\gamma}& (q^2) \,=\, -\frac{4F_A \lambda^{PA}_1}{M_A^2 -q^2} 
+\frac{2F_V\lambda^{PV}_1}{M_V^2}  +\frac{4F_V \lambda^{PV}_2}{M_V^2} +\frac{\sqrt{2}F_AF_V}{(M_A^2-q^2)M_V^2} \times \nonumber \\ &\times \bigg[ -2\lambda^{PVA}_0 -\frac{q^2-M_P^2}{2} (\lambda^{PVA}_2+2\lambda^{PVA}_3) +q^2 (2\lambda^{PVA}_4+\lambda^{PVA}_5) \bigg] \,,  
\end{align} 
\begin{align}
 \mathrm{Im}  \Pi_{_{AA}} (q^2) |_{P\gamma}  &=\, \theta(q^2-M_P^2) \,e^2\, \frac{\left(1-M_P^2/q^2\right)^3}{48\pi}\,q^2 \, |\mathcal{F}^{\,a}_{P\gamma}|^2\, ,
\end{align} 
\begin{align}
& \lambda_2^{PVA}+2\lambda_3^{PVA}-4 \lambda_4^{PVA}-2\lambda_5^{PVA}=-\frac{2\sqrt{2}}{F_A}\left(\lambda_1^{PV}+2\lambda_2^{PV} \right)   ,\quad \mathrm{[cf\,\ref{APV}]}\label{APF}
\end{align}
\begin{eqnarray}
\mathcal{\tilde{F}}^{\,a}_{P\gamma} (q^2) &=&-\frac{\sqrt{2}F_AF_V}{(M_A^2-q^2)M_V^2} \Bigg\{ -\frac{\sqrt{2} M_A^2}{F_A} \left( \lambda_1^{PV}+2\lambda_2^{PV} \right) +\frac{2\sqrt{2}M_V^2}{F_V} \lambda_1^{PA} \nonumber \\ &&+ 2\lambda_0^{PVA}-\frac{M_P^2}{2}\left(\lambda_2^{PVA}+2\lambda_3^{PVA}\right) \Bigg\} \, .
\end{eqnarray}

\subsection{Scalar Form Factors}
\thispagestyle{appendixd}
\subsubsection{Scalar Form Factor to $ \pi \eta$ (Figure \ref{AR0F0})}
%\newpage
\begin{eqnarray}
\bra \eta (p_\eta) \pi^- (p_\pi) | \bar{d} u | 0\ket \,&=& \mathcal{F}^{\,s}_{\pi \eta} (q^2)  \, , \phantom{\frac{1}{2}} \\
\mathcal{F}^{\,s}_{\pi \eta} (q^2) &=&\sqrt{2} \, B_0 \left( 1+4\,\frac{c_m c_d}{F^2}\, \frac{q^2}{M_S^2-q^2} \right) \, , \\
\mathrm{Im}  \Pi_{_{SS}} (q^2) |_{\pi \eta }  &=& \theta(q^2) \frac{1}{16\pi} |\mathcal{F}^{\,s}_{\pi \eta} (q^2)|^2 \, , \\
4\,c_d\,c_m&=&F^2 \, , \phantom{\frac{1}{2}}\label{SGG}\\
\mathcal{\tilde{F}}^{\,s}_{\pi \eta} (q^2) &=&\sqrt{2}B_0\frac{M_S^2}{M_S^2-q^2} \,.
\end{eqnarray}

\subsubsection{Scalar Form Factor to A$\pi$ (Figure \ref{AR1F0})}
\begin{eqnarray}
\bra A^0_{I=0} (p_A,\varepsilon) \pi^- (p_S)  | \bar{d} u | 0\ket &=&
\frac{i}{M_A}  \, q\varepsilon^*  \,\mathcal{F}^{\,s}_{A\pi} (q^2) \, ,\\
\mathcal{F}^{\,s}_{A\pi} (q^2)&=& -\frac{8\, B_0 \,c_m \, \lambda^{SA}_1}{F} \,\frac{M_A^2}{M_S^2-q^2} \, , \\
\mathrm{Im}  \Pi_{_{SS}} (q^2) |_{A\pi}  &=& \theta(q^2-M_A^2) \,\frac{\left(q^2-M_A^2\right)^3 }{64\pi M_A^4 q^2}\, |\mathcal{F}^{\,s}_{A\pi}|^2 \, , \\
\lambda^{SA}_1&=&0 \, , \phantom{\frac{1}{2}}\\
\mathcal{\tilde{F}}^{\,s}_{A\pi} (q^2)&=& 0 \,.\phantom{\frac{1}{2}}
\end{eqnarray}

\subsubsection{Scalar Form Factor to P$\pi$ (Figure \ref{AR1F0})}
\begin{eqnarray}
\bra P^0_{I=0} (p_P) \pi^- (p_\pi) | \bar{d} u | 0\ket \,&=&\mathcal{F}^{\,s}_{P\pi} (q^2)  \, , \phantom{\frac{1}{2}}\\
\mathcal{F}^{\,s}_{P\pi} (q^2) &=&- \frac{4B_0\,d_m}{F} \,+\, \frac{4B_0\,c_m}{F}\frac{q^2-M_P^2}{M_S^2-q^2} \lambda^{SP}_1 \, , \\
\mathrm{Im}  \Pi_{_{SS}} (q^2) |_{P\pi}  &=& \theta(q^2-M_P^2) \frac{1-M_P^2/q^2}{16\pi} |\mathcal{F}^{\,s}_{P\pi} (q^2)|^2 \, , \\
\lambda^{SP}_1&=&-\frac{d_m}{c_m}  \,\label{SGP} ,\\
\mathcal{\tilde{F}}^{\,s}_{P\pi} (q^2) &=&\frac{4B_0d_m}{F} \frac{M_P^2-M_S^2}{M_S^2-q^2} \,.
\end{eqnarray}

\subsubsection{Scalar Form Factor to RR (R=V,A) (Figure \ref{AR2F0})}
\thispagestyle{appendixd}
\begin{align}
\bra R^0_{I=0} (p_1,\varepsilon_1) R^- (p_2,\varepsilon_2) | \bar{d} u | 0\ket &=\,
\frac{1}{M_R^2} ( q\varepsilon_1^*\,q\varepsilon_2^*  - p_1 p_2\, \varepsilon_1^* \varepsilon_2^* )  \mathcal{F}^{\,s}_{RR} (q^2) \,+\, \varepsilon_1^*\,\varepsilon_2^* \,  \mathcal{G}^{\,s}_{RR} (q^2) \, ,
\end{align}
\vspace{-0.3cm}
\begin{align}
\mathcal{F}^{\,s}_{RR} (q^2)&=\, -8\sqrt{2} B_0 \Big[ \lambda^{RR}_6+\frac{c_m}{M_S^2-q^2} \big( \lambda^{SRR}_0-\frac{p_1p_2}{2} \lambda^{SRR}_2-p_1p_2 \lambda^{SRR}_3 \big. \Big. \nonumber \\
& \Big.\big.\qquad\qquad - 2M_R^2\lambda^{SRR}_4 -M_R^2\lambda^{SRR}_5 \big) \Big] \, , \nonumber\\ 
\mathcal{G}^{\,s}_{RR} (q^2)&=\,-8\sqrt{2} B_0\frac{c_m\lambda^{SRR}_1\!\!}{2} \frac{M_R^2}{M_S^2-q^2} \, , 
\end{align}
\vspace{-0.3cm}
\begin{align}
\mathrm{Im}  \Pi_{_{SS}}& (q^2) |_{RR}  \,=\, \theta(q^2-4M_R^2) \frac{\sigma^2_{M_R}}{16\pi} \Bigg\{ \left( 3-\frac{2q^2}{M_R^2}+\frac{q^4}{2M_R^4} \right) |\mathcal{F}^{\,s}_{RR}|^2
  \nonumber \\
& \Bigg. + \left( 3-\frac{q^2}{M_R^2}+\frac{q^4}{4M_R^4}\right) |\mathcal{G}^{\,s}_{RR}|^2+\left( 6-\frac{3q^2}{M_R^2} \right) \mathrm{Re} \{ \mathcal{F}^{\,s}_{RR} {\mathcal{G}^{\,s}_{RR}}^* \}
 \Bigg\} \,,
\end{align} 
\vspace{-0.2cm}
\begin{eqnarray}
\lambda^{SRR}_2+2\lambda^{SRR}_3&=&-\frac{4\lambda^{RR}_6}{c_m} \, ,\phantom{\frac{1}{2}} \nonumber \\
\frac{\lambda^{SRR}_0}{M_R^2}+\frac{\lambda^{SRR}_2}{2}+\lambda^{SRR}_3-2\lambda^{SRR}_4-\lambda^{SRR}_5 &=& -\frac{\lambda^{RR}_6}{c_m} \frac{M_S^2}{M_R^2} \, ,\phantom{\frac{1}{2}}\nonumber \\
\lambda^{SRR}_1&=&0 \, , \label{SRR}\\ \nonumber
\end{eqnarray}
\begin{equation}
\mathcal{\tilde{F}}^{\,s}_{RR} (q^2)\,=\,\mathcal{\tilde{G}}^{\,s}_{RR} (q^2)\,=\,0 \,.
\end{equation}

\subsubsection{Scalar Form Factor to SS (Figure \ref{AR2F0})}
\thispagestyle{appendixd}
\begin{eqnarray}
\bra S^0_{I=0} (p_1) S^- (p_2) | \bar{d} u | 0\ket \,&=&\mathcal{F}^{\,s}_{SS} (q^2)  \, , 
\end{eqnarray}
\vspace{-1cm}
\begin{align}
\mathcal{F}^{\,s}_{SS} (q^2) &=-4\sqrt{2}B_0 \left[ \lambda_3^{SS}+ \frac{3\,c_m\,\lambda^{SSS}_0}{M_S^2-q^2}+\frac{c_m\lambda^{SSS}_1}{2}\frac{q^2+2M_S^2}{M_S^2-q^2} \right]\, , \\
\mathrm{Im} \Pi_{_{SS}} (q^2) |_{SS}  &= \theta(q^2-4M_S^2) \frac{\sigma_{M_S}}{16\pi} |\mathcal{F}^{\,s}_{SS} (q^2)|^2 \, , \\
\lambda^{SSS}_1&=\frac{2\, \lambda^{SS}_3}{c_m} \, ,\label{SSS}\\
\mathcal{\tilde{F}}^{\,s}_{SS} (q^2) &=-\frac{4\sqrt{2}B_0}{M_S^2-q^2} \left[ 3M_S^2 \lambda^{SS}_3+3c_m \lambda^{SSS}_0 \right] \,.
\end{align}

\subsubsection{Scalar Form Factor to PP (Figure \ref{AR2F0})}
\begin{eqnarray}
\bra P^0_{I=0} (p_1) P^- (p_2) | \bar{d} u | 0\ket \,&=&\mathcal{F}^{\,s}_{PP} (q^2)  \, , 
\end{eqnarray}
\vspace{-1cm}
\begin{align}
\mathcal{F}^{\,s}_{PP} (q^2) &=\,-4\sqrt{2}B_0 \left[ \lambda_3^{PP}+ \frac{c_m\,\lambda^{SPP}_0}{M_S^2-q^2}+\frac{c_m\lambda^{SPP}_1}{2}\frac{-q^2+2M_P^2}{M_S^2-q^2} \right]\, , \\
\mathrm{Im} \Pi_{_{SS}} (q^2) |_{PP}  &=\, \theta(q^2-4M_P^2) \frac{\sigma_{M_P}}{16\pi} |\mathcal{F}^{\,s}_{PP} (q^2)|^2 \, , \\
\lambda^{SPP}_1&=\,-\frac{2\, \lambda^{PP}_3}{c_m} \, ,\label{SPP}\\
\mathcal{\tilde{F}}^{\,s}_{PP} (q^2) &=\,-\frac{4\sqrt{2}B_0}{M_S^2-q^2} \left[ (M_S^2-2M_P^2) \lambda^{PP}_3+c_m \lambda^{SPP}_0 \right] \,.
\end{align}

\subsubsection{Scalar Form Factor to SV (Figure \ref{AR2F0})}
\thispagestyle{appendixd}
\begin{eqnarray}
\bra S^0_{I=1} (p_S) V^- (p_V,\varepsilon) | \bar{d} u | 0\ket &=&
\frac{1}{M_V}  \, q\varepsilon^*  \,\mathcal{F}^{\,s}_{SV} (q^2) \, ,
\end{eqnarray}
\vspace{-1cm}
\begin{align}
\mathcal{F}^{\,s}_{SV} (q^2)&=\, -4\sqrt{2}\, B_0 \,c_m \, \lambda^{VSS} \,\frac{M_V^2}{M_S^2-q^2} \, , \\
\mathrm{Im} \Pi_{_{SS}} (q^2) |_{SV}  &=\, \theta(q^2-(M_S+M_V)^2) \,\frac{\lambda^{3/2} \left( q^2, M_S^2, M_V^2 \right) }{64\pi M_V^4 q^2}\, |\mathcal{F}^{\,s}_{SV}|^2 \, , \\
\lambda^{VSS}&=\,0 \, ,\phantom{\frac{1}{2}}\\
\mathcal{\tilde{F}}^{\,s}_{SV} (q^2)&=\, 0 \,.\phantom{\frac{1}{2}}
\end{align}

\subsubsection{Scalar Form Factor to PA (Figure \ref{AR2F0})}
\begin{eqnarray}
\bra P^0_{I=0} (p_P) A^- (p_A,\varepsilon) | \bar{d} u | 0\ket &=&
\frac{i}{M_A}  \, q\varepsilon^*  \,\mathcal{F}^{\,s}_{PA} (q^2) \, ,
\end{eqnarray}
\vspace{-1cm}
\begin{align}
\mathcal{F}^{\,s}_{PA} (q^2)&=\, 4\sqrt{2}\, B_0 \,c_m \, \lambda^{SPA} \,\frac{M_A^2}{M_S^2-q^2} \, ,\\
\mathrm{Im} \Pi_{_{SS}} (q^2) |_{PA}  &=\, \theta(q^2-(M_P+M_A)^2) \,\frac{\lambda^{3/2} \left( q^2, M_P^2, M_A^2 \right) }{64\pi M_A^4 q^2}\, |\mathcal{F}^{\,s}_{PA}|^2 \, , \\
\lambda^{SPA}&=\,0 \, ,\phantom{\frac{1}{2}}\\
\mathcal{\tilde{F}}^{\,s}_{PA} (q^2)&=\,0 \,. \phantom{\frac{1}{2}}
\end{align}
\newpage
\subsubsection{Scalar Form Factor to V$\gamma$ (Figure \ref{AR1F1})}
\begin{equation}
\bra \gamma (p_\gamma, \varepsilon_\gamma )  V^- (p_V,\varepsilon_V) | \bar{d} u | 0\ket \,=\,
\frac{e}{3M_V} ( q\varepsilon_V^*\,q\varepsilon_\gamma^*  - p_V p_\gamma\, \varepsilon_V^* \varepsilon_\gamma^* )  \mathcal{F}^{\,s}_{V\gamma} (q^2) \, , 
\end{equation}
\vspace{-0.7cm}
\begin{align}
\mathcal{F}^{\,s}_{V\gamma} &(q^2)\,=\,
\frac{16B_0\,c_m}{M_S^2-q^2}\lambda^{SV}_3 - \frac{8\sqrt{2} B_0\,F_V}{M_V^2}  \lambda^{VV}_6
- \frac{4\sqrt{2} B_0\,c_m\,F_V}{M_V^2(M_S^2-q^2)} \times
 \nonumber \\
&\times  \Big[ 2\lambda^{SVV}_0- p_V p_\gamma ( \lambda^{SVV}_2+2 \lambda^{SVV}_3 ) -M_V^2(2\lambda^{SVV}_4 +\lambda^{SVV}_5) \big) \Big] \, , 
\end{align}
\vspace{-0.7cm}
\begin{eqnarray}
\mathrm{Im}  \Pi_{_{SS}} (q^2) |_{V\gamma}  &=& \theta(q^2-M_V^2) \frac{(1-M_V^2/q^2)^{3}}{288\pi\,M_V^2} e^2 \, q^4  |\mathcal{F}^{\,s}_{V\gamma}|^2 \,,
\end{eqnarray} 
\vspace{-0.7cm}
\begin{align}
\lambda^{SVV}_2+2\lambda^{SVV}_3&=\,-\frac{4\lambda^{VV}_6}{c_m} \, , \, \mathrm{[cf\,\ref{SRR}]}\nonumber \\
\frac{4\lambda^{SVV}_0}{M_V^2}+\lambda^{SVV}_2\!\!+2\lambda^{SVV}_3\!\!-4\lambda^{SVV}_4\!\!-2\lambda^{SVV}_5\!\! &=\, -\frac{4\,\lambda^{VV}_6}{c_m} \frac{M_S^2}{M_V^2}+\frac{4\sqrt{2} \lambda_3^{SV}}{F_V}   ,\, \mathrm{[cf\,\ref{VSV},\ref{SRR}]}\nonumber \\ \label{SVF}
\end{align}
\vspace{-0.7cm}
\begin{eqnarray}
\mathcal{\tilde{F}}^{\,s}_{V\gamma} (q^2)&=& 0 \,.
\end{eqnarray}

\subsubsection{Scalar Form Factor to S$\gamma$ (Figure \ref{AR1F1})}
\begin{eqnarray}
\bra \gamma \, (p_\gamma, \varepsilon ) S^- (p_S)  | \bar{d} u | 0\ket &=&
 e \, q\varepsilon^*  \,\mathcal{F}^{\,s}_{S\gamma} (q^2) \, , \\
\mathcal{F}^{\,s}_{S \gamma } (q^2)&=&   \frac{8\, B_0 \,c_m}{M_S^2-q^2} \, , \\
\mathrm{Im}  \Pi_{_{SS}} (q^2) |_{S \gamma }  &=&0\,,
\end{eqnarray} 

\subsection{Pseudoscalar Form Factors}
\thispagestyle{appendixd}
\subsubsection{Pseudoscalar Form Factor to V$\pi$ (Figure \ref{BR1F0})}
\begin{eqnarray}
\bra \pi^0 (p_\pi) V^- (p_V,\varepsilon) |i \bar{d}\gamma_5 u | 0\ket &=&
\frac{1}{M_V}  \, q\varepsilon^*  \,\mathcal{F}^{\,p}_{V\pi} (q^2) \, ,
\end{eqnarray}
\vspace{-0.7cm}
\begin{eqnarray}
\mathcal{F}^{\,p}_{V\pi} (q^2)&=& -\frac{2\, B_0}{F} \, \left( \sqrt{2}G_V \frac{M_V^2}{q^2} +4d_m\, \lambda^{PV}_1\frac{M_V^2}{M_P^2-q^2}    \right) \, , \\
\mathrm{Im}  \Pi_{_{PP}} (q^2) |_{V\pi}  &=& \theta(q^2-M_V^2) \,\frac{\left( q^2-M_V^2 \right)^3 }{64\pi M_V^4 q^2}\, |\mathcal{F}^{\,p}_{V\pi}|^2 \, , 
\end{eqnarray} 
\vspace{-0.7cm}
\begin{eqnarray}
-\sqrt{2} G_V + 4d_m \, \lambda^{PV}_1 &=& 0  \, , 
\end{eqnarray}
\vspace{-0.9cm}
\begin{eqnarray}
\mathcal{\tilde{F}}^{\,p}_{V\pi} (q^2)&=&-\frac{2\sqrt{2}B_0 \,G_V}{F} \frac{M_V^2M_P^2}{(M_P^2-q^2)q^2} \,.
\end{eqnarray}

\subsubsection{Pseudoscalar Form Factor to S$\pi$ (Figure \ref{BR1F0})}
\begin{eqnarray}
\bra S^0_{I=0} (p_S) \pi^- (p_\pi) | i \bar{d} \gamma_5 u | 0\ket \,&=&\mathcal{F}^{\,p}_{S\pi} (q^2)  \, , 
\end{eqnarray}
\vspace{-0.3cm}
\begin{align}
\mathcal{F}^{\,p}_{S\pi} (q^2) &=\, \frac{4B_0\,c_m}{F} -\frac{2B_0\,c_d}{F} \frac{q^2-M_S^2}{q^2} \,+\,\frac{4B_0\,d_m}{F}\frac{M_S^2-q^2}{M_P^2-q^2} \lambda^{SP}_1 \, , \\
\mathrm{Im}  \Pi_{_{PP}} (q^2) |_{S\pi}  &=\, \theta(q^2-M_S^2) \frac{1-M_S^2/q^2}{16\pi} |\mathcal{F}^{\,p}_{S\pi} (q^2)|^2 \, , \\
\lambda^{SP}_1&=\,\frac{-2c_m+c_d}{2d_m}  \, ,\label{PGS} \\
\mathcal{\tilde{F}}^{\,p}_{S\pi} (q^2) &=\,\frac{4B_0c_m}{F} \frac{M_P^2-M_S^2}{M_P^2-q^2} +\frac{2B_0c_d}{F} \frac{M_P^2}{M_P^2-q^2} \left( \frac{M_S^2}{q^2}-1 \right) \,.
\end{align}

\subsubsection{Pseudoscalar Form Factor to VA (Figure \ref{BR2F0})}
\thispagestyle{appendixd}
\begin{eqnarray}
\bra V^0_{I=1} (p_V,\varepsilon_V) A^- (p_A,\varepsilon_A) | i \bar{d} \gamma_5 u | 0\ket &=&
\frac{i}{M_V M_A} ( q\varepsilon_V^*\,q\varepsilon_A^*  - p_V p_A\, \varepsilon_V^* \varepsilon_A^* )  \mathcal{F}^{\,p}_{VA} (q^2) \nonumber \\
&& \qquad \quad  +\, i\, \varepsilon_V^*\,\varepsilon_A^* \,  \mathcal{G}^{\,p}_{VA} (q^2) \, ,
\end{eqnarray}
\vspace{-0.5cm}
\begin{align}
\mathcal{F}^{\,p}_{VA} (q^2)&=\, -4\sqrt{2} B_0 \Big[ -2\lambda^{VA}_1+\frac{1}{4\,q^2} \big(-2(q^2+M_V^2+M_A^2)\lambda^{VA}_2+2M_V^2\lambda^{VA}_3 \big.\nonumber \\
& \big. -(q^2+M_V^2-M_A^2)(\lambda^{VA}_4+2\lambda^{VA}_5)\big)+\frac{d_m}{M_P^2-q^2}\big( 2 \lambda^{PVA}_0\big. \nonumber \\
&-p_Vp_A (\lambda^{PVA}_2+2\lambda^{PVA}_3) -M_A^2(2\lambda^{PVA}_4+\lambda^{PVA}_5)-M_V^2\lambda^{PVA}_6 \big.\big) \Big] \, , \nonumber \\
\mathcal{G}^{\,p}_{VA} (q^2)&=\, -4\sqrt{2} B_0 M_A M_V\Big[\frac{1}{2q^2} \big( 2\lambda^{VA}_2+\lambda^{VA}_3\big) +\frac{d_m}{M_P^2-q^2}\lambda^{PVA}_1 \Big]\, , 
\end{align}
\begin{align}
\mathrm{Im}  \Pi_{_{PP}} (q^2)& |_{VA}  = \theta(q^2-(M_V+M_A)^2) \frac{\lambda^{1/2}(q^2,M_V^2,M_A^2)}{16\pi q^2}\! \Bigg\{-\frac{6p_Ap_V}{M_AM_V} \mathrm{Re} \{ \mathcal{F}^{\,s}_{RR} {\mathcal{G}^{\,s}_{RR}}^*\} \nonumber \\
&+  \frac{4M_A^2M_V^2 -q^4+(q^2-M_V^2)^2+(q^2-M_A^2)^2}{2M_A^2M_V^2} |\mathcal{F}^{\,s}_{RR}|^2
  \nonumber \\
& +\frac{10M_A^2M_V^2 -q^4+(q^2-M_V^2)^2+(q^2-M_A^2)^2}{4M_A^2M_V^2} |\mathcal{G}^{\,s}_{RR}|^2
\Bigg. \Bigg\} \,,
\end{align} 
\begin{align}
&\qquad \qquad \lambda^{PVA}_2+2\lambda^{PVA}_3\,=\,\frac{1}{2d_m} \left( 8 \lambda^{VA}_1+2\lambda^{VA}_2+\lambda^{VA}_4+2\lambda^{VA}_5\right) \, ,\qquad \qquad  \nonumber \\
&4\lambda^{PVA}_0+(M_V^2+M_A^2)(\lambda^{PVA}_2+2\lambda^{PVA}_3)-M_A^2(4\lambda^{PVA}_4+2\lambda^{PVA}_5)-2M_V^2 \lambda^{PVA}_6 \,=\, \nonumber\\
&\frac{1}{d_m} \! \left(4M_P^2\lambda^{VA}_1\! +\!(M_P^2\!-\!M_V^2\!-\!M_A^2)\lambda^{VA}_2\! +\!M_V^2\lambda^{VA}_3\!+\!\frac{1}{2} (M_P^2\!-\!M_V^2\!+\!M_A^2)(\lambda^{VA}_4\!+2\lambda^{VA}_5) \!\right)\! , \nonumber\\
&\qquad \qquad 2\lambda^{PVA}_1\,=\,\frac{1}{d_m} \left( 2\lambda^{VA}_2+\lambda^{VA}_3 \right) \, ,\qquad \qquad \qquad \qquad \qquad \vspace{-4.5cm}   \label{PVA}
\end{align}
\begin{align}
\mathcal{\tilde{F}}^{\,p}_{VA} (q^2)&=\frac{\sqrt{2}B_0\,M_P^2}{(M_P^2-q^2)q^2} \!\left[ 2(M_V^2\!+\!M_A^2)\lambda^{VA}_2\! -\!2M_V^2\lambda^{VA}_3 \!+\!(M_V^2\!-\!M_A^2)(\lambda^{VA}_4\!+\!2\lambda^{VA}_5) \right]  , \nonumber \\
\mathcal{\tilde{G}}^{\,p}_{VA} (q^2)&=-2\sqrt{2}B_0 \frac{M_AM_VM_P^2}{(M_P^2-q^2)q^2} \left( 2\lambda^{VA}_2+\lambda^{VA}_3 \right) \, .
\end{align}

\subsubsection{Pseudoscalar Form Factor to PV (Figure \ref{BR2F0})}
\thispagestyle{appendixd}
\begin{eqnarray}
\bra P^0_{I=1} (p_P) V^- (p_V,\varepsilon) |i \bar{d}\gamma_5 u | 0\ket &=&
\frac{1}{M_V}  \, q\varepsilon^*  \,\mathcal{F}^{\,p}_{PV} (q^2) \, ,
\end{eqnarray}
\vspace{-0.5cm}
\begin{align}
\mathcal{F}^{\,p}_{PV} (q^2)&=\, 2\sqrt{2}\, B_0 \, \left( -\frac{M_V^2}{q^2} \lambda^{PV}_1 - \frac{2\,d_m \, M_V^2}{M_P^2-q^2}  \lambda^{VPP}  \right) \, , \\
\mathrm{Im}  \Pi_{_{PP}} (q^2) |_{PV}  &=\, \theta(q^2-(M_P+M_V)^2) \,\frac{\lambda^{3/2} \left( q^2, M_P^2, M_V^2 \right) }{64\pi M_V^4 q^2}\, |\mathcal{F}^{\,p}_{PV}|^2 \, ,  \\
\lambda^{VPP}&=\,\frac{1}{2d_m} \lambda^{PV}_1 \, ,\\
\mathcal{\tilde{F}}^{\,p}_{V\pi} (q^2)&=\,-2\sqrt{2}B_0 \frac{M_V^2M_P^2}{(M_P^2-q^2)q^2}  \lambda^{PV}_1\,.
\end{align}

\subsubsection{Pseudoscalar Form Factor to SA (Figure \ref{BR2F0})}
\begin{eqnarray}
\bra S^0_{I=0} (p_S) A^- (p_A,\varepsilon) | i\bar{d}\gamma_5 u | 0\ket &=&
\frac{i}{M_A}  \, q\varepsilon^*  \,\mathcal{F}^{\,p}_{SA} (q^2) \, ,
\end{eqnarray}
\vspace{-0.7cm}
\begin{align}
\mathcal{F}^{\,p}_{SA} (q^2)&=\, 2\sqrt{2}\, B_0 \, \left( \frac{M_A^2}{q^2} \lambda^{SA}_1 - \frac{2\,d_m \, M_A^2}{M_P^2-q^2}  \lambda^{SPA}  \right) \, ,\\
\mathrm{Im} \Pi_{_{PP}} (q^2) |_{SA} &=\, \theta(q^2-(M_S+M_A)^2) \,\frac{\lambda^{3/2} \left( q^2, M_S^2, M_A^2 \right) }{64\pi M_A^4 q^2}\, |\mathcal{F}^{\,p}_{SA}|^2 \, , \\
\lambda^{SPA}&=\,-\frac{1}{2d_m} \lambda^{SA}_1 \, , \\
\mathcal{\tilde{F}}^{\,p}_{SA} (q^2)&=\,2\sqrt{2}B_0 \frac{M_A^2M_P^2}{(M_P^2-q^2)q^2}  \lambda^{SA}_1\,.
\end{align}

\subsubsection{Pseudoscalar Form Factor to SP (Figure \ref{BR2F0})}
\begin{eqnarray}
\bra S^0_{I=0} (p_S) P^- (p_P) | i \bar{d} \gamma_5 u | 0\ket \,&=&\mathcal{F}^{\,p}_{SP} (q^2)  \, , 
\end{eqnarray}
\vspace{-0.7cm}
\begin{align}
\mathcal{F}^{\,p}_{SP}& (q^2) \,=\, -4\sqrt{2}B_0 \Big[ \lambda_2^{SP}-\frac{q^2+M_S^2-M_P^2}{4\,q^2} \lambda_1^{SP}\nonumber \\
& +\frac{d_m}{2\,(M_P^2-q^2)} \left( 2\lambda_0^{SPP}+(q^2+M_P^2-M_S^2) \lambda_1^{SPP}\right)\Big] \, , %\\
\end{align}
\begin{align}
\mathrm{Im}  \Pi_{_{PP}} (q^2) |_{SP} &=\, \theta(q^2-(M_S+M_P)^2) \frac{\lambda^{1/2}(q^2,M_S^2,M_P^2)}{16\pi q^2} |\mathcal{F}^{\,p}_{SP} (q^2)|^2 \, , \\
\lambda^{SPP}_1&=\,-\frac{1}{2d_m}\lambda^{SP}_1+\frac{2}{d_m}\lambda^{SP}_2 \, , \label{PSP}
\end{align}
\vspace{-0.6cm}
\begin{align}
\mathcal{\tilde{F}}^{\,p}_{SP} &(q^2) \,=\,-\frac{4\sqrt{2}B_0}{M_P^2-q^2} \bigg[ \left(-\frac{M_S^2M_P^2}{4q^2}+\frac{M_P^4}{4q^2}-\frac{3M_P^2}{4}+\frac{M_S^2}{2} \right)\lambda^{SP}_1+ \nonumber \\ & +\left(2M_P^2-M_S^2\right) \lambda^{SP}_2 +d_m\lambda^{SPP}_0 \bigg] \,.
\end{align}

\subsubsection{Pseudoscalar Form Factor to $\pi \gamma$ (Figure \ref{BR0F1})}
\begin{eqnarray}
\bra \gamma \, (p_\gamma, \varepsilon ) \pi^- (p_\pi)  | i \bar{d}\gamma_5 u | 0\ket &=&
 e \, q\varepsilon^*  \,\mathcal{F}^{\,p}_{\pi\gamma} (q^2) \, , \\
\mathcal{F}^{\,p}_{\pi \gamma } (q^2)&=&   \frac{2\sqrt{2} \, B_0 \, F}{q^2} \, , \\
\mathrm{Im}  \Pi_{_{PP}} (q^2) |_{\pi \gamma }  &=&0\, .
\end{eqnarray} 

\subsubsection{Pseudoscalar Form Factor to A$\gamma$ (Figure \ref{BR1F1})}
\thispagestyle{appendixd}
\begin{equation}
\bra \gamma (p_\gamma,\varepsilon_\gamma) A^- (p_A,\varepsilon_A) | i \bar{d} \gamma_5 u | 0\ket \,=\,
\frac{i\,e}{M_A} ( q\varepsilon_\gamma^*\,q\varepsilon_A^*  - p_\gamma p_A\, \varepsilon_\gamma^* \varepsilon_A^* )  \mathcal{F}^{\,p}_{A\gamma} (q^2)  \, ,
\end{equation}
\vspace{-0.5cm}
\begin{align}
\mathcal{F}^{\,p}_{A\gamma} (q^2)&= 
\frac{\sqrt{2} F_A\,B_0}{q^2} \!-\!
\frac{16B_0\,d_m}{M_P^2-q^2}\lambda^{PA}_1\!-\!\frac{4\sqrt{2} B_0\, F_V}{M_V^2}\! \bigg\{\! -2\lambda^{VA}_1 \!+ 
\! \frac{1}{4\,q^2} \!\Big[\!-2(q^2+M_A^2)\lambda^{VA}_2 \nonumber \\
&  -(q^2-M_A^2)(\lambda^{VA}_4+2\lambda^{VA}_5)\Big]+\frac{d_m}{M_P^2-q^2}\Big[ 2 \lambda^{PVA}_0-p_\gamma p_A (\lambda^{PVA}_2+2\lambda^{PVA}_3) \nonumber \\
& -M_A^2(2\lambda^{PVA}_4+\lambda^{PVA}_5) \Big] \bigg\} \, , 
\end{align}
\vspace{-0.7cm}
\begin{eqnarray}
\mathrm{Im}  \Pi_{_{PP}} (q^2) |_{A\gamma}  \,&=& \theta(q^2-M_A^2) \frac{(1-M_A^2/q^2)^{3}}{32\pi\,M_A^2} e^2 \, q^4  |\mathcal{F}^{\,p}_{A\gamma}|^2 \,,
\end{eqnarray} 
\vspace{-0.5cm}
\begin{align}
&\qquad \qquad \lambda^{PVA}_2+2\lambda^{PVA}_3\,=\,\frac{1}{2d_m} \left( 8 \lambda^{VA}_1+2\lambda^{VA}_2+\lambda^{VA}_4+2\lambda^{VA}_5\right) \, ,\qquad \, \mathrm{[cf\,\ref{PVA}]}\qquad  \nonumber \\
& 4\lambda^{PVA}_0\!\!+M_A^2(\lambda^{PVA}_2\!\!+2\lambda^{PVA}_3\!\!-4\lambda^{PVA}_4\!\!-2\lambda^{PVA}_5) = \frac{1}{d_m} \Big(4M_P^2\lambda^{VA}_1 \!+(M_P^2-M_A^2)\lambda^{VA}_2\!  
 &\nonumber\\&
 +\frac{1}{2} (M_P^2+M_A^2)(\lambda^{VA}_4+2\lambda^{VA}_5) \Big)-\frac{M_V^2}{2\sqrt{2}F_Vd_m}\left(  \sqrt{2}F_A+16d_m \lambda^{PA}_1     \right)\, ,\nonumber \\
& \qquad \qquad \qquad \qquad \qquad  \qquad \qquad \qquad \qquad \qquad \qquad\quad \mathrm{[cf\,\ref{VAG},\ref{VPA},\ref{PVA}]}  \label{PAF}
\end{align}
\vspace{-0.7cm}
\begin{align}
\mathcal{\tilde{F}}^{\,p}_{A\gamma} (q^2)&=\,\frac{\sqrt{2}B_0\,M_P^2}{(M_P^2-q^2)q^2} \bigg[ F_A -\frac{F_V}{M_V^2} \left( -2M_A^2  \lambda^{VA}_2+M_A^2( \lambda^{VA}_4+2 \lambda^{VA}_5 ) \right) \bigg] \,.
\end{align}

\subsubsection{Pseudoscalar Form Factor to P$\gamma$ (Figure \ref{BR1F1})}
\thispagestyle{appendixd}
\begin{eqnarray}
\bra \gamma \, (p_\gamma, \varepsilon^* ) P^- (p_P)  | i \bar{d}\gamma_5 u | 0\ket &=&
e \, q\varepsilon  \,\mathcal{F}^{\,p}_{P\gamma} (q^2) \, , \\
\mathcal{F}^{\,p}_{P \gamma } (q^2)&=&   \frac{8\,B_0\, d_m}{M_P^2 - q^2} \, ,\\
\mathrm{Im}  \Pi_{_{PP}} (q^2) |_{P \gamma }  &=& 0\,.
\end{eqnarray} 

\subsection{Form Factors with a Photon}
\thispagestyle{appendixd}
As pointed out in Section 4.3, no new constraints have been obtained from the analysis of form factors with a photon:
\begin{enumerate}
\item $\mathcal{F}^v_{V\gamma}$ and $\mathcal{G}^v_{V\gamma}$ (Eq.~(\ref{VVF})): the 1st constraint is got by adding the 1st and the 3rd constraint of Eq.~(\ref{VVV}) ($\mathcal{F}^v_{VV}$, $\mathcal{G}^v_{VV}$ and $\mathcal{H}^v_{VV}$); the 2nd one subtracting the 1st constraint to the 3rd one of Eq.~(\ref{VVV}); and the 3rd one subtracting the 3rd constraint to the 4th one of  Eq.~(\ref{VVV}).
\item $\mathcal{F}^v_{S\gamma}$ (Eq.~(\ref{VSF})): same constraint than the 1st one of Eq.~(\ref{VSV}) ($\mathcal{F}^v_{SV}$ and $\mathcal{G}^v_{SV}$ ).
\item $\mathcal{F}^a_{\pi\gamma}$ (Eq.~(\ref{AGF})): same constraint than the 1st one of Eq.~(\ref{AVG}) ($\mathcal{F}^a_{V\pi}$ and $\mathcal{G}^a_{V\pi}$).
\item $\mathcal{F}^a_{A\gamma}$ and $\mathcal{G}^a_{A\gamma}$ (Eq.~(\ref{AAF})): the 1st constraint is the same than the 1st one of Eq.~(\ref{AVA}) ($\mathcal{F}^a_{VA}$, $\mathcal{G}^a_{VA}$, $\mathcal{H}^a_{VA}$ and $\mathcal{I}^a_{VA}$); the 2nd one is got subtracting two times the 4th one of Eq.~(\ref{VAA}) ($\mathcal{F}^v_{AA}$, $\mathcal{G}^v_{AA}$ and $\mathcal{H}^v_{AA}$) to the 1st one of Eq.~(\ref{AVA}); and the 3rd one is the same than the 4th one of Eq.~(\ref{AVA}).
\item $\mathcal{F}^a_{P\gamma}$ (Eq.~(\ref{APF})): same constraint than the 1st one of Eq.~(\ref{APV}) ($\mathcal{F}^a_{PV}$ and $\mathcal{G}^a_{PV}$).
\item $\mathcal{F}^s_{V\gamma}$ (Eq.~(\ref{SVF})): the 1st constraint is the same than the 1st one of Eq.~(\ref{SRR}) ($\mathcal{F}^s_{VV}$ and $\mathcal{G}^s_{VV}$); and the 2nd one can be obtained subtracting the 1st one of Eq.~(\ref{VSV}) ($\mathcal{F}^v_{SV}$ and $\mathcal{G}^v_{SV}$) to four times the 2nd one of Eq.~(\ref{SRR}).
\item $\mathcal{F}^s_{S\gamma}$: no constraints.
\item $\mathcal{F}^p_{\pi \gamma}$: no constraints.
\item $\mathcal{F}^p_{A\gamma}$ (Eq.~(\ref{PAF})): the 1st one is the same than the 1st one of Eq.~(\ref{PVA}) ($\mathcal{F}^p_{VA}$ and $\mathcal{G}^p_{VA}$); and the 2nd one is got summing $-M_V^2/(2d_m)$ times the 1st one of Eq.~(\ref{VAG}) ($\mathcal{F}^v_{A\pi}$ and $\mathcal{G}^v_{A\pi}$), $-M_V^2$ times the 1st one of Eq.~(\ref{VPA}) ($\mathcal{F}^v_{PA}$ and $\mathcal{G}^v_{PA}$) and the 2nd one of Eq.~(\ref{PVA}).
\item $\mathcal{F}^p_{P\gamma}$: no constraints.
\end{enumerate}

\appendix
\chapter*{Appendix E \newline \newline Dispersive Relations}
\addcontentsline{toc}{chapter}{Appendix E: Dispersive Relations}
\newcounter{catilinabis}
\renewcommand{\thesection}{\Alph{catilinabis}}
\renewcommand{\theequation}{\Alph{catilinabis}.\arabic{equation}}
\renewcommand{\thetable}{\Alph{catilinabis}}
\renewcommand{\thefigure}{\Alph{catilinabis}.\arabic{figure}}
\setcounter{catilinabis}{5}
\setcounter{equation}{0}
\setcounter{table}{0}
\setcounter{figure}{0}
\setcounter{subsection}{0}

In the purely perturbative calculation (without Dyson resummations) and under the Single Resonance Approximation, the two-point function at next-to-leading order in the $1/N_C$ expansion reads as: 
\begin{eqnarray}
\Pi(t)&=& \frac{D(t)}{\left(M_R^2  -  t\right)^2} \, ,
\end{eqnarray}
where $M_R$ is the mass of the corresponding resonance in the $s$--channel, and $D(t)$ is an analytical  function except for the unitarity logarithmic branch (without poles). 

\begin{figure}
\begin{center}
\includegraphics[angle=-90,clip,scale=0.45]{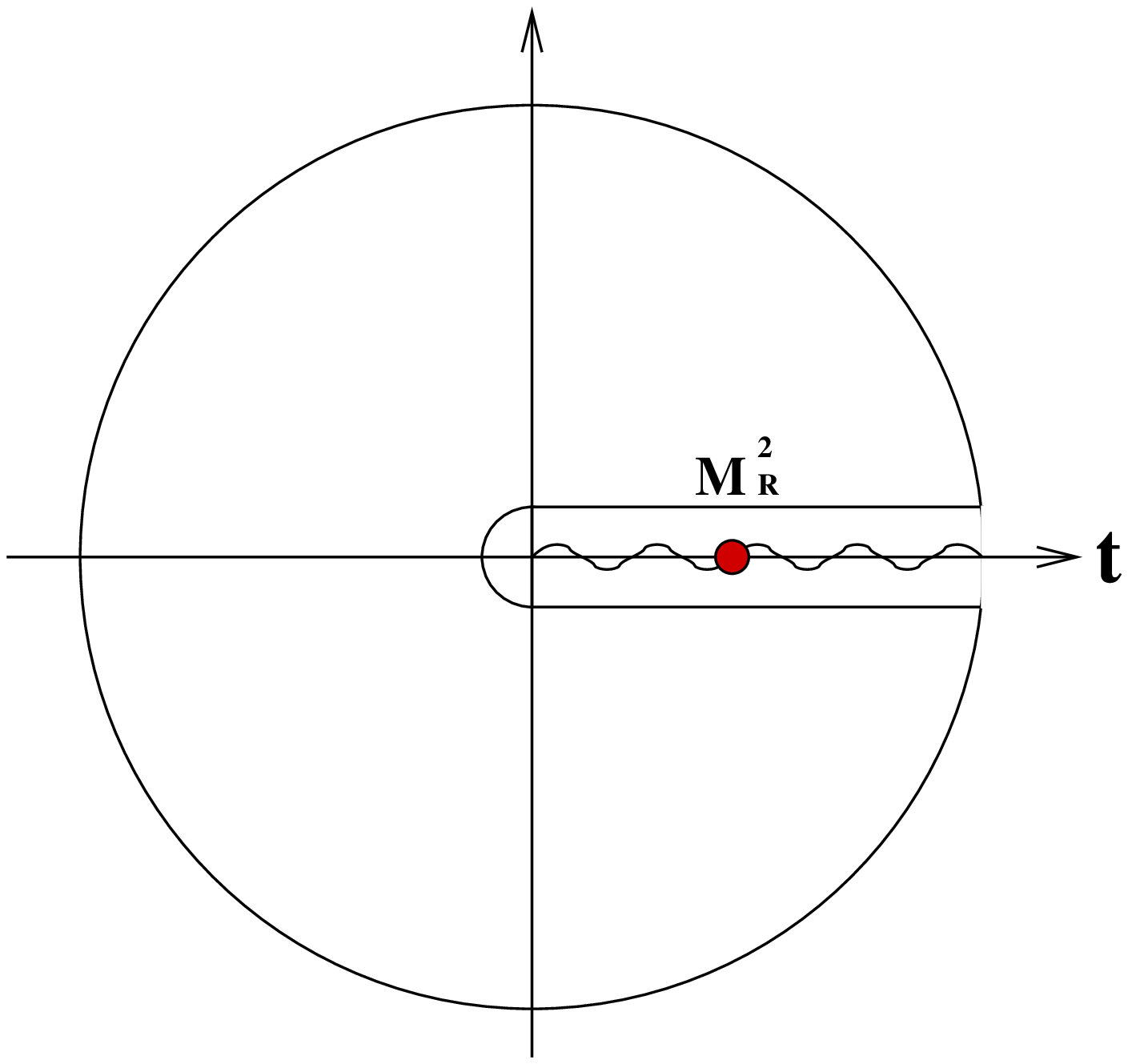}
\caption{\label{fig.circuito}
Integration circuit.}
\end{center}
\end{figure}

In order to recover the correlator, the complex integration in the circuit of Figure~\ref{fig.circuito} is performed:
\begin{eqnarray}
\Pi(q^2) & = & \frac{1}{2\pi i} \, \oint \, \  \mathrm{d}t\, \frac{\Pi(t)}{t -  q^2 }  \, .
\end{eqnarray}
If it is assumed that $|\Pi(t)|\to 0$ when $|t|\to \infty$, the contribution from the external circle of the circuit is zero and it is found that:
\begin{eqnarray}\label{eq.master}
\Pi(q^2)&=& \frac{\overline{D}(q^2)}{(M_R^2 -q^2)^2} -  \frac{\mbox{Re}D'(M_R^2)}{M_R^2-q^2}  +  \frac{\mbox{Re}D(M_R^2)}{\left(M_R^2-q^2\right)^2} \, ,
\end{eqnarray}
with $D'(t)\equiv \frac{d}{dt}D(t)$ and being 
\begin{align}
\label{Dbardef}
\frac{\overline{D}(q^2)}{(M_R^2 -q^2)^2}= \lim_{\epsilon\to 0} \left[ \Int_0^{M_R^2 -\epsilon}\!\!\!\!\!\mathrm{d}t\, \, \frac{1}{\pi}\frac{\mbox{Im}\Pi(t)}{t -q^2} +  \Int_{M_R^2 +\epsilon}^\infty \!\!\!\!\! \mathrm{d}t\,\frac{1}{\pi} \frac{\mbox{Im}\Pi(t)}{t -q^2} - \frac{2}{\pi\epsilon}  \lim_{t\to M_R^2}\left(\frac{\mbox{Im}D(t)}{t-q^2}\right)  \right]  , \quad 
\end{align}
which obeys  $\overline{D}(M_R^2)=0$ and  $\overline{D}\, '(M_R^2)=0$. Notice that in order to recover $\overline{D}(q^2)$ it is not necessary to know Im$D(t)$ at $t=M_R^2$, but just the amplitude in the region $[0,+\infty)-\{M_R^2\}$, where $\Pi(q^2)$ is well defined. 

It is important to remark that, in order to recover the proper asymptotic behaviour of $\Pi(t)$, one must have a spectral function that vanishes at high energies, so the form factors must follow the proper asymptotic behaviours. 

From  Eq.~(\ref{eq.master}),  one notices that, as soon as the value of the real part of $D(t)$ and its first derivative are fixed at $M_R^2$, the whole correlator becomes fixed. This corresponds to providing a renormalization prescription for the corresponding coupling and resonance mass.

The fact that the spectral function vanishes at infinite momentum ensures that there are no terms of the form $\Pi(t)\sim t^{m}\, \ln{(-t)}$, with $m\geq 0$. Furthermore, the polynomial terms $\Pi(t)\sim t^{m}$ with $m\geq 0$ must be also identically zero in order to keep  $\Pi(t)\rightarrow0$ at $|t|\rightarrow \infty$.  Hence, the expression in Eq.~(\ref{eq.master}), is the general expression for the correlator within the Single Resonance Approximation. The inclusion of higher resonances can be performed in a straightforward way. 

This means that although the presence of $\cO(p^4)$ $\chi$PT operators with NLO couplings in $1/N_C$,  $\widetilde{L}_i$,  is not forbidden by the symmetry, the QCD short-distance behaviour imposes that, in our realization,  they do not get renormalized, as suggested in Ref.~\cite{cata}, and they do not contribute to the observable at the end of the day (the polynomial terms $\Pi^(t)\sim t^{m}$ are identically zero). This lack of running in the $\widetilde{L}_i$ related to the analysed currents arisen in the one-loop analysis of the R$\chi$T generating functional with only pseudo-Goldstones, scalar and pseudoscalar resonances after imposing the high energy constraints~\cite{integral}.
\thispagestyle{appendixebis}

\subsection{Diagrammatic Calculation}
For sake of simplicity we will refer now just to the scalar correlator although the extension to other channels is straight-forward. At tree-level order it is found that 
\begin{eqnarray}
\Pi_{_{SS}}(q^2)&=& \frac{16\,B_0^2\,c_m^2}{M_S^2 - q^2} \, .
\end{eqnarray}
The resonance parameters $c_m$ and $M_S$ get renormalized  at next-to-leading order $1/N_C$ ($c_m=c_m^{\, r}+\delta c_m$ and $M_S^2=M_S^{r\,\,2}+\delta M_S^2$) in order to cancel the ultraviolet divergences from the one-loop diagrams:
\begin{align}
\Pi_{_{SS}}(q^2)|_{\mathrm{tree}} &=\frac{16\,B_0^2\,c_m^{r\,2}}{M_S^{r\,\,2} - q^2}\!+\! \frac{32\,B_0^2\,c_m^{r}\,\delta c_m}{M_S^{r\,\,2} - q^2}\!-\! \frac{16\,B_0^2\,c_m^{r\,2}\,\delta M_S^2}{\left(M_S^{2\,\,r} - q^2\right)^2}\! +\!\cO\!\left(\frac{1}{N_C}\right) , \label{tree} \\
\Pi_{_{SS}}(q^2)|_\mathrm{1-loop}&= \frac{D(q^2)|_\mathrm{1-loop}}{\left(M_S^{r\,\,2} - q^2\right)^2} = \frac{\overline{D}(q^2)}{\left(M_S^{r\,\,2}  - q^2\right)^2}\!+\! \frac{c_1 +  \gamma_1\, \lambda_\infty}{M_S^{r\,\,2} - q^2}\!+\! \frac{c_2 +  \gamma_2 \, \lambda_\infty}{\left(M_S^{r\,\,2} -q^2 \right)^2} , \label{loop}
\end{align}
where $\overline{D}(t)$ is provided in terms of the spectral function in Eq.~(\ref{Dbardef}) and $c_{1,2}$ and $\gamma_{1,2}$ are constants determined by the one-loop calculation. Taking into account Eq.~(\ref{eq.master}), one gets   
\begin{eqnarray}
c_1+\gamma_1\, \lambda_\infty &=& -\mbox{Re}\left\{D\,' (q^2=M_S^{r\,\,2})|_\mathrm{1-loop} \right\} \,,\nonumber \\
 c_2+\gamma_2\,\lambda_\infty& =& \mbox{Re}\left\{ D(q^2=M_S^{r\,\,2})|_\mathrm{1-loop}\right\} \,.
\end{eqnarray}
\thispagestyle{appendixebis}

All the relevant  ultraviolet divergences are shown in Eq.~(\ref{loop}).  As mentioned before, the polynomial divergences $\Pi_{_{SS}}(t)\sim \gamma_{-m}\, t^m\,  \lambda_\infty $ cannot produce any contribution at the end of the day, so they exactly cancel at  any energy. Once again,   considering well behaved correlators --and therefore form factors-- at large energies is crucial.

The renormalization procedure through the $c_m$ and $M_S$ counterterms gives
\begin{eqnarray}
32\,B_0^2\, c_m^{r}\delta c_m + \gamma_1\,\lambda_\infty  &=&  0 \, , \nonumber\\
 -16\,B_0\,c_m^{r\,\, 2} \, \delta M_S^{r\,\,2} + \gamma_2\,\lambda_\infty  &=& 0 \, .
\end{eqnarray}

The renormalized amplitude up to next-to-leading order in the $1/N_C$ expansion shows the general structure  
\begin{eqnarray}
\Pi_{_{SS}}(q^2)& =& \frac{\overline{D}(q^2)}{\left(M_S^{r\,\,2}-q^2\right)^2 } +\frac{16\,B_0\,c_m^{r\,\,2} + c_1}{M_S^{r\,\,2}  - q^2} +\frac{c_2}{\left(M_S^{r\,\,2} - q^2 \right)^2}\, .   
\end{eqnarray}
The unknown subraction constants $c_1$ and $c_2$ can be absorved by a redefinition of $c_m^{r}$ and $M_S^{r}$, so 
\begin{eqnarray}
\Pi_{_{SS}}(q^2)\quad & = & \frac{\overline{D}(q^2)}{\left(M_S^{r\,\,2}-q^2\right)^2 } + \frac{16 \, B_0 \, c_m^{r\,\,2} }{M_S^{r\,\,2}  -  q^2} \, ,   
\end{eqnarray}
where $c_m^r$ and $M_S^r$ are now renormalization scale independent.

\subsection{Contribution from High Mass Absorptive Cuts}

Because of the approximation of neglecting intermediate states with two resonances, made in Section~4.4, it is convenient to analyse the  effect on the $\chi$PT couplings of absorptive cuts with higher and higher production thresholds. When the threshold $\Lambda_{th}^2$ is above the resonance mass $M_R^2$, one finds for the low energy limit $q^2\ll \Lambda_{th}^2$,  
\begin{equation}
\frac{\overline{D}(q^2)}{\left(M_R^2-q^2\right)^2} \,=\, \Int_{\Lambda_{th}^2}^\infty \mathrm{d}t \frac{1}{\pi} \frac{\mbox{Im}\Pi(t)}{t-q^2} \,=\, \displaystyle{\sum_{n=0}^\infty}   \left(\frac{q^2}{\Lambda_{th}^2}\right)^n \Int_{1}^\infty \mathrm{d}x  \frac{1}{\pi} \frac{\mbox{Im}\,\Pi(x\cdot \Lambda_{th}^2)}{x^{n+1}}  \, .
\end{equation}
The contributions become smaller and smaller as the value of the production threshold $\Lambda_{th}^2$ is increased, supporting  the approximation in Section~4.4.

On the other hand, in the deep euclidean region $Q^2=-q^2\gg \Lambda_{th}^2$, one gets
\begin{equation}
\left|\frac{\overline{D}(q^2)}{\left(M_R^2-q^2\right)^2}\right| \,\leq \, \frac{1}{Q^2}\, \Int_{\Lambda_{th}^2}^\infty \mathrm{d}t \frac{1}{\pi} \left|\mbox{Im}\Pi(t)\right| \, , 
\end{equation}
which becomes smaller and smaller as $\Lambda_{th}^2$ is increased. 

\thispagestyle{appendixebis}

\appendix
\chapter*{Appendix F \newline \newline Second-order Fluctuation of the Lagrangian}
\addcontentsline{toc}{chapter}{Appendix F: Second-order Fluctuation of the Lagrangian }
\newcounter{catilina}
\renewcommand{\thesection}{\Alph{catilina}}
\renewcommand{\theequation}{\Alph{catilina}.\arabic{equation}}
\renewcommand{\thetable}{\Alph{catilina}}
\setcounter{catilina}{6}
\setcounter{equation}{0}
\setcounter{table}{0}

The expansion around the classical solution of the fields in our lagrangian of Eq.~(\ref{eq:lagr1}) up to second order
(as required for the one loop evaluation) gives:
\begin{eqnarray}
\Delta \mathcal{L}_{\mathrm{R}\chi\mathrm{T}}&=&\Delta \mathcal{L}_{pGB}^{(2)}\,+\,\Delta \mathcal{L}_{\mathrm{kin}}
(\mathrm{S},\mathrm{P}) \,+\,  \Delta \mathcal{L}_{2}(\mathrm{S})\, +\,  \Delta \mathcal{L}_{2}(\mathrm{P})\,+\, 
 \Delta \mathcal{L}_{2}(\mathrm{S},\mathrm{P}) \, ,\quad 
\end{eqnarray}
where
\begin{align}
\Delta \mathcal{L}_{pGB}^{(2)}&=\,-\frac{F^2}{8}\bra 
\chi_+ \Delta^2 \ket \,+\, \frac{F^2}{4} \bra \nabla^\mu \Delta
 \nabla_\mu \Delta \,+\, \frac{1}{4} \left[ u_\mu,\Delta \right] 
 \left[ u^\mu,\Delta \right] \ket \,, \\
\Delta \mathcal{L}_{\mathrm{kin}}(\mathrm{S},\mathrm{P})&=\,\frac{1}{4}\bra \nabla^\mu \es \nabla_\mu \es \ket 
\,-\, \frac{M_S^2}{4}\bra \es \, \es \ket \,+\, \frac{1}{32} \bra 
\big[ [u^\mu,\Delta],S\big] \big[ [u_\mu,\Delta],S\big] \ket \,\nonumber \\ 
& - \,\frac{1}{8} \bra  [\nabla_\mu \Delta,\Delta][S, \nabla^\mu S]
\ket \,+\, \frac{1}{4\sqrt{2}} \bra  [ u_\mu,\Delta]
\Big( [S, \nabla^\mu \es]-[ \nabla^\mu S, \es]\Big) \ket \nonumber \\ 
& +\,\frac{1}{4}\bra 
\nabla^\mu \ep \nabla_\mu \ep \ket \,-\, \frac{M_P^2}{4}\bra \ep \, 
\ep \ket \,+\, \frac{1}{32} \bra \big[ [u^\mu,\Delta],P\big] 
\big[ [u_\mu,\Delta],P\big] \ket \,\nonumber \\ 
& -\, \frac{1}{8} \bra  [\nabla_\mu \Delta,\Delta][P, \nabla^\mu P]
\ket + \!\frac{1}{4\sqrt{2}} \bra  [ u_\mu,\Delta] 
\Big( [P, \nabla^\mu \ep] \!- \![\nabla^\mu P, \ep] \Big) \!\ket ,\!\!\! \\
\Delta \mathcal{L}_{2}(\mathrm{S})&=\,-\frac{i\,c_m}{2\sqrt{2}}\bra 
\es \{\Delta,\chi_-\} \ket \,-\, \frac{c_m}{8} \bra \{S, \Delta\} 
\{\chi_+,\Delta\}\ket - \frac{c_d}{\sqrt{2}} \bra \es \{\nabla_\mu 
\Delta, u^\mu \} \ket\nonumber \\
& + \bra\! \left( c_d  S + \ssa SS \right) \!\! \left( \nabla^\mu 
\Delta \nabla_\mu \Delta +\!\frac{1}{8} \Big\{  \big[\Delta,[u_\mu,\Delta]
\big],u^\mu\Big\} \!\right) \! \ket 
+\!\frac{\ssa}{2} \bra \varepsilon^{2}_{\mathrm{S}}\, u^\mu u_\mu \ket 
 \nonumber \\
&-\, \frac{\ssa}{\sqrt{2}} \bra \left\{ S, \es \right\} \left\{ u_\mu , 
\nabla^\mu \Delta\right\} \ket
\,+\, \ssb \bra S\, \nabla_\mu \Delta\, S \,\nabla^\mu \Delta \ket \,+\,
\frac{\ssb}{2}\bra \es u_\mu \es u^\mu \ket \nonumber \\
&-\sqrt{2}\ssb \bra \es \big( \nabla_\mu \Delta \,S \,u^\mu + u_\mu S \,
\nabla^\mu \Delta \big) \ket  \,+\,
\frac{\ssb}{4} \bra  \big[ [\Delta, u_\mu ],\Delta \big] S \,u^\mu S \ket 
 \nonumber \\&
-\!\frac{\ssc}{8} \bra \{ SS ,\Delta \}  \{\chi_+,\Delta\}  \ket -
 \!\frac{i\, \ssc}{2\sqrt{2}}  \bra \{ S, \es \} \{ \chi_- , \Delta \} \ket
+\!  \frac{\ssc}{2} \bra \varepsilon^{2}_{\mathrm{S}} \,\chi_+ \ket ,\!\!\!\!
\label{problem}\\ \nonumber 
\end{align}
\begin{align}
\Delta \mathcal{L}_{2}(\mathrm{P})&=\,\frac{d_m}{2\sqrt{2}}\bra \ep 
\{\Delta,\chi_+\} \ket \,-\, \frac{i\,d_m}{8} \bra \{ P, \Delta\} 
\{\chi_-,\Delta\}\ket  \nonumber \\
&+\ppa \bra PP \left( \nabla^\mu \Delta \nabla_\mu \Delta +\frac{1}{8} 
\Big\{  \big[\Delta,[u_\mu,\Delta]\big],u^\mu\Big\} \right)  \ket  
\,+\,\frac{\ppa}{2} \bra \varepsilon^{2}_{\mathrm{P}}\, u^\mu u_\mu \ket
\nonumber \\
& - \,\frac{\ppa}{\sqrt{2}} \bra \left\{ P, \ep \right\} \left\{ u_\mu ,
 \nabla^\mu \Delta\right\} \ket
+ \ppb \bra P\, \nabla_\mu \Delta\, P \,\nabla^\mu \Delta \ket +\!
\frac{\ppb}{2}\bra \ep u_\mu \ep u^\mu \ket \nonumber \\
&-\sqrt{2}\ppb \bra \ep \big( \nabla_\mu \Delta \,P \,u^\mu + u_\mu P \,
\nabla^\mu \Delta \big) \ket\,   \ket \,+\,
\frac{\ppb}{4} \bra  \big[ [\Delta, u_\mu ],\Delta \big] P \,u^\mu P \ket  
\nonumber \\&
-\!\frac{\ppc}{8} \bra \{ PP ,\Delta \}  \{\chi_+,\Delta\}  \ket \!-
 \!\frac{i\, \ppc}{2\sqrt{2}}  \bra \{ P, \ep \} \{ \chi_- , \Delta \} \ket
\!+  \!\frac{\ppc}{2} \bra \varepsilon^{2}_{\mathrm{P}} \,\chi_+ \ket ,\!\! 
\label{problem2} \\
\Delta \mathcal{L}_{2}(\mathrm{S},\mathrm{P})&=\,\frac{\spa}{8}\bra 
\{ \nabla_\mu S, P \} \big[ [\Delta, u^\mu ], \Delta \big] \ket
-\frac{\spa}{\sqrt{2}} \bra  \nabla^\mu \Delta \big( \{ \nabla_\mu \es ,
 P\} + \{ \nabla_\mu S,\ep\}  \big)\ket \nonumber \\
&+  \frac{\spa}{4\sqrt{2}} \bra \big[ [ u_\mu, \Delta ], S  \big] \big( 
\{\ep , u^\mu \} - \sqrt{2}\{ P , \nabla^\mu \Delta \} \big) \ket
+\frac{\spa}{2} \bra \{\nabla_\mu \es, \ep \} u^\mu \ket \nonumber \\
& +\frac{\spa}{4\sqrt{2}} \bra \big[ [u_\mu, \Delta ], \es \big] \{ P , 
u^\mu\} \ket
+\frac{\spa}{8} \bra \big[ [ \Delta, \nabla_\mu \Delta ] , S \big] \{ P , 
u^\mu\} \ket \nonumber \\
&-\frac{i\, \spb}{8} \bra \{ S,P \} \big\{\Delta, \{ \chi_- , \Delta \}
 \big\} \ket
+\frac{\spb}{2\sqrt{2}} \bra \{ \Delta, \chi_+ \} \Big( \{\es, P \} +
 \{ S, \ep\} \Big) \ket \nonumber \\
& +\frac{ i\, \spb}{2} \bra \chi_- \{ \es, \ep \} \ket   \, .\label{problem3} 
\end{align}
\thispagestyle{appendixe}

The evaluation of the path integral requires a Gaussian rearrangement of the integration variables. However the second-order fluctuation $\Delta {\cal L}_{R \chi T}$ does not have this structure due to the terms $\bra PP \,\nabla_\mu \Delta \nabla^\mu \Delta \ket$, $\bra P \,\nabla_\mu  \Delta\,P\, \nabla^\mu \Delta \ket$, $ \bra S \nabla_\mu \Delta \nabla^\mu \Delta \ket $, $\bra SS \,\nabla_\mu \Delta \nabla^\mu \Delta \ket$, $\bra S \,\nabla_\mu  \Delta\,S\, \nabla^\mu \Delta \ket$ and $\bra \{ \nabla_\mu \es, P \} \nabla^\mu \Delta \ket$ in Eqs. (\ref{problem}), (\ref{problem2}) and (\ref{problem3}). A way out is provided by a redefinition of the fields that eliminates the unwanted terms:
\begin{align} \label{eq:redefine}
\Delta \, \rightarrow \, & \, \Delta  - \!\frac{c_d}{F^2} 
\left\{ \Delta, S \right\} 
-\! \frac{\tilde{\lam}^{\mathrm{SS}}_1}{F^2} \left\{\Delta ,SS\right\} -
\!\frac{2\tilde{\lam}^{\mathrm{SS}}_2}{F^2} S\, \Delta \, S \, 
- \!\frac{\tilde{\lam}^{\mathrm{PP}}_1}{F^2} \left\{\Delta ,PP\right\} -
\!\frac{2\tilde{\lam}^{\mathrm{PP}}_2}{F^2} P\, \Delta \, P \,, \nonumber \\
\es \, \rightarrow \, & \, \es \,+ \sqrt{2} \spa \{ P, \Delta\} - 
\frac{\sqrt{2}\spa c_d}{F^2} \big\{ P , \{ \Delta, S \} \big\} \, ,
\end{align} 
where the following constants have been defined:
\begin{align}
\tilde{\lam}^{\mathrm{SS}}_1&\equiv \, \lam^{\mathrm{SS}}_1 
-\frac{3}{2}\frac{c_d^2}{F^2} \,, &
\tilde{\lam}^{\mathrm{SS}}_2&\equiv \, \lam^{\mathrm{SS}}_2 
-\frac{3}{2}\frac{c_d^2}{F^2} \,, \nonumber \\
\tilde{\lam}^{\mathrm{PP}}_1&\equiv \, \lam^{\mathrm{PP}}_1 
-{(\spa)}^2 \,, \phantom{\frac{1}{2}} &
\tilde{\lam}^{\mathrm{PP}}_2&\equiv \, \lam^{\mathrm{PP}}_2 
-{(\spa)}^2 \,.\phantom{\frac{1}{2}}
\end{align}
The transformation of the integration measure only yields $\delta^4(0)$ terms which have no effect on the theory~\cite{Politzer:1980me}~\footnote{In dimensional regularization the later result is immediate, as $\delta^d(0)=0$.}.

Performing the transformations given by Eq.~(\ref{eq:redefine}) on $\Delta \mathcal{L}_{\mathrm{R}\chi\mathrm{T}}$ and keeping only terms with up to two resonances we finally obtain:
\begin{align} \label{eq:rchtapp}
\Delta \mathcal{L}_{\mathrm{R}\chi\mathrm{T}}&=\, -\frac{1}{2}\,\Delta_i 
\left( d'_\mu d'^\mu + \sigma \right)_{ij} \Delta_j 
- \frac{1}{2}\,{\es}_{i} \left( d^\mu d_\mu + \ks \right)_{ij} {\es}_{j} 
- \frac{1}{2}\,{\ep}_{i} \left( d^\mu d_\mu + \kp \right)_{ij} {\ep}_{j} 
\nonumber \\
&+ \,{\es}_{i}\, \as_{ij}\, \Delta_j 
+ \,{\ep}_{i}\, \ap_{ij}\, \Delta_j 
+\,{\ep}_i\, \asp_{ij}\, {\es}_j  \phantom{\frac{1}{2}} \nonumber \\ 
&
+ \,{\es}_{k}\, \bs_{\mu \, ki}\, d^\mu_{ij} \Delta_j 
+ \,{\ep}_{k}\, \bp_{\mu \, ki}\, d^\mu_{ij} \Delta_j  
+\, {\ep}_k \, \bsp_{\mu \, ki}\, d^\mu_{ij} {\es}_j   \,\, , 
\phantom{\frac{1}{2}} 
\end{align}
that has the proper Gaussian structure and where the following definitions have been introduced:
\thispagestyle{appendixe}
\begin{align}
d^\mu_{ij}&=\,\delta_{ij}\,\partial^\mu + {\gamma_{ij}^\mu}\big|_{\chi}
 \, ,\\ \nonumber \\ 
d'^\mu_{ij}&=\, d^\mu_{ij} +{\gamma_{ij}^\mu}\big|_{\mathrm{R}} \, , \\
\nonumber \\
{\gamma_{ij}^\mu}\big|_{\chi}&=\,-\frac{1}{2} \bra \Gamma^\mu 
[\lam_i,\lam_j] \ket \, , \\ \nonumber \\
{\gamma_{ij}^\mu}\big|_{\mathrm{R}}&=\,\frac{c_d \spa}{2F^2}
 \bra \{ P, \lam_i\} \{ u^\mu, \lam_j \} \ket 
+ (-\frac{1}{16F^2}+\frac{c_d^2}{8F^4}) \bra [S,\nabla^\mu S]\, 
[\lam_i,\lam_j] \ket   \nonumber \\
& -\frac{1}{16F^2} \bra [P,\nabla^\mu P]\, [\lam_i,\lam_j] \ket    
-\frac{\spa}{16F^2} \bra \big[ S, \{ P, u^\mu \} \big] [\lam_i, 
\lam_j ] \ket \nonumber \\
& +\frac{\ssb \spa}{F^2}  \bra \{ P, \lam_i \} \big( \lam_j S 
\,u^\mu + u^\mu S \,\lam_j  \big)\ket
+\frac{\ssa \spa}{2F^2} \bra \big\{ S, \{ P, \lam_i \} \big\} 
\{ u^\mu, \lam_j \} \ket \nonumber \\
&-\frac{c_d^2 \spa}{2F^4} \bra \{ S, \lam_i \} \big[[P, u^\mu],
 \lam_j \big]\ket -  \Big\{i\leftrightarrow j\Big\}\, , \\ \nonumber \\
\ks_{ij}&=\,\delta_{ij}M_S^2 -\frac{\lambda^{\mathrm{SS}}_1}{2}\bra 
u^\mu u_\mu \{\lambda_i,\lambda_j\} \ket -\lambda^{\mathrm{SS}}_2 
\bra \lambda_i u_\mu \lambda_j u^\mu \ket - \frac{\lambda^{\mathrm{SS}}_3}{2} 
\bra \chi_+ \{\lambda_i , \lambda_j \} \ket \, , \\ \nonumber \\ 
\kp_{ij}&=\,\delta_{ij}M_P^2 -\!\frac{\ppa}{2}\bra u^\mu u_\mu 
\{\lam_i,\lam_j\} \ket -\ppb \bra \lam_i u_\mu \lam_j u^\mu \ket 
-\! \frac{\ppc}{2} \bra \chi_+ \{\lam_i , \lam_j \} \ket  ,\!\!  \\ \nonumber \\
%
%\end{eqnarray}
%
%\begin{eqnarray}
\sigma_{ij}&=\,\frac{1}{16} \bra \chi_+  \{ \hat{\lambda}_i,\hat{\lambda}_j \}   \ket
-\frac{1}{16} \bra [ u_\mu,\hat{\lambda}_i][u^\mu,\hat{\lambda}_j]\ket 
\nonumber \\
&-\frac{c_d}{4F^2} \bra \nabla^2 S \{ \lambda_i , \lambda_j \} \ket
+\frac{c_m}{8F^2} \bra \{ S,\hat{\lambda}_i\} \{ \chi_+, \hat{\lambda}_j \} 
\ket  
+\frac{c_d}{8F^2} \bra \{ S,u^\mu\} \big[ [u_\mu, \hat{\lambda}_i], 
\hat{\lambda}_j\big] \ket \nonumber \\
&
-\frac{c_d \spa}{2F^2}\big(  \bra  \{ \nabla_\mu P, \hat{\lam}_i \} 
\{ u^\mu, \hat{\lam}_j \}  \ket +
 \bra  \{  P, \hat{\lam}_i \} \{\nabla_\mu u^\mu, \hat{\lam}_j \} 
  \ket \big) \nonumber \\
&+\frac{i}{2F^2} \big( \frac{d_m}{4} + c_m \spa \big)\bra \{ P, 
\hat{\lam}_i \} \{ \chi_- , \hat{\lam}_j \} \ket
-\frac{1}{32F^2} \bra \big[ S,[ u^\mu,\lambda_i ]\big]  \big[S, 
[u_\mu,\lambda_j]\big] \ket  \nonumber %\\
\end{align}
\begin{align}
&+\frac{1}{8F^2}  \bra[u^\mu ,\lambda_i ] 
 \big[ u_\mu , \tilde{\lambda}^{\mathrm{SS}}_1 \{ SS, \lambda_j 
 \}+2\tilde{\lambda}^{\mathrm{SS}}_2 S\lambda_j S \big) \big] \ket 
\nonumber \\
&+\frac{1}{8F^2}  \bra[u^\mu ,\lambda_i ] 
\big[\lambda_j, \big( \lambda^{\mathrm{SS}}_1 \{ SS, u_\mu \}+2 
\lambda^{\mathrm{SS}}_2 S u_\mu S \big) \big] \Big) \ket 
\nonumber \\
&-\frac{1}{8F^2}  \bra \{ \chi_+ ,\lambda_i \} \Big( \big(
 \tilde{\lambda}^{\mathrm{SS}}_1
-\lambda^{\mathrm{SS}}_3 \big)
 \{ SS, \lambda_j \}+2\tilde{\lambda}^{\mathrm{SS}}_2 S\lambda_j 
 S \Big) \ket  \nonumber \\
&-\!\frac{c_d^2}{2F^4} \bra \{ \nabla^\mu S, \lambda_i \} \{ \nabla_\mu S ,
 \lambda_j \} \ket
-\!\frac{c_d^2}{4F^4} \bra \{S, \lambda_i  \} \{  \nabla^2 S , 
\lambda_j\} \ket -\!\frac{\tilde{\lambda}^{\mathrm{SS}}_1}{4F^2} \bra \nabla^2 S^2 
\{ \lambda_i , \lambda_j \} \ket \nonumber \\
&-\frac{\tilde{\lambda}^{\mathrm{SS}}_2}{F^2} 
\bra \lambda_i \nabla_\mu \big( \nabla^\mu S \lambda_j  S \big) \ket  +\frac{M_S^2 {(\spa)}^2}{2F^2} \bra \{ P, \lam_i \} \{ P, \lam_j \} 
\ket
 \nonumber \\ 
&
-\frac{\tilde{\lam}^{\mathrm{PP}}_1}{8F^2} \bra \{ \chi_+ ,\lam_i \}\{ PP, 
\lam_j \} \ket
-\frac{\tilde{\lam}^{\mathrm{PP}}_2}{4F^2} \bra \{ \chi_+ ,\lam_i \}P\, 
\lam_j P  \ket \nonumber \\
& -\frac{ \spa \spb}{ 2F^2} \bra \big\{ P, \{ P, \lam_i \} \big\} \{ \chi_+,
 \lam_j \} \ket \,  
+ \, \frac{\ppc}{8F^2} \bra \{ PP , \lam_i \} \{ \chi_+, \lam_j \} \ket 
\nonumber \\
& -\frac{\ssc{(\spa)}^2}{F^2}\bra \{ P, \lam_i \} \{ P, \lam_j \} 
\chi_+ \ket \, 
-\,\frac{1}{32F^2} \bra \big[ P,[ u^\mu,\lam_i ]\big]  \big[P, [u_\mu,\lam_j]
\big] \ket \nonumber \\
& +\frac{c_d^2 {(\spa)}^2}{4F^4} \bra \big[ [u^\mu,P],\lambda_i \big] \big[ 
[u_\mu,P], \lambda_j \big] \ket 
\,- \, \frac{{(\spa)}^2}{4F^2} \bra \big[ [u^\mu,\lambda_i], \{P, \lambda_j\} 
\big] \{P,u_\mu \} \ket \nonumber \\
&-\frac{1}{8F^2} \bra \Big(\ppa \{ u_\mu, PP \}+2\ppb P \,u_\mu P \Big) 
\big[ [ \lam_i , u^\mu ], \lam_j \big] \ket \nonumber \\
&- \, \frac{{(\spa)}^2}{F^2} \bra u^\mu \{ P, \lam_i \} \Big(  \ssa \{ P,
 \lam_j \} u_\mu +
 \ssb u_\mu \{ P, \lam_j \} \Big) \ket \nonumber \\
&+\frac{\tilde{\lam}^{\mathrm{PP}}_1}{8F^2} \bra \big[ u_\mu , \{ PP, 
\lam_i \} \big] [u^\mu ,\lam_j ] \ket 
+\frac{\tilde{\lam}^{\mathrm{PP}}_2}{4F^2} \bra [ u_\mu , P\, \lam_i P 
\,] [u^\mu ,\lam_j ] \ket  \nonumber \\
&-\!\frac{\tilde{\lam}^{\mathrm{PP}}_1}{4F^2} \bra \nabla^2 P^2 \{ \lam_i
 , \lam_j \}\! \ket 
-\!\frac{\tilde{\lam}^{\mathrm{PP}}_2}{F^2} \bra \lam_i \nabla_\mu \big(
 \nabla^\mu P \lam_j  P \big)\! \ket
 -\!\frac{{(\spa)}^2}{2F^2} \bra\! \{ \nabla_\mu P, \lam_i \} \{ \nabla^\mu P, 
 \lam_j \}\! \ket 
\nonumber \\
& +\frac{i\spb}{8F^2} \bra \{ S,P \} \big\{ \lam_i, \{ \chi_-, \lam_j \} 
\big\} \ket 
 \,+ \, \frac{i\,\ssc \spa}{2F^2} \bra \big\{ S, \{ P, \lam_i \} \big\} \{ 
\chi_- ,\lam_j \} \ket \nonumber \\
& -\frac{c_d^2 \spa}{2F^4} \bra \{ P, \lam_i \}  \big\{ u^\mu ,
 \{ \nabla_\mu S, 
\lam_j \} \big\} \ket \, - \, \frac{ \ssb \spa }{ F^2} \bra
 \nabla_\mu \Big( \{ P, \lam_i \} \big(
 \lam_j S u^\mu + u^\mu S \lam_j \big) \Big) \ket \nonumber \\
&-\frac{ \ssa \spa}{2F^2} \bra \{ P, \lam_i \}  \nabla_\mu  \big\{ S , 
\{u^\mu , \lam_j \} \big\} \ket \nonumber \\
&+\frac{\spa}{8F^2} \bra  [ u^\mu ,\lam_i] \Big(  \big[ \lam_j, 
\{ \nabla_\mu S, P \} \big] +
 2\big[ \nabla^\mu S, \{ P, \lam_j \} \big] -2 \big[ S , \{ \nabla^\mu P, 
 \lam_j \} \big] \Big) \ket  \nonumber \\
&
+\!\frac{\spa}{2F^2} \bra \{ u^\mu , \lam_i \} \!\Big( \frac{c_d^2}{F^2} 
\big\{ P, \{ \nabla_\mu S, \lam_j \} \big\} - 
\ssa  \big\{ S, \{ \nabla_\mu P, \lam_j \} \big\}\! \Big)\! \ket 
+  \Big\{i\leftrightarrow j\Big\}  ,\!\! \\ \nonumber \\
\as_{ij}&=\,-\frac{i\,c_m}{2\sqrt{2}F} \bra \chi_- \{\lambda_i,
\hat{\lambda}_j\} \!\ket
-\!\frac{1}{2\sqrt{2}F} \bra \![\nabla^\mu S, \lambda_i ] [ u_\mu,
\hat{\lambda}_j ] \!\ket 
-\!\frac{1}{4\sqrt{2}F} \bra \![S, \lambda_i ] [\nabla^\mu  u_\mu,
\hat{\lambda}_j ]\! \ket\nonumber \\
&+\!\frac{c_d^2}{\sqrt{2}F^3} \bra\! \{u^\mu, \lambda_i \} \{\nabla_\mu S, 
\lambda_j \} \!\ket  
-\!\frac{i \, \lambda^{\mathrm{SS}}_3}{2\sqrt{2} F} \bra\! \{ S , \lambda_i 
\} \{ \chi_- , \hat{\lambda}_j \} \!\ket
-\!\frac{M_S^2 \spa}{\sqrt{2} F} \bra P \{ \lam_i , \hat{\lam}_j \} \!\ket   
\nonumber
\end{align}
%\thispagestyle{appendixe}
%\newpage
\begin{align}
\thispagestyle{appendixe}
&
+ \!\frac{\spb}{2\sqrt{2}F} \bra \!\{ P, \lam_i \} \{ \chi_+ , \hat{\lam}_j 
 \} \!\ket 
+\!\frac{\spa \ssc }{\sqrt{2}F}\bra \!\{\chi_+,\lam_i\}\{P,\hat{\lam}_j \} \!
\ket 
-\!\frac{\spa}{\sqrt{2}F}\bra \nabla^2 P \{ \lam_i, \hat{\lam}_j  \} \!
\ket \nonumber \\
&
 -\!\frac{\spa}{4\sqrt{2}F}\bra\! \{P, u^\mu \} \big[ \lam_i ,
 [u_\mu,\hat{\lam}_j ] \big] \!\ket
+\!\frac{\spa}{\sqrt{2}F} \bra \!\{P, \hat{\lam}_j \} \Big( \ssa 
\{u_\mu u^\mu, \lam_i\} +2\ssb u_\mu \lam_i u^\mu \Big) \!\ket 
\nonumber \\
& +\frac{c_d}{\sqrt{2}F^3} \bra \{u^\mu, \lambda_i \} 
 \left( \tilde{\lambda}^{\mathrm{SS}}_1 \{  \nabla_\mu(SS), \lambda_j 
 \} +2 
\tilde{\lambda}^{\mathrm{SS}}_2 \nabla_\mu(S\lambda_j S) \right)  
\ket  
 \nonumber \\
& +\frac{ \sqrt{2} \lambda^{\mathrm{SS}}_2 \, c_d}{F^3} \bra  \{ 
\nabla^\mu S, \lambda_j \} \Big(  S u_\mu \lambda_i  + \lambda_i u_\mu S 
\Big) \ket 
+\frac{ \lambda^{\mathrm{SS}}_1 \, c_d}{\sqrt{2} F^3} \bra \{ S, \lambda_i 
\} \big\{ u_\mu, \{ \nabla^\mu S, \lambda_j \} \big\} \ket 
 \nonumber \\
& +\!\frac{c_d}{4\sqrt{2}F^3} \bra\! [ S, \lambda_i]\! \big[ u_\mu,\{ \nabla^\mu
 S,\lambda_j\}\big] \!\ket \!
+\!\frac{i\, c_m }{2\sqrt{2} F^3} \bra\! \{ \chi_-,\lambda_i \} \! \Big(\tilde{
\lambda}^{\mathrm{SS}}_1 \{ SS, \lambda_j \}\!+\!2\tilde{\lambda}^{\mathrm{SS}}_2
 S\lambda_j S \Big) \!\ket \nonumber \\
& +\frac{i\, c_m}{2\sqrt{2}F^3} \bra \{ \chi_-, \lam_i\} \Big( 
\tilde{\lam}_1^{\mathrm{PP}} \{PP, \lam_j\}
+2\tilde{\lam}_2^{\mathrm{PP}}P\,\lam_j P \Big) \ket 
 \nonumber \\
&+\frac{c_d}{\sqrt{2}F^3} \bra \{u^\mu, \lam_i \} 
\Big( \tilde{\lam}_1^{\mathrm{PP}} \{ \nabla_\mu (PP),\lam_j\} 
+2\tilde{\lam}_2^{\mathrm{PP}}\nabla_\mu (P\,\lam_j P) \Big) \ket \nonumber \\
&
 +\frac{c_d\spa}{\sqrt{2}F^3} \bra  \{\nabla_\mu P,\lam_i\}\{\nabla^\mu S,
 \lam_j\} \ket \, , \\ \nonumber \\
\ap_{ij}&=\, \!\frac{d_m}{2\sqrt{2}F} \bra \chi_+ \{\lam_i, \hat{\lam}_j \} \!
\ket 
+\!\frac{\spb}{2\sqrt{2}F}\bra \!\{S,\lam_i\}\{\chi_+, \hat{\lam}_j\}\!\ket 
 -\!\frac{\spa}{4\sqrt{2}F}\bra\!\{u_\mu,\lam_i\} \big[S,[u^\mu, \hat{\lam}_j]
\big]\!\ket
\nonumber \\
&
-\frac{1}{4\sqrt{2}F} \bra \Big( [P,\lam_i]  [\nabla^\mu u_\mu, \hat{\lam}_j 
] + 2 [\nabla^\mu P, \lam_i ]
[u_\mu, \hat{\lam}_j ] \Big) \ket \nonumber \\
& +\frac{i\,\spa}{\sqrt{2}F} \spb\bra \{\chi_-,\lam_i\}\{P, 
\hat{\lam}_j\}\ket 
-\frac{i \, \ppc}{2\sqrt{2} F} \bra \{ P , \lam_i \} \{ \chi_- , 
\hat{\lam}_j \} \ket  \nonumber \\
&
+\frac{{(\spa)}^2}{\sqrt{2}F} \bra \{u^\mu, \lam_i\}\{\nabla_\mu P, 
\hat{\lam}_j  \}\ket
+\frac{c_d\spa}{\sqrt{2}F^3}\bra \{\nabla_\mu S, \lam_i\} \{ \nabla^\mu S,
 \lam_j \} \ket \nonumber \\
&-\!\frac{d_m}{2\sqrt{2} F^3} \bra \!\{ \chi_+,\lam_i \}\! \Big(  
\tilde{\lam}^{\mathrm{SS}}_1\{SS,\lam_j\} 
+2\tilde{\lam}^{\mathrm{SS}}_2 S\,\lam_j S 
+\tilde{\lam}^{\mathrm{PP}}_1\{ PP , \lam_j \} +2 
\tilde{\lam}^{\mathrm{PP}}_2 P \,\lam_j P \Big) \!\ket\,  \nonumber \\
& -\frac{c_d {(\spa)}^2}{\sqrt{2}F^3} \bra \{u^\mu, \lam_i\} 
 \big\{P,\{\nabla_\mu S,\lam_j\} \big\} \ket
+\frac{c_d}{4\sqrt{2}F^3} \bra  [P,\lam_i] \big[ u_\mu,
\{\nabla^\mu S,\lam_j\}\big] \ket
\nonumber \\
& +\frac{c_d }{\sqrt{2} F^3}\bra \{\nabla^\mu S,\lam_j\} \Big(
 \ppa \big\{ u_\mu ,\{P,\lam_i\} \big\}
+2\ppb \left( P \,u_\mu \lam_i +\lam_i u_\mu P \right) \Big) \ket  
 \, , \\ \nonumber \\
\asp_{ij}&=\,\frac{i\,\spb}{2} \bra \chi_- \{ \lam_i,\lam_j\} \ket  
\, , \\ \nonumber \\
{{\bs}^{\mu}_{ij}}&=\, -\frac{c_d}{\sqrt{2}F} \bra u^\mu \{ \lambda_i ,
\hat{\lambda}_j \} \ket
-\frac{1}{4\sqrt{2}F}\bra [S,\lambda_i][u^\mu,\hat{\lambda}_j]\ket 
-\frac{\lambda^{\mathrm{SS}}_1}{\sqrt{2}F} \bra \{ S, \lambda_i \} \{ 
u^\mu, \hat{\lambda}_j \} \ket 
 \nonumber \\
&- \!\frac{\sqrt{2}\lambda^{\mathrm{SS}}_2}{F} \bra S \!\Big( u^\mu 
\lambda_i \hat{\lambda}_j + 
\hat{\lambda}_j \lambda_i u^\mu \Big) \ket 
-\frac{\spa}{\sqrt{2}F}\bra \nabla^\mu P \{\lam_i, \hat{\lam}_j \}\!
\ket \nonumber 
\thispagestyle{appendixe}
\end{align}
\begin{align}
\thispagestyle{appendixe}
&+ \!\frac{c_d }{\sqrt{2} F^3} \bra\! \{ u^\mu , \lambda_i \}\! \Big( 
\tilde{\lambda}^{\mathrm{SS}}_1\! \{ SS, \lambda_j \} \!+\!2
\tilde{\lambda}^{\mathrm{SS}}_2 \!S\lambda_j S 
\!+\! \tilde{\lam}_1^{\mathrm{PP}} \!\{PP,\lam_j\} 
\!+\!2\tilde{\lam}_2^{\mathrm{PP}}\!P\lam_jP \Big) \!\ket   ,\!\!\! \\ \nonumber \\
{{\bp}^{\mu}_{ij}}&=\,-\frac{\spa}{\sqrt{2}F} \bra \nabla^\mu S
 \{\lam_i,\hat{\lam}_j \} \ket 
-\frac{1}{4\sqrt{2}F}\bra [P,\lam_i][u^\mu,\hat{\lam}_j]\ket  
-\frac{\ppa}{\sqrt{2}F} \bra \{ P, \lam_i \} \{ u^\mu, \hat{\lam}_j \} 
\ket \nonumber \\ 
& - \frac{\sqrt{2}\ppb}{F} \bra P  \Big( u^\mu \lam_i \hat{\lam}_j +
\hat{\lam}_j \lam_i u^\mu\Big) \ket  
 +\frac{{(\spa)}^2}{\sqrt{2}F} \bra\{u^\mu,\lam_i\}\{P, \hat{\lam}_j\}
  \ket \, , \\ \nonumber \\
{{\bsp}^{\mu}_{ij}}&=\, \frac{\spa}{2} \bra u^\mu \{\lam_i,\lam_j\} \ket \, ,
\end{align}
and the following definitions have been used,
\begin{equation}
\hat{\lambda}_i \,\equiv \, \lambda_i - \frac{c_d}{F^2} 
\{ \lambda_i , S \} \,, \qquad \qquad
\nabla_\mu \left( A \,\lambda_i\, B \right) \, \equiv \, 
\nabla_\mu A \, \lambda_i \, B \, + \, A \, \lambda_i \, \nabla_\mu B  \,,
\end{equation}
where $A$ and $B$ are any chiral tensor or resonance field.
\par
As commented in the text we can write Eq.~(\ref{eq:rchtapp}) as:
\begin{equation}
\Delta {\cal L}_{\mathrm{R} \chi \mathrm{T}} \, = \, 
- \, \frac{1}{2} \, \eta \, \left( \, \Sigma_{\mu} \, \Sigma^{\mu} \, + \, 
\Lambda \, \right) \, \eta^{\top} \; , 
\end{equation}
where $\eta$ collects the fluctuation fields, 
$\eta=\left(\Delta_i,{\es}_j,{\ep}_k\right)$, $i,j,k = 0,...,8$, 
$\eta^{\top}$ is its transposed and $\Lambda$ and $\Sigma_\mu$ are defined as:
\thispagestyle{appendixe}
\begin{eqnarray}
\left( \Lambda \right)_{ij}  &=& \left( \begin{array}{ccc}
\sigma+ 
\frac{1}{4}{\bs_{\mu}}^{\top} \bsm  
& -\as^{\top}+\frac{1}{2}\tilde{d^{\mu}_-} {\bs_{\mu}}^{\top}
&-\ap^{\top}+\frac{1}{2}\tilde{d^{\mu}_-} {\bp_{\mu}}^{\top}
 \\
 \\
+ \frac{1}{4}{\bp_{\mu}}^{\top} \bpm 
& + 
\frac{1}{4}{\bp_{\mu}}^{\top} \bspm 
& - 
 \frac{1}{4}{\bs_{\mu}}^{\top} {\bspm}^{\top} 
\\ \\
\\
 -\as+\frac{1}{2} \bar{d^{\mu}_+} \bs_{\mu}
&\ks+ \frac{1}{4}\bsm {\bs_{\mu}}^{\top}  
& -\asp^{\top}+\frac{1}{2}\hat{d^{\mu}} {\bsp_{\mu}}^{\top}
\\
\\
 + \frac{1}{4}{\bsp_{\mu}}^{\top}
  \bpm 
& + \frac{1}{4}{\bsp_{\mu}}^{\top} 
\bspm 
&  + 
\frac{1}{4}\bsm{\bp_{\mu}}^{\top} 
\\ \\ \\
-\ap+\frac{1}{2} \bar{d^{\mu}_+} \bp_{\mu} 
&-\asp+\frac{1}{2} \hat{d^{\mu}} \bsp_{\mu} 
& \kp+ \frac{1}{4}\bpm {\bp_{\mu}}^{\top}  \\ \\
 - \frac{1}{4}\bspm \bs_{\mu} 
& + \frac{1}{4}\bpm 
{\bs_{\mu}}^{\top} 
& + \frac{1}{4}\bspm
  {\bsp_{\mu}}^{\top} 
 \end{array} \right)_{ij} \; .\nonumber \\ 
\end{eqnarray}
\begin{eqnarray}
\left( \Sigma_\mu \right)_{ij} &=&\delta_{ij} \, \partial_\mu\, + \, 
\left( Y_{\mu} \right)_{ij} \, ,\\
& & \nonumber \\
\left( Y_{\mu} \right)_{ij} & = & \left( \begin{array}{ccc}
\gamma'_{\mu} & \frac{1}{2} {\bs_{\mu}}^{\top} &
\frac{1}{2} {\bp_{\mu}}^{\top}
\\\\ -\frac{1}{2} \bs_{\mu} & \gamma_{\mu} & \frac{1}{2} {\bsp_{\mu}}^{\top} 
\\\\
-\frac{1}{2} \bp_{\mu}  & -\frac{1}{2} \bsp_{\mu}& \gamma_{\mu}\end{array} 
\right)_{ij}\;\;\;,\label{eq:y}\\ \nonumber \\ 
& & \nonumber 
\end{eqnarray}\thispagestyle{appendixe}
Here some new expressions have been defined:
\begin{eqnarray} \label{eq:gamos}
\gamma^{\mu} &=&{\gamma^\mu}\big|_{\chi} \, , \nonumber \\
\gamma'^{\mu} &=& {\gamma^\mu}\big|_{\chi} + {\gamma^\mu}\big|_{\mathrm{R}} 
\, ,\nonumber \\ 
\hat{d^{\mu}} \, X &=& \partial^{\mu} \, X \, + \, \left[ \, \gamma^{\mu}  \, ,
X \, \right] \, , \nonumber  \\
\tilde{d^{\mu}_{\pm}} X &=& \hat{d^{\mu}} \,  X \, \pm \, 
\left(\gamma'^{\mu}-\gamma^{\mu}\right)\,X \,,\nonumber  \\
\bar{d^{\mu}_{\pm}} X &=& \hat{d^{\mu}} \, X \, \pm \, X\,
\left(\gamma'^{\mu}-\gamma^\mu \right) \, .
\end{eqnarray}
\newpage
\thispagestyle{appendixe}

\appendix
\chapter*{Appendix G \newline \newline $\beta$-function Coefficients}
\addcontentsline{toc}{chapter}{Appendix G: $\beta$-function Coefficients}
\newcounter{catilinatris}
\renewcommand{\thesection}{\Alph{catilinatris}}
\renewcommand{\theequation}{\Alph{catilinatris}.\arabic{equation}}
\renewcommand{\thetable}{\Alph{catilinatris}.\arabic{table}}
\setcounter{catilinatris}{7}
\setcounter{equation}{0}
\setcounter{table}{0}

\pagestyle{appendixg}
The divergent part of the R$\chi$T lagrangian shown in Chapter~5, at one loop, can be expressed in a basis of operators that satisfy the same symmetry requirements that our starting lagrangian of Eq.~(\ref{eq:lagr1}). At one loop our bare lagrangian reads:
\begin{equation} \label{appendix-main}
{\cal L}_{1} \, = \,   \sum_ {i=1}^{18} \, \alpha_i \, 
{\cal O}_i \, + \, \sum_{i=1}^{66} \, \beta_i^R \, {\cal O}_i^R \, + \, 
\sum_{i=1}^{379} \, \beta_i^{RR} \, {\cal O}_i^{RR} \; .
\end{equation}
The notation of Section~5.3.3 is followed. The couplings in the lagrangian $ {\cal L}_{L=1}$ read:
\begin{eqnarray} \label{appendix-coef}
\alpha_i & = & \mu^{D-4} \, \left( \, \alpha_i^r(\mu) \, + \, 
\frac{1}{(4\pi)^2} \, \frac{1}{D-4} \, \gamma_i  \right) \; , \nonumber \\
\beta_i^R & = & \mu^{D-4} \, \left( \, \beta_i^{R,r}(\mu) \, + \, 
\frac{1}{(4\pi)^2} \, \frac{1}{D-4} \, \gamma_i^R  \right) \; , \nonumber \\
\beta_i^{RR} & = & \mu^{D-4} \, \left( \, \beta_i^{RR,r}(\mu) \, + \, 
\frac{1}{(4\pi)^2} \, \frac{1}{D-4} \, \gamma_i^{RR}  \right) \; ,
\end{eqnarray}
where  $\gamma_i$, $\gamma_i^R$ and $\gamma_i^{RR}$ are the
divergent coefficients that constitute the $\beta$-function of our
lagrangian. $\gamma_i^R$ and $\gamma_i^{RR}$ are given in Tables~\ref{apendixf-a} and \ref{apendixf-b}, while $\gamma_i$ were shown in Table 5.1. 

We indicate with an asterisk all the operators whose $\beta$-function coefficient vanishes once the short-distance constraints of Eqs.~(\ref{eq:largecd}) and (\ref{eq:largecd2}) are considered.

\begin{center}
\tabletail{\hline}	 
\topcaption{\label{apendixf-a}Operators with one resonance and their $\beta$-function 
coefficients.}
\tabletail{\hline}
\begin{supertabular}{|>{$}r<{$} | >{$}c<{$} | >{$}p{10cm}<{$} |}
\hline & & \\
i & \cO_i^R &\qquad \qquad \qquad \qquad \qquad \qquad \quad \gamma_i^R  \\ & & \\ \hline \hline

   1&   \bra  S   u \cdot u \ket &   - 3    N    F^{-4}    c_d^3    {M_S^2} + 2    N  F^{-2}  c_d    {\ssa}    {M_S^2} + 4    N  F^{-2}  c_d    {\ssb}    {M_S^2} - N
          F^{-2}   c_d    {M_S^2}    (\spa)^2 -N   F^{-2} c_d{M_P^2}    (\spa)^2\\ \hline

 2  &   \bra  S \ket \bra u \cdot u \ket &   - 3    F^{-4}    c_d^3    {M_S^2} + 2 F^{-2}   c_d    {\ssa}    {M_S^2} -F^{-2} c_d    {M_S^2}    (\spa)^2 +F^{-2} c_d
             {M_S^2} -F^{-2} c_d {M_P^2}    (\spa)^2 \\ \hline

   3&   \bra  u_\mu  \ket \bra u^\mu    S \ket &   - 2    F^{-4}    c_d^3    {M_S^2} + 4   F^{-2} c_d    {\ssa}    {M_S^2} + 4  F^{-2}  c_d
             {\ssb}    {M_S^2} + 2  F^{-2}  c_d    {M_S^2}    (\spa)^2 - F^{-2} c_d    {M_S^2} \\ \hline

  4&   \bra  S   {\chi_+} \ket &   - 2    N    F^{-4}    c_d^2    c_m    {M_S^2} - 2    N  F^{-2}  d_m    {M_S^2}    {\spa} + 1/2    N  F^{-2}  c_d
         {M_S^2} - 2    N  F^{-2}  c_m    {M_S^2}    (\spa)^2-N   F^{-2} c_m{M_P^2}    (\spa)^2 \\ \hline

   5&   \bra  S \ket \bra {\chi_+} \ket &   - 2    F^{-4}    c_d^2    c_m    {M_S^2} - 2 F^{-2}   d_m    {M_S^2}    {\spa} + 1/2 F^{-2}   c_d    {M_S^2}
          - 2  F^{-2}  c_m    {M_S^2}    (\spa)^2 -F^{-2} c_m {M_P^2}    (\spa)^2 \\
\end{supertabular}
\end{center}

\begin{center}
\tablecaption{\label{apendixf-b} Operators with two resonances, scalars and pseudoscalars,
 and their $\beta$-function 
coefficients.}
\tabletail{\hline}
\begin{supertabular}{|>{$}r<{$} | >{$}c<{$} | >{$}p{10cm}<{$} |}
\hline & & \\
i & \cO_i^{RR} &\qquad \qquad \qquad \qquad \qquad \qquad \quad \gamma_i^{RR}  \\ & & \\ \hline \hline

   1&   \bra   SS \ket &  N    F^{-4}    c_d^2    {M_S^4} + N F^{-2}   {M_S^4}    (\spa)^2 +N   F^{-2} {M_P^2M_S^2}    (\spa)^2\\ \hline

     2&   \bra   S\ket^2 &  F^{-4}    c_d^2    {M_S^4} +F^{-2} {M_S^4}    (\spa)^2 +F^{-2}  {M_P^2}  M_S^2  (\spa)^2 \\ \hline

  3&   \bra   SS   u \cdot u \ket &   - 4    N    F^{-4}    c_d^2    {\ssa}    {M_S^2} - 4    N    F^{-4}    c_d^2    {\ssb}    {M_S^2} - 2    N
         F^{-4}    c_d^2    {M_S^2}    (\spa)^2 - N    F^{-4}    c_d^2    {M_S^2} - 4    N  F^{-2}  {\ssa}    {M_S^2}    (\spa)^2 + 1/2    N  F^{-2}  {\ssa}
         {M_S^2} + N  F^{-2}  (\ssa)^2    {M_S^2} + 2    N F^{-2}   {\ssb}    {M_S^2}    (\spa)^2 + 4    N F^{-2}   (\ssb)^2    {M_S^2} + 1/16    N F^{-2}   {M_S^2} -2N   F^{-2} {\ssa}{M_P^2}    (\spa)^2\\ \hline

  4&   \bra   u_\nu S u^\nu    S \ket &   - 2    N    F^{-4}    c_d^2    {\ssa}    {M_S^2} - 4    N    F^{-4}    c_d^2    {\ssb}    {M_S^2} + 3/2    N
         F^{-4}    c_d^2    {M_S^2} + 4    N  F^{-2}  {\ssa}    {\ssb}    {M_S^2} + N   F^{-2} {\ssa}    {M_S^2}    (\spa)^2 - 4    N F^{-2}   {\ssb}    {M_S^2}    (\spa)^2 - N
           F^{-2}  {\ssb}    {M_S^2} + 3/4    N F^{-2}   {M_S^2}    (\spa)^2-2N   F^{-2}\ssb {M_P^2}    (\spa)^2 \\ \hline

   5&   \bra   u_\mu  \ket \bra  u^\mu    SS \ket &   - 4    F^{-4}    c_d^2    {\ssa}    {M_S^2} - 4    F^{-4}    c_d^2    {\ssb}
             {M_S^2} - 4    F^{-4}    c_d^2    {M_S^2}    (\spa)^2 + 2    F^{-4}    c_d^2    {M_S^2} + 4 F^{-2}   {\ssa}    {\ssb}    {M_S^2} + 2 F^{-2}   {\ssa}
         {M_S^2}    (\spa)^2 - 2  F^{-2}  {\ssa}    {M_S^2} + 2  F^{-2}  (\ssa)^2    {M_S^2} + 2  F^{-2}  {\ssb}    {M_S^2}    (\spa)^2 + F^{-2}{\ssb}    {M_S^2} - 1/8  F^{-2}  {M_S^2}
          \\
\end{supertabular}

%\end{center}
%\newpage

%\begin{center}
%\tabletail{\hline}
%\begin{supertabular}{|>{$}r<{$} | >{$}c<{$} | >{$}p{10cm}<{$} |}
%\\
\hline
       346& i\,  \bra   S   u \cdot u   P   {\chi_-}   \ket + \mathrm{h.c.} &   - 12    N    F^{-6}      d_m    c_d^3    (\spa)^2 - 1/2    N    F^{-6}      d_m
         c_d^3 - 19    N    F^{-6}      c_d^3    c_m    {\spa} + 11    N    F^{-6}      c_d^4    {\spa} + 2    N    F^{-6}
         c_d^4    {\spb} + 6    N    F^{-4}      d_m    c_d    {\ssa}    (\spa)^2 - 1/2    N    F^{-4}      d_m    c_d    {\ssa} + 4    N
         F^{-4}      d_m    c_d    {\ssb}    (\spa)^2 + 1/2    N    F^{-4}      d_m    c_d    {\ssb} + 1/2    N    F^{-4}      d_m    c_d
         (\spa)^2 - 1/8    N    F^{-4}      d_m    c_d - 3    N    F^{-4}      {\ppa}    c_d    c_m    {\spa} + 5/2    N    F^{-4}
         {\ppa}    c_d^2    {\spa} - N    F^{-4}      {\ppa}    c_d^2    {\spb} + 2    N    F^{-4}      {\ppb}    c_d    c_m    {\spa} - 2    N
         F^{-4}      {\ppb}    c_d^2    {\spa} - N    F^{-4}      {\ppc}    c_d^2    {\spa} + 7    N    F^{-4}      c_d    c_m    {\ssa}    {\spa}
          + 6    N    F^{-4}      c_d    c_m    {\ssb}    {\spa} + 5/4    N    F^{-4}      c_d    c_m    {\spa} - 4    N    F^{-4}      c_d
         c_m    (\spa)^3 - 13/2    N    F^{-4}      c_d^2    {\ssa}    {\spa} + N    F^{-4}      c_d^2    {\ssa}    {\spb} - 5    N
         F^{-4}      c_d^2    {\ssb}    {\spa} - 2    N    F^{-4}      c_d^2    {\ssb}    {\spb} + 3    N    F^{-4}      c_d^2    {\ssc}    {\spa}
          - 11/8    N    F^{-4}      c_d^2    {\spa} - N    F^{-4}      c_d^2    (\spa)^2    {\spb} + 11/2    N    F^{-4}
         c_d^2    (\spa)^3 + 1/2    N    F^{-4}      c_d^2    {\spb} + 1/4    N  F^{-2}    {\ppa}    {\ssa}    {\spa} - N F^{-2}     {\ppa}    {\ssa}
         {\spb} - 3/2    N F^{-2}     {\ppa}    {\ssb}    {\spa} + N F^{-2}     {\ppa}    {\ssc}    {\spa} - 3/16    N F^{-2}     {\ppa}    {\spa} + 2    N  F^{-2}    {\ppa}
         (\spa)^2    {\spb} - 1/4    N F^{-2}     {\ppa}    {\spb} + 3/2    N  F^{-2}    {\ppb}    {\ssa}    {\spa} - N  F^{-2}    {\ppb}    {\ssc}    {\spa} + 1/8
         N  F^{-2}    {\ppb}    {\spa} - N F^{-2}     {\ppb}    (\spa)^2    {\spb} - N  F^{-2}    {\ppc}    {\ssa}    {\spa} + N  F^{-2}    {\ppc}    {\ssb}    {\spa} - 2    N F^{-2}
             {\ssa}    {\ssc}    {\spa} + 9/16    N F^{-2}     {\ssa}    {\spa} + 6    N   F^{-2}   {\ssa}    (\spa)^2    {\spb} - 11/4    N F^{-2}     {\ssa}    (\spa)^3
          - 1/4    N  F^{-2}    {\ssa}    {\spb} + N   F^{-2}   (\ssa)^2    {\spa} - 1/8    N  F^{-2}    {\ssb}    {\spa} - N  F^{-2}    {\ssb}    (\spa)^2    {\spb}
          + 1/2    NF^{-2}      {\ssb}    (\spa)^3 + 1/2    N F^{-2}     {\ssc}    (\spa)^3 + 1/64    N F^{-2}     {\spa} - 1/16    N  F^{-2}
         (\spa)^3 + 1/2    NF^{-2}      (\spa)^4    {\spb} - 1/16    N F^{-2}     {\spb} \\
\end{supertabular}

\end{center}

\end{document}